\documentclass[a4paper]{book}
\usepackage[english]{layout} 
\usepackage{fullpage}
\usepackage{boxedminipage}
\usepackage{xcolor}
\usepackage{amsmath}
\usepackage{amssymb}
\usepackage{yfonts}
\usepackage{eufrak}
\usepackage{epsfig}
\usepackage{graphicx,subfigure}
\usepackage{hyperref}
\usepackage[toc,page]{appendix} 
\usepackage[nottoc, notlof, notlot]{tocbibind}
\usepackage{multirow}
\usepackage{tabularx}
\usepackage{tabulary}
\usepackage{mcite}
\usepackage[usenames,dvipsnames]{pstricks}
\usepackage{epsfig}
\usepackage{pst-grad} 
\usepackage{pst-plot} 
\usepackage{cite} 
\usepackage{fancybox} 
\usepackage{alltt} 
\usepackage{undertilde} 
\usepackage{feynmp} 
\usepackage{slashed} 
\usepackage{bm}
\usepackage{bbm}
\renewcommand\Re{\operatorname{Re}}

\usepackage{rotating}
\usepackage{lscape}

\newcommand{\bea}{\begin{eqnarray}}
\newcommand{\eea}{\end{eqnarray}}
\newcommand{\n}{\nonumber \\}

\newcommand{\delm}{\partial_\mu} 
\newcommand{\delmm}{\partial^\mu} 
\newcommand{\deln}{\partial_\nu} 
\newcommand{\delnn}{\partial^\nu} 
\newcommand{\epsilonbar}{\bar{\epsilon}} 
\newcommand{\lag}{\mathcal{L}} 
\newcommand{\psibar}{\bar{\psi}} 
\newcommand{\sigmabar}{\bar{\sigma}} 
\newcommand{\lambdabar}{\bar{\lambda}} 
\newcommand{\alphadot}{\dot{\alpha}} 
\newcommand{\betadot}{\dot{\beta}} 
\newcommand{\gammadot}{\dot{\gamma}} 
\newcommand{\phibar}{\bar{\phi}} 
\newcommand{\thetabar}{\bar{\theta}} 
\newcommand{\Qbar}{\bar{Q}} 
\newcommand{\tttt}{\theta\cdot\theta \thetabar\cdot\thetabar}
\newcommand{\phidagger}{\phi^{\dagger}}
\newcommand{\Phidagger}{\Phi^{\dagger}}
\newcommand{\chibar}{\bar{\chi}}
\newcommand{\Dbar}{\bar{D}}

\newcommand{\lagr}{\mathcal{L}}
\newcommand{\mP}{\mathcal{P}}
\newcommand{\mC}{\mathcal{C}}
\newcommand{\hPhi}{\hat{\Phi}}
\newcommand{\tQ}{\tilde{Q}}
\newcommand{\tL}{\tilde{L}}
\newcommand{\hDelta}{\hat{\Delta}}
\newcommand{\et}{\mbox{ and }}
\newcommand{\Tr}{{\rm Tr}}
\newcommand{\fr}{\sc FeynRules}
\newcommand{\mk}{\sc Mathematica}
\newcommand{\insurge}{\sc InSuRGE}
\newcommand{\asperge}{\sc ASperGe}

\newcommand{\tchi}{\tilde{\chi}}

\newcommand{\pPhi}{\mathbf{\Phi}}
\newcommand{\pPsi}{\mathbf{\Psi}}
\newcommand{\vV}{\mathbf{V}}
\newcommand{\mV}{\mathcal{V}}
\newcommand{\singlet}{{\utilde{\mathbf{1}}}}
\newcommand{\doublet}{{\utilde{\mathbf{2}}}}
\newcommand{\triplet}{{\utilde{\mathbf{3}}}}
\newcommand{\octet}{\utilde{\mathbf{8}}}
\newcommand{\nc}{\newcommand}
\nc{\postscript}[2]{\setlength{\epsfxsize}{#2\hsize}\centerline{\epsfbox{#1}}}
\nc{\be}{\begin{equation}}   \nc{\ee}{\end{equation}}
\nc{\beq}{\begin{equation}}   \nc{\eeq}{\end{equation}}
\nc{\baa}{\begin{array}}      \nc{\eaa}{\end{array}}
\nc{\bi}{\begin{itemize}}    \nc{\ei}{\end{itemize}}
\nc{\ben}{\begin{enumerate}}  \nc{\een}{\end{enumerate}}
\nc{\bce}{\begin{center}}     \nc{\ece}{\end{center}}
\nc{\non}{\nonumber}
\nc{\sh}{\hat s}
\nc\bpm{\begin{pmatrix}}      \nc\epm{\end{pmatrix}} 
\def\bsp#1\esp{\begin{split}#1\end{split}}

\newcommand{\ie}{{\textit i.e.}}

\nc\e{\varepsilon} 
\nc\Nbar{{\bar N}}
\nc\lbar{{\bar l}}
\nc\qbar{{\bar q}}
\nc\xibar{{\bar \xi}}
\nc\Psibar{{\bar \Psi}}
\nc\Lbar{{\bar L}}
\def\d{\mathrm{d}}

\definecolor{mygreen}{rgb}{0.16862745098039217,0.4392156862745098,0.03529411764705882} 

\newcommand*\xbar[1]{%
  \hbox{%
    \vbox{%
      \hrule height 0.5pt 
      \kern0.2ex
      \hbox{%
        \kern-0.1em
        \ensuremath{#1}%
        \kern-0.1em
      }%
    }%
  }%
} 

\usepackage{vmargin}					
\usepackage{setspace}					
\usepackage{algorithm}				
\usepackage{algorithmic}			
\usepackage{latexsym}					
\usepackage{xspace}						
\usepackage{multicol}					
\usepackage{array}						
\usepackage{fancyhdr}
\begin{document}

\newenvironment{vcenterpage}
{\newpage\vspace*{\fill}}
{\vspace*{\fill}\par\pagebreak}

\thispagestyle{empty}

\setmarginsrb{10mm}{0mm}{10mm}{0mm}{0mm}{0mm}{0mm}{0mm}

\begin{tabular}{ p{3cm} p{9cm} p{3cm}}
		\begin{minipage}{3cm}
			\includegraphics[width=3cm]{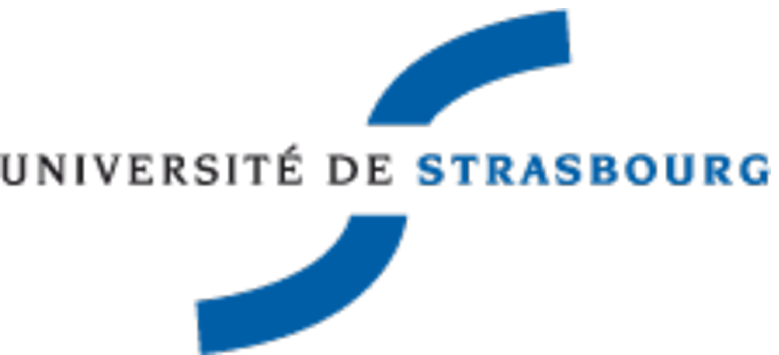} 
		\end{minipage}
	&
		\begin{minipage}{9cm}
			\begin{center}
				\textbf{UNIVERSIT\'E DE STRASBOURG}
			\end{center}
		\end{minipage}
	&
		\begin{minipage}{3cm}
			\includegraphics[width=3cm]{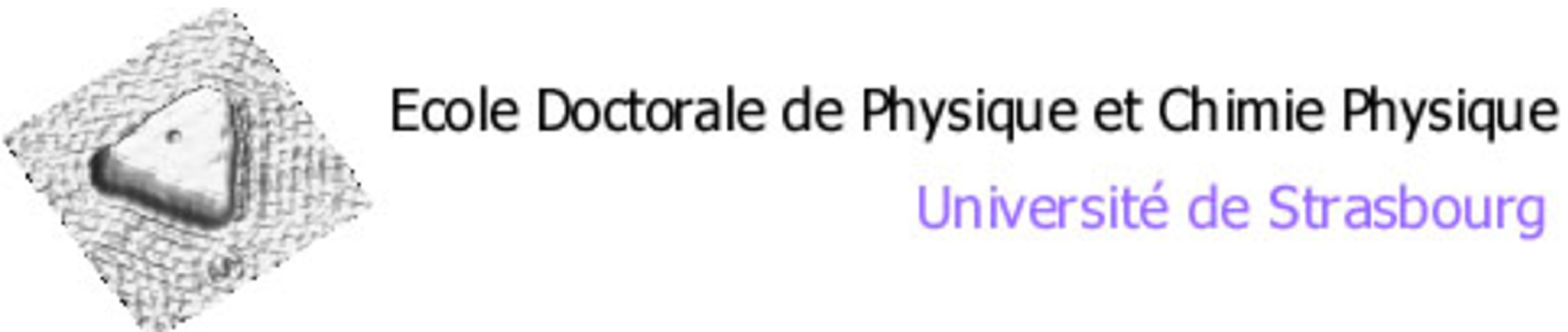} 
		\end{minipage}
\end{tabular}

\vspace{0.5cm}

\begin{minipage}{16cm}
	\begin{center}
		\textbf{\LARGE{\'Ecole Doctorale de Physique et de Chimie Physique}}
	\end{center}
\end{minipage}
	
\vspace{1cm}
	
\begin{minipage}{16cm}
	\begin{center}
		$\ $ D\'epartement de Recherche Subatomique
	\end{center}
\end{minipage}
	
\vspace{1cm}
	
\begin{minipage}{16cm}
	\begin{center}
		\LARGE \textbf{TH\`ESE} \normalsize pr\'esent\'ee par : \\ \vspace{0.2cm}
		\large \textbf{Adam ALLOUL}\\ \vspace{0.2cm}
		soutenue le : 20 september 2013
	\end{center}
\end{minipage}
	
\vspace{1cm}
	
\begin{minipage}{16cm}
	\begin{center}
		pour obtenir le grade de : Docteur de l'universit\'e de Strasbourg \\ \vspace{0.2cm}
		Discipline/ Sp\'ecialit\'e : Physique th\'eorique
	\end{center}
\end{minipage}
	
\vspace{0.8cm}

\hspace{0.65truecm}\fbox{
	\begin{minipage}{16cm}
		\vspace{0.5cm}
		\begin{center}
			\Large \textbf{Top-down and bottom-up excursions beyond the Standard Model:}\\ \vspace{0.2cm}
			\large \textbf{The example of left-right symmetries in supersymmetry}
		\end{center}
		\vspace{0.3cm}
	\end{minipage}
}

\vspace{1cm}
		 
\textsc{\large TH\`ESE dirig\'ee par : } $\ $ \vspace{0.2cm}\\ 
\begin{tabular}{l p{1.5cm} p{9.5cm}}
	\hspace{1cm}M. \textsc{Rausch de Traubenberg} Michel & $\ $ &  Professeur, universit\'e de Strasbourg\\
\end{tabular}

\vspace{0.3cm}

\textsc{RAPPORTEURS :}  $\ $ \vspace{0.2cm}\\
\begin{tabular}{p{4.8cm} p{3.69truecm} p{9.5cm}}
	\hspace{1cm}M. \textsc{Maltoni}  Fabio & $\ $ &  Professeur, Universit\'e Catholique de Louvain\\
	\hspace{1cm}M. \textsc{Orloff}  Jean & $\ $ &  Professeur, Clermont Universit\'e, Universit\'e Blaise \mbox{Pascal}\\
\end{tabular}

\vspace{0.3cm}

\textsc{JURY : }  $\ $ \vspace{0.2cm} \\
\begin{tabular}{p{4.8cm} p{3.69cm} p{9.5cm}}
	\hspace{1cm}M. \textsc{Grojean}  Christophe & $\ $ &  Professeur, CERN / Universitat Autonoma de Barcelona,  Pr\'esident du jury\\\\
	\hspace{1cm}M. \textsc{Fuks}  Benjamin & $\ $ & Docteur, CERN / Universit\'e de Strasbourg \\
	\hspace{1cm}M. \textsc{Roy}  Christelle & $\ $ &  Professeur, CNRS\\
\end{tabular}

\vspace{0.3cm}

		 
\setlength{\voffset}{0pt}
\thispagestyle{empty}

\setlength{\voffset}{0pt} 				
\setlength{\topmargin}{0pt}				
\setlength{\headheight}{30pt}			
\setlength{\headsep}{30pt}				
\setlength{\textheight}{620pt}		
\setlength{\footskip}{30pt}				
\setlength{\hoffset}{0pt} 				
\setlength{\oddsidemargin}{20pt}	
\setlength{\evensidemargin}{0pt}	
\setlength{\textwidth}{420pt}			
\setlength{\marginparsep}{10pt}		
\setlength{\marginparwidth}{40pt}	
\parskip=3pt						
\pagestyle{fancy}


\chapter*{Remerciements}
\markright{\MakeUppercase{Remerciements}}
M\^eme si s'engager dans une th\`ese rel\`eve de la seule d\'etermination du candidat, en pratique cette d\'ecision a une port\'ee bien plus grande. Cela revient en effet \`a imposer \`a sa famille, ses amis, son entourage un rythme de vie d\'etermin\'e, au moins partiellement, par l'avanc\'ee du travail de recherche. Les frustrations, les sauts d'humeur, les joies, les \'ecueils, les avanc\'ees et les d\'eceptions d\'ebordent souvent du cadre du laboratoire et s'invitent partout dans nos vies priv\'ees pesant, par cons\'equent, sur nos relations avec nos proches. J'aimerais donc profiter de ces quelques lignes pour remercier toutes les personnes qui ont \'et\'e l\`a pour moi, qui ont cru en moi, qui m'ont aid\'e, soutenu et parfois m\^eme prot\'eg\'e.\\

Dans l'ordre chronologique des choses, je commencerai \'evidemment par remercier ceux sans qui je ne serai m\^eme pas de ce monde, mes parents. Leur clairvoyance quand ils ont d\'ecid\'e de nous scolariser, ma soeur et moi, dans une \'ecole priv\'ee en Alg\'erie, leurs conseils, leur disponibilit\'e, leur amour, tous les sacrifices auxquels ils ont d\^u consentir pour que nous ayons une chance de r\'eussir font que je leur serai \'eternellement reconnaissant. Ils ont toujours \'et\'e l\`a pour moi et les entendre me t\'emoigner leur fiert\'e m'\'emeut au plus haut point.\\

J'aimerais aussi t\'emoigner ma gratitude \`a tous mes enseignants. Vous avez su, \`a coups de frustrations et d'encouragements, me faire aimer les \'etudes, me faire travailer et finalement vous avez contribu\'e \`a ma r\'eussite. J'aimerais t\'emoigner ma gratitude tout particuli\`erement \`a MM. Gilbert Moultaka et Cyril Hugonie, mes deux enseignants de Master 2 \`a Montpellier sans l'appui desquels je n'aurais jamais r\'eussi \`a d\'ecrocher cette th\`ese. Je r\'eserverai aussi une ligne sp\'ecialement \`a M. Dyakonov qui a encadr\'e mon stage de M1. Sa fa\c con particuli\`ere d'aborder la physique mais aussi de corriger mes erreur m'auront marqu\'e \`a jamais!\\

La r\'eussite d'une th\`ese d\'epend directement des relations encadrant - \'etudiant et du sujet de travail. J'ai eu la chance de tomber entre les mains de deux chercheurs exceptionnels qui ont su d\`es le d\'epart me mettre \`a l'aise et \'etablir une atmosph\`ere de travail id\'eale. Merci \`a toi Michel d'avoir \'et\'e l\`a tous les jours, sans exception (ou presque :) ). Ta rigueur scientifique et ton engagement dans ma th\`ese ont \'et\'e d\'eterminants pour ma r\'eussite. Tu as \'et\'e un vrai guide pour moi, tu m'as beaucoup appris et j'esp\`ere sinc\`erement qu'un jour j'arriverai \`a ton niveau de ma\^itrise des concepts physiques, ayant abandonn\'e le r\^eve de pouvoir te d\'epasser niveau blagues! Merci \`a toi aussi Benjamin! Tu m'as accueilli d\`es le premier jour comme un ami, tu m'as propos\'e le tutoiement tr\`es rapidement et tr\`es vite on a trinqu\'e ensemble. Avec toi on se sent tout de suite \`a l'aise et c'est remarquable! J'aimerais aussi te remercier d'avoir cru en moi et de m'avoir toujours pouss\'e \`a rendre un travail complet et bien fait. Ta rigueur et ton intransigeance ont fait que je me suis donn\'e \`a fond et m\^eme si parfois c'\'etait dur, le r\'esultat \'etait toujours tr\`es bon gr\^ace \`a toi! J'en profite pour souhaiter beaucoup de r\'eussite \`a mon successeur, Damien Tant. Accroche-toi, travaille et tout se passera pour le mieux!\\

\`A mon arriv\'ee \`a Strasbourg, j'ai eu la chance de ne pas me retrouver seul. En effet, gr\^ace \`a toi Fred j'ai pu rencontrer Jon et rapidement nous avons form\'e un joyeux trio \'ecumant les bars de Strasbourg, les rues de Dambach-La-Ville et cuisinant comme des chefs pour des d\^iners inimitables! Ces sorties ont agi comme de vraies soupapes de s\'ecurit\'e et l'amiti\'e que vous m'avez t\'emoign\'ee a \'et\'e un r\'eel filet de s\'ecurit\'e! \`A cette bande de joyeux se sont ajout\'ees d'autres personnes, au fil du temps et des ann\'ees. Ainsi, Paul, Nuria et Julie nous ont rejoint, tour \`a tour, amenant avec eux plein de vie, d'humour et de bons sentiments. Une amie qui m\'eriterait un paragraphe \`a elle toute seule est Ana-Maria! Toujours bouillonnante, pleine de vie, exhub\'erante et chaleureuse elle a \'egay\'e toutes nos soir\'ees mais surtout, surtout elle m'a fait rencontrer la femme de ma vie (disons-le franchement!). Merci Ana, Merci :)\\

Pendant cette p\'eriode de stage, j'ai aussi eu la chance de rencontrer deux femmes formidables: ma voisine Hafida et No\'emie. Gr\^ace \`a elles, les soir\'ees \`a la r\'esidence universitaires \'etaient toujours ponctu\'ees de franches rigolades mais je pense que cette p\'eriode a \'et\'e marqu\'ee par deux faits majeurs: La premi\`ere soir\'ee qu'on a pass\'ees ensemble alors qu'on se connaissait \`a peine, nos convives passant de mon appartement \`a celui de ma voisine et vice-ver\c ca et le visionnage terrifiant du film "Paranormal Activity". Malheureusement, les conditions ont fait que vous \^etes parties toutes les deux mais nos liens d'amiti\'e restent tr\`es fort et pour longtemps j'esp\`ere, hein voisine!\\

Me voici donc en octobre 2010, fra\^ichement embauch\'e en tant que doctorant \`a l'Institut Pluridisciplinaire Hubert Curien au m\^eme temps que deux autres acolytes: Guillaume et Emmanuel.\\
Avec toi Guillaume, le contact a \'et\'e super facile. Depuis la premi\`ere mission qu'on a pass\'ee ensemble \`a Lyon, je t'ai toujours consid\'er\'e et je te consid\`ere toujours comme un tr\`es bon ami. Ensemble, nous en avons v\'ecu des choses! Des moments tr\`es durs mais aussi des moments de franches rigolade autour de la piscine en Hollande! Malheureusement, la pression pour toi est devenue trop grande et t'as d\^u arr\^eter ta th\`ese avant la fin. Je comprends tr\`es bien les raisons qui t'ont pouss\'e vers cette solution et tu sais bien que nous sommes bien d'accord, toi et moi. J'esp\`ere seulement que notre amiti\'e restera aussi solide qu'elle ne l'a \'et\'e.\\
Manu, avec toi ce fut un peu plus difficile mais nous avons fini par nous comprendre. Ensemble, nous nous sommes accompagn\'es dans la r\'edaction de nos manuscrits et je pense que cela a facilit\'e grandement notre entente. Maintenant que chacun d'entre nous poursuit son chemin s\'epar\'ement, j'esp\`ere seulement qu'on pourra garder contact et continuer \`a faire \'evoluer notre amiti\'e.\\

\`A l'IPHC, j'ai aussi eu l'occasion de rencontrer un autre homme d'exception, un homme au grand coeur, \`a la g\'en\'erosit\'e sans limites (litt\'eralement), tr\`es discret mais efficace et dont le travail soign\'e m'a toujours impressionn\'e. J'aimerais donc te remercier Eric Conte pour ton aide inestimable! Je te l'ai dit \`a plusieurs reprises mais sans toi, je n'aurais s\^urement pas r\'eussi \`a terminer ma th\`ese dans les temps! Sans toi, je serais aujourd'hui m\^eme s\^urement dans une entrprise priv\'ee! Merci! Et parce que je n'ai pas le droit de faire figurer ton nom sur la page de garde, j'en profite aussi pour te remercier d'avoir accept\'e de faire partie de mon jury.\\

Maintenant, j'aimerais prendre le temps de remercier une personne qui a pris une place tr\`es particuli\`ere dans ma vie. Tu es arriv\'ee \`a l'improviste en novembre 2010, chez moi pour ma pendaison de cr\'emaill\`ere. Depuis, je pense qu'on peut dire que tu n'es pratiquement plus ressortie :) En fait depuis, il y a eu un sacr\'e remue-m\'enage dans ma vie, un vrai chamboulement puisque d\`es le mois de mars nous vivions officiellement ensemble. J'aimerais te remercier plus particuli\`erement parce que ta pr\'esence m'a apport\'e l'\'equilibre n\'ecessaire pour mener une vie saine. Avec toi, je me suis m\^eme mis \`a aller faire mes courses au march\'e!! Ta patience, ta chaleur et ta tendresse m'ont aid\'e \`a garder la t\^ete froide et \`a ne jamais d\'esp\'erer. Ta pr\'esence \`a mes c\^ot\'es a fait que tu t'es souvent retrouv\'ee en premi\`ere ligne quand j'avais des soucis ou quand le moral n'\'etait pas au top mais tu as toujours r\'epondu pr\'esente, tu n'as jamais rechign\'e \`a m'apporter tout le r\'econfort dont j'avais besoin m\^eme si je ne t'ai jamais demand\'e ton avis! Merci, mille fois merci et comme je te le dis toujours "\c ca finira par s'arranger".

Permettez-moi de remercier plus g\'en\'eralement toutes celles et ceux que j'ai rencontr\'ees durant ma vie de doctorant et qui, de pr\`es ou de loin, ont contribu\'e \`a la r\'eussite de mes travaux de recherche. Je pense notamment \`a (dans un ordre totalement al\'eatoire) Mathieu Planat; Sophie Schlaeder; Fran\c cois Schmidt; Marie-Anne Bigot-Sazy; Alexandre Penot; Laurent; Muriel; Camille; Thierry le bruxellois; Claude Duhr; mon meilleur ami Amir; Khodor; l'artiste, doctorant et directeur du Bureau des Doctorants Harold Barquero; C\'ecile Bopp; J\'er\'emy Andr\'ea ainsi que toute l'\'equipe CMS de Strasbourg; Marcus Slupinski; ma famille toute enti\`ere et particuli\`erement Jimmy qui a fait le d\'eplacement sp\'ecialement de New York pour assister \`a ma soutenance de th\`ese. Je voudrais aussi remercier tout le personnel administratif de l'IPHC et de l'Universit\'e de Strasbourg qui ont fait tout ce qui \'etait en leur pouvoir pour all\'eger les d\'emarches administratives. \\

Enfin, je voudrais remercier les membres du jury qui ont accept\'e d'examiner \`a la loupe mon travail de recherche. Je suis tr\`es honor\'e d'avoir pu recevoir de votre part mon titre de Docteur de l'Universit\'e de Strasbourg et je veillerai \`a ne jamais d\'ecevoir la confiance que vous avez plac\'ee en moi. Je voudrais aussi remercier tout particuli\`erement le Professeur Maltoni pour sa g\'en\'erosit\'e et sa disponibilit\'e ainsi que la directrice de mon laboratoire M$^{\rm mme}$ Roy qui faisait partie de mon jury et dont les f\'elicitations appuy\'ees m'ont beaucoup touch\'ees.




\begin{vcenterpage}

\chapter*{R\'esum\'e}
\markright{\MakeUppercase{R\'esum\'e}}

\paragraph{}
Une tr\`es grande effervescence secoue le monde de la physique des particules depuis le lancement du grand collisionneur de hadrons (LHC) au CERN. Cette \'enorme machine capable de faire se collisionner des protons \`a des \'energies \'egales \`a 14 TeV promet de lever le voile sur la physique r\'egissant les int\'eractions \`a ces \'echelles d'\'energies. Ces r\'esultats sont d'autant plus attendus que l'on a acquis la certitude que le Mod\`ele Standard de la physique des particules est incomplet et devrait, en fait, \^etre interpr\'et\'e comme la th\'eorie effective d'une th\'eorie plus fondamentale. Toutefois, depuis le lancement des exp\'eriences au LHC avec des \'energies de 7 puis de 8 TeV aucun signe de nouvelle physique n'a \'et\'e d\'ecouvert. Par contre, un \'enorme bond en avant a \'et\'e franchi avec la d\'ecouverte d'une particule scalaire de masse \'egale \`a 125 GeV et dont les propri\'et\'es sont relativement proches de celles du boson de Higgs telles que pr\'edites par le Mod\`ele Standard. C'est dans ce contexte de forte \'emulation internationale que mon travail de th\`ese s'est inscrit.\\

Dans un premier temps, nous avons voulu explorer la ph\'enom\'enologie associ\'ee au secteur des neutralinos et charginos du mod\`ele supersym\'etrique sym\'etrique gauche-droit. Cette \'etude peut \^etre motiv\'ee par plusieurs raisons notamment le fait que leur caract\`ere supersym\'etrique apporte une solution au probl\`eme dit de la hi\'erarchie mais implique aussi l'unification des constantes de jauge ainsi que l'explication de la mati\`ere noire. L'introduction de la sym\'etrie entre les fermions gauchers et les fermions droitiers permet, quant \`a elle, d'expliquer naturellement, via le m\'ecanisme dit de la balan\c coire, la petitesse de la masse des neutrinos mais aussi de r\'epondre \`a plusieurs autres questions non solubles dans le cadre du Mod\`ele Standard. Nous concentrant uniquement sur le secteur des charginos et neutralinos les plus l\'egers, nous avons montr\'e que ces mod\`eles peuvent \^etre facilement mis en \'evidence dans les \'ev\`enements multi-leptoniques en ce sens que les signatures qu'ils induisent sont tr\`es diff\'erentes compar\'ees \`a celles du Mod\`ele Standard et de sa version supersym\'etrique.\\
Dans un second temps, nous avons voulu explorer la ph\'enom\'enologie associ\'ee aux particules doublement charg\'ees. La d\'ecouverte de telles particules repr\'esenterait une preuve incontestable de nouvelle physique mais soul\`everait beaucoup de questions notamment le fait de savoir quelle th\'eorie d\'ecrit le mieux leurs int\'eractions. Dans notre analyse nous sommes partis du postulat qu'une telle particule \'etait d\'etect\'ee au LHC et avons essay\'e de donner quelques cl\'es  qui permettraient de comprendre la th\'eorie qui d\'ecrit le mieux cette d\'ecouverte. Pour ce faire nous avons consid\'er\'e des particules doublement charg\'ees scalaires, fermioniques ou vectorielles se transformant trivialement, dans la fondamentale ou l'adjointe du groupe $SU(2)_L$ et avons \'ecrit, pour chaque cas, le Lagrangien effectif d\'ecrivant les int\'eractions de ce nouveau champ avec ceux du Mod\`ele Standard. Nous concentrant uniquement sur les \'ev\`enements o\`u la multiplicit\'e de leptons dans l'\'etat final est au moins \'egale \`a trois, nous avons montr\'e que les limites exp\'erimentales pouvaient \^etre facilement contourn\'ees et qu'ainsi des particules doublement charg\'ees avec une masse sup\'erieure \`a 100 GeV n'\'etaient pas encore totalement exclues. Nous avons aussi analys\'e les observables cin\'ematiques associ\'ees \`a chacun des cas envisag\'es et avons conclu que, en l'absence de tout autre signal de nouvelle physique, il fallait combiner plusieurs variables cin\'ematiques pour pouvoir discriminer de mani\`ere claire entre les diff\'erentes possibilit\'es.\\

Un autre volet, compl\'ementaire au pr\'ec\'edent, de ma th\`ese a consist\'e \`a d\'evelopper des modules informatiques dans le cadre du programme {\sc FeynRules}. J'ai ainsi particip\'e au d\'eveloppement d'une routine capable de calculer, automatiquement, les \'equations du groupe de renormalisation au niveau de deux boucles associ\'ees \`a toute th\'eorie supersym\'etrique renormalisable. Un autre travail dans lequel j'ai pris une part importante a consist\'e \`a d\'evelopper un g\'en\'erateur de spectre dans {\sc FeynRules}. L'id\'ee a \'et\'e de doter ce dernier d'un ensemble de routines capables d'extraire les matrices de masse associ\'ees \`a n'importe quel Lagrangien automatiquement puis d'exporter ces matrices sous la forme d'un code source en {\sc C++} capable de diagonaliser ces matrices et de retourner dans un fichier {\sc SLHA} les matrices de m\'elange ainsi que le spectre en masse.

\paragraph{Mots clés : }
Physique au-del\`a du Mod\`ele Standard; Supersym\'etrie; Construction de mod\`eles; Approches top-down et bottom-up; Approches mod\`ele ind\'ependantes; D\'eveloppement d'outils informatiques.
\end{vcenterpage}

\begin{vcenterpage}

\chapter*{Abstract}
\markright{\MakeUppercase{Abstract}}

\paragraph{}
The field of high-energy physics has been living a very exciting period of its history with the Large Hadron Collider (LHC) at CERN collecting data. Indeed, this enormous machine able to collide protons at a center of mass energy of 14 TeV promises to unveil the mystery around the physics at such energy scales. From the physicists side, the expectations are very strong as it is nowadays a certitude that the Standard Model of particle physics is incomplete and should, in fact, be interpreted as the effective theory of a more fundamental one. Unfortunately, the 7 and 8 TeV runs of the LHC did not provide any sign of new physics yet but there has been at least one major discovery in 2010, namely the discovery of a scalar particle with a mass of 125 GeV and which properties are very close to those of the Standard Model Higgs boson. Since then, many questions have come up as we now want to understand if it really is the Standard Model Higgs boson or if it exhibits any deviations. It is in this peculiar context that my research work was carried.\\

In a first project, we, my supervisors, our collaborator and I, have wanted to explore the phenomenology associated with the neutralinos and charginos sector of the left-right symmetric supersymmetric model. Such an analysis can be motivated by several reasons such as the fact that the supersymmetric nature of these models provides a natural explanation for the infamous hierarchy problem, implies the unification of the gauge coupling constants at very high energy and provides a natural candidate for dark matter. In addition to these nice features, the left-right symmetry introduces a natural framework for explaining the smallness of neutrino masses but also helps in addressing several other unresolved issues in the Standard Model framework. Only focusing on the lightest charginos and neutralinos decaying into one or more light leptons, we have shown in our study that these models can be easily discovered in multi-leptonic final states as they lead to signatures very different from those induced by the Standard Model or its supersymmetric version. \\
In a second project, we have explored the phenomenology associated with doubly-charged particles. The discovery of such particles would be an irrefutable proof of new physics but would also raise the problem of knowing which model describes best their properties. Starting from the hypothesis that a particle carrying a two-unit electric charge is discovered at the LHC, we have carried an analysis which aim was to provide some key observables that would help in answering the latter question. To do so, we have adopted a model-independent approach where the Standard Model field content is extended minimally to contain a scalar, fermionic or vector multiplet transforming either as a singlet, doublet or triplet under $SU(2)_L$. The hypercharge of the latter field is chosen so that the highest electric charge carried by its components is equal to two and the Standard Model Lagrangian is extended to account for the new interactions. In our results, we have shown that the experimental constraints could be evaded easily so that the new fields can have a mass at least equal to 100 GeV. In addition, we have shown that, in the hypothesis that no other signal of new physics exists, only a combination of several kinematical distributions can help in distinguishing between the cases we have considered.\\

Another part of my thesis, complementary to the phenomenology work, has consisted in developping computer programs that might be helpful for phenomenological studies. Working in the framework of the {\sc Mathematica} package {\sc FeynRules}, I took part in the development of a routine able to extract automatically the analytical expressions of the renormalization group equations at the two-loop level for any renormalizable supersymmetric model. I have also been involved in the development of another module of {\sc FeynRules} able to extract automatically the analytical expressions for the mass matrices associated to any model implemented in {\sc FeynRules} and to export these equations in the form of a {\sc C++} source code able to diagonalize the matrices and store the mixing matrices as well as the spectrum in an SLHA-compliant file.

\paragraph{Keywords : }
Beyond the Standard Model phenomenology; Supersymmetry; Model building; Top-down and bottom-up approaches; Model-independent approach; Development of computer programs.
\end{vcenterpage}

\pagenumbering{arabic}
\tableofcontents
\pagestyle{fancy}

\chapter{R\'esum\'e d\'etaill\'e}
En physique des hautes \'energies, on peut tranquilement affirmer que toutes les particules et int\'eractions connues (sauf la gravit\'e) sont bien d\'ecrites par le Mod\`ele Standard (SM). En effet, ce dernier dont la construction th\'eorique repose sur des principes de sym\'etrie s'est av\'er\'e tr\`es robuste puisque toutes les exp\'eriences qui ont \'et\'e conduites jusqu'\`a aujourd'hui ont point\'e tout au plus vers de simples extensions de ce mod\`ele. En tout cas, aucune d'entre elles n'a conclu \`a un r\'eel besoin de changement de paradigme. Cette affirmation pose toutefois de s\'erieux probl\`emes.\\

Si aucune extension du Mod\`ele Standard ne devait \^etre d\'ecouverte, ceci serait en confrontation directe avec les r\'esultats cosmologiques indiquant clairement que la mati\`ere dite ordinaire ne repr\'esente en fait qu'environ 5$\%$ de la masse de l'Univers.\\

Le Mod\`ele Standard de la physique des particules n'explique que les int\'eractions faibles, \'electro-magn\'etiques et fortes ignorant de fait l'int\'eraction gravitationnelle. Aux \'echelles d'\'energies jusqu'\`a aujourd'hui explor\'ees, n\'egliger les effets de cette derni\`ere par rapport aux autres est une approximation tout \`a fait l\'egitime dans le monde des particules fondamentales. Le probl\`eme appara\^it \`a l'\'echelle d'\'energie de Planck ($10^{19} {\rm GeV}$) o\`u il est pr\'edit que l'int\'eraction gravitationnelle devienne aussi forte que les autres int\'eractions. \`A de telles \'energies nous ne savons donc pas comment d\'ecrire les int\'eractions.\\

Un autre probl\`eme du Mod\`ele Standard r\'eside dans le m\'ecanisme de brisure de la sym\'etrie \'electrofaible. En suivant le m\'echanisme de Higgs-Brout-Englert \cite{Englert:1964et,Higgs:1964ia}, toutes les particules du Mod\`ele Standard sont cens\'ees acqu\'erir leurs masses \`a travers leurs int\'eractions avec un champ scalaire fondamental, le boson de Higgs. R\'ecemment, les colllaborations {\sc ATLAS}\cite{:2012gk} et {\sc CMS}\cite{:2012gu} ont indiqu\'e toutes les deux avoir d\'ecouvert une particule scalaire avec une masse autour de 125 GeV exhibant les m\^emes propri\'et\'es que le boson de Higgs. Cette d\'ecouverte nous permet \'evidemment de mieux comprendre le m\'ecanisme de brisure de la sym\'etrie \'electrofaible mais le probl\`eme de la hi\'erarchie s'en trouve raviv\'e. En effet, si les mesures confirmaient que c'est bien le boson de Higgs tel que pr\'edit par le Mod\`ele Standard, sa masse est cens\'ee recevoir des corrections quantiques quadratiquement divergentes induisant un probl\`eme de naturalit\'e.\\

Quelques r\'esultats exp\'erimentaux montrent aussi clairement qu'on a besoin, tout au moins, d'une extension du Mod\`ele Standard. L'observation de l'oscillation des neutrinos est peut-\^etre l'un des arguments les plus forts puisqu'elle implique que ceux-ci ont une masse non nulle. La mesure du moment magn\'etique anomale du muon a, quant \`a elle, montr\'e une d\'eviation l\'eg\`erement sup\'erieure \`a 3 sigma. Ceci n'est \'evidemment pas suffisant pour \'etablir une d\'ecouverte mais c'est un r\'esultat qui interpelle.\\

L'accumulation \`a travers les ann\'ees de tous ces indices (parmi d'autres) a contribu\'e \`a faire na\^itre au sein de la communit\'e des physiciens des hautes \'energies de grandes attentes vis-\`a-vis du grand collisionneur de hadrons (LHC) du CERN. Du c\^ot\'e th\'eorique, le LHC a induit une forte activit\'e de recerche o\`u th\'eoriciens et ph\'enom\'enologistes ont travaill\'e ensemble \`a imaginer de nouveaux mod\`eles et \`a produire des pr\'edictions que l'on pourra comparer aux r\'esultats exp\'erimentaux. La r\'esistance du Mod\`ele Standard face aux r\'esultats exp\'erimentaux jouant le r\^ole de guide en ceci qu'elle place de fortes restrictions sur les th\'eories r\'ealistes puisque toutes les observations doivent pouvoir \^etre expliqu\'ees dans le cadre de ce nouveau mod\`ele. Parmi les nouveaux mod\`eles ou th\'eories les plus connues, on peut citer par exemple les th\'eories de Grande Unification (GUTs)\cite{Raby:2006sk,Georgi:1974yf,Lucas:1996bc} dans lesquelles toutes les int\'eractions de jauge du Mod\`ele Standard s'unifient; les th\'eories o\`u la dimensionnalit\'e de l'espace-temps est \'etendue \`a un nombre sup\'erieur \`a 4 \cite{PerezLorenzana:2004na,PerezLorenzana:2005iv,Hewett:2002hv}; les th\'eories des cordes dans lesquelles chaque champ est consid\'er\'e comme la manifestation d'un certain mode de vibration d'une unique corde; la supersym\'etrie qui \'etend les sym\'etries de l'espace temps pour lier des champs de diff\'erentes statistiques $\dots$ etc. \\

Un vieux r\^eve en physique th\'eorique est de pouvoir r\'ealiser l'unification de toutes les int\'eractions de jauge, c'est-\`a-dire, \^etre capable d'expliquer avec la m\^eme th\'eorie les int\'eractions faibles, \'electro-magn\'etiques et fortes. Ce r\^eve provient de la volont\'e de trouver les sym\'etries profondes qui gouvernent notre Univers et au m\^eme temps de la conviction que plus le nombre de param\`etres libres est petit, plus pr\'edictive est la th\'eorie. Cette conviction est confort\'ee par l'observation de l'\'evolution de la valeur des constantes de couplage de jauge dans le m\^eme sens. Une c\'el\`ebre tentative d'unification a \'et\'e faite par Georgi et Glashow en 1974 \cite{Georgi:1974sy} o\`u ils consid\`erent le plongement du groupe de jauge du SM dans le groupe $SU(5)$. En choisissant le bon m\'ecanisme de brisure, le groupe $SU(5)$ peut en effet se briser dans $SU(3)_c\times SU(2)_L\times U(1)_Y$ par contre leur mod\`ele pr\'edisant un temps de demi-vie du proton trop rapide, il a d\^u \^etre abandonn\'e dans sa forme \mbox{originelle}.\\

Depuis, plusieurs autres tentatives ont \'et\'e propos\'ees o\`u des groupes comme $E_6$ ou $SO(10)$, plus gros que $SU(5)$, \'etaint consid\'er\'es. Consid\'erer des groupes plus gros n\'ecessitant forc\'ement des m\'ecanismes de brisure de sym\'etrie plus complexes, ces mod\`eles ne sont pas minimaux comme le premier. Un exemple int\'eressant dans le cadre de ce manuscrit est de consid\'erer le groupe $SO(10)$ comme \'etant le groupe d'unification. Le m\'ecanisme de brisure se fait alors en plusieurs \'etapes faisant appara\^itre, lors de la premi\`ere \'etape, deux groupes $SU(2)$ \cite{Chang:1984uy}
$$ SO(10) \to SU(3)\times SU(2)\times SU(2)\times U(1) \times \mP \to \dots$$
o\`u $\mP$ est le groupe de parit\'e et les points repr\'esentent le reste des \'etapes menant au groupe de jauge du SM. Ces deux groupes $SU(2)$ peuvent \^etre interpr\'et\'es comme correspondant \`a $SU(2)_L {\rm et} SU(2)_R$ impliquant alors une sym\'etrie entre les fermions gauchers et les fermions droitiers.\\

La supersym\'etrie est certainement la plus populaire des extensions du Mod\`ele Standard. Elle est en fait la seule extension non-triviale du groupe de Poincar\'e (th\'eor\`emes de Haag-Lopuszanski-Sohnius et Colman-Mandula) reliant les degr\'es de libert\'e bosoniques et fermioniques. Parmi les avantages qu'induit la supersym\'etrie, les plus connus sont le fait qu'elle propose une solution au probl\`eme dit de la hi\'erarchie et le fait que les constantes de couplage de jauge s'unifient \`a tr\`es haute \'energie. La r\'ealisation minimale de la supersym\'etrie en physique des particules, c'est-\`a-dire le Mod\`ele Standard Supersym\'etrique Minimal (pour une revue sur le sujet voir, par example \cite{Csaki:1996ks}) qui est obtenue en "supersymm\'ertrisant" le Mod\`ele Standard est certainement l'un des mod\`eles les plus \'etudi\'es. Cette c\'el\'ebrit\'e est d\^ue \`a la (relative) simplicit\'e de ce mod\`ele (encore cette recherche de minimalit\'e) mais aussi \`a ses particularit\'es ph\'enom\'enologiquement int\'eressantes telle que sa capacit\'e \`a pr\'edire l'existence d'une particule massive \'electriquement neutre int\'eragissant faiblement avec les particules du Mod\`ele Standard et donc pouvant \^etre un bon candidat pour la mati\`ere noire. Le co\^ut de cette c\'el\'ebrit\'e est qu'aujourd'hui ce mod\`ele a \'et\'e tellement bien \'etudi\'e tant th\'eoriquement qu'exp\'erimentalement que l'espace des param\`etres encore autoris\'e est fortement r\'eduit, surtout dans la version contrainte de ce mod\`ele (cMSSM). La derni\`ere contrainte exp\'erimentale vient de la d\'ecouverte m\^eme de la particule scalaire au LHC qui pose de s\'erieux probl\`emes de naturalit\'e \`a ce mod\`ele et contribue donc fortement \`a r\'eduire cet espace des param\`etres et nous poussant, par la m\^eme occasion, \`a aller explorer des mod\`eles moins minimaux.\\

Durant mes trois ann\'ees de th\`ese, mon travail s'est divis\'e en deux parties. Une premi\`ere facette de mon travail a consist\'e \`a r\'ealiser des \'etudes ph\'enom\'enologiques autour du mod\`ele supersym\'etrique sym\'etrique gauche-droit et des particules doublement charg\'ees. Un second volet, compl\'ementaire au premier, a consist\'e \`a d\'evelopper des outils informatiques utiles pour les \'etudes ph\'enom\'enologiques. Nous allons d\'evelopper ces deux volets dans les quelques paragraphes qui suivent.


\section{Travail en ph\'enom\'enologie}
\subsection{\'Etude du mod\`ele supersym\'etrique sym\'etrique gauche-droit}
En essayant de joindre la motivation d'avoir une th\'eorie pr\'edisant l'unification des couplages de jauge ainsi que la supersym\'etrie, mes directeurs de th\`ese, notre collaboratrice et moi-m\^eme avons men\'e une \'etude ph\'enom\'enologique sur le mod\`ele supersym\'etrique sym\'etrique gauche-droit. Ces mod\`eles caract\'eris\'es par un groupe de jauge plus large que ceux du Mod\`ele Standard et de son \'equivalent supersym\'etrique, pr\'edisent un grand nombre de nouvelles particules fondamentales scalaires et fermioniques induisant donc une ph\'enom\'enologie tr\`es riche. Dans le papier que nous avons publi\'e r\'ecemment \cite{Alloul:2013xx}, nous avons investigu\'e, dans le cadre d'une approche \textit{"top-down"}, la ph\'enom\'enologie que les charginos et neutralinos de ce mod\`eles induiraient au LHC. Pour ce faire, nous avons commenc\'e par la construction du mod\`ele lui-m\^eme. En effet, dans la th\'eorie il subsiste parfois des confusions quant \`a la d\'efinition des matrices g\'en\'erant les repr\'esentations pour les champs appartenant \`a la repr\'esentation fondamentale de $SU(2)$. Plus pr\'ecis\'ement, le groupe de jauge de ce mod\`ele \'etant
$$ SU(3)_c\times SU(2)_L \times SU(2)_R \times U(1)_{\rm B-L}$$
o\`u B-L est la diff\'erence entre les nombres baryonique et leptonique, les champs de jauge sont, en termes de superchamps vectoriels
\bea
V_3 &=& ( \octet, \singlet, \singlet, 0 ) \equiv (\tilde{g}^a,g_\mu^a)\n
V_{2L} &=& ( \singlet, \triplet, \singlet, 0 ) \equiv (\tilde{W}^k_L,W^k_{L\mu}) , \n
V_{2R} &=& ( \singlet, \singlet, \triplet, 0 ) \equiv (\tilde{W}^k_R,W^k_{R\mu}) , \n
V_{1} &=& ( \singlet, \singlet, \singlet, 0 ) \equiv (\tilde{B},B_{\mu}). \nonumber
\eea
et le contenu en mati\`ere en termes de superchamps chiraux est
\bea
(Q_L)^{fmi} = \begin{pmatrix} u^{fm}_L \\ d_L^{fm} \end{pmatrix} = (\triplet,\doublet,\singlet,\frac13), && (Q_R)_{fmi'} = \begin{pmatrix} u_{Rfm}^c &      d_{Rfm}^c\end{pmatrix} = (\utilde{\mathbf{\bar{3}}}, \singlet, \doublet^{*}, -\frac13),\n
(L_L)^{fi} = \begin{pmatrix} \nu_L^f \\ l_L^f \end{pmatrix} = (\singlet,\doublet,\singlet,-1), && (L_R)_{fi'} = \begin{pmatrix} \nu^c_{Rf} & l_{R_f}^c \end{pmatrix} = (\singlet,\singlet,\doublet^*,1),\n
\delta_{1L} = (\singlet, \triplet, \singlet, -2) = \begin{pmatrix} \delta^1_{1L} \\ \delta^2_{1L} \\ \delta^3_{1L}\end{pmatrix}, && \delta_{1R} =(\singlet, \singlet, \triplet, -2) = \begin{pmatrix} \delta^1_{1R} \\ \delta^2_{1R} \\ \delta^3_{1R}\end{pmatrix}, \n
\delta_{2L} = (\singlet, \triplet, \singlet, +2) = \begin{pmatrix} \delta^1_{2L} \\ \delta^2_{2L} \\ \delta^3_{2L}\end{pmatrix}, && \delta_{2R} =(\singlet, \singlet, \triplet, +2) = \begin{pmatrix} \delta^1_{2R} \\ \delta^2_{2R} \\ \delta^3_{2R}\end{pmatrix},\n
\Phi_{a=1,2} = (\singlet, \doublet, \doublet, 0) = \begin{pmatrix} \phi^0_a & \phi^+_a \\ \phi^-_a & \phi^{'0}_a \end{pmatrix}, && S = (\singlet, \singlet, \singlet, 0).\nonumber
\eea
Ici les indices $f,m, i~{\rm et}~i'$ correspondent \`a des indices de saveur, couleur, $SU(2)_L$ et $SU(2)_R$ respectivement; $Q_L$ et $L_L$ repr\'esentent les superchamps chiraux contenant les quarks et leptons gauchers du Mod\`ele Standard; $Q_R$ et $L_R$ quant \`a eux sont des superchamps chiraux contenant des fermions de chiralit\'e droite se transformant trivialement sous $SU(2)_L$ mais pas sous $SU(2)_R$. Les superchamps $\delta_{1\{L,R\}}$ et $\delta_{2\{L,R\}}$ contiennent les champs de Higgs n\'ecessaires \`a la brisure de la sym\'etrie $SU(2)_L\times SU(2)_R$ alors que les superchamps chiraux $\Phi_a$ sont n\'ecessaires pour la brisure de la sym\'etrie \'electrofaible. Les superchamps triplets de $SU(2)$ peuvent se r\'e\'ecrire sous forme matricielle plus adapt\'ee pour la construction de Lagrangiens:\\
$$\Delta = \frac{1}{\sqrt2}\sigma_a \delta^a $$
Le contenu en champs fix\'e, le Lagrangien associ\'e \`a ce mod\`ele s'\'ecrit simplement
\bea \lagr = \lagr_{gauge} + \lagr_{chiral} + \lagr_{int} + V_D + V_F + \lagr_{soft}, \nonumber\eea
avec
\begin{itemize}
	\item le lagrangien de jauge donn\'e par
	$$ \lagr_{gauge} = -\frac{1}{4} V_{k}^{\mu\nu}V^k_{\mu\nu} + \frac{i}{2}(\tilde{V}^k \sigma^\mu D_\mu \bar{\tilde{V}}_k - D_\mu \tilde{V}^k \sigma^\mu \bar{\tilde{V}}_k) $$
	o\`u k est un indice de jauge; $V$ un champ vectoriel; $V_{k}^{\mu\nu}$ le tenseur de champs; $\tilde{V}$ un fermion de jauge et $\sigma^\mu = (\sigma^0, \sigma^i)$ o\`u $\sigma^0$ est la matrice identit\'e de taille $2\times 2$ et $\sigma^i$ sont les matrices de Pauli. 
	\item Le terme $\lagr_{chiral}$ correspond aux termes cin\'etiques des champs de mati\`ere et leurs int\'eractions avec les champs de jauge, il est donn\'e par
	$$ \lagr_{chiral} = D_\mu \phidagger D^\mu \phi + \frac{i}{2} ( \psi \sigma^\mu D_\mu \psibar - D_\mu \psi \sigma^\mu \psibar) + (i g\sqrt2 \bar{\tilde{V}}^k \cdot \psibar_i T_k \phi^i + {\rm h.c.} ) $$
	o\`u $\phi$ est un champ scalaire; $D^\mu$ une d\'eriv\'ee covariante; $\psi$ un champ fermionique; $T_k$ sont les matrices g\'en\'erant les repr\'esentations des groupes de jauge et h.c. correspond au terme herm\'etique conjug\'e. 
	\item Le terme $\lagr_{int}$ correspond \`a la partie du Lagrangien d\'ecrivant les int\'eractions entre les champs de mati\`ere, il provient directement du superpotentiel qui s'\'ecrit, dans le cadre de notre mod\`ele, comme suit
\end{itemize}
	\bea
	W &=& (\tQ_L)^{mi} y_Q^1 (\hPhi)_i{}^{i'} (\tQ_R)_{mi'} + (\tQ_L)^{mi}y_{Q}^2 (\hPhi_2)_i{}^{i'} (\tQ_R)_{mi'} + (\tL_L)^iy_L^1(\hPhi)_i{}^{i'}(\tL_R)_{i'} \n
	&+& (\tL_L)^iy_L^2 (\hPhi_2)_i{}^{i'} (\tL_R)_{i'} + (\hat{\tL}_L)_i y_L^3 (\Delta_{2L})^i{}_j(\tL_L)^j + (\hat{\tL}_R)_{i'} y_L^4 (\Delta_{1R})^{i'}{}_{j'} (\tL_R)^{j'} \n
	&+& (\mu_L + \lambda_L S) \Delta_{1L} \cdot \hDelta_{2L} + (\mu_R + \lambda_R S)\Delta_{1R}\cdot\hDelta_{2R} + (\mu_3 + \lambda_3 S) \Phi_1 \cdot \hPhi_2 \n
	&+& \frac13 \lambda_s S^3 + \mu_s S^2 + \xi_S S.\nonumber
	\eea
	o\`u $\tQ_L, \tQ_R, \tL_L \mbox{ et } \tL_R$ sont les composantes scalaires des superchamps $Q_L$, $Q_R$, $T_L$ et $T_R$ respectivement. Les autres quantit\'es sont d\'efinies comme suit
	\bea 
	(\hat{\tL}_L)_i &=& \epsilon_{ij}(\tL_L)^j, ~~~ (\hat{\tL}_R)^{i'} = \epsilon^{i'j'}(\tL_R)_{j'}, \n
	(\hDelta_{2L})_i{}^j &=& \epsilon_{ik}\epsilon^{jl} (\Delta_{2L})^k{}_l, ~~~ (\hDelta_{2R})_{i'}{}^{j'} = \epsilon_{i'k'}\epsilon^{j'l'} (\Delta_{2R})^{k'}{}_{l'}, \n
	\Delta_{1L} \cdot \hDelta_{2L} &=& {\rm Tr}(\Delta^t_{1L} \hDelta_{2L}) = (\Delta_{1L})^i{}_j (\hDelta_{2L})_i{}^j, ~~~ \Delta_{1R} \cdot \hDelta_{2R} = {\rm Tr}(\Delta^t_{1R} \hDelta_{2R}) = (\Delta_{1R})^{i'}{}_{j'} (\hDelta_{2R})_{i'}{}^{j'}, \n
	(\hPhi_{1,2})_i{}^{i'} &=& \epsilon^{i'j'}\epsilon_{ij}(\Phi_{1,2})^j{}_{j'},~~~ \Phi_1\cdot\hPhi_2 = {\rm Tr}(\Phi^t_1 \hPhi_2) = (\Phi_1)^i{}_{i'} (\hPhi_2)_i{}^{i'}.\nonumber
	\eea
\begin{itemize}
	\item Les termes $V_D \mbox{ et } V_F$ sont les termes $D$ et $F$, respectivement, du potentiel scalaire.
	\item Le Lagrangien de brisure douce de la supersym\'etrie est lui dict\'e, en partie, par la forme du superpotentiel et son expression est \'egale \`a
\end{itemize}
	\bea
	\lagr_{soft} &=& -\frac12 \big[ M_1 \tilde{B}\cdot\tilde{B} + M_{2L} \tilde{W}^k_L \cdot \tilde{W}_{Lk} + M_{2R} \tilde{W}^k_R \cdot \tilde{W}_{Rk} + M_3 \tilde{g}^a \cdot \tilde{g}_a + {\rm h.c.} \big] \n
	             &-& \Big[\tilde{Q}^\dagger m_{Q_L}^2 \tilde{Q}_L + \tilde{Q}_R m_{Q_R}^2 Q_R^\dagger + \tL_L^\dagger m_{L_L}^2 \tL_L + \tL_R m_{L_R}^2 \tL_R^\dagger - (m_\Phi^2)^{ff'} {\rm Tr}(\Phi^\dagger_f \Phi_{f'}) \n
	&+& m^2_{\Delta_{1L}}{\rm Tr}(\Delta^\dagger_{1L} \Delta_{1L}) +  m^2_{\Delta_{2L}}{\rm Tr}(\Delta^\dagger_{2L} \Delta_{2L}) + m^2_{\Delta_{1R}}{\rm Tr}(\Delta^\dagger_{1R} \Delta_{1R}) +  m^2_{\Delta_{2R}}{\rm Tr}(\Delta^\dagger_{2R} \Delta_{2R}) + m^2_S S^\dagger S \Big]\n
	&-& \Big[ \tQ_L T_Q^1 \hPhi_1 \tQ_R + \tQ_L T_Q^2 \hPhi_2 \tQ_R + \tL_L T_L^1\hPhi_1\tL_R + \tL_L T_L^2\hPhi_2\tL_R + \hat{\tL}_LT^3_L\Delta_{2L}\tL_L + \tL_R T^4_L \Delta_{1R} \hat{\tL}_R + {\rm h.c.} \Big]\n
	&-& \Big[ T_L S \Delta_{1L} \cdot \hDelta_{2L} + T_R S \Delta_{1R} \cdot \hDelta_{2R} + T_{3}S \Phi_1 \cdot \hPhi_2 + {\rm h.c.} \Big]\nonumber
	\eea

\begin{figure}
\begin{center}
    \includegraphics[width=.49\columnwidth]{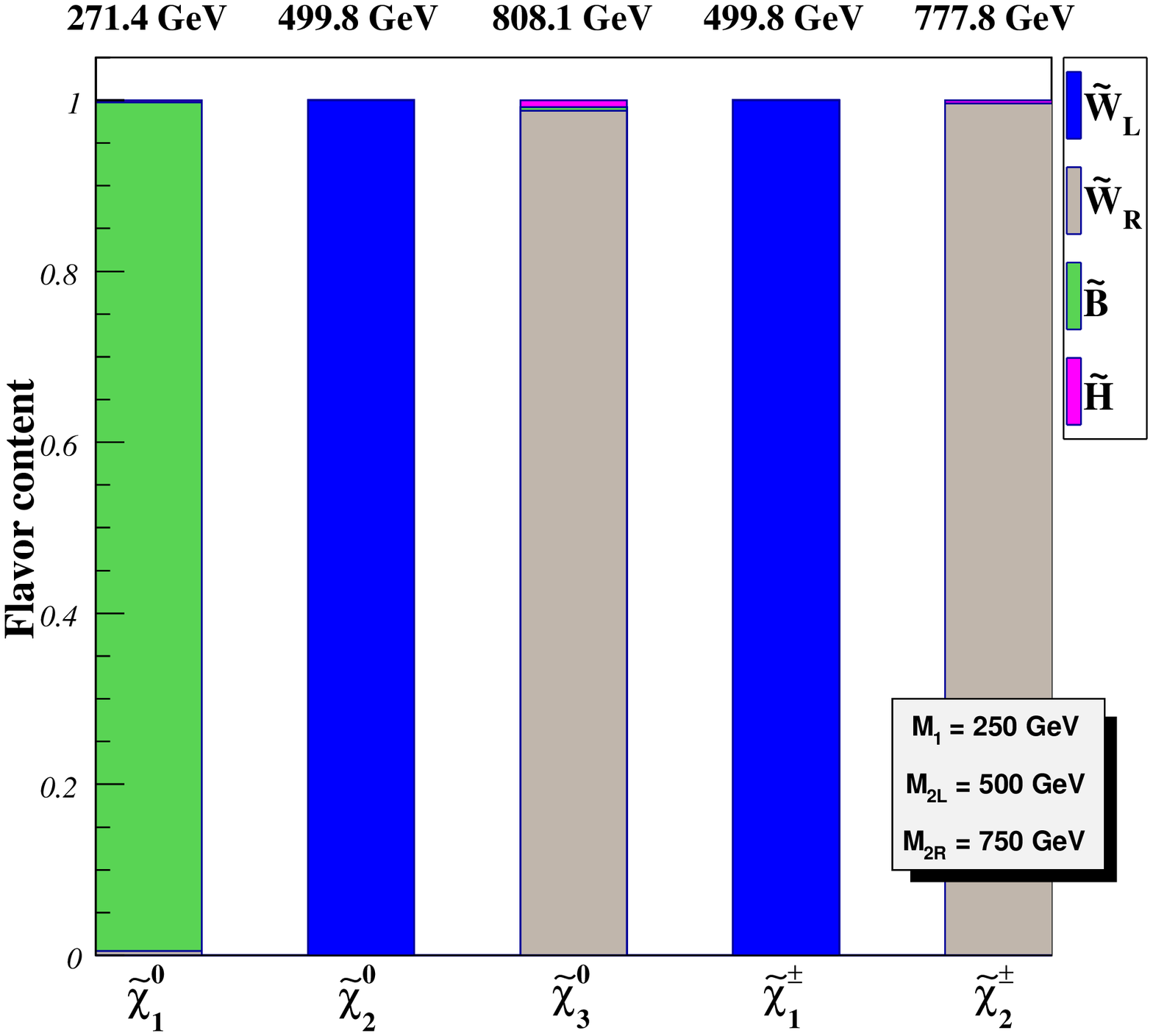}
    \includegraphics[width=.49\columnwidth]{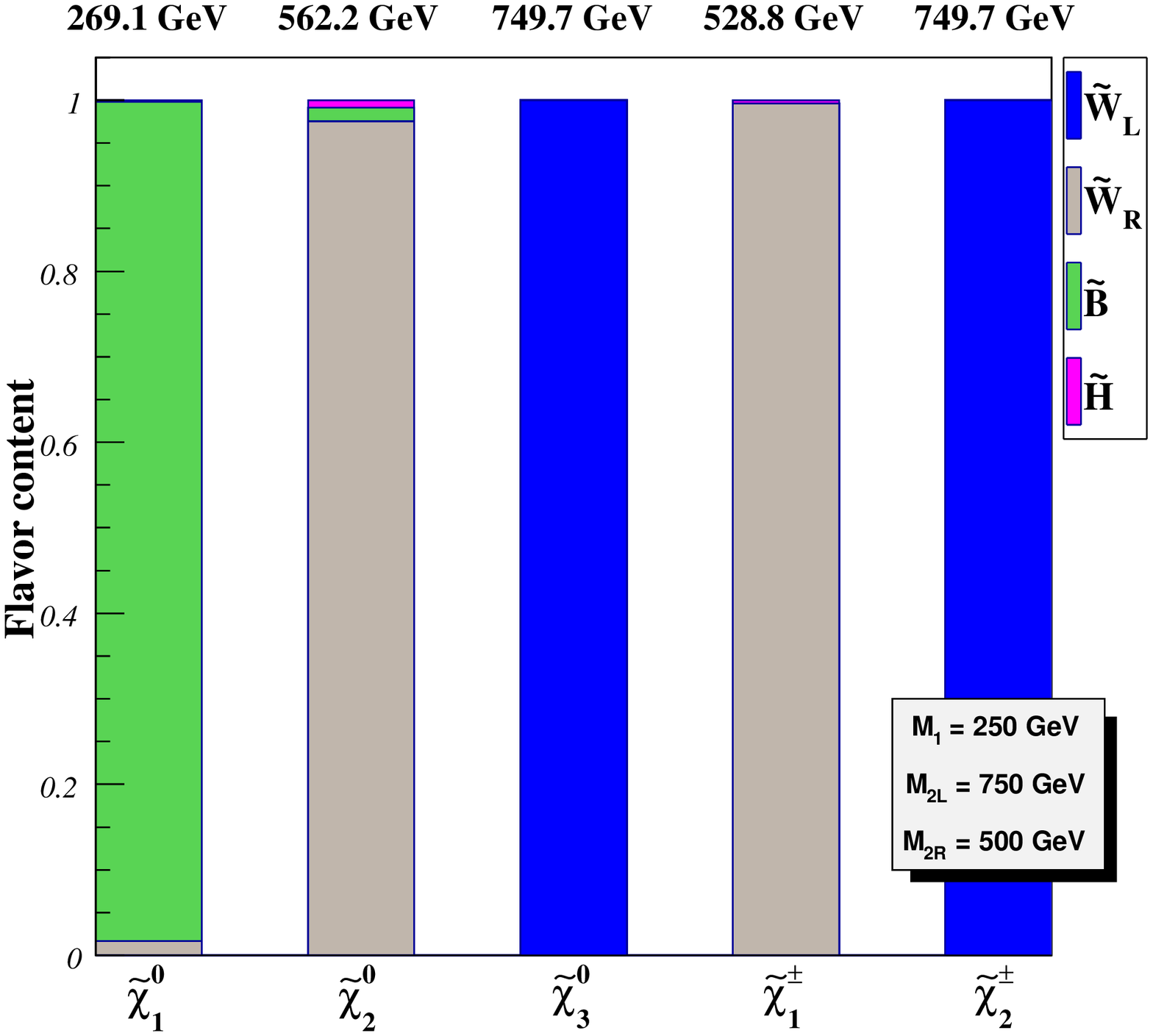}\\
    \includegraphics[width=.49\columnwidth]{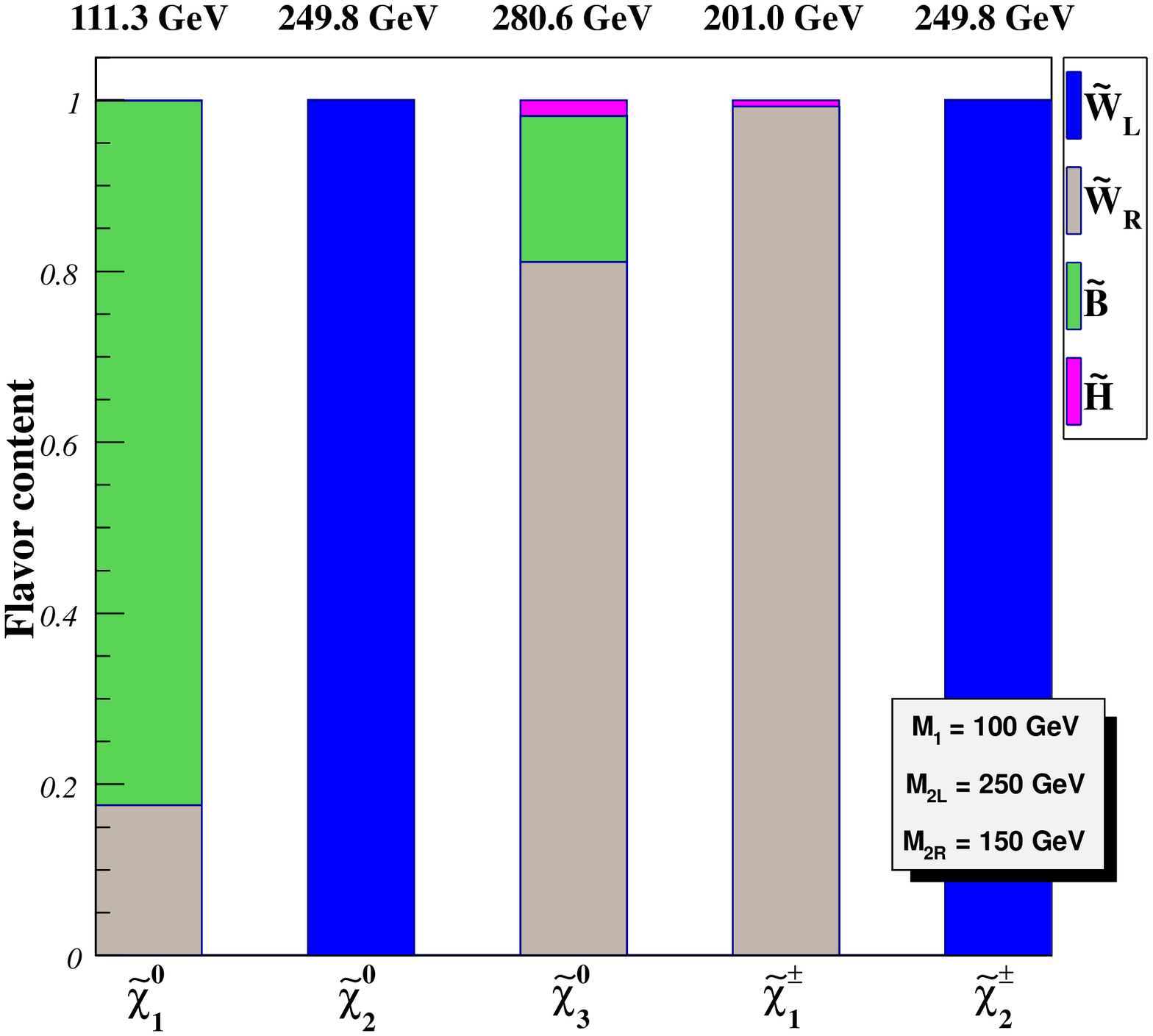}
    \includegraphics[width=.49\columnwidth]{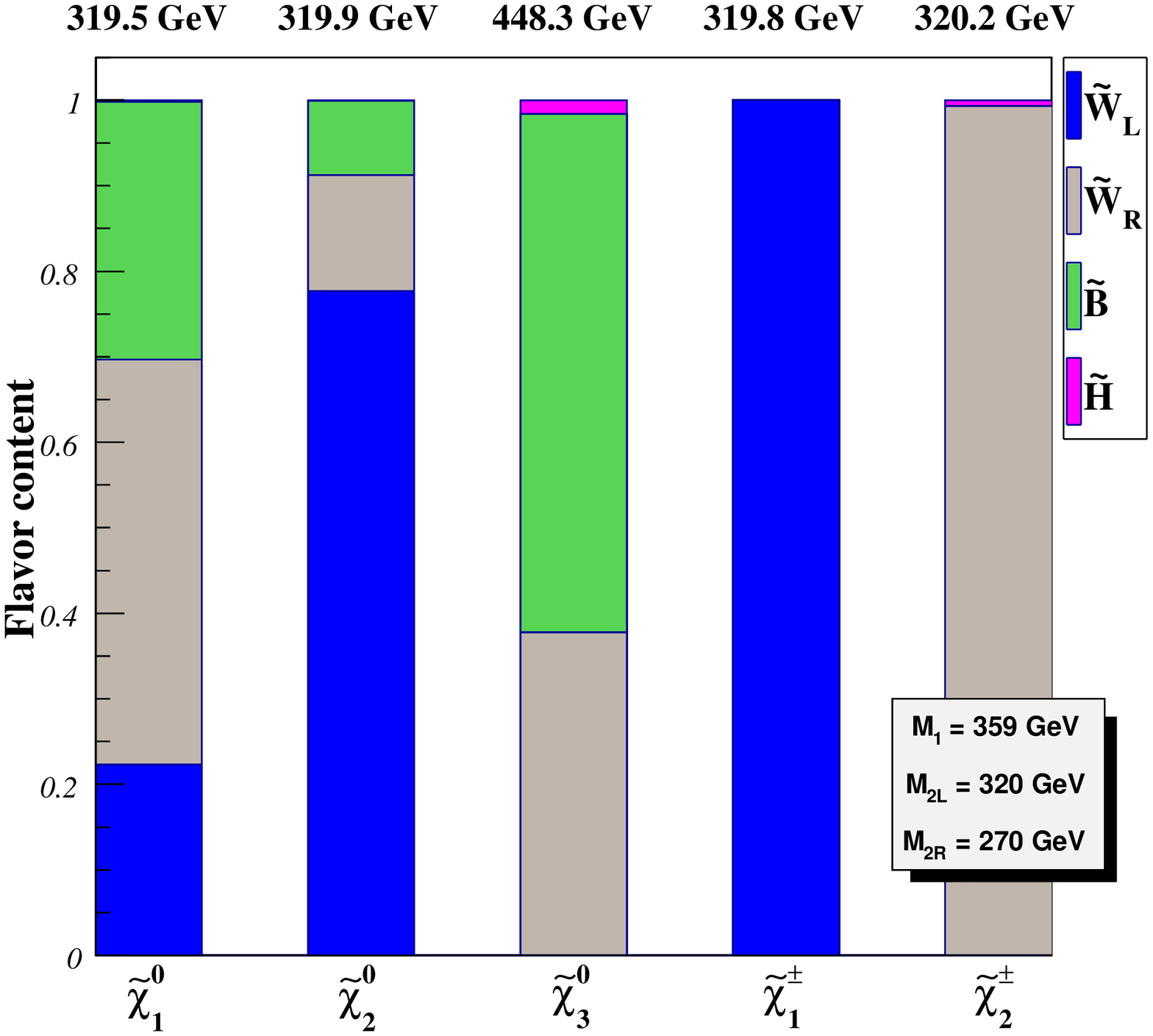}
\end{center}
\caption{\footnotesize Dans ces figures sont montr\'ees les d\'ecompositions des \'etats propres de masse en fonction des \'etats propres de jauge. $M_1, M_{2L} \mbox{ et } M_{2R}$ sont les masses des jauginos associ\'es aux groupes $U(1)_{B-L}, SU(2)_L$ et $SU(2)_R$ respectivement}
\label{fig:lrsusy fr}
\end{figure}

La construction du mod\`ele est maintenant termin\'ee et nous pouvons passer \`a l'\'etude du secteur des charginos et neutralinos nous int\'eressant. Pour ce faire, nous avons d'abord construit quatre sc\'enarios d'\'etude de telle sorte \`a ce que dans deux cas les \'etats propres de jauge sont aussi \'etats propres de masse et deux autres cas o\`u les \'etats propres de masse sont des vrais m\'elanges des \'etats propres de jauge. La figure \ref{fig:lrsusy fr} r\'esume ces quatres configurations et donne aussi les masses des diff\'erents \'etats.

Ensuite, nous avons impl\'ement\'e le mod\`ele dans le programme {\sc FeynRules} afin de profiter des facilit\'es qu'offre l'interface {\sc UFO}. Le mod\`ele {\sc UFO} g\'en\'er\'e, nous utilisons le simulateur Monte Carlo {\sc MadGraph 5} afin de calculer les largeurs de d\'esingt\'egrations des neutralinos et charginos les plus l\'egers mais aussi pour g\'en\'erer les \'ev\'enements Monte Carlo partoniques simulant la production de ces particules au LHC. Ces \'ev\'evenements sont ensuite trait\'es \`a l'aide du programme {\sc Pythia 8} pour simuler proprement les d\'esint\'egrations, hadronisations et le ``parton showering'' (la simulation des ph\'enom\`enes de la QCD est tr\`es importante pour le LHC). Enfin, nous devons aussi prendre en compte le ``jet-clustering'' \`a l'aide du programme {\sc FastJet}\footnote{Nous consid\'erons un d\'etecteur parfait.} et utilisons le programme {\sc MadAnalysis 5} pour analyser les \'ev\'enements ainsi produits et pouvoir les comparer au bruit de fond du Mod\`ele Standard et de sa version supersym\'etrique.\\

Les r\'esultats de notre analyse montrent que le meilleur canal pour d\'etecter des \'ev\'enements provenant du mod\`ele supersym\'etrique sym\'etrique gauche-droit est celui o\`u la configuration de l'\'etat final contient au moins un lepton charg\'e l\'eger. En effet, nous avons montr\'e qu'imposer des restrictions sur les variables cin\'ematiques telles que l'\'energie transverse manquante soit au moins \'egale \`a 100 GeV, l'impulsion transverse du lepton le plus \'energ\'etique au moins \'egale \`a 80 GeV et l'impulsion transverse du second lepton plus \'energ\'etique au moins \'egale \`a 70 GeV garantit une pr\'edominance du signal par rapport aux pr\'edictions du Mod\`ele Standard. De plus, ces m\^emes restrictions ont \'et\'e utilis\'ees pour comparer notre mod\`ele avec le Mod\`ele Standard Supersym\'etrique Minimal et montrer que, l\`a aussi, les diff\'erences entre ces deux mod\`eles \'etaient assez grandes pour pouvoir les distinguer facilement.\\

\subsection*{Les particules doublement charg\'ees au LHC}
Dans le mod\`ele pr\'esent\'e ci-dessus, le secteur du Higgs a ceci de particulier qu'il contient des particules doublement charg\'ees. En effet, si on reprend les composantes scalaires\footnote{Le m\^eme raisonnement tient pour les composantes fermioniques} des champs se transformant dans l'adjointe de $SU(2)$ sous leur forme matricielle et qu'on d\'etermine la charge \'electrique de chacune des composantes, l'on trouve
\bea \Delta_{1 \{L,R\}} = \begin{pmatrix} \frac{\Delta^{-}_{1\{L,R\}}}{\sqrt2} & \Delta^{0}_{1\{L,R\}} \\ \Delta^{--}_{1\{L,R\}} & -\frac{\Delta^{-}_{1\{L,R\}}}{\sqrt2} \end{pmatrix},~~~\Delta_{2 \{L,R\}} =  \begin{pmatrix} \frac{\Delta^{+}_{2\{L,R\}}}{\sqrt2} & \Delta^{++}_{2\{L,R\}} \\ \Delta^{0}_{2\{L,R\}} & -\frac{\Delta^{+}_{2\{L,R\}}}{\sqrt2} \end{pmatrix} \nonumber \eea
o\`u les exposants indiquent les charges \'electriques. Si de telles particules devaient \^etre produites au LHC, les traces qu'elles laisseraient seraient facilement mises en \'evidence de par leur charge \'electrique. Cependant, une telle d\'etection ne signifierait en aucun cas qu'on a d\'ecouvert le mod\`ele d\'ecrit ci-dessus puisque les particules doublement charg\'ees sont pr\'edites par plusieurs extensions du Mod\`ele Stadard. Ainsi la question de savoir quel mod\`ele d\'ecrit le mieux ces particules se poserait imm\'ediatement et seule une analyse pr\'ecise des propri\'et\'es de ces particules permetteraient de donner les cl\'es pour les comprendre. \\

Dans chaque extension du Mod\`ele Standard pr\'edisant l'existence d'une particule doublement charg\'ee, cette derni\`ere a des nombres quantiques diff\'erents. Par exemple, dans le mod\`ele sym\'etrique gauche-droit non supersym\'etrique, les particules doublement charg\'ees sont uniquement scalaires et se transforment dans la $\singlet$ ou la $\triplet$ de $SU(2)_L$ alors que dans la version supersym\'etrique de ce mod\`ele l'on a des particules scalaires et fermioniques. Dans d'autres extensions du Mod\`ele Standard, les particules doublement charg\'ees peuvent m\^eme \^etre des vecteurs. Pour pouvoir rester le plus g\'en\'erique possible, dans notre publication \cite{Alloul:2013zz}, mes collaborateurs et moi-m\^eme avons justement construit des mod\`eles effectifs partant du contenu en champ et du Lagrangien du Mod\`ele Standard et les \'etendant afin qu'ils contiennent une nouvelle particule doublement charg\'ee ainsi que ses int\'eractions avec les autres particules du Mod\`ele Standard. Ainsi nous avons d\'efini neuf cas correspondant \`a une particule charg\'ee scalaire, fermionique ou vecteur appartenant \`a un multiplet se transformant sous $SU(2)_L$ comme un singlet, un doublet ou un triplet. Les hypercharges de ces multiplets sont choisies de telle sorte \`a ce que la particule doublement charg\'ee ait la charge \'electrique la plus \'elev\'ee. Dans le cas d'un multiplet fermionique se transformant dans la $\doublet$ de $SU(2)_L$, il n'est pas interdit que sa composante dont la charge \'electrique est \'egale \`a 1 se m\'elange avec les leptons du Mod\`ele Standard; nous distinguerons dans ce cas les deux cas extr\^emes o\`u il y a un m\'elange et o\`u le m\'elange est inexistant. Enfin, par souci de minimalit\'e nous avons choisi d'inclure dans le Lagrangien des couplages non-renormalisables au lieu d'augmenter le contenu en champs et avons interdit les d\'esint\'egrations \`a l'int\'erieur d'un m\^eme multiplet.\\

Une premi\`ere partie th\'eorique de notre travail a consist\'e \`a calculer analytiquement les expressions des largeurs de d\'esint\'egration de ces particules doublement charg\'ees ainsi que les expressions des sections efficaces de leur production. Ceci nous a permis de dresser le tableau \ref{tab:lim mass dc} ci-dessous dans lequelles sont report\'ees les masses maximales admises pour chacun des cas consid\'er\'es de telle sorte \`a ce que la section efficace de production de ces particules menant \`a un \'etat final avec au moins trois leptons l\'egers charg\'es soit au moins \'egale \`a 1 fb pour une \'energie dans le centre de masse de 8 TeV. Dans la seconde colonne de ce tableau est donn\'e, pour chaque cas, le nombre de leptons charg\'es l\'egers maximal que l'on peut produire.
\begin{table}[!h]
\centering
\begin{tabular}{c|c|c|c||c|c|c|}
\cline{2-7}
& \multicolumn{3}{|c||}{Masse maximale [GeV]} & \multicolumn{3}{|c||}{Nombre maximale de leptons} \\
\cline{2-7}
& Singlet & Doublet & Triplet & Singlet & Doublet & Triplet \\
\hline
\hline
Scalars & 330 & 257 & 350 & 4 & 4 & 5\\
Fermions (3 Gen) & 555 & 661 & 738 & 4 & 4 & 4\\
Fermions (4 Gen) & - & 525 & 648 & - & 6 & 5 \\
Vectors & 392 & 619 & 495 & 4 & 4 & 4\\
\hline
\end{tabular}
\caption{\footnotesize\label{tab:lim mass dc} Dans ce tableau sont pr\'esent\'es les r\'esultats r\'esumant pour tous les cas simplifi\'es que nous avons consid\'er\'e dans notre \'etude la masse maximale que l'on peut atteindre pour que la section efficace li\'ee \`a la production de ces particules et leur d\'esint\'egration en leptons charg\'es l\'egers soit au moins \'egale \`a 1 fb. La seconde colonne correspond quant \`a elle au nombre maximal de leptons que l'on peut esp\'erer dans chaque cas.}
\end{table}

\begin{figure}
\centering\begin{tabular}{c c c}
\includegraphics[width=.32\columnwidth]{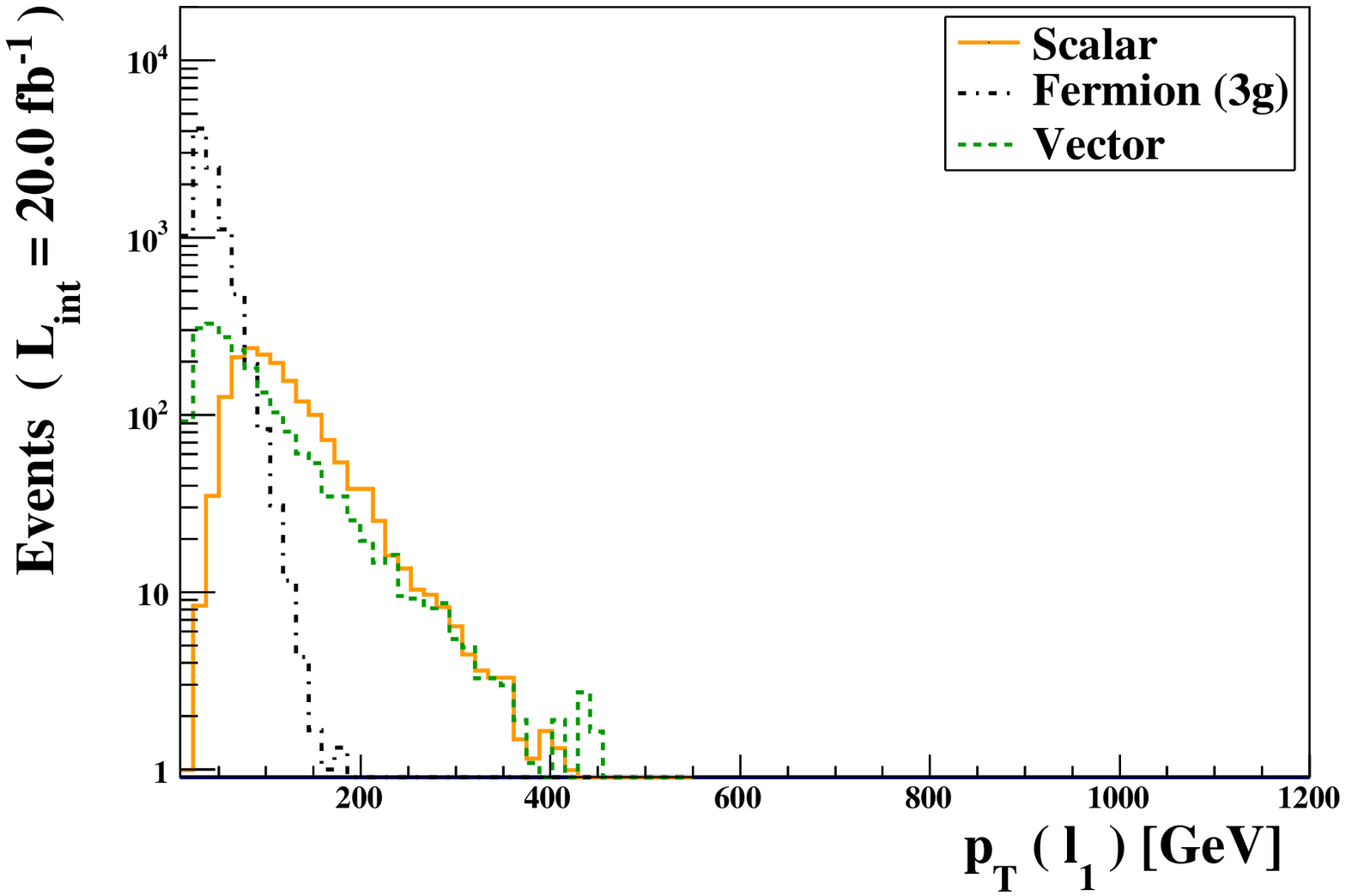} & \includegraphics[width=.32\columnwidth]{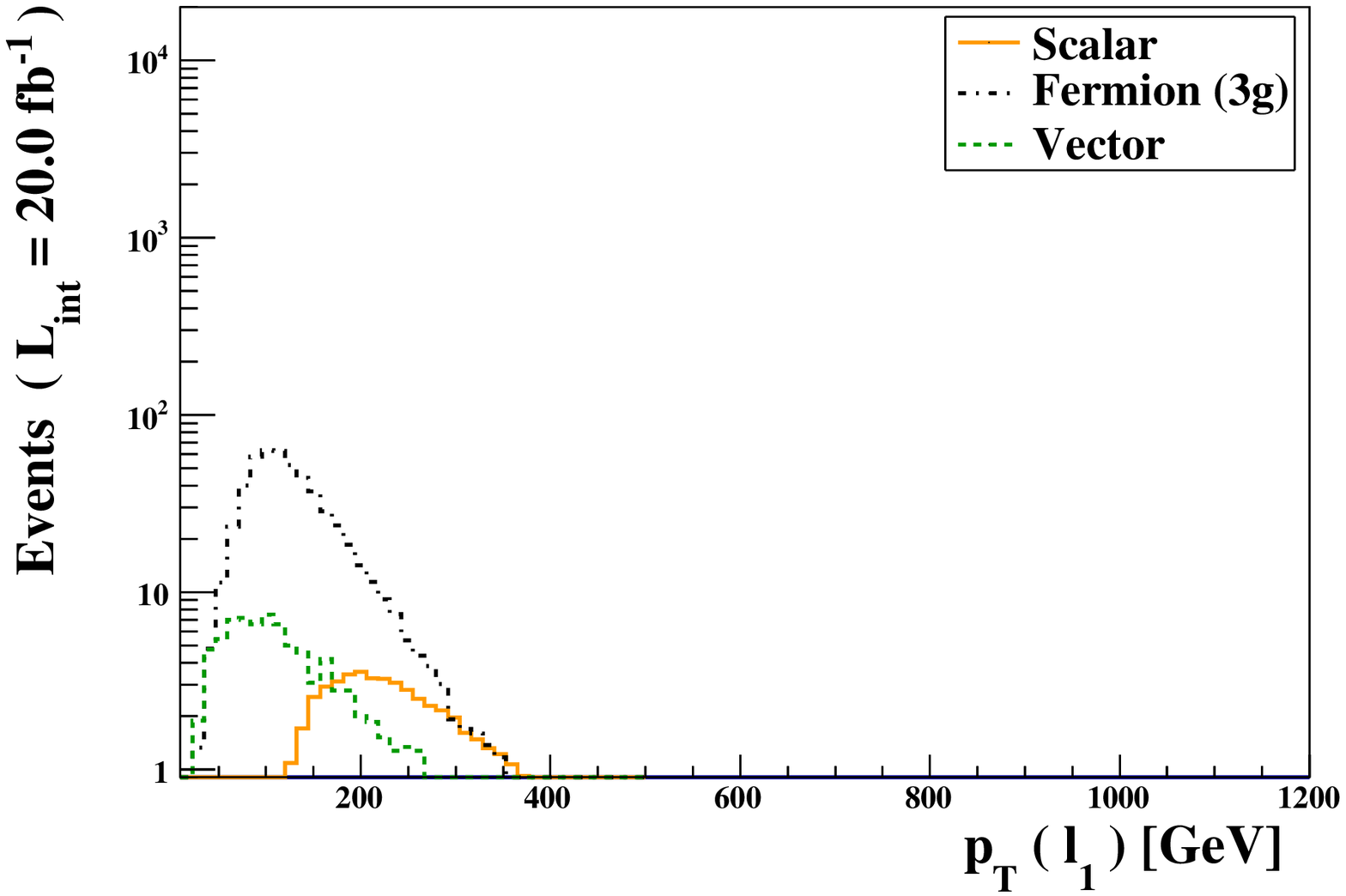} & \includegraphics[width=.32\columnwidth]{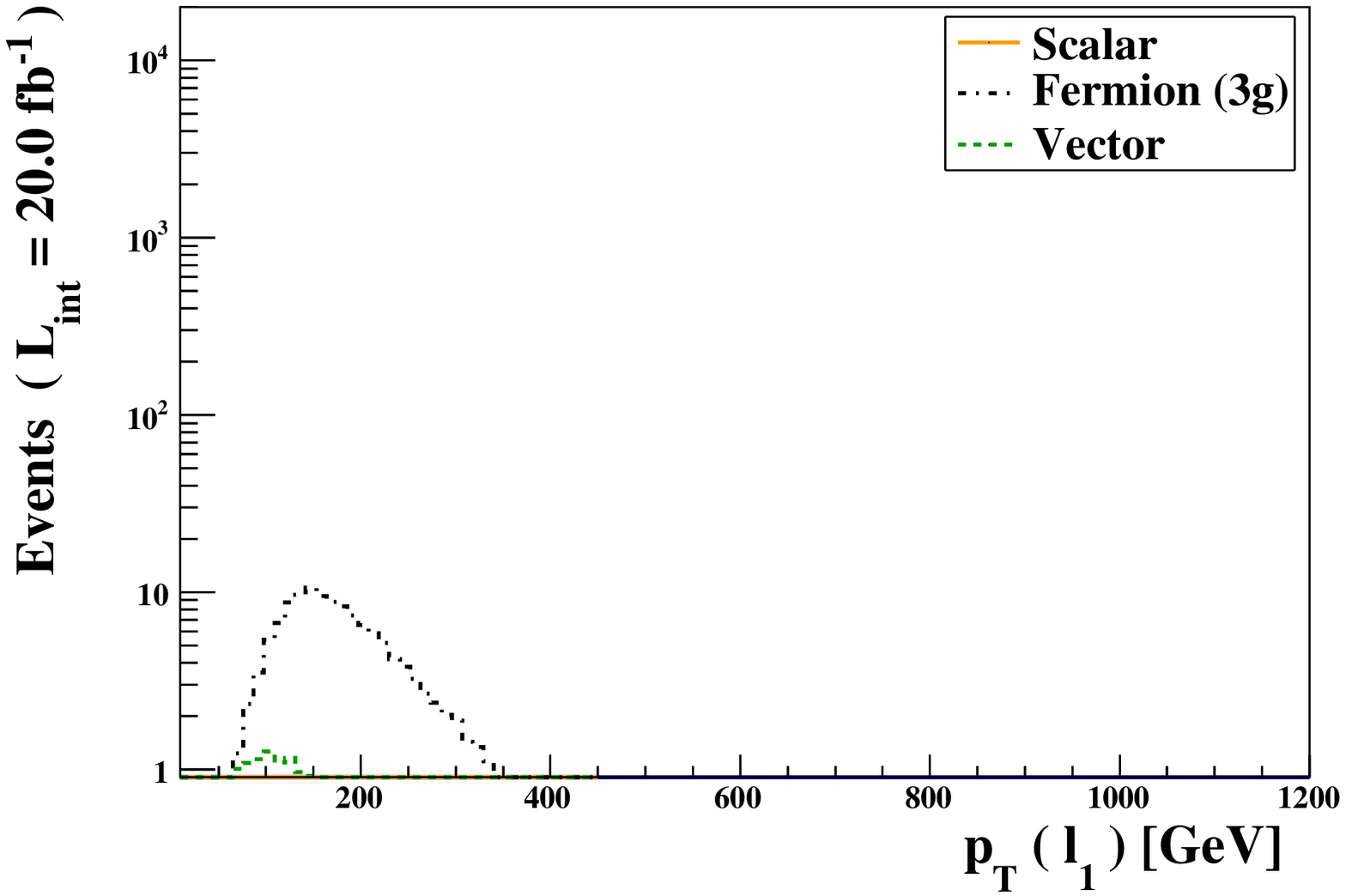}\\
\includegraphics[width=.32\columnwidth]{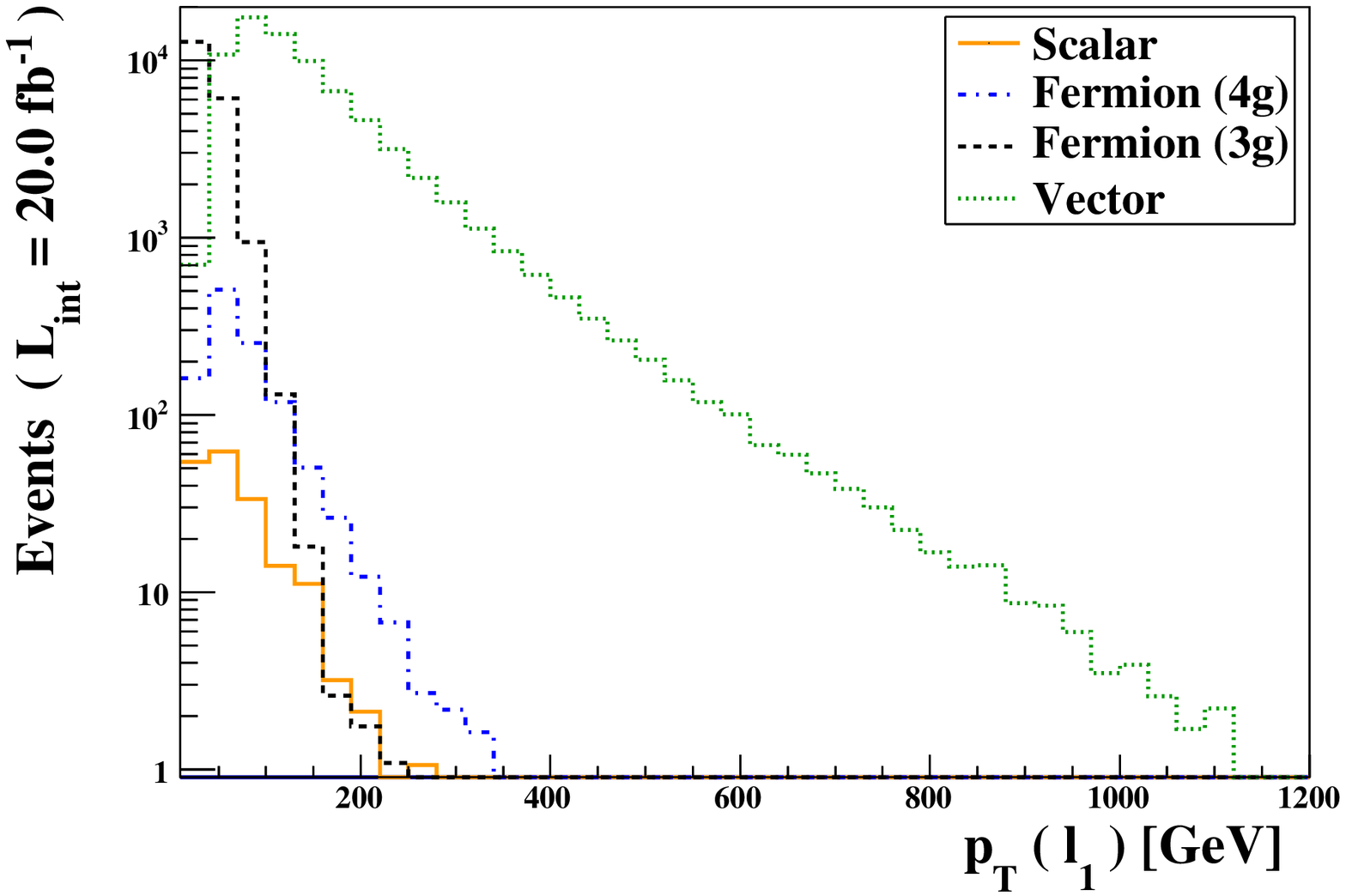} & \includegraphics[width=.32\columnwidth]{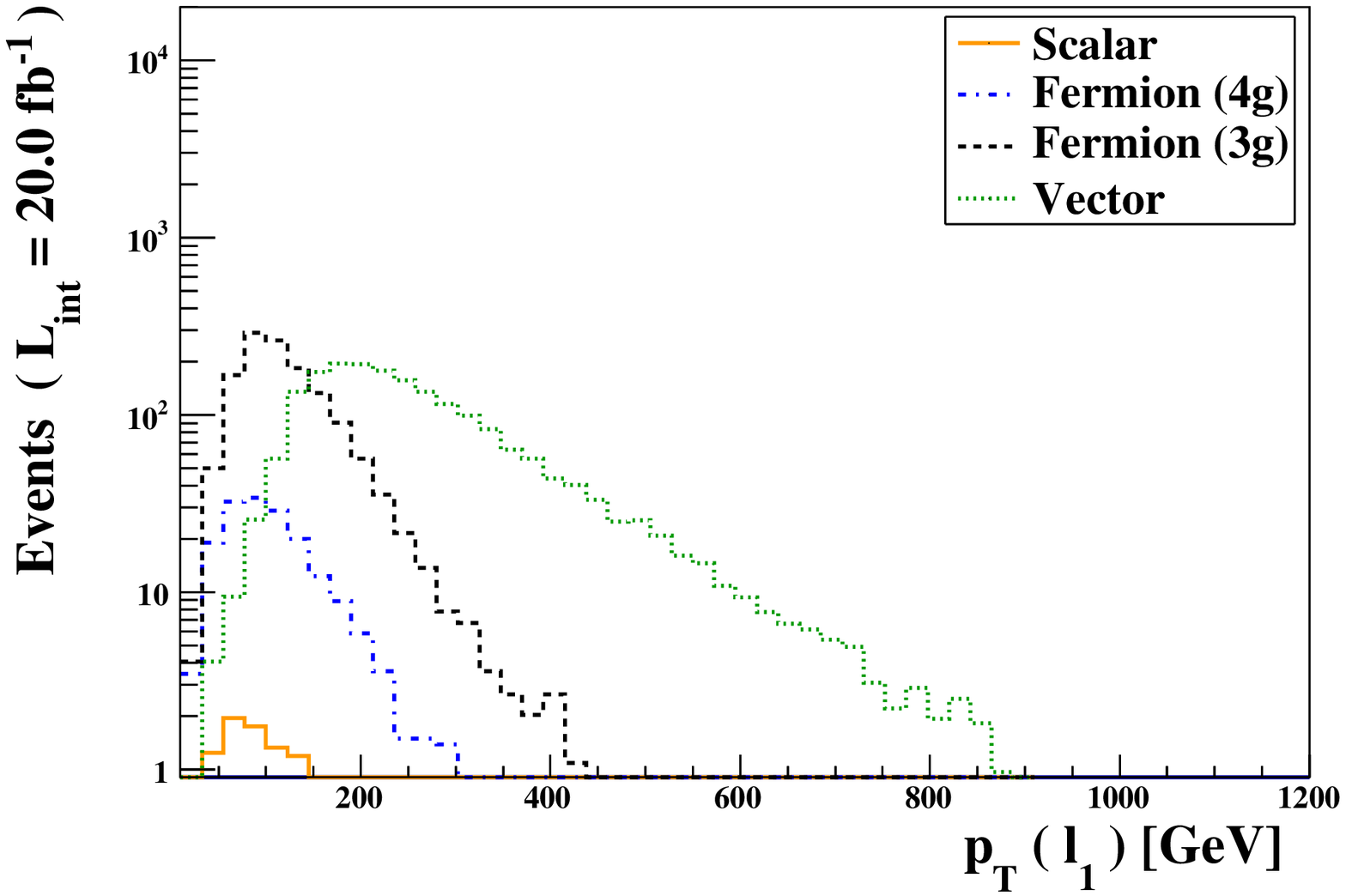} & \includegraphics[width=.32\columnwidth]{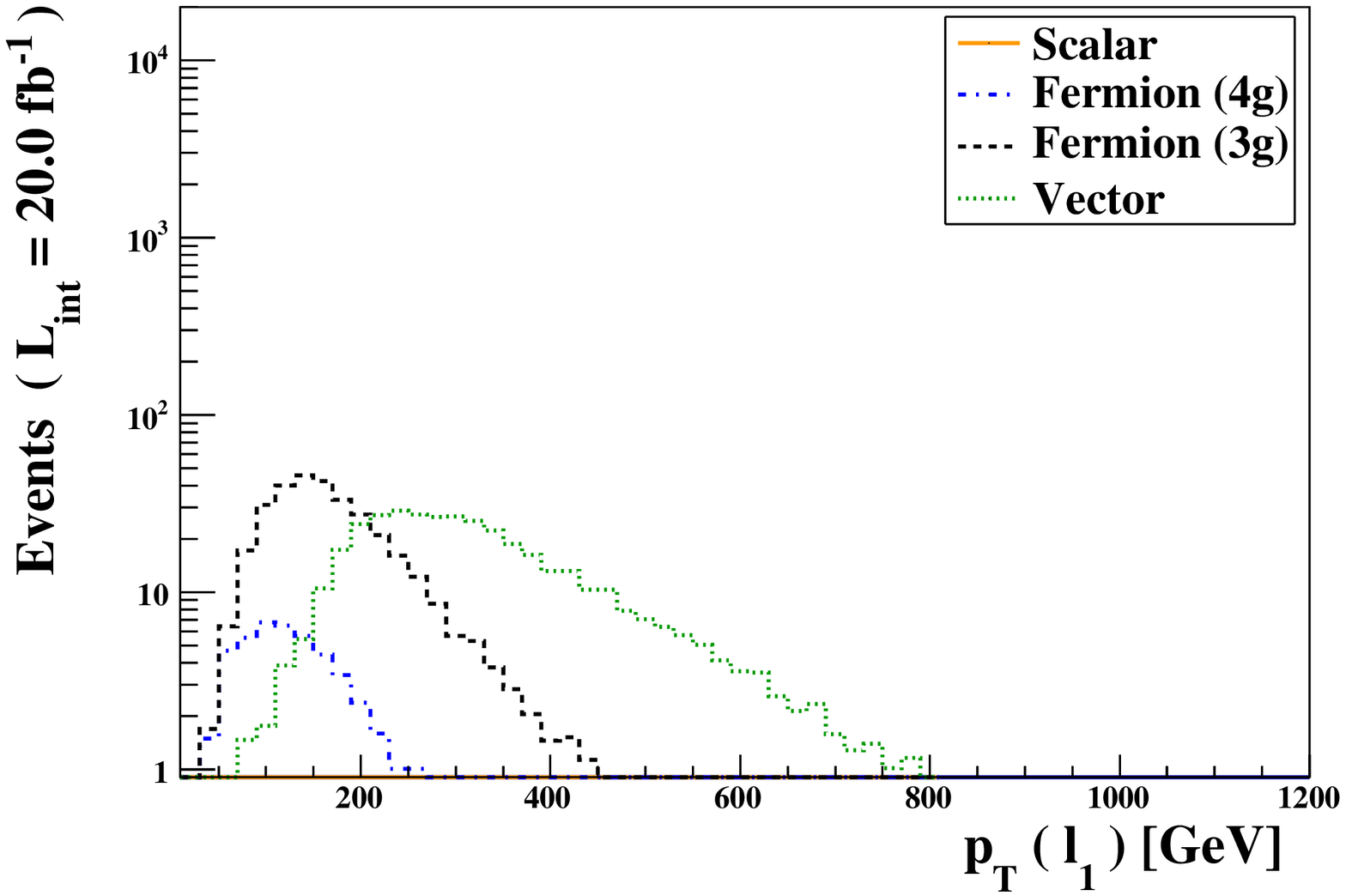}\\
\includegraphics[width=.32\columnwidth]{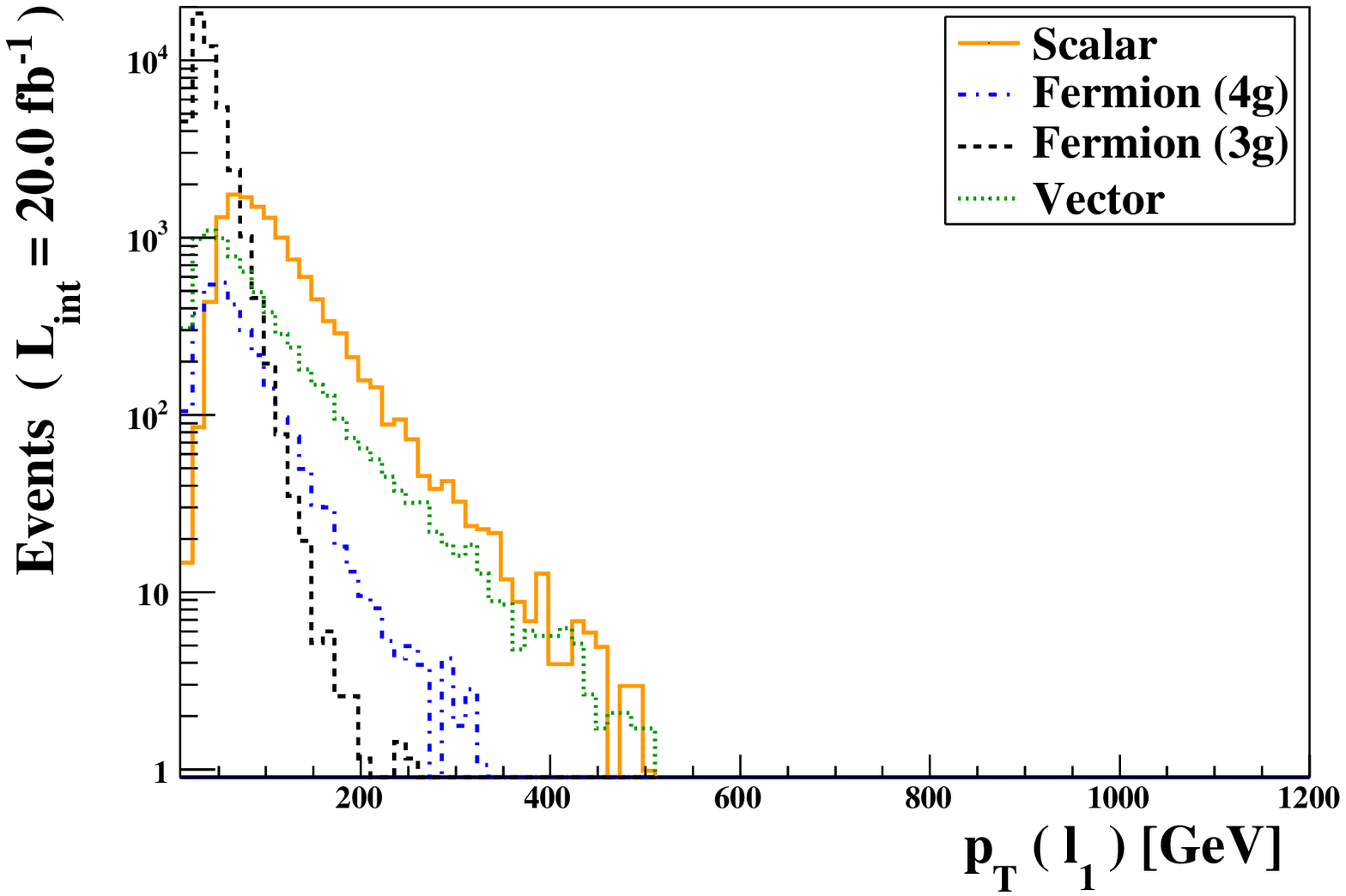} & \includegraphics[width=.32\columnwidth]{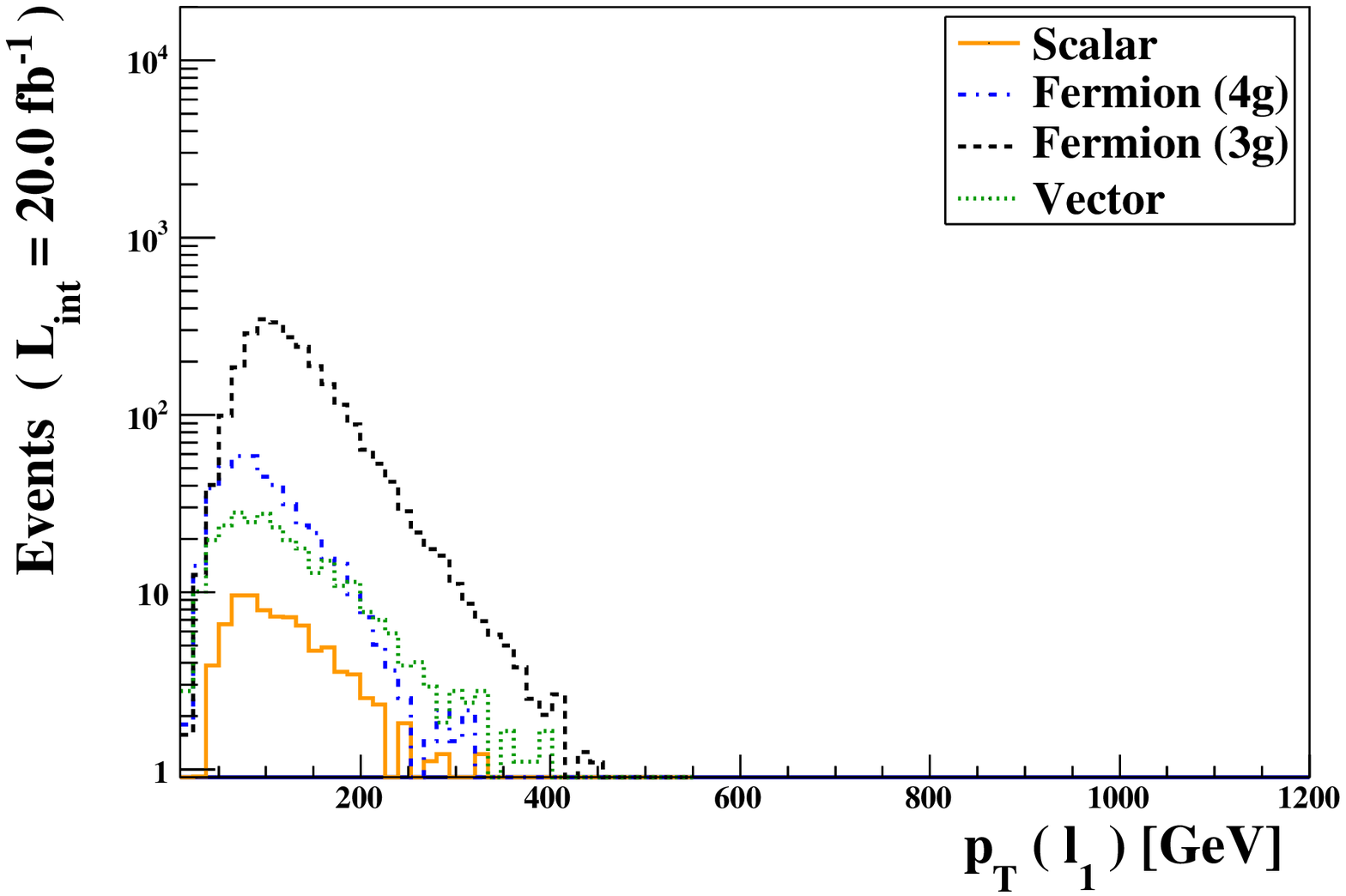} & \includegraphics[width=.32\columnwidth]{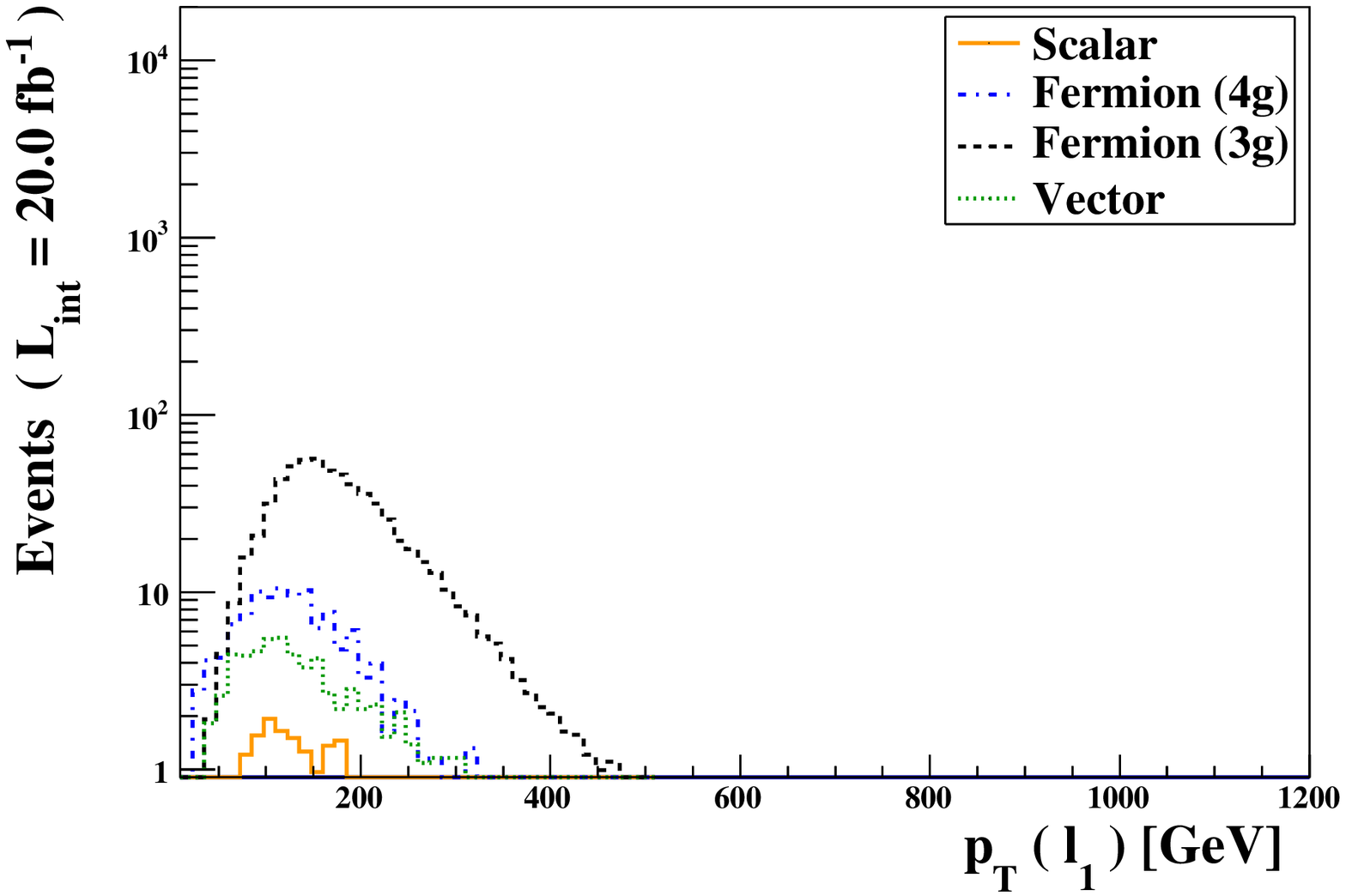}\\
\end{tabular}
\caption{\footnotesize\label{fig:graph db} \'Evolution de la distribution de l'impulsion transverse quand les repr\'esentations et les masses des nouveaux multiplets varient.}
\end{figure}

Les distributions cin\'ematiques \'etant les seules observables nous permettant d'avoir acc\`es aux propri\'et\'es des particules produites au LHC, nous avons proc\'ed\'e, dans une seconde partie de notre \'etude \`a une simulation Monte Carlo pour chacun des cas consid\'er\'es. Pour ce faire, nous avons fix\'e tous les couplages \`a 0.1 et les masses des nouveaux multiplets ont \'et\'e fix\'ees \`a 100, 250 et 350 GeV successivement. Pour cette analyse, par contre, aucune simulation du bruit de fond n'a \'et\'e op\'er\'ee mais, forts de notre pr\'ec\'edente \'etude, nous savions qu'avec au moins trois leptons l\'egers charg\'es dans l'\'etat final, celui-ci \'etait sous contr\^ole. Enfin, nous avons arr\^et\'e notre simulation Monte Carlo \`a la hadronisation des leptons ``tau'' (effectu\'ee par {\sc Pythia 6}). \\ 

\`A l'aide du programme {\sc MadAnalysis 5}, nous avons analys\'e plusieurs variables cin\'ematiques (impulsions transverses, \'energie transverse manquante, distances angulaires) et avons pu conclure que, en l'absence de toute autre indication de nouvelle physique, seule une analyse combin\'ee de plusieurs variables pouvait aider \`a distinguer entre les diff\'erents cas. Dans la figure \ref{fig:graph db} o\`u de gauche \`a droite sont pr\'esent\'es les cas singlet, doublet et triplet et de haut en bas les masses 100, 250 et 350 GeV on peut voir par exemple la distributions de l'impulsion du lepton le plus \'energ\'etique.  Dans cet exemple, si on consid\`ere le cas o\`u la nouvelle particule a une masse de 100 GeV et se comporte comme un doublet sous $SU(2)_L$ (graphique au centre de la premi\`ere ligne) on voit que les distributions pour les diff\'erents cas ne sont pas fondamentalement diff\'erentes. En revanche, si la particule doublement charg\'ee a une masse de 250 GeV et se transforme comme un triplet sous $SU(2)_L$ (graphique tout \`a droite de la ligne du milieu) les diff\'erents cas m\`enent \`a des distributions clairement distinguables.\\
 
Une limite claire de notre travail est introduite par le fait que nous n'avons men\'e aucune \'etude sur les effets des d\'etecteurs sur les distributions cin\'ematiques. Il serait donc int\'eressant de pouvoir continuer ce travail dans cette direction l\`a mais aussi de le sp\'ecialiser dans le cas de mod\`eles non simplifi\'es.

\section{D\'eveloppement d'outils informatiques}
Les \'etudes en ph\'enom\'enologie aujourd'hui reposent en grande partie sur l'utilisation de programmes informatiques. Ceux-ci, en automatisant certains calculs tr\`es lourds voire impossibles \`a faire \`a la main, sur une feuille de papier, nous facilitent grandement la vie, nous permettent d'\^etre tr\`es r\'eactifs aux r\'esultats du LHC et nous permettent aussi de publier des r\'esultats plus pr\'ecis (calculs NLO, NNLO \dots etc). C'est donc naturellement que durant ma th\`ese, j'ai \'et\'e impliqu\'e dans le d\'eveloppement de deux outils informatiques automatisant, pour l'un, l'extraction des \'equations du groupe de renormalisation \`a deux boucles pour tout mod\`ele supersym\'etrique renormalisable et, pour l'autre, l'extraction et la diagonalisation des matrices de masses pour n'importe quel mod\`ele de physique au-del\`a du Mod\`ele Standard et la sauvegarde de ces r\'esultats dans un fichier compatible avec le format SLHA. Dans les quelques paragraphes qui suivent, je pr\'esenterai un peu plus en d\'etails ces deux programmes mais ces deux programmes ayant \'et\'e inclus dans le programme {\sc FeynRules}\cite{Christensen:2008py,Christensen:2009jx,Duhr:2011se,Alloul:2013yy} et ayant moi-m\^eme contribu\'e au d\'eveloppemenet de ce dernier, je commencerai par une petite introduction \`a ce programme.\\

{\sc FeynRules} donc est un programme informatique \'ecrit dans le langage de programmation {\sc Mathematica} capable d'extraire automatiquement les r\`egles de Feynman associ\'ees \`a tous les vertexes d'un mod\`ele donn\'e \`a partir d'un nombre d'informations minimal et de les exporter ensuite, \`a travers des interfaces d\'edi\'ees, vers plusieurs g\'en\'erateurs Monte Carlo diff\'erents .

En pratique, il est demand\'e \`a l'utilisateur de fournir dans un fichier mod\`ele structur\'e les informations sur le groupe de jauge de son mod\`ele, le contenu en champ ainsi que tous leurs nombres quantiques, les d\'efinitions des param\`etres du Lagrangien et le Lagrangien lui-m\^eme. Quelques commandes suffisent ensuite, \`a partir d'une session {\mk}, \`a extraire automatiquement toutes les r\`egles de Feynman associ\'ees \`a ce mod\`ele. Les interfaces d'exportation impl\'ement\'ees dans {\fr} permettent ensuite de traduire ces vertexes dans un langage que les g\'en\'erateurs {\sc CalcHep, FeynArts, MadGraph 5, Sherpa, UFO} et {\sc Wizhard} peuvent comprendre. Ceci a pour int\'er\^et, bien \'evidemment, de simplifier la t\^ache d'impl\'ementation de chaque mod\`ele dans chaque g\'en\'erateur de spectre (les fichiers mod\`eles suivent des conventions diff\'erentes) et de minimiser le risque d'erreurs. Le gain en temps est inestimable.\\

Le programme {\fr} a, au cours des ann\'ees, beaucoup \'evolu\'e offrant de plus en plus de fonctionnalit\'es. Il est maintenant possible de d\'efinir les mod\`eles supersym\'etriques dans le formalisme du superespace; d'exporter dans le format {\sc UFO} les mod\`eles ce qui a pour avantage de contourner les contraintes et limitations habituellement impos\'ees par les autres formats de mod\`eles. On peut aussi citer le fait que {\fr} supporte maintenant les champs de Rarita-Schwinger dont le spin est \'egale \`a 3/2 et calcule aussi automatiquement et analytiquement les largeurs de d\'esint\'egration de toutes les particules. \\

Le premier module que j'ai d\'evelopp\'e dans {\fr}, intitul\'e {\sc InSuRGE} pour ``model-Independent Supersymmetry Renormalization Group Equations'' consiste en un ensemble de routines capables d'extraire \`a partir d'un mod\`ele impl\'ement\'e dans le formalisme des superchamps dans {\fr} les formules analytiques des \'equations du groupe de renormalisation \`a deux boucles et ce pour n'importe quel mod\`ele supersym\'etrique renormalisable. Ce module qui a fait l'objet d'une publication dans la r\'ef\'erence \cite{Brooijmans:2012yi} a \'et\'e test\'e pour le MSSM, le NMSSM et le LRSUSY et les r\'esultats sont en accord avec la lit\'erature.\\

Le second module au d\'eveloppement duquel j'ai particip\'e apporte \`a {\sc FeynRules} la capacit\'e \`a extraire automatiquement et analytiquement les matrices de masse associ\'ees \`a tout mod\`ele (supersym\'etrique ou non) impl\'ement\'e dans {\fr}. Les formules analytiques sont ensuite export\'ees sous la forme d'un code source {\sc C++} pour que le programme {\sc ASperGe}\cite{Alloul:2013fw} pour {\sc Automated Spectrum Generation} les diagonalise et extraie les valeurs propres et matrices de m\'elange. Les r\'esultats sont ensuite sauvegard\'es dans un fichier texte compatible avec le format SLHA.

Pour \^etre plus pr\'ecis, il est demand\'e \`a l'utilisateur de d\'eclarer dans le fichier mod\`ele {\fr} les m\'elanges entre les champs \'etats propres de jauge et les champs \'etats propre de masse qui en r\'esultent. Ainsi, le m\'elange des vecteurs de jauge B et W$^1$\footnote{l'exposant correspond \`a la premi\`ere composante du champ vecteur $W$. }, par exemple, associ\'es aux groupes $U(1)_Y \mbox{ et } SU(2)_L$ respectivement donnant lieu au photon et au boson vecteur $Z$ doit \^etre d\'eclar\'e comme
\begin{verbatim}
   Mix["weakmix"] == {
     MassBasis    -> {A, Z}, 
     GaugeBasis   -> {B, Wi[3]}, 
     MixingMatrix -> UG, 
     BlockName    -> WEAKMIX}, 
\end{verbatim}
o\`u le terme \verb?weakmix? sert \`a identifier le m\'elange en question; \verb?MassBasis? sert \`a identifier la base des \'etats propres de masse; \verb?GaugeBasis? sert \`a identifier les \'etats propres de jauge; \verb?MixingMatrix? sert \`a identifier le nom de la matrice de m\'elange et, enfin, \verb?BlockName? est utilis\'e par le code num\'erique {\sc ASperGe} pour sauvegarder les r\'esultats au format SLHA\footnote{Il existe plusieurs autres options qui ont \'et\'e d\'etaill\'ees dans la publication\cite{Alloul:2013fw}. }. Ce fichier est ensuite lu par {\fr} et si l'utilisateur demandait le calcul de la matrice de masse il aurait pour r\'esultat la matrice
\bea
\frac{1}{4}\begin{pmatrix}
	g'^2 v^2 & g' g v^2\\
	-g' g v^2 &g^2 v^2
\end{pmatrix}
\eea
o\`u $g'$ est la constante de couplage associ\'ee au groupe $U(1)_Y$, $g$ celle associ\'ee au groupe $SU(2)_L$ et $v$ la valeur moyenne dans le vide du boson de Higgs. Si l'interface vers le programme {\sc ASperGe} \'etait utilis\'ee, ce dernier retournerait une masse nulle pour le photon et une masse d'environ 91 GeV pour le boson $Z$.

\chapter{Introduction}
In high energy particle physics, one can safely state that all the known particles and interactions (but gravity) are well described by the Standard Model (SM) of particle physics\footnote{At the time of writing.}. Indeed, the latter whose theoretical construction relies on symmetry principles, has shown very robust through the decades as all the experiments that have been conducted until today (proton colliders, electron colliders, neutrino experiments, astroparticle experiments) have at most shown hints towards its extension. None of them however pointed clearly towards a real change of paradigm. The first statement poses however at least two deep theoretical issues.\\

If no extension to the Standard Model was found this would imply the question to know how to interpret the cosmological data indicating clearly that the ``known particles" only represent a tiny fraction of $5\%$ of the mass of the Universe.\\ 

The Standard Model of particle physics only accounts for the weak, the electromagnetic and the strong interactions and gravity is just ignored. At the energy scales that have been probed until now, neglecting the effects of the latter with respect to the other interactions can be regarded as a safe approximation in the realm of fundamental particles. The problem is that gravity is expected to become more and more important with increasing energy until the Plank
scale ($10^{19}$~GeV) where it is expected to induce effects at least as important as those induced by the other three fundamental forces. At such scales we do not know how to describe the interactions.\\

Another problem of the Standard Model resides in the mechanism of electroweak symmetry breaking. Following the Higgs-Brout-Englert mechanism \cite{Englert:1964et,Higgs:1964ia} all the Standard Model particles are supposed to acquire their mass through their interaction with a fundamental scalar, the Higgs boson. Recently, both {\sc ATLAS}\cite{:2012gk} and {\sc CMS}\cite{:2012gu} reported the discovery of a scalar field with a mass around 126 GeV exhibiting the same properties as the Higgs boson. Though this discovery will help us in better understanding the mechanism of electroweak symmetry breaking it also revives the problem of the hierarchy. Indeed if this new scalar field is the one predicted by the Standard Model, its mass is supposed to receive quadratically divergent corrections which induces thus a naturality problem.  \\

Some experimental results point also clearly towards extensions of the Standard Model. The observation of neutrino oscillations is maybe the clearest of the latter as it implies that neutrinos have a mass which is not taken into account in the SM. The measure of the anomalous magnetic moment of the muon has shown a deviation slightly higher than $3~\sigma$. This is not enough to claim a discovery but clearly something is happening there.\\

The accumulation over the years of all these hints (amongst others) has contributed to rise in the high energy physics community high expectations regarding the results the Large Hadron Collider (LHC) experiment could bring. On the theory side the latter have contributed to triggering an intense research activity where theorists and phenomenologists have worked together in order to imagine and build new theories and make predictions testable at the LHC. However the resistance of the SM to almost all the experiments conducted until now places stringent constraints on what is usually called ``Beyong the Standard Model (BSM) theories". Indeed, if the physics at the TeV scale remains to be discovered yet, the low energy physics is very well known due, for example, to electroweak precise measurements\cite{Drees:2001xw}. The new theories are thus constrained to include as a low energy effective theory, the Standard Model. Amongst the famous attempts to construct such theories one can for example cite\footnote{alphabetical order} the Grand Unified theories (GUTs)\cite{Raby:2006sk,Georgi:1974yf,Lucas:1996bc} where all the SM gauge interactions unfiy, extradimension theories\cite{PerezLorenzana:2004na,PerezLorenzana:2005iv,Hewett:2002hv} where space-time dimensionality is extended to $D>4$, string theory where particles turn out to be different excitation modes of a single string, supersymmetry which extends the space-time symmetries to link fields of different statistics $\dots$ and so on.\\

An early dream in theoretical particle physics is to achieve the unification of all gauge interactions, that is, being able to explain with a same theory the weak, the electromagnetic and the strong interactions. This dream proceeds by the search for the deep symmetries that govern our universe and at the same time the thought that the less number of free parameters we have, the more predictive is the theory. On the experimental side, the running of the gauge coupling constants, {\ie},
their evolution with energy, has been measured and has shown that they evolve in the same direction. A famous attempt to realize such unification was made by Georgi and Glashow in 1974\cite{Georgi:1974sy} where they achieved the embedding of the SM gauge group in $SU(5)$. By a correct symmetry breaking, the latter is then broken into the SM gauge group
$$ SU(5) \to SU(3)_c \times SU(2)_L \times U(1)_Y$$ 
respecting thus the low energy limit constraint. However, their model predicts a too fast proton decay and is thus ruled out in its original form. 

This first attempt can be thought of as minimal and other models have considered even larger groups such as $E_6$\cite{Gursey:1975ki} or $SO(10)$\cite{Chang:1984uy,Lucas:1996bc}. Larger groups implying more envolved symmetry breaking patterns, these theories reveal a non-minimal character. An example of interest to us, in the context of this manuscript, is to consider $SO(10)$ as being the unification group. The symmetry breaking pattern of this group into the SM group reveals then, that at a certain energy, two $SU(2)$ gauge groups appear\cite{Chang:1984uy}
$$ SO(10) \to SU(3)\times SU(2)\times SU(2)\times U(1) \times \mP \to \dots$$
where $\mP$ is parity and the dots stand for the subsequent symmetry breakings leading to the SM gauge group. These two $SU(2)$ gauge groups can then be interpreted as $SU(2)_L$ and $SU(2)_R$ leading thus a symmetry between left-handed and right-handed fermions.\\

Supersymmetry is certainly the most popular extension of the Standard Model. It is the only possible non-trivial extension of the Poincar\'e group (Haag-Lopuszanski-Sohinus and Coleman-Mandula theorems) linking bosonic and fermionic degrees of freedom. Amongst its attractive features, the most cited ones are its ability to cure the hirearchy problem in an elegant fashion and the fact that gauge coupling constants unify at very high scale. Its minimal realization in particle physics,
the Minimal Supersymmetric Standard Model (for a review, see for example \cite{Csaki:1996ks}) which is achieved by ``supersymmetrizing" the Standard Model of particle physics, is certainly one of the most studied models in particle physics. This fame is due to both the simplicity of the model (still that early dream of minimality) and the attractive phenomenological signatures it leads to such as its ability to predict a neutral stable particle weakly interacting with the other fundamental particles, {\ie} dark matter. The cost of this fame has however contributed to narrowing down significantly the parameter space of the model, especially in the case of the so-called constrained-MSSM. The last experimental constraint comes from the discovery of the Higgs-like boson which reduces significantly the available parameter space but most of the other experimental constraints drop when considering non minimal models.\\

Joining both motivations for potential Grand Unified Theories and supersymmetry, I have been envolved in a phenomenological study on the left-right symmetric supersymmetric models. These models, exhibiting a larger group than that of both the SM and the MSSM predict a plethora of new scalar and fermionic fundamental fields and lead hence to a very rich phenomenology. In a recent paper Mariana Frank, my promoters and I have published\cite{Alloul:2013xx}, we have investigated, in a \textit{top-down} approach, the phenomenology charginos and neutralinos of left-right symmetric supersymmetric particles would yield at the Large Hadron Collider. Focusing on final states where at least one charged light lepton appears we have shown that the signal from these models can be easily extracted from the Standard Model background by analyzing events where at least three charged leptons are produced and by imposing requirements on some kinematical variables.\\

In another analysis we have published in \cite{Alloul:2013zz}, we have adopted a rather complementary approach. Starting from the observation that left-right symmetric models and other high energy completions of the Standard Model can predict particles carrying a two-unit electric charge we have decided to investigate the signatures these exotic particles would leave at the LHC. Such an approach, often dubbed \textit{bottom-up} approach, allows then to take into account several different models in a simplified effective theory in order to lead a prospective analysis. In our paper, for example, in order to take into account the various models predicting such particles, we have allowed the latter to be either a scalar, a fermionic or a vector field transforming as a singlet, a doublet or a triplet under the weak gauge group. We have then shown that to every case one could associate a distinctive behaviour allowing thus for a proper discrimination between the models.\\

Finally, the predictions we have made in our analyses would not have been possible without the use of automated tools. Indeed, the combination of both the complexity of the calculations due to more and more sophisticated models and the necessity to be responsive in a field in constant progress constrain the phenomenologists to make use of automated tools. In this context, I have been involved in the development of two modules of {\sc FeynRules}\cite{Christensen:2008py,Christensen:2009jx,Duhr:2011se,Alloul:2013yy}, namely {\sc InSuRGE}\cite{Brooijmans:2012yi}\footnote{The acronym stands for {\sc model Independent Supersymmetric Renormalization Group Equations}.} and {\sc ASperGe}\cite{Alloul:2013fw}\footnote{The acronym stands for {\sc Automated Spectrum Generation}.}. The first one is an ensemble of routines able to extract automatically the renormalization group equations for any renormalizable supersymmetric theory at the two-loop level. The second one allows to derive automatically the mass matrices of any quantum field theory implemented in {\sc FeynRules} and to generate automatically a {\sc C++} source code able to diagonalize these mass matrices and return the mass spectrum of the theory.\\

In this manuscript, I describe the above mentionned projects in the following order. In chapter \ref{chap: SM}, I present a brief and succint review on the construction of the Standard Model and on some of its successes and shortcomes. The next chapter is then dedicated to the construction of supersymmetry and its minimal realization in particle physics, the Minimal Supersymmetric Standard Model. The aim of this chapter is twofold as it will serve to derive some results that will be useful for later use but also as an argumentation in favor of non-minimal supersymmetric models.\\

Chapter \ref{chap:lrsusy} will be dedicated to the study of the left-right symmetric supersymmetric model. In this chapter, after reviewing briefly the non-supersymmetric case, we will tackle the core subject and start building the model piece-by-piece. Once the Lagrangian obtained, we will perform an analysis of the signatures the charginos and neutralinos of this model would yield at the LHC. The conclusion of this chapter will be threefold as in addition to discussing the results, I will also use it to motivate the work presented in the two next chapters  \ref{chap: spec gen} and \ref{chap: doubly charged}. In the former, as a member of the {\sc FeynRules} collaboration, I briefly describe the {\sc FeynRules} package and then move to the description of the two modules {\sc InSuRGE} and {\sc ASperGe}. In chapter \ref{chap: doubly charged}, I present the \textit{bottom-up} analysis we have conducted on the doubly-charged particles. Finally, in chapter \ref{chap: conclusion}, are gathered concluding remarks and some open questions that might deserve some more interest.\\

\chapter{The Standard Model of particle physics}\label{chap: SM}

\unitlength = 1mm
\section{Construction and successes}
Standard Model of particle physics\cite{Glashow:1961tr,Weinberg:1967tq,Glashow:1970gm,Weinberg:1971nd,Gross:1973ju,Gross:1974cs,Politzer:1974fr,Marciano:1977Su} is one of the most tested and accurate theories in physics. Its construction relies on the use of symmetries to describe the interactions amongst particles. The symmetry group is a direct product between the Poincar\'e group $P$ and the gauge group $G = SU(3)_c \times SU(2)_L \times U(1)_Y$
\bea \label{eq:group} P \times G .\eea
The former describes the space-time symmetries (leading for example to spin) while the latter describes the internal symmetries responsible for the electroweak and strong interactions. Particles are classified following the representations to which they belong and form two big classes: fermions and bosons. The former which further split into leptons and quarks have half integer spin and describe the elementary constituents of matter while the latter which are
spin-1 vector gauge bosons and are the ones carrying the gauge interactions. The Higgs boson, which is a spin-0 particle, occupies a peculiar place in this picture. Indeed by minimizing the scalar potential, the Higgs scalar acquires a vacuum expectation value and electroweak symmetry gets broken to electromagnetism
$$ SU(2)_L \times U(1)_Y \longrightarrow U(1)_{e.m}. $$
The field content of the theory is given in table \ref{table:field sm} and before writing the Lagrangian, we introduce our notations.
\begin{itemize}
\item The gauge group $SU(3)_c \times SU(2)_L \times U(1)_Y$ will be associated to the gauge coupling constants $g_s, g \et g'$ respectively. The structure constants will be noted $f_{abc}$.
\item The field strength for a generic gauge field $F_\mu^a$ will be written
$$ F^a_{\mu\nu} = \partial_\mu F_\nu^a - \partial_\nu F_\mu^a + gf^{a}{}_{bc} F^b_{\mu} F^c_{\nu} .$$
Note that the last term is only present for non abelian gauge groups.
\item A generic covariant derivative is defined as follows
$$ D_\mu = \partial_\mu - i g_a T^a v_\mu $$
where $v_\mu$ is a vector field, $g$ a gauge coupling constant and $T^a$ the matrix generating the representation in which lies the field under consideration. For a field belonging to the fundamental representations of both $SU(3)_c \et SU(2)_L$ and has hypercharge $Y$, the covariant derivative becomes
$$ D_\mu =  \partial_\mu + i g_s T^a G_{a\mu} + i g \frac{\sigma^a}{2}W_{a\mu} + i g' Y B_\mu $$
where $T^a = \frac{\lambda^a}{2}$, $\lambda^a$ are Gell-Mann matrices generating the fundamental representation of $SU(3)_c$ and $\sigma^a$ are the Pauli matrices (see appendix \ref{annex:susy}). 
\end{itemize}
\small{\setlength{\extrarowheight}{3pt}
\begin{table}
\begin{center}
\begin{tabular}{| c | c c c|}
\hline
& Field & Spin & $SU(3)_c \times SU(2)_L \times U(1)_Y$ \\
\hline
\multirow{5}{*}{Matter} &  $L $ & $\frac12$ & ($\singlet,\utilde{\mathbf{2}}$,-$\frac12$) \\
& $\bar{l}_R $  & $\frac12$ & $(\singlet,\singlet,-1)$ \\
\cline{2-4}
& $Q_L$ & $\frac12$ & ($\utilde{\mathbf{3}},\utilde{\mathbf{2}},\frac16$) \\
& $\bar{u}_R$ & $\frac12$ & ($\utilde{\mathbf{3}},\singlet,\frac23$) \\
& $\bar{d}_R$ & $\frac12$ & ($\utilde{\mathbf{3}},\singlet,-\frac13$) \\
\hline
Higgs boson & $\phi$ & 0 & $(\singlet,\utilde{\mathbf{2}}$,$\frac12$) \\
\hline
\multirow{3}{*}{Vector Bosons} & $B^\mu$ & 1 & $(\singlet,\singlet,0)$ \\
& $W^\mu$ & 1 & ($\singlet, \utilde{\mathbf{3}}$,0) \\
& $G_a^\mu$ & 1 & ($\utilde{\mathbf{8}},\singlet, 0$)\\
\hline
\end{tabular}
\end{center}
\caption{\footnotesize Field content of the Standard Model of particle physics in terms of gauge eigenstates.\label{table:field sm}}
\end{table}
Finally, the Lagrangian of the Standard Model reads
\bea 
\label{lag: sm}
\lagr_{SM} &=& - \frac14 G_{\mu\nu}^a G_{a}^{\mu\nu} - \frac14 W^a_{\mu\nu}W_a^{\mu\nu} - \frac14 B_{\mu\nu}B^{\mu\nu} \n
&+& i \big( L^i \sigmabar^\mu D_\mu \bar{L}_{i} + l_R^i \sigma^\mu D_\mu \bar{l}_{R,i} \big) \n
&+& i \big( Q_{L}^i  \sigmabar^\mu D_\mu \bar{Q}_{L,i} +   u_{R}^i \sigma^\mu D_\mu\bar{u}_{R,i} +  d_{R}^i\sigma^\mu D_\mu  \bar{d}_{R,i}  \big)\n
&-& (y_e)^i{}_{j} \bar{L}_{Li} \cdot \phi ~l_R^{j} - (y_d)^i{}_{j} \bar{Q}_{Li} \cdot \phi~ d_{R}^{j}  - (y_u)^i{}_{j} Q^\dagger_{Li} \cdot \phi^\dagger~ u_{R}^{j} \n
&+& D_\mu\phi^\dagger D^\mu\phi - \mu \phi^\dagger\cdot \phi - \lambda (\phi^\dagger \cdot \phi)^2.
\eea
where $a$ refers to gauge indices, $i,j$ to generation indices and the dot stands for an $SU(2)$ invariant product. Here, we have used Weyl fermions that is two-component fermions. In this notation, $\bar{u}_R$ is a right-handed fermion and
$$ (\bar{u}_R)^\dagger = u_R $$
is a left-handed fermion. To avoid any confusion about the chirality of the field we use the ``bar" sign to distinguish chiralities, its presence indicating a right-handed chirality.  This notation is introduced here because of its utility in supersymmetry, however when moving to phenomenology using Dirac fermions turns out to be more practical. In the latter notation, instead of 
$$ u_{R}^i  \sigmabar^\mu D_\mu \bar{u}_{R,i} $$
we would have written
$$ \bar{\mathcal{U}}^i_R \gamma^\mu D_\mu \mathcal{U}_{R,i} $$
where $\gamma^\mu$ are the Dirac matrices defined in function of the Pauli $\sigma$ matrices:
$$ \gamma^\mu = \begin{pmatrix} 0 & \sigma^\mu \\ \sigmabar^\mu & 0 \end{pmatrix}; $$
we define $\mathcal{U}_R$ as being the Dirac fermion whose right-handed component is $\bar{u}_R$ while its left-handed component is zero
$$ \mathcal{U}_R = \begin{pmatrix} 0 \\ \bar{u}_R \end{pmatrix}. $$
Finally, the confusion about the meaning of the conjugation is lifted by the number of components of the field on which it acts.
\paragraph{The bosonic sector} As stated above, by minimizing the scalar potential of the Standard Model one finds that the Higgs scalar acquires a vacuum expectation value and the electroweak symmetry is broken. We write thus:
$$ \langle \phi \rangle = \langle \begin{pmatrix} \phi^0 \\ \phi^- \end{pmatrix} \rangle = \frac{1}{\sqrt2}\begin{pmatrix} v + h + i A^0 \\ 0 \end{pmatrix}. $$
where $h$ (resp. $A^0$) is a $\mC\mP$ even (resp. odd) real scalar. After diagonalizing the mass matrices, we find that the gauge bosons $W^\pm$ and $Z$ are the admixtures of the gauge eigenstates $B_\mu \et W_\mu$
$$ W^{\pm}_\mu = \frac{1}{\sqrt2}(W_\mu^1 \mp i W_\mu^2),~~~~ Z_\mu = \cos\theta_w W_\mu^3 - \sin\theta_w B_\mu. $$
Their masses are given by
$$ M_Z = \frac12 \sqrt{g^2 + g'^2} v, ~~~~ M_{W^\pm} = \cos \theta_w M_Z $$
with the electroweak mixing angle (or Weinberg angle) $\theta_w$ defined as
$$ \tan\theta_w = \frac{g'}{g}. $$
The photon $A_\mu$ rises also from the admixture of the gauge bosons eigenstates 
$$ A_\mu = \cos\theta_w B_\mu + \sin\theta_w W^3_\mu$$
and is obviously massless. This angle and the masses of the gauge bosons having been measured (see ref.\cite{Beringer:1900zz} and figure \ref{lep results}) at different experiments allow us to give an approximate value for the vev of the Higgs field of 246 GeV. \\
\begin{figure}[!t]
\hskip-1truecm
\subfigure[]{%
\label{fig:mw lep}   \includegraphics[scale=0.4]{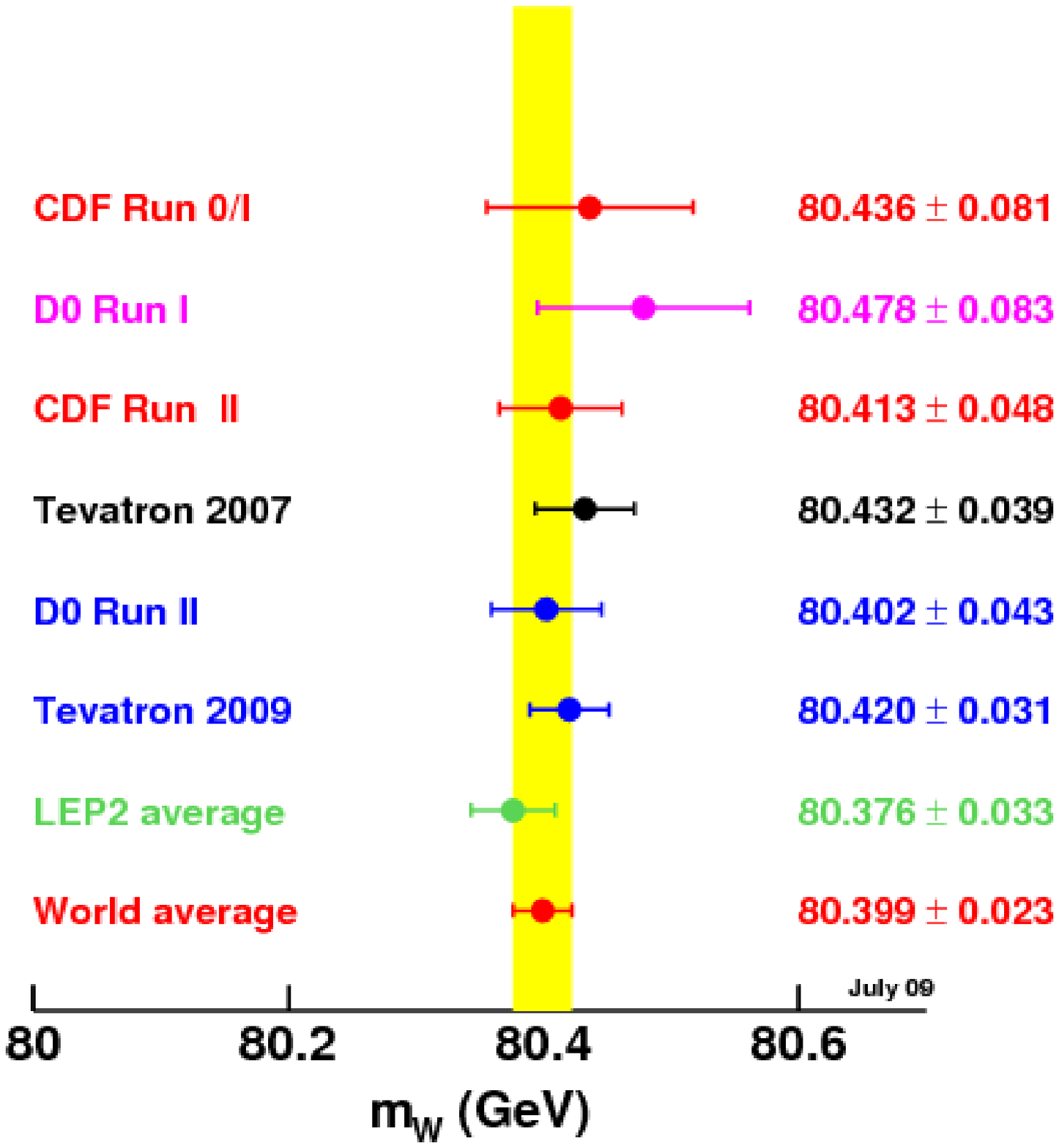} }%
\subfigure[]{%
\label{fig:mz lep}   \includegraphics[scale=0.4]{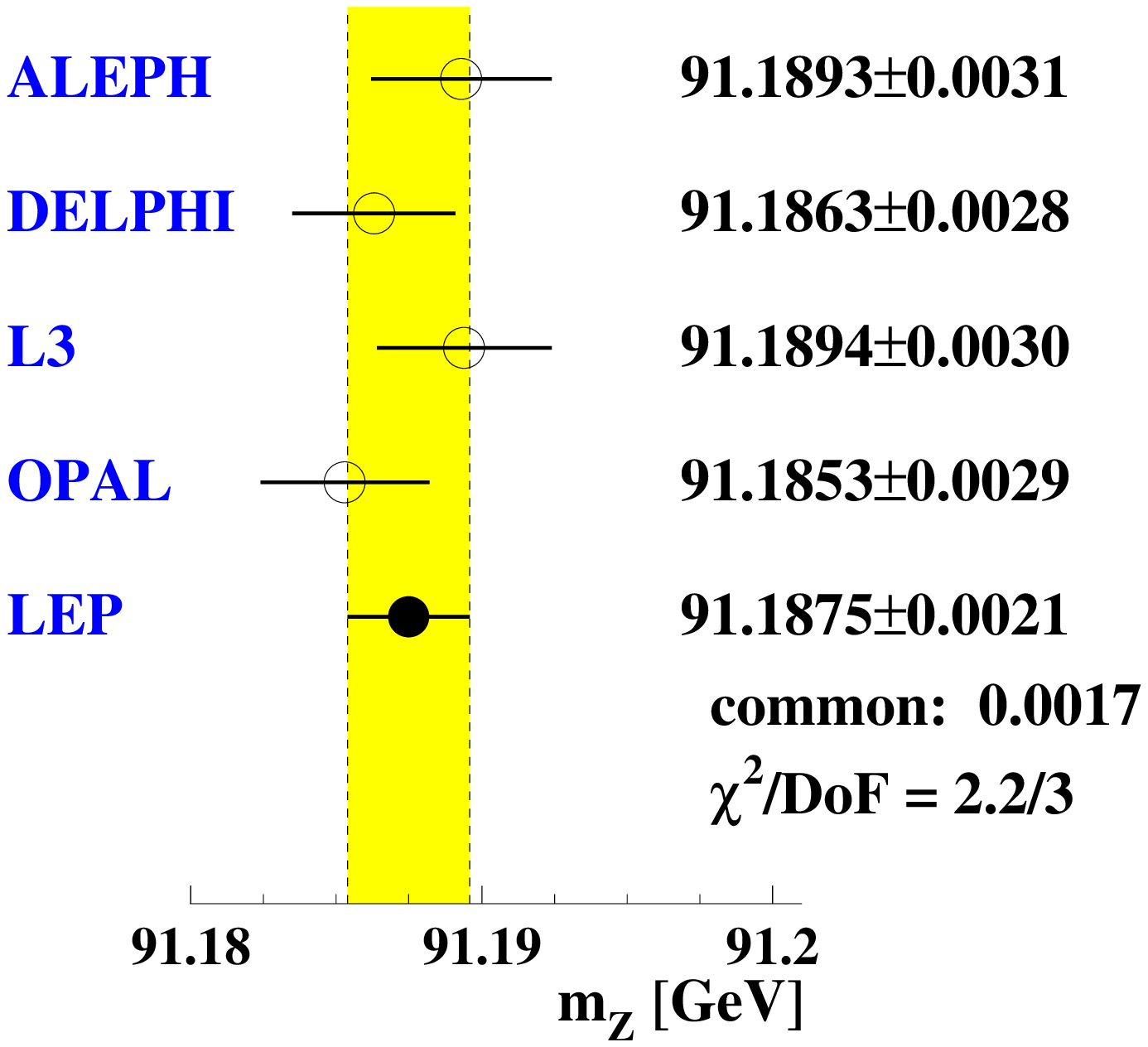}}\\ 

\caption{\footnotesize\label{lep results} Latest LEP results for the W and Z bosons masses compared to SM predictions (in yellow). The plot for the W mass (left panel) was taken from ref.\protect\cite{Melnitchouk:2012qy} and the one for the Z boson mass (right panel) was taken from ref.\protect\cite{Z-Pole}.  }
\end{figure}
\paragraph{Fermion sector} In the quark sector, when the Higgs field acquires a vacuum expectation value, the Yukawa couplings
$$- (y_d)^i{}_{j} \bar{Q}_{Li} \cdot \phi~ d_{R}^{j}  - (y_u)^i{}_{j} Q^\dagger_{Li} \cdot \phi^\dagger~ u_{R}^{j} + {\rm h.c.} $$
induce a mass term. To get the physical states one diagonalizes the matrices $y_d \et y_u$ by four unitary rotations $V^{u,d}_{L,R}$ such that:
$$ M^f = V_L^f y^f V_R^f (\frac{v}{\sqrt2}) $$ 
where $f= u,d$. The Cabibbo Kobayashi Maskawa ($CKM$) matrix is a $3\times 3$ unitary matrix and can be parametrized by three mixing angles and a $\mC\mP$-violating phase.

Under the light of the results from neutrino experiments, it is nowadays acknowledged that neutrinos do oscillate and consequently they have a tiny mass and do mix together. This implies that the field content of the Standard Model has to be extended to include right-handed neutrinos as gauge singlets so that one can write mass terms for these particles. The mixing in the neutrino sector is then described by the Pontecorvo-Maki-Nakagawa-Sakata ($PMNS$) $V^{PMNS}$ matrix\footnote{for a review, the reader can have a look at the paper \cite{Mohapatra:2006gs}}. By convention, down-type (\textit{i.e.} charged) leptons are considered mass eigenstates and only the neutrinos have to be rotated. It is noteworthy that the elements of these two mixing matrices, the $PMNS$ and the $CKM$, are not predicted by the SM and are thus free parameters to be measured (see \cite{Beringer:1900zz}).\\

The Standard Model of particle physics proved very successful and many of its predictions revealed true. For example, the charm\cite{Glashow:1970gm}, bottom \cite{Kobayashi:1973fv} and top quarks\cite{Kobayashi:1973fv} and also the gluons\cite{Fritzsch:1972jv,Fritzsch:1973pi,Politzer:1973fx} were predicted theoretically before their experimental discovery. One can also cite the W and Z gauge bosons masses which are in very good agreement with experimental measurements (see. fig.\ref{lep results}) but also the Higgs boson discovery which is at the moment of writing still compatible with Standard Model's predictions\cite{:2012gu,:2012gk}.

\section{Towards extensions of the Standard Model}
Though very successfull in describing the fundamental interactions at low energy ($\lesssim 140$ GeV), there exists strong theoretical motivations as well as some tension between predictions and experimental observations pushing us to consider the Standard Model of particle physics as an effective theory of a more fundamental one. Below is a small list of the known problems and open questions\footnote{For a review of the Standard Model's flaws see ref\cite{Murayama:2007ek}.}.\\

\paragraph{Gravity, unification and all that} Though successful in unifying electromagnetic and weak interactions in the so-called electroweak force, the SM does not provide a unified framework in which all the strong, weak and electromagnetic interactions unify. Besides, renormalization group equations which govern the evolution with respect to energy of the gauge coupling constants show clearly their non-unification (see fig.\ref{fig:sm rge}) which may be seen as a problem in this theory. On the other hand and more importantly, gravity, the weakest of all known interactions, is totally ignored in this framework.
\begin{figure}
\begin{center}
\includegraphics[scale=0.35]{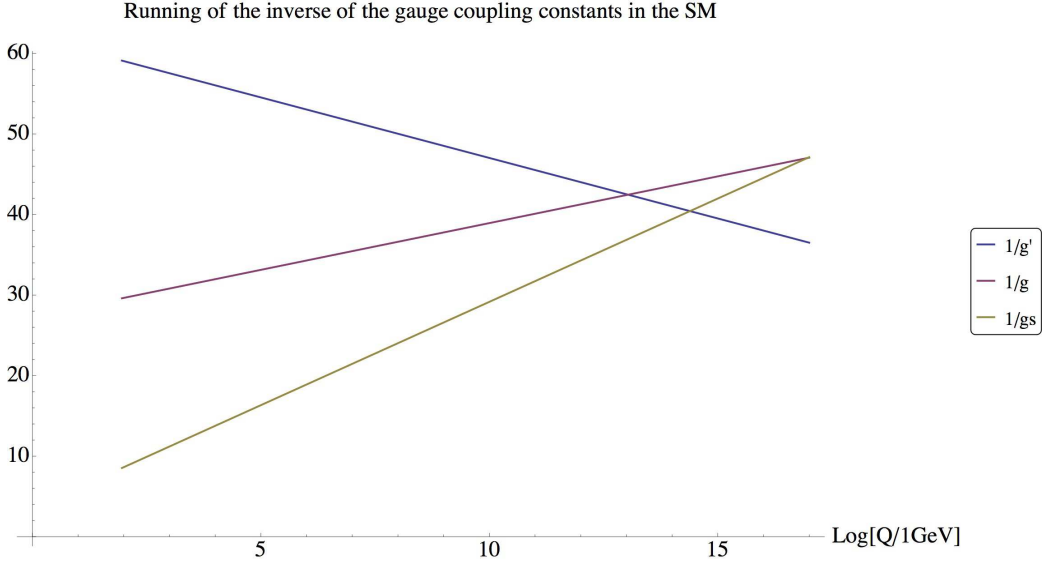}
\caption{{\footnotesize\label{fig:sm rge} Running of the gauge coupling constants of the Standard Model at the one-loop level. Here $Q$ is the energy in GeV, and the running starts at $M_Z = 91.18$ GeV.}}
\end{center}
\end{figure}

\paragraph{Families} In the zoology of elementary particles, leptons and quarks are known to regroup in three\footnote{Actually, a fourth generation of quarks and leptons is not ruled out by experiment but this goes beyond the scope of this manuscript.} different families. The Standard Model does not answer to the question why three.
\paragraph{Neutrino masses} The various neutrino experiments (solar, atmospheric, reactor and accelerator) have provided strong evidences in favor of oscillating neutrinos. This fact implies that these peculiar particles are massive but with very tiny masses which is in contradiction with the Standard Model in which they are considered massless. Moreover, even though one wanted to introduce by hand a mass term for these particles, one would have to add the right-handed neutrinos to the Standard Model field content. This issue requires, at the minimum, an extension of the Standard Model which was built at a time where neutrinos were thought to be massless. 
\paragraph{Dark matter and dark energy} It is nowadays recognized that the ordinary matter (baryonic) only represents a tiny fraction ($\sim 4 - 5\%$) of the mass of the Universe; the remaining $95\%$ is called generically ``dark matter" and ``dark energy". The former which represents around $25\%$ of the mass of the Universe  has the property of interacting very weakly\footnote{in the case where dark matter is a Weakly Interacting Massive Particle (WIMP). Other candidates do exist, for a review one can have a look at ref.\cite{Drees:2012ji}.} with the baryonic matter and is hence very difficult to detect. The only possible candidate in the Standard Model to explain this\footnote{One should keep in mind that dark matter could be made of different particles} mysterious particle is the neutrino but its properties do not satisfy all the cosmological constraints (\textit{e.g.}, relic density). Dark energy concentrates the most significant amount of mass as it represents $70\%$ of the total mass of the Universe. Neither is it explained.

\paragraph{The muon anomalous magnetic moment} Though the Standard Model succeeded in passing most of the experimental tests, there is at least one measurement where the tension between theory and experiment is strong. The anomalous magnetic moment of the muon measured at the Brookhaven National Laboratory (BNL) was shown to be \cite{Teubner:2012qb}
$$\frac{(g-2)_\mu}{2}_{BNL} = (11659208.9 \pm 6.3 )\cdot 10^{-10} $$
which represents a discrepancy of $3.3 \sigma$ with respect to the predicted value
$$\frac{(g-2)_\mu}{2}_{theo} = (11659182.8.9 \pm 4.9 )\cdot 10^{-10}. $$
Of course no conclusion can be drawn from this result and complementary measurements have to be run however one might consider this fact as a hint toward new physics.

\begin{figure}[!t]
\begin{center}
\begin{fmffile}{test1}
\begin{fmfgraph*}(40,25)
\fmfleft{i}\fmflabel{$H$}{i}
\fmfright{o}\fmflabel{$H$}{o}
\fmf{dashes}{i,v1}
\fmf{dashes}{v2,o}
\fmf{fermion,left,label=$f$,tension=0.4}{v1,v2,v1}
\fmfdot{v1,v2}
\fmfv{lab=$y_f$,lab.side=left,l.a=60}{v2}
\fmfv{lab=$y_f$,lab.side=right,l.a=120}{v1}
\end{fmfgraph*}
\hskip2truecm
\begin{fmfgraph*}(40,25)
\fmfleft{i1}\fmflabel{$H$}{i1}
\fmfright{o1}\fmflabel{$H$}{o1}
\fmf{dashes}{i1,v3}
\fmf{dashes}{o1,v3}
\fmfdot{v3,v3}
\fmf{dashes,label=$S$}{v3,v3}
\end{fmfgraph*}
\end{fmffile}
\end{center}
\caption{\footnotesize\label{higgs corrections} One-loop radiative corrections to the Higgs squared mass du to a Dirac fermion $f$ and a scalar $S$.}
\end{figure}

\paragraph{Hierarchy problem} Another naturality problem and probably the most famous of all Standard Model's flaws is the so-called hierarchy problem. In a few words one can summarize it as the question to know why the $\mu$ parameter (see eq\eqref{lag: sm}) which has a dimension of mass in the Standard Model's Lagrangian and which value is (at tree-level)
$$ |\mu| =  \frac{M_H}{\sqrt2} \simeq 90 ~\text{GeV, (with }M_H = 125 \text{ GeV)} $$
 has to be so small compared to the Planck scale ($10^{19}$ GeV). This question is related to the fact that in the framework of the Standard Model there is absolutely no symmetry protecting scalar masses from diverging. \\
In practice, one can, for example, compute the radiative corrections to the Higgs mass parameter $\mu$. This term receives contributions from the quartic Higgs couplings $\lambda$ as well as from the Yukawa couplings Higgs-fermion-fermion that we denote generically $y_f$ (see fig.\ref{higgs corrections}). These processes induce a correction of order
\bea
(\lambda - y_f^2) \Lambda^2 \eea
where $\Lambda$ is a certain cut-off which value is of crucial importance as the correction depends on its square. If one considers the Standard Model valid up to the Planck scale than $\Lambda$ should be of order $10^{19}$ GeV! This implies that, to have a value of $\mu$ compatible with the experimental constraints, one needs either $\lambda = y_f^2$ or a huge amount of fine-tuning. The first possibility establishing the equality between quartic boson couplings and the square of a boson-fermion coupling should be the consequence of a symmetry to be well motivated; we will see that supersymmetry does exactly the job (see next chapter). The second possibility is exactly the hierarchy problem also dubbed as fine-tuning problem.

\chapter{Introduction to supersymmetry}\label{chap:susy}

The Standard Model of particle physics allowed to unify in a simple framework electromagnetic and weak interactions and is actually one of the most tested theories in physics. Though very successful, its theoretical issues coupled with some experimental results are powerful indicators that this model has to be considered as an effective theory of a more fundamental one. Actually, theorists have been trying to build a realistic quantum field theory able to overcome these issues since early '70s. String theory, technicolour, extra-dimensional space-time are some of the alternatives that have been considered\cite{Lykken:2005up,delAguila:2004sj,Mohapatra:1990jc}. Supersymmetry is also one of those beyond the Standard Model theories that have been studied. Its properties allowing to solve (amongst others) in an elegant way the hierarchy problem have made of it one of the most famous and studied theories in the last decades.\\

Indeed, though a number of no-go theorems concluding that direct symmetries between different spin particles are in conflict with quantum field theory\cite{Coleman:1967ad}, the introduction of supersymmetry allows to achieve this purpose through the use of Lie super-algebra instead of Lie algebra. As a consequence to this symmetry between fermions and bosons, scalar and fermionic divergent loops do cancel each other as they involve the same couplings but with a minus sign for fermionic loops. The hierarchy problem is thus solved. As another theoretical motivation for supersymmetry one could also cite the fact that its locally invariant version naturally encompasses gravity. Finally, from a phenomenology point of view, the minimal realization of supersymmetry in particle physics, namely the Minimal Supersymmetric Standard Model (MSSM), though exhibiting only a global supersymmetry provides a natural candidate for dark matter and predicts the unification of the gauge coupling constants.\\

From the experimental side, though not excluded, supersymmetry has not been discovered yet, that is, not a single scalar particle carrying the same quantum numbers than those of one of the known fermions has been observed yet. This implies, at least, that supersymmetry is a broken symmetry and the superpartners of the Standard Model particles are just too heavy for the energy range that has been probed at collider expriments. \\
 
In this chapter we introduce all the ingredients to build a supersymmetric theory. In this scope, We present in the first section the supersymmetric algebra which will allow us to build our first renormalizable Lagrangian in the second section. After introducing the superspace formalism where all supersymmetric quantities express naturally, we will build the most general renormalizable supersymmetric Lagrangian. In the following two sections, we present some of the supersymmetry breaking mechanisms as well as the renormalization group equations. Finally, in section \ref{sec:mssm}, we present the minimal realization of supersymmetry, the Minimal Supersymmetric Standard Model (MSSM), and argue in favour of the study of non-minimal models.\\

This chapter is aimed to present the main results in supersymmetry therefore calculations will not be carried out explicitely. However, a list of references is given when necessary and I used as guidelines the following two references \cite{MichBenj,Sohnius:1985qm} as well as the conventions listed in the appendix \ref{annex:susy}.

\section{Supersymmetric algebra}
\paragraph{Special conventions for this section}
In mathematics, usually,  conventions used for writing Lie algebra (that is commutation relations) do not involve an explicit $i$ whereas, in physics, they do. For example, let $\tilde{A}_{i|i=1\dots n}$ be three operators obeying the Lie algebra and $f_{ab}{}^c$ the structure constant of this algebra. The commutation relation reads
\bea
\text{In mathematics} &&[\tilde{A}_a, \tilde{A}_b] = f_{ab}{}^c \tilde{A}_c , \n
\text{In physics} &&[\tilde{A}^*_a, \tilde{A}^*_b] = i f_{ab}{}^c \tilde{A}^*_c . \nonumber \eea
Thus if we have a unitary representation of the Poincar\'e group acting on a Hilbert space, operators $\tilde{A}$ will be hermitian while $\tilde{A}^*$ will be anti-hermitian. In this section, we will use mathematician's conventions to show the results but to recover the physicist's notations it will be enough to apply the replacements:
\bea M^{\mu\nu} &\rightarrow& iM^{\mu\nu} \n
P^\mu &\rightarrow& iP^\mu. \eea

Let us consider now a collection of fermionic $\Psi^i$ and bosonic fields $\Phi^a$ whose dynamics are described by the Lagrangian $\lagr$. We consider two symmetries generated by $B_A^1$ and $B_A^2$ on the one hand, and by $F_I^1$ and $F_I^2$ on the other hand. The two first generators act on fields without changing their nature
$$ \delta_A \Phi^a = (B_A^1)^a{}_b \Phi^b ~~~,~~~ \delta_A \Psi^i = (B_A^2)^i{}_i \Psi^j $$
and are called bosonic generators. The two other generators $F_I^1 \mbox{ and } F_I^2$ act on the fields by shifting their spin by 1/2, that is by changing their nature:
$$ \delta_I \Phi^a = (F_I^1)^a{}_i\Psi^i~~~,~~~\delta_I \Psi^i = (F_I^2)^i{}_a\Phi^a $$
and are by consequence called fermionic.\\
These conserved symmetries induce two conserved charges $\tilde{B}_A$ and $\tilde{F}_I$ whose actions on the fields are:
$$ [\Phi^a, \tilde{B}_A] = \delta_A \Phi^a ,~~~~ [\Psi^i , \tilde{B_A}] = \delta_A \Psi^i, $$
$$ [\Phi^a, \tilde{F_I}] = \delta_I \Phi^a ,~~~~ \{\Psi^i , \tilde{F_I}\} = \delta_I \Psi^i, $$
where the $\delta$ were defined previously and $\{ ,\}$ is the anti-commutator defined as
$$ \{\Psi^i , \tilde{F_I}\} = \Psi^i \tilde{F_I} +\tilde{F_I} \Psi^i.  $$
We can show that the bosonic and the fermionic charges $\tilde{B}_A \et \tilde{F}_I$, obey the algebra
$$ [ \tilde{B}_A , \tilde{B}_B] = f_{AB}{}^C \tilde{B}_C ,~~~ \{\tilde{F}_I, \tilde{F}_J\} = Q_{IJ}{}^C \tilde{B}_C, ~~~ [\tilde{B}_A , \tilde{F}_I] = R_{AI}{}^J \tilde{F}_J$$
where $f_{AB}{}^C,Q_{IJ}{}^C$ and $R_{AI}{}^J$ are some constants. This is exactly the so called Lie superalgebra. The latter is defined as a $\mathbb{Z}_2-$graded vector space $\mathfrak{g} = \mathfrak{g}_0 \oplus \mathfrak{g}_1$ where $\mathfrak{g}_0$ operators are said bosonic and $\mathfrak{g}_1$'s operators fermionic.\\
Now, if we consider $\{B_i, i=1,\dots,n\}$ ($n$ being the dimension of $\mathfrak{g}_0$) a basis of $\mathfrak{g}_0$ and $\{F_i, i=1,\dots,m\}$ ($m$ being the dimension of $\mathfrak{g}_1$) a basis of $\mathfrak{g}_1$ , the algebra $\mathfrak{g}$ reads
\begin{itemize}
\item $\mathfrak{g}_0$ is a Lie algebra: $[\tilde{B}_i,\tilde{B}_j] = f_{ij}{}^k \tilde{B}_k$.
\item $\mathfrak{g}_1$ is a representation of $\mathfrak{g}_0$, \ie, $[\tilde{B}_i, \tilde{F}_a] = R_{ia}{}^b \tilde{F}_b$. It is thus a non trivial extension of $\mathcal{g}_0$. 
\item Composing two fermionic transformations leads a bosonic one $\{\tilde{F}_a, \tilde{F}_b\} = Q_{ab}{}^c \tilde{B}_c$
\item The Jacobi relations are also satisfied
\bea
\big[\tilde{B}_i, [\tilde{B}_j, \tilde{B}_k]\big] + \big[\tilde{B}_j,[\tilde{B}_k,\tilde{B}_i]\big] + \big[\tilde{B}_k, [\tilde{B}_i,\tilde{B}_j]\big] &=& 0 ,\n
\big[\tilde{B}_i, [\tilde{B}_j, \tilde{F}_a]\big] + \big[\tilde{B}_j,[\tilde{F}_a,\tilde{B}_i]\big] + \big[\tilde{F}_a, [\tilde{B}_i,\tilde{B}_j]\big] &=& 0 ,\n
\big[\tilde{B}_i, \{\tilde{F}_a, \tilde{F}_b\}\big] - \{[\tilde{B}_i,\tilde{F}_a],\tilde{F}_b\} - \{\tilde{F}_a, [\tilde{B}_i,\tilde{F}_b]\} &=& 0 ,\n
\big[\tilde{F}_a, \{\tilde{F}_b, \tilde{F}_c\}\big] - \big[\{\tilde{F}_a,\tilde{F}_b\},\tilde{F}_c\big] + \big[\tilde{F}_b, \{\tilde{F}_a,\tilde{F}_c\}\big] &=& 0 .\n
\eea
\end{itemize}

Supersymmetry is a non trivial extention of the Poincar\'e algebra. It is generated by the generators of the Poincar\'e algebra
$$\mathfrak{g}_0 = \{M_{\mu \nu},P_\mu , ~\mu,\nu = 0,\dots,3\}$$
for the bosonic part and by
$$\mathfrak{g}_1 = (\frac12,0) \oplus (0,\frac12) = \{Q_\alpha, \alpha=1,2\} \oplus \{\bar{Q}^{\dot{\alpha}}, \dot{\alpha} = 1,2\},$$
where $(Q_\alpha, \bar{Q}^{\dot{\alpha}})$ is a Majorana spinor
$$ (Q_\alpha)^\dagger = \bar{Q}_{\dot{\alpha}}.$$
The algebra is described by the \mbox{(anti-)commutators} 
\bea \label{susy algebra}
\big[M_{\mu \nu},M_{\rho \sigma}\big] &=& \eta_{\nu \sigma}M_{\rho \mu} - \eta_{\nu \sigma}M_{\rho \nu} + \eta_{\nu\rho} M_{\mu \sigma} - \eta_{\mu\rho} M_{\nu\sigma}, \big[M_{\mu\nu},P_{\rho}\big] = \eta_{\nu\rho} P_\mu - \eta_{\mu\rho}P_\nu\n
\big[M_{\mu\nu},\bar{Q}^{\alphadot}\big] &=& \bar{\sigma}_{\mu\nu}^{\alphadot}{}_{\dot{\beta}}\bar{Q}^{\dot{\beta}},~~~~~ \{Q_\alpha, \bar{Q}_{\dot{\alpha}}\} = -2i\sigma^\mu_{\alpha \dot{\alpha}}P_\mu, ~~~~~\big[M_{\mu \nu},Q_\alpha \big] = \sigma_{\mu \nu \alpha}{}^\beta Q_\beta, \n
\{Q_\alpha, Q_\beta\} &=&  \{\Qbar_{\alphadot}, \Qbar_{\dot{\beta}}\} = \big[P_\mu,P_\nu \big] = \big[P_\mu , Q_\alpha\big] = 0.
\eea

We clearly see that the combination of two supersymmetric transformations ($Q$) induces a translation in space-time which is a bosonic operator.\\
Some remarks are in order here. These transformations induce naturally that a supersymmetric multiplet must contain both fermions and bosons. Moreover, the operator $P^\mu P_\mu$ commutes with all the generators of the algebra and is thus a Casimir operator. This ensures that the members of a same multiplet will carry the same mass. A natural question arises as a consequence of these two assertions: how many fermionic (bosonic, respectively) degrees of freedom are there in a supersymmetric multiplet? Let us assume that we have $N$ supersymmetry generators $Q^I_\alpha~,~\bar{Q}_I^{\dot{\alpha}}$ with $ (Q^I_\alpha)^\dagger=\bar{Q}_{I \dot{\alpha}}$ and
\bea \label{central charges} \{Q^I_\alpha, \bar{Q}_{J \dot{\alpha}} \} = -2i\delta^I_J \sigma^\mu_{\alpha \dot{\alpha}} P_\mu,~~~ \{Q^I_\alpha , Q^J_\beta \} = 0,~~~ \{\bar{Q}_{I \dot{\alpha}} , \bar{Q}_{J \dot{\beta}} \} = 0. \eea
Let us also introduce the operator ``fermionic number" $(-1)^N$ whose eigenvalues are $-1$ when acting on a fermion and $+1$ when acting on a boson. We have
\bea {\rm Tr}[(-1)^N \{Q^I_\alpha, \bar{Q}_{J\dot{\alpha}} \}] = -2 i \delta^I{}_J \sigma^\mu_{\alpha \alphadot}{\rm Tr}[(-1)^N P_\mu] \nonumber \eea
which, by virtue of the properties of the trace and the relation 
$$(-1)^N Q^I_\alpha = -Q^I_\alpha (-1)^N $$
induces
$$ -2 i \delta^I{}_J \sigma^\mu_{\alpha \alphadot}{\rm Tr}[(-1)^N P_\mu] = 0. $$
This relation must hold for any value of the momentum so that the only solution left is 
$${\rm Tr}[(-1^N)] = 0.$$
In plain words, this means that we have as many fermions as bosons. This is known as the \textit{``fermion=boson"} rule. The number of fermionic and bosonic degrees of freedom depends on the number of supersymmetry generators we allow. However, to build realistic field theories or at least renormalizable ones, one must constrain the highest spin in a supersymmetric multiplet. For renormalizable theories, the highest spin in the multiplet must be smaller or equal to one. In this case the number of supersymmetry generators cannot be larger than four. The latter case, $N=4$, is very interesting as it has been proven to be renormalizable and finite and is know as super Yang-Mills theory \cite{Sohnius:1985qm,n=4-1,*n=4-2,*n=4-3,*n=4-4}. The multiplet in this case contains four bosonic and four fermionic degrees of freedom. If we chose to ignore the renormalizability condition than we can include gravity as it is carried by a spin-2 particle. In this case the number of generators cannot be larger than eight in which case the multiplet contains $128+128$ degrees of freedom. As for $N > 8$ cases, multiplets will contain fields with spin $\geq \frac52$ which cannot be consistently coupled to gravity.\\

Before concluding this section, I would like to state that equation \eqref{central charges} is not the most general. For facility, we have made the assumption that there is no central charge in the theory that is we did not consider the case
$$ \{Q^I_\alpha , Q^J_\beta \} = \epsilon_{\alpha\beta} Z^{IJ} $$
where $Z^{IJ}$ is the central charge. In the sequel we will only focus on $N=1$ supersymmetric theories as they are the most interesting phenomenologically and restrain ourselves to the renormalizable ones.

\section{Building a supersymmetric renormalizable Lagrangian}\label{sec:susy}

Now that the supersymmetric algebra is established, we can proceed to the building of a Lagrangian for supersymmetric theories. \\

We shall start the discussion by describing the very first linear realization of supersymmetry introduced by Wess and Zumino in 1973 \cite{Wess:1974tw}. We will argue afterwards that in the superspace formalism supersymmetric quantities express naturally. In this formalism, we shall deduce the final form of the most general and renormalizable supersymmetric Lagrangian. \\

\subsection{The Wess-Zumino Lagrangian} \label{sec: wess zumino}
\paragraph{Matter free Lagrangian}
Consider a multiplet consisting of two complex scalar fields $\phi$ and $F$ and a left-handed Weyl spinor $\psi$. Let an infinitesimal supersymmetric transformation be:
\bea \label{chiral transfo} \delta_\epsilon \phi = \sqrt2 \epsilon \cdot \psi ~~~~~,~~~~~ \delta_\epsilon \psi = -i \sqrt2 \sigma^\mu \bar{\epsilon} \partial_\mu \phi -\sqrt2 F\epsilon ~~~~~,~~~~~ \delta_\epsilon F = -i \sqrt2 \partial_\mu\psi \sigma^\mu \bar{\epsilon}. \eea 
where $\epsilon$ is a left handed spinor (Grassman variable). Combining two supersymmetric transformations leads to:\\
\bea [\delta_{\epsilon 1}, \delta_{\epsilon 2}] \phi &=& \delta_{\epsilon 1} (\delta_{\epsilon 2} \phi) - \delta_{\epsilon 2} (\delta_{\epsilon 1}\phi), \n
&=& -2 i [\epsilon_2 \sigma^\mu \bar{\epsilon}_1 - \epsilon_1 \sigma^\mu \bar{\epsilon}_2] \delm \phi. \n
\big[\delta_{\epsilon 1}, \delta_{\epsilon 2}\big] \psi_\alpha &=& \delta_{\epsilon 1} (\delta_{\epsilon 2} \psi_\alpha) - \delta_{\epsilon 2} (\delta_{\epsilon 1}\psi_\alpha), \n
&=&  2i [ \epsilon_1 \sigma^\mu \epsilonbar_2  - \epsilon_2 \sigma^\mu \epsilonbar_1] \delm \psi . \n
\big[\delta_{\epsilon 1}, \delta_{\epsilon 2}\big] F &=& \delta_{\epsilon 1} (\delta_{\epsilon 2} F) - \delta_{\epsilon 2} (\delta_{\epsilon 1}F), \n
&=& -2i (\epsilon_2 \sigma^\mu \epsilonbar_1 - \epsilon_1 \sigma^\mu \epsilonbar_2) \delm F. \eea

A few remarks are in order here. First, to carry out properly the calculations one must use the properties of the Pauli matrices and the formulas given in appendix \ref{annex:susy}. Secondly, we see that for both scalars $\phi$ and $F$ the combination of two supersymmetric transformations induces a space-time translation. The transformation laws given above preserve thus supersymmetry and we are now tempted to build our first Lagrangian:
\bea \label{chiral free lag} \lag_0 = \delm \phi^\dagger \delmm \phi + i \bar{\psi}\bar{\sigma}^\mu \delm \psi + F^\dagger F, \notag \eea
which is essentially the sum of the free Lagrangians for the fields $\phi$ and $\psi$. The variational equation for the field $F$ is simply $F=0$ it is nevertheless essential for the transformation properties as it helps in restoring the equality between the number of scalar and fermionic degrees of freedom when off-shell as can be seen from the count of the degrees of freedom in the chiral multiplet
\begin{center}
\begin{tabular}{c c c c}
Field & $\phi$ & $\psi$ & $F$ \\
Off-shell & 2 & 4 & 2 \\
On-shell & 2 & 2 & 0 \\
\end{tabular}
\end{center}
Finally, using the transformation laws given in \eqref{chiral transfo}, we can show that the variation of the Lagrangian:
\bea \delta_\epsilon \lag_0 &=& (\partial_\phi \lag) \delta_\epsilon \phi + (\partial_{\phi^\dagger} \lag) \delta_\epsilon \phi^\dagger \n
&+& (\partial_{\delm\psi} \lag) \delta_\epsilon \delm\psi + (\partial_{\psibar} \lag) \delta_\epsilon \psibar \n
&+& (\partial_F \lag) \delta_\epsilon F + (\partial_{F^\dagger} \lag) \delta_\epsilon F^\dagger, \notag\eea
is simply equal to
\bea \delta_\epsilon \lag_0 &=& -\sqrt2 \delm \big[ \deln \phi^\dagger \epsilon \sigma^\nu \sigmabar^\mu \psi - \delmm \phi^\dagger \epsilon \cdot \psi - \psibar \cdot \epsilonbar \delmm \phi + i \psibar \sigmabar^\mu \epsilon F \big], \eea
which is a total derivative and thus implies that this free Lagrangian is invariant under supersymmetric transformations. 

\paragraph{Vector free Lagrangian}
Let us now consider the vector multiplet $(A_\mu, \lambda, D)$ associated to an abelian gauge group and where $A_\mu$ is a vector, $(\lambda, \lambdabar)$ a Majorana spinor and $D$ a real auxiliary field. These fields obey the following supersymmetric transformation laws:
\bea \label{vector transfo} \delta A_\mu = i(\epsilon \sigma_\mu \lambdabar - \lambda \sigma_\mu \epsilonbar),~~~~~ \delta \lambda = iD\epsilon + \frac12 \sigma^\mu \sigmabar^\nu \epsilon F_{\mu \nu} ,~~~~~ \delta D = \epsilon \sigma^\mu \delm \lambdabar + \delm \lambda \sigma^\mu \epsilonbar. \eea

Following the same idea as in the previous paragraph, we can show that the free Lagrangian for this multiplet is simply the sum of the free Lagrangians for $A_\mu$ and $\lambda$. An additional term is required for the auxiliary field $D$ but, as for the $F$ term in $\lag_0$,
\bea \label{vector free lag} \lag_1 = -\frac14 F^{\mu \nu} F_{\mu \nu} + i \lambdabar \sigmabar^\mu \delm \lambda + \frac12 D^2. \eea
The equation of motion for $D$ will be simply $D=0$.
\paragraph{Interactions in the matter sector}
There are three main ingredients to construct a Lagrangian for interacting fields
\begin{enumerate}
\item The product of two chiral multiplets is a chiral multiplet.
\item The $F$-component of a chiral multiplet transforms as a total derivative.
\item A Lagrangian density must have dimension 4 in mass.
\end{enumerate}
Let us thus consider a collection of $n$ chiral multiplets $\Phi^i = (\phi^i, \psi^i , F^i)_{i = 1,\mathellipsis,n}$ transforming as in \eqref{chiral transfo} and the holomorphic function $W(\phi)$ which we define to be:
\bea \label{superpotential} W(\phi) = \frac16 \lambda_{ijk} \phi^i \phi^j \phi^k + \frac12 \mu_{ij}\phi^i \phi^j + \alpha_i \phi^i \eea
where $\lambda_{ijk},m_{ij} \mbox{ and } \alpha_i$ are complex. This is called the \textbf{superpotential} and we can show that if we define
$$ W_i = \frac{\partial W}{\partial \phi^i}, ~~~ W_{ij} = \frac{\partial^2 W}{\partial \phi^i \partial \phi^j}, $$
then 
\bea \mathcal{L}_{int} = -W_i F^i - \frac12 W_{ij} \psi^i \cdot \psi^j + {\rm hc} , \eea
where hc stands for the hermitian conjugate, is invariant up to a total derivative:
$$ \delta \mathcal{L}_{int} = i \sqrt{2} \delm [ W_i \psi^i \sigma^\mu \epsilonbar - W^{*i} \epsilon \sigma^\mu \psibar_i ]. $$
We can now redefine our matter Lagrangian as 
\bea \label{int lagr} \mathcal{L} = \mathcal{L}_0 + \mathcal{L}_{int} \eea 
and start eliminating the auxiliary fields $F^i \mbox{ and } F^\dagger_i $. Their equations of motion are given by:
$$ \frac{\partial \mathcal{L}_{int}}{\partial F^i} = - W_i + F^\dagger_i = 0 ~~~~,~~~~\frac{\partial \mathcal{L}_{int}}{\partial F^\dagger_i} = - W^{*i} + F^{i} = 0. $$
Inserting these two results into our Lagrangian, we get:
$$ \mathcal{L} = \delm \phi^\dagger \delmm \phi + i \psibar \sigmabar^\mu \delm \psi - \frac12 W_{ij} \psi^i \cdot \psi^j - \frac12 W^{*ij}\psibar_i \cdot \psibar_j - W^{*i}W_i. $$ 
The scalar potential is thus
$$ \label{f terms} V_F = W^{*i} W_i $$
which appears to be a sum of positive terms. It is thus always positive and bounded from below.
\subsection[Supersymmetry in superspace formalism]{Supersymmetry in superspace formalism \protect \footnote{for a more formal description of the superspace formalism, the reader can refer to \cite{Sohnius:1985qm,Salam:1974yz}.}}
Superspace formalism is a very practical and usefool tool to reexpress supersymmetric quantities in a compact and easy form. It has been introduced by A. Salam and J. Strathdee in 1974 \cite{Salam:1974yz} as a new approach to build supersymmetric theories. In their paper they extend the usual space-time with coordinates $x^\mu$ by the adjunction of the (anti-commuting complex number) Majorana 	spinor $(\theta_\alpha, \bar{\theta}_{\dot{\alpha}})$ and reexpress all the supersymmetric quantities. In this section we will see how this can be achieved.

\subsubsection{Supercharges and covariant derivatives}
In Minkowski space, if $\phi(x)$ is an ordinary quantum field depending only on the coordinates $x^\mu$ than we can think of it as having been translated from $x^\mu=0$ 
\bea \phi(x) = e^{x\cdot P}\phi(0) e^{-x\cdot P} \nonumber \eea
In superspace, we introduce the Majorana spinor $(\theta^\alpha, \thetabar_{\alphadot})$ as the superpartners of the space-time coordinates. With these elements, the supersymmetry algebra can be integrated to the super-Poincar\'e group with group elements defined by
$$ G(x,\theta,\thetabar) = e^{x^\mu P_\mu + i(\theta^\alpha Q_\alpha + \Qbar_{\alphadot} \thetabar^{\alphadot})}. $$ 
We then try to evaluate the right action and the left action of the generators on a point in superspace,
$$G(x, \theta, \thetabar) G(0,\epsilon, \epsilonbar) ~~~ G(0, \epsilon, \epsilonbar) G(x,\theta,\thetabar).$$
To do so we first have to calculate the commutator
$$ [\theta^\alpha Q_\alpha + \Qbar_{\alphadot} \thetabar^{\alphadot}, \epsilon^\beta Q_\beta + \Qbar_{\dot{\beta}} \epsilonbar^{\dot{\beta}} ] $$
where $(\epsilon, \epsilonbar)$ is the parameter of a supersymmetric transformation and $(Q,\Qbar)$ the supersymmetry generators. We use the results of \eqref{susy algebra} and the properties
$$ \{Q,\theta\} = \{Q, \epsilon\} = \{ \theta, \epsilon \} = 0, $$
to find
$$ [\theta^\alpha Q_\alpha + \Qbar_{\alphadot} \thetabar^{\alphadot}, \epsilon^\beta Q_\beta + \Qbar_{\dot{\beta}} \epsilonbar^{\dot{\beta}} ] = -2i (\theta \sigma^\mu \epsilonbar - \epsilon \sigma^\mu \thetabar) \delm $$
which is, by no surprise, a translation in superspace. Finally, with the help of the Baker-Campbell-Hausdorff formula
$$ e^A e^B = e^{A+B+\frac12 [A,B]} $$
if $[[A,B],A] = [[A,B],B] = 0 $, we find
\bea G(x,\theta,\thetabar) G(0.\epsilon,\epsilonbar) = e^{x^\mu - i(\epsilon \sigma^\mu \thetabar - \theta \sigma^\mu \epsilonbar)P_\mu + i(\theta + \epsilon)\cdot Q + i \Qbar \cdot (\thetabar + \epsilonbar) }, \n
G(0,\epsilon, \epsilonbar) G(x,\theta,\thetabar) =  e^{x^\mu + i(\epsilon \sigma^\mu \thetabar - \theta \sigma^\mu \epsilonbar)P_\mu + i(\theta + \epsilon)\cdot Q + i \Qbar \cdot (\thetabar + \epsilonbar) }, \eea
showing that if we note
$$ x^\mu \rightarrow x^\mu + \delta_{L,R} x^\mu ~~~,~~~ \theta^\alpha \rightarrow \theta^\alpha + \delta_{L,R} \theta^\alpha ~~~,~~~ \thetabar^{\alphadot} \rightarrow \thetabar^{\alphadot} + \delta_{L,R} \thetabar^{\alphadot} $$
then the transformation laws read\\
\begin{center}
\begin{tabular}{c c c c c} 
$\delta_L x^\mu = i(\epsilon \sigma^\mu \thetabar - \theta \sigma^\mu \epsilonbar)$  &,& $\delta_L \theta^\alpha = \epsilon^\alpha $ &,& $\delta_L \thetabar^{\alphadot} = \epsilonbar^{\alphadot} ,$\\
$\delta_R x^\mu = -i(\epsilon \sigma^\mu \thetabar - \theta \sigma^\mu \epsilonbar)$ &,& $\delta_R \theta^\alpha = \epsilon^\alpha $ &,& $\delta_R \thetabar^{\alphadot} = \epsilonbar^{\alphadot}. $ \end{tabular}
\end{center}
Finally, if we define $Q_{L,R} \mbox{ and } \Qbar_{L,R} $ as the supercharges corresponding to the left and the right action respectively so that:
$$ \delta_{L,R} x^\mu = [i(\epsilon \cdot Q_{L,R} + \Qbar_{L,R} \cdot \epsilonbar), x^\mu] ~,~ \delta_{L,R} \theta^\alpha = [i(\epsilon \cdot Q_{L,R} + \Qbar_{L,R} \cdot \epsilonbar), \theta^\alpha]  ~,~ \delta_{L,R} \thetabar^{\alphadot} = [i(\epsilon \cdot Q_{L,R} + \Qbar_{L,R} \cdot \epsilonbar), \thetabar^{\alphadot}], $$
we show that
\bea Q_{R \alpha} = -i (\partial_\alpha - i\sigma^\mu_{\alpha \alphadot} \thetabar^{\alphadot} \delm) ~~&,&~~ \Qbar_{R \alphadot} = -i (-\bar{\partial}_{\alphadot} + i\theta_\alpha \sigma^\mu_{\alpha \alphadot} \delm), \n
 Q_{L \alpha} = -i (\partial_\alpha + i\sigma^\mu_{\alpha \alphadot} \thetabar^{\alphadot} \delm) ~~&,&~~ \Qbar_{L \alphadot} = i (\bar{\partial}_{\alphadot} + i\theta_\alpha \sigma^\mu_{\alpha \alphadot} \delm).\eea
The definitions of the derivatives $\partial_\alpha \mbox{ and } \bar{\partial}_{\alphadot}$ are given in appendix \ref{annex:susy}.\\
We also define the supercharges $Q = Q_L, \Qbar = \Qbar_L$ and the covariant derivatives 
$$ D = iQ_R, \bar{D} = -i \Qbar_R$$
and check the algebra:
\bea
\{Q_\alpha, \Qbar_{\alphadot} \} &=& -\{D_\alpha, \bar{D}_{\alphadot}\} = 2i \sigma^\mu_{\alpha \alphadot} \delm ,\n
\{D_\alpha, Q_\alpha\} &=& \{\bar{D}_{\alphadot}, Q_\alpha\} = \{D_\alpha, \Qbar_{\alphadot}\} = \{\bar{D}_{\alphadot}, \Qbar_{\alphadot}\} = 0 \nonumber. \eea
\subsubsection{General superfield}
In this new formalism, superfields will depend explicitely on $\theta \mbox{ and } \thetabar$, we thus expand them in powers of the latters 
\bea \label{general superfield} \Phi &=& z(x) + \theta \cdot \xi(x) + \thetabar \cdot \bar{\zeta}(x) + \theta \cdot \theta f(x) + \thetabar \cdot \thetabar g(x) \n
&+& \theta \sigma^\mu \thetabar A_\mu(x) + \thetabar \cdot \thetabar \theta \cdot \omega(x) + \theta \cdot \theta \thetabar\cdot\bar{\rho}(x) + \theta \cdot \theta \thetabar \cdot \thetabar d(x)
\eea
where $z(x), f(x), g(x) \mbox{ and } d(x)$ are complex scalar fields, $\xi(x), \omega(x)$ are left-handed spinors while $\bar{\zeta} \mbox{ and } \bar{\rho}(x)$ are right-handed spinors and $A_\mu(x)$ is a complex vector field. All in all, we have 16 complex bosonic degrees of freedom and 16 complex fermionic ones.\\
To get the transformation laws of every component, we simply have to calculate 
$$ \delta \Phi(x,\theta,\thetabar) = i(\epsilon \cdot Q + \Qbar \cdot \epsilonbar) \Phi(x,\theta,\thetabar) $$
and then regroup terms in powers of the Grassman variablethe Grassman variables dix \ref{annex:susy}, one gets:
\bea \label{general transfo rules}
\delta z &=& \epsilon \cdot \xi + \epsilonbar \cdot \bar{\zeta}, \n
\delta \xi &=& 2 \epsilon f + \sigma^\mu \epsilonbar (A_\mu + i \delm z), \n
\delta \bar{\zeta} &=& 2 g \epsilonbar - \sigmabar^\mu \epsilon (A_\mu + i \delm z) ,\n
\delta f &=& \frac{i}{2} \delm \xi \sigma^\mu \epsilonbar + \epsilonbar \cdot \bar{\rho} ,\n
\delta g &=& - \frac{i}{2} \epsilon \sigma^\mu \delm \bar{\zeta} + \epsilon \cdot \omega, \n
\delta A_\mu &=& -\frac{i}{2} \epsilon \cdot \delm \xi - i \epsilon \sigma_{\nu \mu} \delnn \xi + \frac{i}{2} \epsilonbar \cdot \delm \bar{\zeta} - i \epsilonbar \sigmabar_{\nu\mu } \delnn \bar{\zeta} - \epsilonbar \sigmabar_\mu \omega - \bar{\rho} \sigmabar_\mu \epsilon, \n
\delta \omega &=& -i \sigma^\mu \epsilonbar \delm g + \frac{i}{2} \epsilon \delm A^\mu + \frac{i}{2} \epsilon \sigma^{\mu \nu} F_{\mu \nu} + 2 \epsilon d, \n
\delta \bar{\rho} &=& i \epsilon \sigma^\mu \delm f - \frac{i}{2} \epsilonbar \delm A^\mu + \frac{i}{2} \sigmabar^{\mu \nu} \epsilonbar F_{\mu \nu} + 2 \epsilonbar d, \n
\delta d &=& \frac{i}{2} \delm \omega \sigma^\mu \epsilonbar - \frac{i}{2} \epsilon \sigma^\mu \delm \bar{\rho}. \eea

\subsubsection{Chiral superfield}
It is important to remark that $\bar{D}_{\alphadot}$ anticommutes with $\epsilon, \epsilonbar, Q \mbox{ and } \Qbar $. This implies that the constraint 
\bea \label{chiral constraint} \bar{D}_{\alphadot} \Phi(x,\theta,\thetabar) = 0 \eea
is preserved by a supersymmetric transformation. Moreover, if one remarks that 
$$ \bar{D}_{\alphadot} y^\mu = 0 \mbox{ and } \bar{D}_{\alpha}\theta = 0 $$
where $y^\mu =  x^\mu - i \theta \sigma^\mu \thetabar$, then the constraint \eqref{chiral constraint} defines a chiral superfield $\Phi(x,\theta,\thetabar)$ such that
\bea \Phi(x,\theta,\thetabar) = \Phi(y,\theta) = \phi(y) + \sqrt{2} \theta \cdot \psi(y) - \theta \cdot \theta F(y) .\eea
This chiral superfield is thus composed of two complex scalar fields and a left-handed spinor. Following the same path as for the general case and after some algebra, one finds the transformation laws for these fields
\bea 
\delta \phi &=& \sqrt{2} \epsilon \cdot \psi, ~~~ \delta \psi = -\sqrt{2} \epsilon F - i \sqrt2 \sigma^\mu \epsilonbar \delm \phi,~~~ \delta F = -i \sqrt2 \delm \psi \sigma^\mu \epsilonbar \eea
which reproduces equation \eqref{chiral transfo} 
The next step in our way to constructing a supersymmetric Lagrangian is to reproduce the results of \eqref{int lagr}. We investigate the product $\Phi^\dagger \Phi$ by looking at its $\tttt$ component which transforms as a total derivative under supersymmetric transformations. We expand $\Phi$ around $x^\mu$ 
\bea  \Phi(y,\theta) &=& \phi(x) - \sqrt{2} \theta \cdot \psi(x) - \theta \cdot \theta F(x) - i \theta \sigma^\mu \thetabar \delm \phi(x),\n
 &+& \frac{i}{\sqrt{2}} \theta \cdot \theta \delm \psi(x) \sigma^\mu \thetabar - \frac14 \tttt \square \ \phi(x). \nonumber \eea
and find that the Lagrangian reads
\bea
\mathcal{L}_0 = \Phidagger \Phi_{|\tttt}= \delm \phidagger \delmm \phi + \frac{i}{2}(\psi \sigma^\mu \delm \psibar - \delm\psi\sigma^\mu \psibar) + F^\dagger F ,\eea
which is that of the matter free Wess-Zumino Lagrangian. Following the same path as in section \ref{sec: wess zumino}, we adapt the superpotential of \eqref{superpotential} by promoting the fields there to superfields:
\bea \label{superfield superpotential} W(\Phi) = \frac16 \lambda_{ijk} \Phi^i \Phi^j \Phi^k + \frac12 \mu_{ij}\Phi^i \Phi^j + \alpha_i \Phi^i \nonumber \eea
where $\{ \Phi^i\}_{|i=1\dots n}$ is a collection of chiral superifleds. We deduce that $W(\Phi)_{|\theta \theta}$ transforms as a total derivative and is thus a Lagrangian density. After carrying out the algebra, we find that the superpotential can be reexpressed as
\bea W(\Phi) = W(\phi) + \sqrt2 \theta \psi^i \frac{\partial W(\phi)}{\partial \phi^i} - \theta \cdot \theta \Big[ F^i \frac{\partial W(\phi)}{\partial \phi^i} + \frac12 \psi^i \cdot \psi^j \frac{\partial^2 W(\phi)}{\partial \phi^i \partial \phi^j} \Big].\eea
Finally, we obtain the superspace formulation of the Wess-Zumino Lagrangian:
\bea \label{wz superspace} \mathcal{L}_{WZ} &=& \Phidagger_i\Phi^i_{|\tttt} + W(\Phi)_{|\theta \theta} + W^*(\Phidagger)_{|\thetabar \thetabar}. \eea
As for the scalar potential, it appears after elimination of the auxliary fields that:
$$ V(\phi, \phidagger) = \frac{\partial W}{\partial \phi^i}\frac{\partial W^*}{\partial \phidagger_i} .$$

\subsubsection{Vector superfield}

To get the vector superfield, one must first apply a reality condition to the general superfield of \eqref{general superfield}. After renaming it to $V$ and regrouping terms together, one gets
\bea V &=& V^\dagger \n
V &=& C(x) + i \theta\cdot \chi(x) - i \thetabar \cdot \chibar(x) \n
&+& \frac{i}{2} \theta\cdot\theta (M(x) + i N(x)) - \frac{i}{2} \thetabar\cdot\thetabar (M(x) -i N(x)) + \theta \sigma^\mu \thetabar v_\mu(x) \n
&+& i \theta \cdot \theta \thetabar\cdot(\lambdabar(x) - \frac{i}{2} \sigmabar^\mu \delm \chi(x)) - i \thetabar\cdot\thetabar\theta\cdot(\lambda(x) - \frac{i}{2} \sigma^\mu \delm \chibar(x)) \n
&+& \frac12 \tttt(D(x) + \frac12 \square C(x)), \eea
where $M, N, D \mbox{ and } C$ are real scalar fields, $(\chi,\chibar) \mbox{ and } (\lambda, \lambdabar)$ are Majorana fermions and $v_\mu$ is a real vector field. \\
The superfield we get contains too many degrees of freedom which we try to eliminate via a gauge transformation. Let $\Phi$ be a chiral superfield and let us apply the following gauge transformation
\bea \label{gauge transformation} V \rightarrow V + \Phi + \Phi^\dagger \eea
which preserves the reality of $V$. It induces:
\bea C &\rightarrow& C + \phi + \phidagger, ~~ \chi \rightarrow \chi - i \sqrt{2} \psi, ~~ M + i N \rightarrow M + i N + 2 i F, \n
v_\mu &\rightarrow& v_\mu - i \delm(\phi - \phidagger), ~~ \lambda \rightarrow \lambda, ~~ D \rightarrow D .\nonumber \eea
The fields $\lambda$ and $D$ are gauge invariant and if we choose the chiral superfield such that
$$ C = -(\phi + \phidagger),~~ \chi = i \sqrt2 \psi, ~~ M+iN = -2iF $$
then we can eliminate the first components of $V$ and recover the vector superfield in the Wess-Zumino gauge:
\bea V_{WZ} = \theta \sigma^\mu \thetabar v_\mu + i \theta \cdot \theta \thetabar \cdot \lambdabar -i \thetabar\cdot\thetabar \theta \cdot \lambda + \frac12 \tttt D. \eea
The transformation laws for the vector superfield in this gauge are problematic as they lead to non-vanishing transformation laws for the fields that we cancelled:
\bea
C = 0 &\rightarrow& C' = 0 \n
\chi = 0 &\rightarrow& \chi' = - i \sigma^\mu \epsilonbar v_\mu \n
M + i N = 0  &\rightarrow&  M' + i N' = 2 \epsilonbar \cdot \lambdabar \nonumber. \eea
To have the transformation laws of the vector superfield in this gauge, we use the transformation laws of the general superfield \eqref{general transfo rules} to define a gauge transformation depending on the fields:
\bea \lambda = \omega + \frac{i}{2} \sigma^\mu \delm \chibar, ~~~ D = \frac12 d, ~~~ v_\mu = A_\mu .\nonumber \eea
It is then easy to find
\bea 
\delta v_\mu = i(\epsilon \sigma_\mu \lambdabar - \lambda \sigma^\mu \epsilonbar),~~ \delta \lambda = \sigma^{\mu \nu} \epsilon F_{\mu \nu} + i \epsilon D,~~ \delta D = \delm \lambda \sigma^\mu \epsilonbar + \epsilon \sigma^\mu \delm \lambdabar. \eea

To extract a Lagrangian density for the vector superfields, one might want to investigate the powers of $V_{WZ}$. We find that the powers of $V_{WZ}$ vanish for $n\geq3$
\bea V^2_{WZ} = \frac12 \tttt v^\mu v_\mu,~~ V^3_{WZ} = 0 \nonumber \eea
and no kinetic term appears for the fields $v_\mu \mbox{ and } \lambda_\mu$. We then define the spinorial multiplet
\bea  W_\alpha = -\frac14 \bar{D}\cdot\bar{D} D_\alpha V, ~~ \bar{W}_{\alphadot} = -\frac14 D \cdot D \bar{D}_{\alphadot} V \eea
which is clearly a chiral supermultiplet ($\bar{D}_{\alphadot} (\bar{D} \cdot \bar{D}) = 0$) and invariant under the gauge transformation \eqref{gauge transformation}. To have its expression in terms of components, one must first introduce the new coordinate $y^\mu = x^\mu - i \theta \sigma^\mu \thetabar \mbox{ and } y^{\dagger\mu}= x^\mu + i \theta \sigma^\mu \thetabar$ and reexpress the covariant derivatives $D \mbox{ and } \bar{D}$:
\bea \Dbar_{\alphadot} y^\mu = 0 &\Rightarrow& \Dbar_{\alphadot} = \bar{\partial}_{\alphadot}, \n
D_\alpha y^\mu = -2i\sigma^\mu \thetabar &\Rightarrow& D_\alpha = \partial_\alpha - 2i\sigma^\mu \thetabar \partial_{y_\mu} \nonumber. \eea
Finally, we use the vector superfield in the Wess-Zumino gauge
$$ V(x,\theta,\thetabar) = V(y + i \theta \sigma^\mu \thetabar,\theta, \thetabar) = \theta \sigma^\mu \thetabar v_\mu + i \theta \sigma^\nu \thetabar \theta \sigma^\mu \thetabar \deln v_\mu + i \theta \cdot \theta \thetabar \cdot \lambdabar - \thetabar \cdot \thetabar \theta \cdot \lambda + \frac12 \tttt D $$
and get
\bea
W_\alpha &=& -\frac{i}{2} (\sigma^\mu \sigmabar^\nu \theta)_\alpha F_{\mu \nu} - i \lambda_\alpha + \theta_\alpha D - \theta \cdot \theta (\sigma^\mu \delm \lambda)_\alpha. \n
\bar{W}_{\alphadot} &=& \frac{i}{2} (\thetabar \sigmabar^\nu \sigma^\mu)_{\alphadot} F_{\mu \nu} + i \lambdabar_{\alphadot} + \thetabar_{\alphadot} D - \thetabar \cdot \thetabar (\delm \lambda \sigma^\mu)_{\alphadot}. \eea

To construct a Lagrangian density out of the spinorial supermultiplet, we use the property of chiral superfields telling that their \textit{F-component} transforms as a total derivative under supersymmetric transformations. We compute then:
$$(W^\alpha W_\alpha)_{|\theta \theta} = -\frac12 F^{\mu \nu}F_{\mu \nu} - \frac{i}{4}(\lambda \sigma^\mu \delm \lambdabar - \delm \lambda \sigma^\mu \lambdabar) + \frac12 D^2 - \frac{i}{4} \epsilon^{\mu\nu\rho\sigma}F_{\mu\nu}F_{\rho\sigma}$$
and $(\bar{W}_{\alphadot} \bar{W}^{\alphadot})_{|\thetabar \thetabar}$ which is just the complex conjugate of $(W^\alpha W_\alpha)_{|\theta \theta}$ and find that it returns the Wess Zumino free Lagrangian for vector fields:
\bea
\mathcal{L}_V &=& \frac14 (W^\alpha W_\alpha)_{|\theta \theta} + \frac14 (\bar{W}_{\alphadot} \bar{W}^{\alphadot})_{|\thetabar \thetabar}, \n
\mathcal{L}_V &=& -\frac14 F^{\mu \nu}F_{\mu \nu} + \frac{i}{2}(\lambda \sigma^\mu \delm \lambdabar - \delm \lambda \sigma^\mu \lambdabar) + \frac12 D^2. \eea

\subsubsection{Non-abelian gauge invariance and supersymmetry}
On our way to the most general renormalizable Lagrangian for supersymmetric theories, we need now to consider generalizing the formalism of non-abelian gauge theories to the supersymmetric case. To do so, let us consider a set of chiral superfields $\Phi^i$ transforming according to an arbitrary unitary representation $\mathcal{R}$ of a gauge group $G$, and a set of vector supermultiplets $V^a$ belonging to the adjoint representation of $G$. In the case of a theory invariant under global transformations of the symmetry group $G$, the transformations of the chiral multiplet are
\bea \label{global invariance}\Phi^{'i} = [e^{i \Lambda^a T_a}]^i{}_j \Phi^j \eea
or infinitesimally
$$ \delta\phi^i = i \Lambda^a (T_a)^i{}_j \Phi^j$$
where the matrices $T_a$ are the hermitian generators in the representation $\mathcal{R}$ of the gauge group $G$
\bea [T_a, T_b] = i f_{ab}{}^c T_c, ~~~ {\rm Tr}(T_aT_b) = \tau_{\mathcal{R}} \delta_{ab}, \eea
$f_{ab}{}^c \mbox{ and } \tau_{\mathcal{R}}$ are real constant parameters. Parameters $\Lambda^a$ are real constants but since constants are also superfields, they can be considered as chiral superfields so that equation\eqref{global invariance} is a superfield equation. This transformation leaves unchanged the kinetic term in \eqref{wz superspace} but imposes the superpotential $W(\Phi)$ to only contain group invariant quantities. The superfield in the linear term, for example, must be a singlet under the group $G$. \\

When considering local gauge invariance, the parameters $\Lambda^a(x)$ are no longer superfields. Hence, for the last two transformations laws to be consistent with supersymmtry, one must allow the parameters $\Lambda^a$ to be complete left-handed chiral superfields, but then the kinetic terms for chiral superfields are no longer invariant under \eqref{global invariance} since $\Lambda^{a\dagger} \neq \Lambda^a$. To restore invariance one must consider the vector supermultiplet
\bea V = V^a T_a \eea
where $V^{a\dagger} = V^a$. It transforms under gauge transformations as 
\bea e^{2gV} \rightarrow e^{2gV'} = e^{-2gi\Lambda} e^{2gV} e^{2gi\Lambda^\dagger} \eea
where $g$ is the gauge coupling constant and $\Lambda = \Lambda^a T_a$. This transformation is chosen to make the kinetic terms
\bea \mathcal{L}_{kin} = [\Phi^\dagger e^{-2gV} \Phi]_{\tttt} \eea
locally gauge invariant. In this case, we need also to redefine the spinorial chiral multiplet as follows
\bea W_\alpha = -\frac14 \bar{D}\cdot\bar{D} e^{2gV} D_\alpha e^{-2gV}, ~~ \bar{W}_{\alphadot} = -\frac14 D\cdot D e^{-2gV} \bar{D}_{\alphadot} e^{2gV} \eea
which in components read
\bea 
W_\alpha &=& i g (\sigma^\mu \sigmabar^\nu \theta)_\alpha F_{\mu\nu} + 2 g \theta \cdot \theta (\sigma^\mu D_\mu \lambdabar)_\alpha + 2ig \lambda_\alpha - 2g\theta_\alpha D ,\n
\bar{W}_{\alphadot} &=& i g (\thetabar \sigmabar^\mu\sigma^\nu)_{\alphadot} F_{\mu\nu} + 2g \thetabar\cdot\thetabar (D_\mu \lambda \sigma^\mu)_{\alphadot} -2ig\lambdabar_{\alphadot} -2g\thetabar_{\alphadot} D,\eea
where we have introduced the following quantities
\bea
F^0_{\mu \nu} &=& \delm v_\nu - \deln v_\mu,\n
F_{\mu\nu} &=& F^0_{\mu\nu} + g[v_\mu,v_\nu] ,\n
D_\mu \lambdabar &=& \delm \lambdabar - ig[v_\mu, \lambdabar],\n
D_\mu \lambda &=& \delm \lambda - i g[v_\mu, \lambda] \nonumber\eea
and let $v_\mu = v_\mu^a T_a$ to be a real vector field, $(\lambda,\bar{\lambda})\mbox{ with } \lambda = \lambda^a T_a$ a Majorana fermion and $g$ the coupling constant associated with the gauge group. We can show that the transformation laws under a supersymmetric transformation read
\bea W_\alpha \rightarrow e^{-2ig\Lambda} W_\alpha e^{2ig\Lambda},~~ \bar{W}_{\alphadot} \rightarrow e^{-2ig\Lambda^\dagger} \bar{W}_{\alphadot} e^{2ig\Lambda^\dagger}. \eea

Finally, gathering all the terms together, it is clear that the Lagrangian
\bea \label{general lagrangian} 
\mathcal{L} &=& [\Phi^\dagger e^{-2gV} \Phi]_{|\tttt} + \frac{1}{16g^2\tau_{\mathcal{R}}} {\rm Tr}(W^\alpha W_\alpha)_{|\theta\theta} + \frac{1}{16g^2\tau_{\mathcal{R}}} {\rm Tr}(\bar{W}_{\alphadot}\bar{W}^{\alphadot})_{|\thetabar\thetabar}\n
&+& W(\Phi)_{|\theta\theta}+ W^{*}(\Phidagger)_{|\thetabar\thetabar} \eea
is the most general renormalizable Lagrangian in $N=1$ supersymmetry. We see here how compact the Lagrangian is in superspace formalism. If we assume no gauge singlet field is present in the theory, this Lagrangian reads in components
\bea 
\mathcal{L} &=& -\frac14 F^a_{\mu\nu}F_a^{\mu\nu} + \frac{i}{2} (\lambda^a \sigma^\mu D_\mu \lambdabar_a - D_\mu \lambda^a \sigma^\mu \lambdabar_a) + \frac12 D_aD^a \n
&+& D_\mu \phidagger_i D^\mu \phi^i - \frac{i}{2} (D_\mu \psibar_i \sigmabar^\mu \psi^i - \psibar_i \sigmabar^\mu D_\mu \psi_i) + F^\dagger_i F^i\n
&-& gD^a\phidagger_i T_a \phi^i + i\sqrt2 g \lambdabar^a\cdot \psibar_i T_a \phi^i - i \sqrt2 g \phidagger_i T_a \psi^i\cdot \lambda^a \n
&-& \mu_{ij} (\phi^i F^j + \frac12 \psi^i \cdot \psi^j) - \frac12 \lambda_{ijk}(\phi^i\phi^jF^k + \phi^i \psi^j\cdot\psi^k) + {\rm hc}
\eea
where we have used
\bea
W(\Phi) &=& \frac12 \mu_{ij}\Phi^i\Phi^j + \frac16 \lambda_{ijk} \Phi^i\Phi^j\Phi^k, \n
D_\mu \phi^i &=& (\delm - i gv_\mu )\phi^i, ~~ D_\mu \psi^i = (\delm - i gv_\mu )\psi^i, \n
D_\mu \phidagger_i &=& \delm\phidagger_i + i g \phidagger_i v_\mu , ~~ D_\mu \psibar_i = \delm\psibar_i + i g\psibar_i v_\mu. \nonumber \eea
The auxiliary fields $D \mbox{ and } F$ can be eliminated through their equations of motion, which leads to the final form
\bea
\label{general renormalizable lagrangian}
\mathcal{L} &=&  -\frac14 F^a_{\mu\nu}F_a^{\mu\nu} + \frac{i}{2} (\lambda^a \sigma^\mu D_\mu \lambdabar_a - D_\mu \lambda^a \sigma^\mu \lambdabar_a) + D_\mu \phidagger D^\mu \phi \n
&-& \frac{i}{2} (D_\mu \psibar \sigmabar^\mu \psi - \psibar \sigmabar^\mu D_\mu \psi) + i\sqrt2 g \lambdabar^a\cdot \psibar T_a \phi - i \sqrt2 g \phidagger T_a \psi\cdot \lambda^a \n
&-& \frac12 \frac{\partial^2W(\phi)}{\partial \phi^i \partial \phi^j} \psi^i\cdot\psi^j - \frac12 \frac{\partial^2 W^{*}}{\partial \phidagger_i \partial \phidagger_j} \psibar^i\cdot\psibar^j - V(\phi,\phidagger) \eea
where the scalar potential is defined as follows
\bea
\label{scalar potential}
V(\phi,\phidagger) &=& F^\dagger_i F^i + \frac12 D^aD_a ,\n
V(\phi,\phidagger) &=& \frac{\partial W}{\partial \phi^i}\frac{\partial W^{*}}{\partial \phidagger_i} + \frac12 g^2 (\phidagger T^a \phi)(\phidagger T_a \phi).
\eea
\paragraph{Remark}
In more general theories, where we allow for non-renormalizable couplings such as supergravity, the Lagrangian \eqref{general lagrangian} can be extended to
\bea \label{non renormalizable lagrangian} 
\mathcal{L} &=& \Re\big[K(e^{-2gV}\Phi,\Phidagger)\big]_{|\tttt} + \frac{1}{16g^2}\big[h_{ab}(\Phi)(W^{a \alpha} W^b_\alpha)_{|\theta\theta} + h_{ab}^{*}(\Phidagger)(\bar{W}^a_{\alphadot} \bar{W}^{b \alphadot})_{|\thetabar\thetabar}\big]\n
&+& W(\Phi)_{|\theta\theta} + W^{*}(\Phidagger)_{|\thetabar\thetabar} \eea
where we have introduced
\begin{itemize}
\item a vector superfield $K(e^{-2gV}\Phi,\Phidagger)$ depending on $\Phi$, $\Phidagger$ and $V$ called the K\"ahler potential,
\item a holomorphic function $h_{ab}(\Phi)$ depending on the chiral superfield $\Phi$ called the kinetic gauge function.
\end{itemize}

\section{Supersymmetry breaking}\label{sec:susy breaking}
As we have seen in the beginning of this chapter, supersymmetry implies that all members of a same multiplet have the same mass which is not phenomenologically viable. Indeed, not a single scalar particle with exactly the same quantum numbers than, for example, the electron has been detected so far. One has thus to find a way to induce supersymmetry breaking and this must happen in such a way that all supersymmetric particles are heavier than their Standard Model counterpart\footnote{Except in some cases where it is possible to have a scalar superpartner of a Standard Model particle lighter than the latter.}. \\

In this section, we will describe qualitatively some mechanisms available to induce supersymmetry breaking. First, we will see that the spontaneous breaking of supersymmetry induces the apparition of a massless fermion field, the Nambu-Goldstone mode associated with this breaking. We will proceed then to calculate the mass matrices in supersymmetry and describe the O'Raifeartaigh and Fayet-Iliopoulos mechanisms. Finally, we will give the main ideas behind supergravity and gauge mediated supersymmetry breaking. For a review on the subject, one might have a look at the two following reviews \cite{Intriligator:2007cp,Chung:2003fi}.


\subsection{Goldstino}	
From equation \eqref{scalar potential}, it is clear that to break supersymmetry spontaneously it is sufficient that one of the auxiliary fields develops a vacuum expectation value (vev)
\bea \label{breaking vacuum}\langle F_i \rangle \neq 0 \mbox{ and/or } \langle D_a \rangle \neq 0 \eea
where the indices $i \mbox{ and } a$ run over all the chiral and vector supermultiplets, respectively. The spontaneous breaking of a global continuous symmetry always implies a massless Nambu-Goldstone mode with the same quantum numbers as the broken symmetry. In our case, the broken generator is the fermionic charge $Q_\alpha$ so that the Nambu-Goldstone particle ought to be a massless neutral Majorana fermion \cite{Salam:1974zb}. To see how this massless fermion rises in supersymmetry, let us assume that supersymmetry is broken and thus some of the $F$ and $D$ terms acquire a vacuum expectation value. The interaction terms for left-handed fermions in the lagrangian \eqref{general renormalizable lagrangian}
\bea
\mathcal{L}_{int} =  - i \sqrt2 g \phidagger T_a \psi\cdot \lambda^a - \frac12 \frac{\partial^2W(\phi)}{\partial \phi^i \partial \phi^j} \psi^i\cdot\psi^j \eea
can be re-written as
\bea \label{goldstino mass matrix} -\frac12 \begin{pmatrix}\psi^i & i\sqrt2 \lambda_a \end{pmatrix} \begin{pmatrix} W_{ij} & gT^{bk}{}_i \phidagger_k \\ g T^{ak}{}_j \phidagger_j & 0 \end{pmatrix} \begin{pmatrix} \psi^j \\ i \sqrt2 \lambda_b \end{pmatrix}. \eea
We consider the following two constraints
\begin{enumerate}
\item Minimization of the scalar potential $ \frac{\partial V}{\partial \phi^k} = F^\dagger_i W^{i}_k + g \phidagger_j T^{aj}_k D_a = 0, $
\item Gauge invariance of the superpotential implying that $ \delta_\omega W^* = \frac{\partial W^{*}}{\partial \phidagger_i} \delta_\omega \phidagger_i = - i F^i \omega_a T^{aj}{}_i \phidagger_j $ vanishes. We thus have $ F^i T^{aj}{}_i \phidagger_j = 0.$
\end{enumerate}
Equations above imply that the state $\Psi_G$ 
\bea \Psi_G = \frac{1}{\sqrt2} \langle F^\dagger_i \rangle \psi^i + \frac{i}{2} \langle D^a \rangle \lambda_a \eea
is an eigenstate of the mass matrix \eqref{goldstino mass matrix} with a mass equal to 0.\\
Supersymmetry breaking vacuum is thus characterized by the condition \eqref{breaking vacuum} and the existence of a massless fermion, the goldstino. There is also a useful sum rule that governs the tree-level squared masses of particles in theories with or without supersymmetry breaking vacuum. Assuming that $\mathcal{M}_0$ is the mass matrix for scalar particles, $\mathcal{M}_{\frac12}$ the mass matrix for fermions and $\mathcal{M}_1$ the mass matrix for vector fields, we define the supertrace as
\bea \label{super trace} {\rm sTr}(\mathcal{M}^2) = {\rm Tr}(\mathcal{M}^2_0) -2 {\rm Tr}(\mathcal{M}^2_{\frac12}) + 3 {\rm Tr}(\mathcal{M}^2_1). \eea
We proceed to the analysis of this equation and assume a spontaneously broken supersymmetry.


\paragraph{Scalars mass matrices}
We deduce from the lagrangian \eqref{general renormalizable lagrangian} that the scalar particles get their masses from the quadratic terms in the scalar potential. The mass matrix can be written as
\bea \label{scalar mass matrix} \mathcal{M}^2_0 = \begin{pmatrix}
\langle \frac{\partial^2 V}{\partial \phi^i \partial \phidagger_j} \rangle & \langle \frac{\partial^2 V}{\partial \phidagger_j \phidagger_i} \rangle\\
\langle \frac{\partial^2 V}{\partial \phi^i \partial \phi^j} \rangle & \langle\frac{\partial^2 V}{\partial \phi^j \phidagger_i}\rangle
\end{pmatrix}. \eea
If we replace $V$ by its value \eqref{scalar potential}, note the Casimir
$$ C(i) \delta_k{}^j = {\rm Tr}(T^a T^a)_k{}^j $$
where $T$ matrices are the generators in the representation $\mathcal{R}$ of the gauge group $G$ in which lies the field $\phi$ and assume \footnote{Phenomenologically, allowing for gauge anomalies \textit{i.e.} ${\rm Tr}(T_a)\neq 0$ for a $U(1)$ group introduces quadratic divergences and gravitational anomalies which is not desirable \cite{Derendinger:1990tj}.  }
$$ {\rm Tr}(T_a) = 0,$$
it is easy to show that
\bea {\rm Tr}(\mathcal{M}_0^2) = 2 W_{ik}W^{*ik} + 2 g^2 C(i) \langle \phi^i \phidagger_i \rangle. \eea


\paragraph{Vector fields mass matrices}
The lagrangian \eqref{general renormalizable lagrangian} is here also the starting point for building the vector fields mass matrices. Mass terms for these fields originate from the covariant derivatives $D_\mu$ acting on the scalar fields and read
\bea \mathcal{M}^2_1 = \begin{pmatrix} 2g^2 \langle \phidagger_k \rangle T_a{}^k{}_i T^{bi}{}_l \langle \phi^l \rangle \end{pmatrix}. \eea
The trace of this matrix simply reads
\bea {\rm Tr}(\mathcal{M}^2_1) = 2g^2 C(i) \langle \phidagger_i \phi^j \rangle. \eea
\paragraph{Fermion mass matrices}
Using the result \eqref{goldstino mass matrix}, we deduce immediately the mass matrix for fermions
\bea \label{fermion mass matrix}
\mathcal{M}_{\frac12} =\begin{pmatrix} \langle W_{ij} \rangle & i \sqrt2 gT^{bk}{}_i \langle \phidagger_k \rangle \\ i \sqrt2 g T^{ak}{}_j \langle \phidagger_j \rangle  & 0 \end{pmatrix}\eea
in the basis $\begin{pmatrix} \psi^i & \lambda_a \end{pmatrix} $.\\
Squaring the matrix and taking its trace returns
\bea {\rm Tr}(\mathcal{M}^2_{\frac12}) = \langle W^{*ij}W_{ji} \rangle + 4 g^2 C(i) \langle \phidagger_i \phi^i \rangle .\eea


\paragraph{The mass formula} Finally, the mass formula \eqref{super trace} becomes
\bea s{\rm Tr}(\mathcal{M}^2) = 0 \eea
for any supersymmetric theory with an arbitrary vacuum, provided the absence of chiral anomalies.


\paragraph{The Fayet-Iliopoulos mechanism} As seen in \eqref{breaking vacuum}, a way to break supersymmetry is to consider $\langle F \rangle = 0 \mbox{ while } \langle D \rangle \neq 0 $. Such a pattern, is known as the Fayet-Iliopoulos mechanism and was presented in ref. \cite{Fayet:1974jb} in 1974. In this paper, they introduce a new term in the lagrangian
\bea \label{fayet iliopoulos term} \mathcal{L}_{F-I} = \xi D \eea
where $D$ is the auxiliary field of a gauge supermultiplet $V = (A_\mu, \lambda,D)$ associated to an $U(1)$ gauge group. To analyze this supersymmetry breaking mechanism, let us consider the following superpotential
\bea W = \mu \Phi_+\Phi_-\eea
where $\Phi_{\pm} = (\phi_\pm, \psi_\pm,F_\pm)$ are two chiral superfields with charge $\pm e$ under the $U(1)$ gauge group. The scalar potential of this model is
\bea V(\phi_+,\phi_-) &=& \frac12 (A_+^2 + B^2_+) + \bm{\frac12(A_-^2 + B^2_-)(\mu^2 - e\xi)} \n 
&+& \frac{e^2}{8}(A_+^2 + B^2_+ - A_-^2 - B_-^2)^2 + \frac12 \xi^2 \eea
where we have split the two complex scalars  $\phi_\pm = \frac{(A_{\pm} + i B_{\pm})}{\sqrt2}$. It is thus a sum of positive terms except for the term in bold font whose sign depends on the difference $\mu^2 - e\xi $.
\begin{itemize}
\item A first possibility is to have $\mu^2 > e\xi$. In this case the scalar potential is positive and supersymmetry is broken. However, the minimum is obtained for $\langle A_+ \rangle = \langle A_- \rangle = \langle B_+\rangle = \langle B_- \rangle = 0$ which preserves the $U(1)$ gauge symmetry. In this configuration, the mass matrices give rise to two massive complex scalars, a goldstino and two fermions. The masses are the following
\bea
m_{A_1} &=& m_{B_1} = \sqrt{\mu^2 - e\xi},~~  m_{A_2} = m_{B_2} = \sqrt{\mu^2 + e\xi}, \n
m_{\psi_2}&=& m_{\psi_3} = m,~~  m_{\psi_1} = 0. \eea
This spectrum is obviously not acceptable phenomenologically as it implies that the superpartner of the electron, for example, is lighter than the electron.
\item In the other case, $\mu^2 < e\xi$, the minimum of the scalar potental is obtained for $\langle A_-\rangle = \langle B_- \rangle = 0$ and, after redefinition of the field $\phi_+$, $\langle B_+ \rangle = 0$ and $\langle A_+ \rangle = \frac{\sqrt{2(e\xi - \mu^2)}}{e}$. This setting, induces also the apprition of a goldstino, two massive fermion fields, two scalar fields and two pseudo-scalar fields (of which a Goldstone boson) with the masses
\bea
m_{A_1} &=& \sqrt{2(e\xi-\mu^2)},~~m_{B_1} =0,~~  m_{A_2} = m_{B_2} = \sqrt{2}\mu ,\n
m_{\psi_2}&=& \sqrt{2(e\xi-\mu^2)},~~  m_{\psi_3} = 0. \eea
Neither in this configuration, no viable phenomenology can be found.
\end{itemize}
Finally, we conclude that Fayet-Iliopoulos supersymmetry breaking mechanism, though successful, cannot accomodate a realistic phenomenology as it predicts automatically a scalar partner to the SM fermions lighter than the fermion itself. This result generalizes to other forms of the superpotential. Nevertheless, it is interesting to remark that in the second case, the minimum breaks supersymmetry together with gauge invariance.


\paragraph{The O'Raifeartaigh mechanism} The other way to break supersymmetry is to consider $\langle F \rangle \neq 0 \mbox{ and } \langle D \rangle = 0$ and this is what O'Raifeartaigh did, one year after Fayet and Iliopoulos, in his paper \cite{O'Raifeartaigh:1975pr}. In the latter, he comes to the conclusion that in order to break supersymmetry without involving any vector superfield ($\langle D_a \rangle = 0$), one needs at least three chiral superfields. One of them would develop a vacuum expectation value while the others not. Let us investigate the phenomenological implications of such a model by first writing down the superpotential
\bea W = \lambda \Phi^3 + m \Phi^1 \Phi^2 + g \Phi^1 \Phi^1 \Phi^3 \eea
where $\Phi^1, \Phi^2 \mbox{ and }\Phi^3$ are three chiral superfields. We also define the real scalars $A$ and $B$ such that
$$ \phi^j = \frac{1}{\sqrt2}(A^j + i B^j)$$
and the real auxiliary fields $f$ and $g$ 
$$ F^j = \frac{1}{\sqrt2}(f^j - i g^j).$$
From \eqref{scalar potential}, we deduce the equations of motion for the real and imaginary parts of the auxiliary fields
\bea\label{Fi values}
   f^1 = mA^2 + \sqrt2 g (A^1A^3 - B^1B^3) &,& g^1  = mB^2 + \sqrt2 g (A^1B^3 - B^1A^3),\n
   f^2 = mA^1 &,& g^2 = mB^1,\n
   f^3 = \sqrt2 \lambda + \frac{g}{\sqrt2}( (A^1)^2 - (B^1)^2) &,& g^3 = \sqrt2 gA^1B^1.
\eea
We observe that if $\langle F^2\rangle =0 $ then $\langle F^1\rangle$ is simulteneously zero but $\langle F^3\rangle = \lambda = 0 $ is not possible which induces the breaking of supersymmetry. The question is now to know whether such a mechanism can lead to a viable phenomenology. To do so, let us start first by plugging $F_i$ values \eqref{Fi values} into the scalar potential
\bea V &=& F^1F^\dagger_1 + F^2F^\dagger_2 + F^3F^\dagger_3 ,\n
     V &=& \frac{m^2}{2} ( (A^1)^2 - (B^2)^2) + g^2 ( (A^1A^2)^2 + (A^1B^3)^2 + (B^1B^3)^2) \n
       &+& \sqrt2 mg ( A^1(A^3A^2 + B^3 B^2) + B^1(A^3B^2 - A^2B^3)) \n
       &+& (A^1)^2 ( \lambda g + \frac{m^2}{2}) + (B^1)^2 ( \frac{m^2}{2} - \lambda g) \frac{g^2}{4}((A^1)^2 + (B^1)^2)^2. \eea
We impose $m^2 \geq 2\lambda g$ to ensure that the scalar potential is bounded from below. We remark also that if $\langle A^1 \rangle = \langle A^2 \rangle = \langle B^1 \rangle = \langle B^2 \rangle = 0 $ then the scalar field $\phi^3$ can acquire an arbitrary vacuum expectation value. For simplicity, we set 
$$ \langle B^3 \rangle = 0,~~~  \langle A^3 \rangle = \frac{\mu}{\sqrt2 g} $$
and derive the scalar mass matrix by generalizing \eqref{scalar mass matrix} to the basis $\begin{pmatrix} A^1 & A^2 & A^3 \end{pmatrix}$
\bea
\mathcal{M}_S^2 = \begin{pmatrix} \mu^2 + m^2 + 2\lambda g & m\mu & 0\\ m\mu & m^2 & 0\\ 0  & 0 & 0 \end{pmatrix}.\eea
This scalar squared mass matrix has a massless eigenvector and two massive scalars
$$ m^2_{A^1} = \lambda g + m^2 + \frac{\mu^2}{2} + \frac12 \sqrt{(2\lambda g + \mu^2)^2 + 4m^2\mu^2}, ~~m_{A^2}^2 =  \lambda g + m^2 + \frac{\mu^2}{2} - \frac12 \sqrt{(2\lambda g + \mu^2)^2 + 4m^2\mu^2}.$$
The pseudo-scalars squared mass matrix is obtained the same way, its eigenvalues are related to the previous ones by substituting $\lambda \rightarrow - \lambda$,
$$ m^2_{B^1} = -\lambda g + m^2 + \frac{\mu^2}{2} + \frac12 \sqrt{(-2\lambda g + \mu^2)^2 + 4m^2\mu^2}, ~~m_{B^2}^2 =  -\lambda g + m^2 + \frac{\mu^2}{2} - \frac12 \sqrt{(-2\lambda g + \mu^2)^2 + 4m^2\mu^2}.$$
As to the fermion fields, starting from \eqref{fermion mass matrix} we find in the basis $\begin{pmatrix} \psi^1 & \psi^2 & \psi^3 \end{pmatrix}$
\bea 
\mathcal{M}_\psi = \begin{pmatrix}\mu & m & 0 \\ m & 0 & 0 \\ 0 & 0 & 0 \end{pmatrix}, \eea
with the eigenvalues
$$m_{\psi^1} = \frac12 \sqrt{\mu^2 + 4 m^2} + \frac{\mu}{2},~~ m_{\psi^2} = \frac12 \sqrt{\mu^2 + 4m^2} - \frac{\mu}{2},~~ m_{\psi^3} = 0, $$
exhibiting the goldstino which turns out to be the fermion from the chiral supermultiplet $\Phi^3$.\\
All in all, we see clearly that we have the hierarchy
\bea m_{A^1} > m_{\psi^1} > m_{B^1} \eea 
leading thus to a phenomenology that is not compatible with experimental observations.\\


\subsection{Non-renormalization theorems}
There exists other ways to break supersymmetry but they all imply the coexistence of two radically different energy scales, the electroweak scale and, for example, the Planck scale. It is therefore important to study the behaviour of supersymmetric theories in such configurations. Indeed, in the Standard Model of particle physics, such a discrepancy in the energy scales leads to the well-known hierarchy problem. Here, we want to know how the parameters in the supersymmetric Lagrangian behave under renormalization.\\

\paragraph{Scalar masses}
In the Standard Model of particle physics, the photon mass is strictly equal to 0 because of the $U(1)_{em}$ symmetry of the model. We say that the photon mass is protected by the $U(1)$ symmetry. Fermions' masses are protected by the chiral symmetry. It is however broken but only leads to relatively small masses with no quadratic divergencies. Finally, there is no symmetry to protect scalar masses from quadratic divergencies unless one introduces supersymmetry. In the latter, the $U(1)$ and chiral symmetry are still there but there is also a symmetry between scalars and fermions (in unbroken supersymmetry). Scalar fields ``inherit" thus of the chiral symmetry and get their masses protected from quadratic divergencies.\\
To see how it works, as divergencies arise from scalar self-couplings, it is sufficient to consider a theory with only one chiral superfield $\Phi=(\phi, \psi, F)$ and no vector superfield obeying to the lagrangian \eqref{general renormalizable lagrangian}. Then, simply by splitting the complex scalar field into its real degrees of freedom
$$\phi = \frac{(A+ iB)}{\sqrt2}$$
and writing down the Feynman diagrams renormalizing their masses (fig.\ref{scalar mass renormalization}), we immediately see that quadratic divergencies cancel out and we are only left with logarithmic ones which are much easier to absorb in counterterms.\\

\begin{figure}
\unitlength = 1mm
\begin{fmffile}{test45}
\begin{fmfgraph*}(30,15)
\fmfleft{i}\fmflabel{$A$}{i}
\fmfright{o}\fmflabel{$A~~~~+$}{o}
\fmf{dashes}{i,v1,o}
\fmf{dashes,tension=0.8,label=$A/B$}{v1,v1}
\end{fmfgraph*}
\hskip2cm
\begin{fmfgraph*}(30,15)
\fmfleft{i1}\fmflabel{$A$}{i1}
\fmfright{o1}\fmflabel{$A~~~~+$}{o1}
\fmf{dashes}{i1,v1}
\fmf{dashes}{v2,o1}
\fmf{dashes,left,tension=0.4,label=$A/B$}{v1,v2}
\fmf{dashes,left,tension=0.4,label=$A/B$}{v2,v1}
\end{fmfgraph*}
\hskip2cm
\begin{fmfgraph*}(30,15)
\fmfleft{i1}\fmflabel{$A$}{i1}
\fmfright{o1}\fmflabel{$A~~~~ \sim ~~\ln \Lambda $}{o1}
\fmf{dashes}{i1,v1}
\fmf{dashes}{v2,o1}
\fmf{plain,left,tension=0.4,label=$\psi$}{v1,v2}
\fmf{plain,left,tension=0.4,label=$\psibar$}{v2,v1}
\end{fmfgraph*}
\end{fmffile}

\caption{\footnotesize \label{scalar mass renormalization} Feynman diagrams renormalizing the mass of the scalar $A$. The symbol $A/B$ neans that either the field $A$ or $B$ is in the loop and $\Lambda$ is the cut-off}
\end{figure}

\paragraph{Other parameters in the lagrangian}
Now we move on to the parameters of the lagrangian. In general, one writes
\bea \mathcal{L}_{renormalized} = \mathcal{L} + \mathcal{L}_{counterterms} \eea
Historically, Grisaru, Siegel and Rocek were the first to study this problem in 1977 \cite{Grisaru1979429}. They established the non-renormalization theorem stating that
\begin{itemize}
\item the superpotential needs not to be renormalized,
\item neither the Fayet-Iliopoulos terms \eqref{fayet iliopoulos term} provided that the $U(1)$ generator is traceless.
\item Gauge kinetic function needs to be renormalized at the one-loop level.
\item K\"ahler potential must be renormalized to all orders.
\end{itemize}
N. Seiberg, based on string theory arguments by Witten in 1985, gave a quite elegant proof of this theorem using the inner symmetries of the supersymmetric algebra \cite{Seiberg:1993vc,Quevedo:2010ui}. Such theorems are of very big importance in supersymmetric theories as they imply for example, that if a particle is massless at tree-level, than it will remain massless at all orders of perturbation theory as long as supersymmetry is conserved. Moreover, these theorems stabilize the mass spectra and allow the coexistence of extremely different energy scales \cite{Sohnius:1985qm}.\\
These results are very important in supersymmetric theories as they allow us to consider supersymmetry breaking mechanisms originating from more fundamental theories. The following paragraphs will give a brief description of the ideas behind these mechanisms but not the detailed calculations which are beyond the scope of this manuscript.\footnote{If interested by the calculational details, the reader can refer to the book \cite{MichBenj} where calculations are carried out explicitely.}

\subsection{Supergravity mediated supersymmetry breaking}
In this mechanism\cite{Chamseddine:1982jx,Barbieri:1982eh,Ibanez:1982ee,Hall:1983iz,Ohta:1982wn}, we suppose the coexistence of two sectors, the visible sector represented by the chiral superfields $\Phi^i$ and the hidden sector represented by the chiral superfield $Z$. The physics in the hidden sector is governed by supergravity (locally invariant supersymmetry) while the matter sector only exhibits a global supersymmetry. The superpotential is
\bea W(\Phi^i,Z) = W_m(\Phi^i) + W_h(Z). \eea
At the minimum of the potential, we suppose that only the scalar $z$ acquires a vev. This induces the spontaneous breaking of supergravity and the apparition of a massless fermion field (the goldstino). The latter is eaten by the gravitino which becomes then massive (super-Higgs mechanism).
Finally, supersymmetry breaking in the matter sector is induced through the coupling of the $Z$ with $\phi$ via the K\"ahler potential which generate the so-called soft-breaking terms. If the superpotential reads
$$ W(\Phi) = \alpha_i \Phi^i + \frac12 \mu_{ij} \Phi^i \Phi^j + \frac16 \lambda_{ijk} \Phi^i \Phi^j \Phi^k $$ 
then the soft breaking potential simply reads
\bea V_{soft} = \alpha_i \eta_0 \phi^i+ \frac12 B_0 \mu_{ij} \phi^i \phi^j + \frac16 A_0 \lambda_{ijk} \phi^i \phi^j \phi^k + m_0 \phi^i\phi^{\dagger}_j \eea
where $\eta_0, B_0 \mbox{ and } A_0$ are the linear, bilinear and trilinear universal soft breaking terms and $m_0$ is the universal scalar and gauginos masses. The gauge kinetic function $h_{ab}(\Phi)$ induces universal mass terms $m_{1/2}$ for the gaugino masses which completes the set of soft supersymmetry breaking terms.
\subsection{Gauge mediated supersymmetry breaking}
Here we assume that a gauge group, for example $SU(5)$, is broken in a hidden sector and a singlet field acquires a vacuum expectation value. The latter field, propagates the breaking of the gauge group to the visible sector through its interactions with a messenger superfield. Finally, mass terms for the fields in the visible sector and a splitting in mass between the components of a same supermultiplet are generated which induces supersymmetry breaking.\\

The two mechanisms described here are not the only ones. Other solutions to the problem of supersymmetry breaking have been considered such as the anomaly gauge mediated supersymmetry breaking (AMSB) or extra-dimensional supersymmetry breaking (XMSB) \cite{Martin:1997ns}. In any case, it is always supposed that supersymmetry breaking is induced by some mechanism at high scale (generally unification scale or Planck scale) inducing, in the low energy limit, a new Lagrangian called the soft supersymmetry breaking Lagrangian which reads:
\bea
\lagr_{soft} = \frac16 h^{ijk}\phi_i\phi_j\phi_k - \frac12 b^{ij} \phi_i\phi_j - a^i \phi_i - \frac12 (m^2)_i{}^j \phibar^i \phi_j - \frac12 M \lambda\cdot\lambdabar +{\rm h.c.}
\eea
These new parameters are related to the high energy ones through the renormalization group equations.

\section{Renormalization group equations}\label{sec:rge-formulas}
Gauge couplings unification and supersymmetry breaking mechanisms are powerful hints for Grand Unified Theory (GUT) or some other organizing principle such as string theory. From a phenomenology point of view, that is if one only considers the low energy theory, these mechanisms induce a large number of free parameters\footnote{We will see later that, in the minimal realization of supersymmetry, the number of free parameters (more than 100) is essentially due to supersymmetry breaking parameters.} making thus supersymmetry a non-predictive theory. A possible way out is to study, with the renormalization group equations, the relation between the high-energy principles where supersymmetry breaking takes place and the low energy theory where experiments are held. By doing so, the number of free parameters is reduced significantly and supersymmetry becomes very predictive.\\


Another motivation to study the renormalization group equations (RGEs) is the fact that they have a generic form and can be applied for any renormalizable supersymmetric theory. Their analytic expressions have been known for some time now (see for example \cite{Jones:1974pg,Jones:1983vk,Falck:1985aa,Derendinger:1990tj,Martin:1993zk,Martin:1993yx,Martin:1997ns}) but I thought it might be useful to provide them here also. In the next section, where the simplest phenomenological realization of supersymmetry is described, an example of application of these equations is given with an emphasis on the calculational details.\\

Let us then set the context: following the same notations as above, we consider a general $N=1$ renormalizable supersymmetric theory with gauge group $G$. The latter is supposed to be a direct product of subgroups $G_a$. The chiral superfields $(\Phi_i)_{i=1\dots n}$ contain a complex scalar $\phi_i$ and a two-component fermion $\psi_i$ which lie in the representation $R$ of the gauge group. Matrices $T^a$ are the generators in the representation $\mathcal{R}$ of the gauge group $G$ and we introduce the quadratic Casimir invariant $C( R )$ and the Dynkin index $\tau_R$:
\bea
&& (T^a T^a)^j{}_i \equiv C( R ) \delta^j{}_i \n
&& {\rm Tr}(T^A T^B) \equiv \tau_R \delta^{AB} \eea 
In the case of a non-abelian Lie gauge group, we have
$$
	\tau_R = \left\{ 
	  \begin{array}{l l}
	    \frac12 & \quad \text{for the fundamental representation}\\
	     2 & \quad \text{for the adjoint representation}
	  \end{array} \right.
$$
In the case of an $SU(n)$ group, the quadratic Casimir invariant is simply zero but the Dynkin index is defined as follows
\bea \tau_R = C ~ Y^2 \eea
where $Y$ is the $U(1)$ charge and $C$ is a constant whose value depends on the unification scheme chosen. \\

Finally, the equations that are reported here are taken from \cite{Martin:1993zk} and are thus expressed in the dimensional reduction scheme (DRED) with modified minimal substraction ($\xbar{\text{DR}}$) which does not violate supersymmetry (for a pedagogical introduction, see \cite{Capper:1979ns}) unlike the dimensional regularization scheme (DREG) \cite{'tHooft:1972fi} with modified minimal substraction ($\xbar{\text{MS}}$).

Indeed, as shown by Delbourgo and Prasad in \cite{Delbourgo:1974az}, ``the action associated with supersymmetric theories in arbitrary dimensions $2l$ is not an invariant (and the associated spinor current is not conserved) except for $l=2$.". The problem of this scheme is that if we extend the number of space-time parameters to $2l$ then the number of spinor parameters is extended to $2^l$ which results in either having a number of fields depending on the number $l$ in the Lagrangian or in having a supersymmetry-violating action. Siegel in his paper \cite{Siegel:1979wq} introducing dimensional reduction proposed to first carry-out the supersymmetric algebra in 4 dimensions and then vary the number of space-time coordinates from 4 to $d$ with $d = 4 - 2\epsilon<4$ (dimensional reduction). Though this procedure revealed some possible inconsistencies \cite{Siegel:1980qs}, Avdeev and Vladimirov \cite{Avdeev:1982xy} concluded that dimensional regularization via dimensional reduction ``secures manifest supersymmetry of the quantum super Yang-Mills theories to all orders and a consistency of calculations up to five loops''. 

We note that the running couplings computed in $\xbar{\text{MS}}$ will differ from those computed in $\xbar{\text{DR}}$ by finite one-loop corrections, and the $\beta$-functions will be different for the two schemes starting at the two-loop level. A ``dictionnary" has been provided by Martin and Vaughn \cite{Martin:1993yx} for translating couplings between the two schemes including all finite one-loop radiative corrections.\\

The equations will be grouped into three categories: gauge coupling constants and gaugino masses, superpotential parameters and their counter-parts in the soft-supersymmetry breaking lagrangian and scalar masses. Furthermore, when not mentionned clearly, we will use the letters of the beginning of the alphabet $(a, b, c)$ to denote gauge indices while the letters of the middle of the alphabet $(i, j, k\dots)$ will refer to indices running over all the chiral superfields.

\subsection{Gauge coupling constants and gaugino masses}
The renormalization group equations for the gauge coupling constants and the gaugino masses will be noted generically
\bea \frac{d}{dt}p_i = \frac{1}{16\pi^2}\beta_{p_i}^{(1)} + \frac{1}{(16\pi^2)^2}\beta_{p_i}^{(2)}\eea
where $t \equiv \ln E$, $p_i$ is the parameter to renormalize and $\beta^{(1)}_{p_i} \et \beta^{(2)}_{p_i}$ are the one-loop and two-loop beta-functions associated to the parameter $p_i$.\\

The $\beta$-functions for the gauge coupling constants are
\bea
\beta_{g_a}^{(1)} &=& g_a^3 \big[ \tau_{aR} - 3C( G_a ) \big]; \n
\beta_{g_a}^{(2)} &=& g_a^3 \bigg[ -6 g_a^2 \big(C(G_a)\big)^2 + 2g_a^2C(G_a) \tau_{aR} + 4\sum_{b}g_b^2\tau_{a R }C_b( R ) \bigg] - g_a^3 \lambda^{ijk}\lambda_{ijk} \frac{C(k)}{d(G)}.
\eea
where
\begin{itemize}
\item the $\sum_b$ is a sum over subgroups, 
\item $\lambda_{ijk} \equiv (\lambda^{ijk})^*$
\item $d(G)$ is the dimension of the adjoint representation
\item $\tau_{aR}$ is the Dynkin index summed over all chiral multiplets
\item $\tau_{aR}C_b( R )$ is the sum of the Dynkin indices weighted by the quadratic Casimir invariant.
\end{itemize}
The $\beta$-functions for the gaugino masses are 
\bea 
\beta_{M_a}^{(1)} &=& g_a^2 \big[2\tau_{aR} - 6 C( G_a )\big] M_a; \n
\beta_{M_a}^{(1)} &=& g_a^2 \bigg[ -24 g_a^2(C(G_a))^2 M_a + 8g_a^2C(G_a)\tau_{aR}M_a + 8 \sum_{b} g_b^2 \tau_{aR} C_b( R )(M_a + M_b) \bigg] \n
&+& 2g_a^2 \bigg[ h^{ijk} - M_a \lambda^{ijk} \bigg] \lambda_{ijk} \frac{C_a(k)}{d(G_a)}.
\eea

\subsection{Superpotential parameters and soft supersymmetry breaking parameters}
The renormalization group equations for the superpotential parameters are noted
\bea
\frac{d}{dt} \lambda^{ijk} &=& \lambda^{ijp} \bigg[ \frac{1}{16\pi^2} \gamma_p^{(1)k} +  \frac{1}{(16\pi^2)^2} \gamma_p^{(2)k} \bigg] + (k \leftrightarrow i) + (k \leftrightarrow j),\n
\frac{d}{dt} \mu^{ij} &=& \mu^{ip} \bigg[ \frac{1}{16\pi^2} \gamma_p^{(1)j} +  \frac{1}{(16\pi^2)^2} \gamma_p^{(2)j} \bigg] + (j \leftrightarrow i),\n
\frac{d}{dt} \alpha^{i} &=& \alpha^{p} \bigg[ \frac{1}{16\pi^2} \gamma_p^{(1)i} +  \frac{1}{(16\pi^2)^2} \gamma_p^{(2)i} \bigg].
\eea
where the anomalous dimensions $\gamma$ read
\bea \label{eq:anomdim}
\gamma_i^{(1)j} &=& \frac12 \lambda_{ipq}\lambda^{jpq} - 2\delta^j_i \sum_a g^2_a C_a(i), \n
\gamma_i^{(2)j} &=& -\frac12 \lambda_{imn}\lambda^{npq}\lambda_{pqr}\lambda^{mrj} + \sum_a g_a^2 \lambda_{ipq}\lambda^{jpq}\bigg[2C_a( p )- C_a(i)\bigg]\n
&+&2\delta_i^j \bigg[\sum_a g_a^4 C_a(i) \tau_{aR}  + 2 \sum_a \sum_b g_a^2g_b^2C_a(i)C_b(i) - 3\sum_a g_a C(G_a) C_a(i) \bigg].
\eea
In these equations, $C( i ) $ always refers to the quadratic Casimir invariant of the representation carried by the chiral superfield $i$ while $S( R)$ refers to the total Dynkin index summed over all of the chiral superfields. \\

Turning on to the RGEs associated with the trilinear supersymmetry breaking parameters, we write
\bea
\frac{d}{dt} h^{ijk} &=& \frac{1}{16\pi^2}\big[\beta_{h}^{(1)}\big]^{ijk} +  \frac{1}{(16\pi^2)^2}\big[\beta_{h}^{(2)}\big]^{ijk}
\eea
with $\big[\beta_{h}^{(1)}\big]^{ijk} \et \big[\beta_{h}^{(2)}\big]^{ijk}$ defined as follows
\bea
\big[\beta_{h}^{(1)}\big]^{ijk} &=& \frac12 h^{ijl}\lambda_{lmn}\lambda^{mnk} + \lambda^{ijl}\lambda_{lmn}h^{mnk} \n
&-& 2\sum_a\bigg( h^ijk - 2M_a \lambda^{ijk}\bigg)g_a^2C_a(k) + (k \leftrightarrow i) + ( k \leftrightarrow j)\n
\big[\beta_{h}^{(2)}\big]^{ijk} &=& -\frac12 h^{ijl}\lambda_{lmn}\lambda^{npq}\lambda_{pqr}\lambda^{mrk} - \lambda^{ijl}\lambda_{lmn}\lambda^{npq}\lambda_{pqr}h^{mrk} - \lambda^{ijl}\lambda_{lmn}h^{npq}\lambda_{pqr}\lambda^{mrk};\n
&+& \sum_a g_a^2\bigg[h^{ijl}\lambda_{lpq}\lambda^{pqk} + 2\lambda^{ijl}\lambda_{lpq}h^{pqk} - 2M_a\lambda^{ijl}\lambda_{lpq}\lambda^{pqk} \bigg] \bigg[2C( p ) - C(k) \bigg] \n
&+& \sum_a g_a^4\bigg[2h^{ijk}-8M_a\lambda^{ijk}\bigg] \bigg[C_a(k) \tau_{aR} - 3C(G_a) C_a(k)\bigg]  \n
&+& 2\sum_a\sum_b g_a^2g_b^2\bigg[2h^{ijk}-8M_a\lambda^{ijk}\bigg]C_a(k)C_b(k) + (k \leftrightarrow i) + (k \leftrightarrow j).
\eea

The equation for the bilinear supersymmetry breaking parameters reads
\bea 
\frac{d}{dt} b^{ij} &=& \frac{1}{16\pi^2}\big[\beta_{b}^{(1)}\big]^{ij} +  \frac{1}{(16\pi^2)^2}\big[\beta_{b}^{(2)}\big]^{ij}
\eea
where
\bea
\big[\beta_{b}^{(1)}\big]^{ij} &=& \frac12b^{il}\lambda_{lmn}\lambda^{mnj} + \frac12 \lambda^{ijl}\lambda_{lmn}b^{mn} + \mu^{il}\lambda_{lmn}h^{mnj}\n
& - &2\sum_a\bigg[b^{ij} - 2M_a \mu^{ij}\bigg]g_a^2 C_a(i) + (i \leftrightarrow j);\n
\big[\beta_{b}^{(2)}\big]^{ij} &=& -\frac12 b^{il}\lambda_{lmn}\lambda^{pqn}\lambda_{pqr}\lambda^{mrj} - \frac12\lambda^{ijl}\lambda_{lmn}b^{mr}\lambda_{pqr}\lambda^{pqn} \n
&-&\frac12\lambda^{ijl}\lambda_{lmn}\mu^{mr}\lambda_{pqr}h^{pqn} - \mu^{il}\lambda_{lmn}h^{npq}\lambda_{pqr}\lambda^{mrj} \n
&-&\mu^{il}\lambda_{lmn}\lambda^{npq}\lambda_{pqr}h^{mrj} + 2\sum_a \lambda^{ijl}\lambda_{lpq}(b^{pq} - \mu^{pq}M_q)g_a^2C_a( p) \n
&+&\sum_a g_a^2 \bigg[b^{il}\lambda_{lpq}\lambda^{pqj}+2\mu^{il}\lambda_{lpq}h^{pqj} - 2\mu^{il}\lambda_{lpq}\lambda^{pqj}M_a\bigg]\bigg[2C_a( p )- C_a(i) \bigg]\n
&+&\sum_a g_a^4 \bigg[2b_{ij} - 8\mu^{ij}M_a\bigg]\bigg[C_a(i)\tau_{aR} - 3C(G_a) C_a(i) \bigg]\n
&+&\sum_a\sum_b g_a^2g_b^2 \bigg[2b_{ij} - 8\mu^{ij}M_a\bigg]C_a(i)C_b(i).
\eea
Finally, the renormalization group equations for the linear soft breaking parameter and the vacuun expectation values $v_i$ have been taken from ref.\cite{Fonseca:2011vn} and read
\bea
\frac{da^i}{dt}  = \frac{1}{16\pi^2} [\beta^{(1)}_{a}]^i +  \frac{1}{(16\pi^2)^2} [\beta^{(1)}_{a}]^i\n
\eea
with 
\bea
[\beta^{(1)}_a]^i &=& \frac12 \lambda^{iln}\lambda_{pln}a^p + \alpha^p \lambda_{pln}h^{iln} + \mu_{ik}\lambda_{kln}b^{ln} + 2\lambda^{ikp}(m^2)^l_p\mu_{kl} + h^{ikl}b_{kl}; \n
\big[\beta^{(2)}_a\big]^i &=& 2 \sum_a g^2_a C_a(l)\lambda^{ikl}\lambda_{pkl} a^p - \frac12 \lambda^{ikq}\lambda_{qst} \lambda^{lst}\lambda_{pkl}a^p - 4\sum_a g_a^2 C_a(l)(\lambda^{ikl}M_a - h^{ikl})\lambda_{pkl}\alpha^p \n
&-& \bigg[\lambda^{ikq}\lambda_{qst}h^{lst}\lambda_{pkl} + h^{ikq}\lambda_{qst}\lambda^{lst}\lambda_{pkl}\bigg]\alpha^p - 4\sum_a g_a^2C_a(l) \lambda_{jnl}\bigg[\mu^{nl}M_a - b^{nl}\bigg]\mu^{ij} \n
&-& \bigg[\lambda_{jnq}h^{qst}\lambda_{lst}\mu^{nl} + \lambda_{jnq}\lambda^{qst}\lambda_{lst}b^{nl}\bigg] \mu^{ij} + 4\sum_a g_a^2 C_a(l) \bigg[ 2\lambda^{ikl} \mu_{kl}|M_a|^2 - \lambda^{ikl}b_{kl}M_a \n
&-& h^{ikl}\mu_{kl} M_a^\dagger + h^{ikl}b_{kl} + \lambda^{ipl}(m^2)^k_p\mu_{kl} + \lambda^{ikp}(m^2)^l_p \mu_{kl} \bigg] - \bigg[\lambda^{ikq} \lambda_{qst}h^{lst}b_{kl} \n
&+& 8\delta^j_i \sum_a \bigg[g_a^4C_a(i){\rm Tr}[\tau_{aR}m^2] - C_a(G) |M_a|^2 \bigg] \n
&+& h^{ikq}\lambda_{qst}\lambda^{lst}b_{kl} + h^{ikq}h_{qst}\lambda^{lst}\mu_{kl} + \lambda^{ipq}(m^2)^k_p\lambda_{qst}\lambda^{lst}\mu_{kl} + \lambda^{ikq}\lambda_{qst}\lambda^{pst}(m^2)^l_p\mu_{kl} \n
&+& \lambda^{ikp}(m^2)^q_p\lambda_{qst}\lambda^{lst}\mu_{kl} + 2\lambda^{ikq}\lambda_{qsp}(m^2)^p_t\lambda^{lst}\mu_{kl} + \lambda^{ikq}h_{qst}h^{lst}\mu_{kl} \bigg].
\eea
\bea
\frac{d}{dt}v^i = v^p \bigg[\frac{1}{16\pi^2} \gamma_p^{(1)i} + \frac{1}{(16\phi^2)^2}\gamma_p^{(2)i}\bigg]
\eea
where the anomalous dimensions $\gamma^{(1)i} \et \gamma^{(2)i}$ are those of equations \eqref{eq:anomdim}.\\
\subsection{Scalar squared masses}
The equations for renormalizing the squared scalar masses are 
\bea
\frac{d}{dt} (m^2)^j_i = \frac{1}{16\pi^2} \big[\beta^{(1)}_{m^2}\big]^j_i + \frac{1}{(16\pi^2)^2} \big[\beta^{(2)}_{m^2}\big]^j_i
\eea
with
\bea
\big[\beta^{(1)}_{m^2}\big]^j_i &=& \frac12 \lambda_{ipq}\lambda^{pqn}(m^2)^j_n + \frac12\lambda^{jpq}\lambda_{pqn}(m^2)^n_i + 2 \lambda_{ipq}\lambda^{jpq}(m^2)^q_r + h_{ipq}h^{jpq} \n
&-& 8\delta^j_i\sum_a |M_a|^2g_a^2 C_a(i) + \sum_ag_a^2(T^A_a)^j_i {\rm Tr}[T^A_am^2];\n
\big[\beta^{(2)}_{m^2}\big]^j_i &=&  -\frac12(m^2)^l_i\lambda_{lmn}\lambda^{mrj}\lambda_{pqr}\lambda^{pqn} - \frac12 (m^2)^j_i\lambda^{lmn}\lambda_{mri}\lambda^{pqr}\lambda_{pqn} - \lambda_{ilm}\lambda^{jnm}(m^2)^l_r\lambda_{npq}\lambda^{rpq} \n
&-& \lambda_{ilm}\lambda^{jnm}(m^2)^r_n\lambda_{rpq}\lambda^{lpq} - \lambda_{ilm}\lambda^{jnr}(m^2)^l_n\lambda_{pqr}\lambda^{pqm} - 2\lambda_{ilm}\lambda^{jln}\lambda_{npq}\lambda^{mpr}(m^2)^q_r \n
&-& \lambda_{ilm}\lambda^{jln}h_{npq}h^{mpq} - h_{ilm}h^{jln}\lambda_{npq}\lambda^{mpq} - h_{ilm}\lambda^{jln}\lambda_{npq}h^{mpq} - \lambda_{ilm}h^{jln}h_{npq}\lambda^{mpq}\n
&+&\sum_a g_a^2\bigg[(m^2)^l_i\lambda_{lpq}\lambda^{jpq} + \lambda_{ipq}\lambda^{lpq}(m^2)^j_l + 4\lambda_{ipq}\lambda^{jpl}(m^2)^q_l + 2h_{ipq}h^{jpq} - 2h_{ipq}\lambda^{jpq}M_a\n
&&~~~~~~ - 2\lambda_{ipq}h^{jpq}M_a^\dagger + 4\lambda_{ipq}\lambda^{jpq}|M_a|^2 \bigg] \bigg[C_a℗ + C_a(q) - C_a(i)\bigg]\n
&-& 2 \sum_ag_a^2(T^A_a)^j_i(T^A_am^2)^l_r \lambda_{lpq}\lambda^{rpq} + 8 \sum_a\sum_b g_a^ag_b^2(T^A_a)^j_i {\rm Tr}[T^A_aC_b( R )m^2]\n
&+& \delta^j_i \sum_a\bigg[ 24 g_a^4 |M_a|^2 \big(C_a(i)\tau_{aR} - 3 C(G_a) C_a(i)\big) \n
&&~~~~~~~~~~ +\sum_b g_a^2g_b^2C_a(i)C_b(i)\big(32 |M_a|^2 + 8M_aM_b^\dagger + 8M_bM_a^\dagger\big) \bigg].
\n
\eea

Note that in the above equations, the traces are over all of the chiral superfields and the $C( r)$ are the quadratic Casimir invariants for the irreducible representations of chiral superfields in the traces. The terms which explicitely involve $T^{Aj}_i$ vanish for non-abelian groups.

\section{The minimal supersymmetric standard model}\label{sec:mssm}
The easiest way to embody supersymmetry in particle physics consists in associating to every Standard Model particle, a supersymmetric partner. In particular, to every fermion one would associate a scalar (the sfermion)  and to every boson a fermion whose name is made up from the addition of the boson name and the suffixe \textit{-ino}. We keep thus the same gauge group $SU(3)_c \times SU(2)_L \times U(1)_Y$ and the particle content in terms of superfields reads
\begin{itemize}
\item Gauge sector
	\begin{itemize}
	\item $V_1 = (\singlet,\singlet,0)$ associated to $U(1)$;
	\item $V_2 = (\singlet,\triplet,0)$ associated to $SU(2)$;
	\item $V_3 = (\octet,\singlet,0)$ associated to $SU(3)$.
	\end{itemize}
\item Matter sector
	\begin{itemize}
	\item Left-handed quarks $Q^i = \begin{pmatrix} u_L \\ d_L \end{pmatrix} = (\triplet,\doublet,\frac16)$;
	\item Left-handed up-type anti-quarks $U^i = (\triplet,\doublet,-\frac23)$;
	\item Left-handed down-type anti-quarks $D^i =  (\triplet,\singlet,\frac13)$;
	\item Left-handed leptons $L^i = \begin{pmatrix} \nu_L \\ e_L \end{pmatrix} = (\singlet,\doublet,-\frac12)$;
	\item Left-handed charged anti-leptons $E^i = (\singlet,\singlet,\singlet)$.
	\end{itemize}
\end{itemize}
In the sequel, the supersymmetric partners of the Standard Model particles will be denoted with a tilde symbol; e.g. the scalar superpartner of the quark will be written $ \tilde{Q}_L \mbox{ or } \tilde{u}_L$ if we were speaking about the up-type scalar quark. The number between brackets are the quantum numbers of the fields under the gauge group. The index $i=1,2,3$ is a flavor index running over the three families of quarks and leptons. Auxiliary fields were omitted for simplicity. The superpotential being a holomorphic function, the Higgs sector is composed of two chiral superfields $H_u$ and $H_d$ to give masses to both up-type and down-type fermions as well as to avoid chiral anomalies.
\begin{itemize}
\item $H_u = (H_u,\tilde{H}_u) = (\singlet,\doublet,\frac12)$ gives masses to the up-type fermion;
\item $H_d = (H_d,\tilde{H}_d) =(\singlet,\doublet,-\frac12)$ gives masses to the down-type fermions.
\end{itemize}
The most general superpotential for such a particle content is straightforward to obtain. It suffices to build gauge group invariant products. For example, no linear term is allowed as no gauge singlet exists. The superpotential is
\bea \label{rpv superpot} W(\Phi) &=& \lambda'' U D D + \lambda' Q\cdot L D + \lambda L \cdot L E + \mu' H_u \cdot L \n
&-& y_e L\cdot H_d E - y_d Q \cdot H_d D + y_u Q \cdot H_u U + \mu H_u \cdot H_d \eea
where we have omitted all the indices for clarity and noted by a dot the $SU(2)$ invariant product. \\

The terms in the first line of the superpotential are problematic as they induce lepton and/or baryon number violating processes. Such processes have not been observed experimentally and, more importantly, the parameters $\lambda'' \mbox{ and } \lambda'$ induce a too short proton lifetime if not suppressed. It is thus common to suppose the existence of a discrete symmetry, namely the R-parity which is a multiplicative quantum number defined as $(-1)^{3B - L + 2S}$ where $B$ is the baryon number, $L$ the lepton number and $S$ the spin of the particle. The conservation of the latter implies that at every vertex, the product of the $R$-parity number for all the fields involved is equal to +1. 
It is also remarkable that all Standard Model's particles have $R$-parity equal to $+1$ while their supersymmetric counterparts have $-1$ $R$-parity. This implies that supersymmetric particles always appear by pair at vertices. Phenomenologically, when $R$-parity is conserved, the lightest supersymmetric particle, dubbed the LSP, is also a stable particle. If the LSP is electrically neutral it can be considered as a good candidate for dark matter. Finally, the superpotential of the $R$-parity conserving minimal supersymmetric standard model (MSSM) is
\bea \label{nrpv superpot} W = y_e L\cdot H_d E + y_d Q \cdot H_d D + y_u Q \cdot H_u U + \mu H_u \cdot H_d. \eea
Its lagrangian can be deduced simply from \eqref{general lagrangian}. One must however be careful to take into account the behaviour of the various fields under the gauge group, for example the term $(\Phidagger e^{-2gV} \Phi)_{\tttt}$ reads for left-handed quarks
\bea Q^\dagger e^{-2 \frac16 g' V_1} e^{-2g V_2} e^{-2g_sV_3} Q \eea
where $g',g,g_s$ are the coupling constants associated to the gauge groups $SU(3)_c, SU(2)_L \mbox{ and } U(1)_Y$ respectively.\\

As seen in section \ref{sec:susy breaking}, supersymmetry has to be broken to account for the experimental constraints on the mass spectrum and the hierarchies between particles. Certainly the most studied and explored mechanisms to induce supersymmetry breaking in the MSSM are those inspired by supergravity with a minimal (diagonal) K\"ahler potential. These scenarios imply an organizing principle at very high scale leading to only three free parameters: the universal scalar masses $m_0$, the universal gaugino masses $m_{\frac12}$ and the universal trilinear soft-breaking couplings $A_0$. The value of the bilinear soft breaking coupling $B$ is constrained by the minimization of the scalar potential to ensure a correct electroweak symmetry breaking at low scale. as well as supersymmetry breaking but its sign remains as a free parameter.  In this case, we often speak of constrained minimal supersymmetric standard model (cMSSM).\\
Such mechanism is very interesting from a theorist point of view as it reduces drastically the number of free parameters in the theory. Indeed, at the low scale, the minimal supersymmetric standard model counts more than 100 free parameters mostly due to the soft breaking terms while at the high scale we only have the 3 parameters above plus the sign of $\mu$ and the ratio of vacuum expectation values of the Higgs fields $\tan\beta = \frac{\langle H_u \rangle}{\langle H_d \rangle}$
given, in general, at the $Z$-boson scale. The link between these two scales is ensured by the renormalization group equations flow.\\
\subsection{Example of calculation of the renormalization group equations}
In this subsection, we will apply the generic formulas for the renormalization groupe equations given in section\ref{sec:rge-formulas} to obtain the analytic formula of the two-loop RGEs for the gauge coupling constant $g$ and the Yukawa coupling $y_u$ introduced above. 
\paragraph{The gauge coupling constant $g$}\mbox{}\\
The general formula is:
$$ \frac{d}{dt}g_i = \frac{1}{16\pi^2}\beta_{g_i}^{(1)} + \frac{1}{(16\pi^2)^2}\beta_{g_i}^{(2)} $$
with
\bea
\beta_{g_a}^{(1)} &=& g_a^3 \big[ \tau_{aR} - 3C( G_a ) \big]; \n
\beta_{g_a}^{(2)} &=& g_a^3 \bigg[ -6 g_a^2 \big(C(G_a)\big)^2 + 2g_a^2C(G_a) \tau_{aR} + 4\sum_{b}g_b^2\tau_{a R }C_b( R ) \bigg] - g_a^3 \lambda^{ijk}\lambda_{ijk} \frac{C(k)}{d(G)}.\nonumber
\eea
At the one-loop level, only quantities coming from group theory are present. To calculate them, it is sufficient to only consider the chiral superfields transforming non trivially under $SU(2)_L$ that is $Q, L, H_u \et H_d$. We have, thus:

$$ \beta_{g}^{(1)} = g^3 \Big[ \underbrace{3\times 3 \times \frac12}_{Q}~~~+~~~\underbrace{3 \times \frac12}_{L}~~~+~~~ \underbrace{1\times\frac12}_{H_u}~~~+~~~\underbrace{1\times\frac12}_{H_d}~~~-~~~\underbrace{3 \times 2}_{3C(G_a)}\Big] = 1$$
where the factor $\frac12$ is the Dynkin index in the fundamental representation of $SU(2)_L$ and $2$ is the quadratic Casimir invariant for $SU(2)_L$ in the adjoint representation. Some remarks:
\begin{itemize}
    \item The left-handed quarks come under three different families and three different colors hence the factor $3\times 3$.
    \item We have three families of leptons, hence the factor $3$.
    \item Finally, the Higgs fields only contribute with a factor of one.
\end{itemize}
At the two-loop level, calculations are a little more tricky so let us split the calculation into four steps:
\begin{enumerate}
\item The term $ -6 g_a^2 \big(C(G_a)\big)^2$ simply returns, in our case, $$ - 6 g^2 (2)^2 = -24 g^2 $$
\item The second term, $2g_a^2C(G_a) \tau_{aR}$ is just the quadratic Casimir invariant times the total Dynkin index summed over all chiral superfields. It involves the same reasoning than above and returns
$$ 2g^2C(G_a) \tau_{aR} = 28 g^2 $$
\item the last term in the brackets $4\sum_{b}g_b^2\tau_{a R }C_b( R ) $ is the sum of the Dynkin Indices of all the chiral superfields weighted by the quadratic Casimir invariant in the representation of every field. There is also a sum over the gauge groups:
\bea
    4\tau_{a R }\sum_{b}g_b^2C_b( R ) &=& 4 g'^2 \times\frac35\times\frac12 \times\Big[ \underbrace{\frac14}_{H_u}~~~+~~~ \underbrace{\frac14}_{H_d} ~~~+~~~ \underbrace{3\times\frac14}_{L}~~~+~~~ \underbrace{3\times3\times\frac{1}{36}}_{Q} \Big] \n
&&+ 4g^2 \times \frac12\times\Big[ \underbrace{\frac34}_{H_u}~~~+~~~ \underbrace{\frac34}_{H_d} ~~~+~~~ \underbrace{3\times\frac34}_{L}~~~+~~~ \underbrace{3\times3\times\frac{3}{4}}_{Q} \Big] \n
&&+ 4g_s^2 \times\frac12\times \Big[\underbrace{3\times3\times\frac{4}{3}}_{Q} \Big] \n
4\tau_{a R }\sum_{b}g_b^2C_b( R ) &=& \frac95 g'^2 + 21 g^2 + 24 g_s^2 \nonumber
\eea
where the overal factor $\frac35$ multiplying $g'$ in the first line is the normalization of the hypercharge, the factor $\frac34$ in the second is the quadractic Casimir invariant for the fundamental representation of $SU(2)_L$ and $\frac43$ is the one for the fundamental representation of $SU(3)_c$.
\item To get the last term, \textit{i.e.} the one involving Yukawa couplings, one should pay attention to the fact that an implicit sum running over all chiral superfields is understood for the indices $i,j,k$. The only non-zero terms will be those involving the Yukawa couplings present in the superpotential \eqref{nrpv superpot} 
\bea
  g_a^3 \lambda^{ijk}\lambda_{ijk} \frac{C(k)}{d(G)} &=& 2\times \frac13 g^3 \Bigg\{\lambda^{L,H_d,E}\lambda_{L,H_d,E} \Big(\underbrace{\frac34}_{H_d} + \underbrace{3\times\frac34}_{L} \Big) + 3\times \lambda^{Q,H_d,D} \lambda_{Q,H_d,D} \Big( \underbrace{3\times\frac34}_{Q} +  \underbrace{\frac34}_{H_d} \Big) \n
&&+ 3\times \lambda^{Q,H_u,U} \lambda_{Q,H_u,U} \Big( \underbrace{3\times\frac34}_{Q} + \underbrace{\frac34}_{H_u} \Big)
 \Bigg\} \n
  g_a^3 \lambda^{ijk}\lambda_{ijk} \frac{C(k)}{d(G)} &=& g^3 \Big[ 2\times{\rm Tr}(y_e y^\dagger_e) + 6\times{\rm Tr}(y_d y^\dagger_d) + 6\times{\rm Tr}(y_u y^\dagger_u) \Big] \nonumber
\eea
 where the overall factor $2$ is added to account for the permutations, the dimension of the adjoint representation $d(G)$ is equal to 3 in the case of $SU(2)_L$ hence the other overall factor $\frac13$,  $\frac34$ is the quadratic Casimir invariant for the fundamental representation in our case and, finally the factor of $3$ multiplying the quark Yukawa couplings are added to take into account the fact that these particles carry a color index.
\end{enumerate}
Adding all the contributions we get:
\bea 
\frac{d}{dt}g &=& \frac{g^3}{16\pi^2} + \frac{9g'^2g^3}{1280\pi^4} + \frac{3g_s^2g_w^3}{32\pi^4}+\frac{25g^5}{256\pi^4}\n
&& -\frac{3}{128\pi^4}{\rm Tr}(y_dy^\dagger_d) - \frac{3}{128\pi^4}{\rm Tr}(y_uy_u^\dagger) - \frac{1}{128\pi^4}{\rm Tr}(y_ey_e^\dagger) \nonumber
\eea

\paragraph{The Yukawa coupling $y_u$}\mbox{}\\
We turn now to the Yukawa coupling $y_u$ whose running with respect to energy is given by the general formula:
\bea
\frac{d}{dt} \lambda^{ijk} &=& \lambda^{ijp} \bigg[ \frac{1}{16\pi^2} \gamma_p^{(1)k} +  \frac{1}{(16\pi^2)^2} \gamma_p^{(2)k} \bigg] + (k \leftrightarrow i) + (k \leftrightarrow j)\n
\mbox{with}\n
\gamma_i^{(1)j} &=& \frac12 \lambda_{ipq}\lambda^{jpq} - 2\delta^j_i \sum_a g^2_a C_a(i)\n
\gamma_i^{(2)j} &=& -\frac12 \lambda_{imn}\lambda^{npq}\lambda_{pqr}\lambda^{mrj} + \sum_a g_a^2 \lambda_{ipq}\lambda^{jpq}\bigg[2C_a( p )- C_a(i)\bigg]\n
&+&2\delta_i^j \bigg[\sum_a g_a^4 C_a(i) \tau_{aR}  + 2 \sum_a \sum_b g_a^2g_b^2C_a(i)C_b(i) - 3\sum_a g_a C(G_a) C_a(i) \bigg]\nonumber
\eea
We choose to present here only the calculation of the one-loop anomalous dimension $\gamma^{(1)}$ as the calculation of the second one involves quantities that we have already met. The Yukawa coupling $y_u$ involves the superfields $Q, H_u \et U$ which allows us to re-write the above formula as follows:

\bea
\frac{d}{dt} \lambda^{Q H_u U} &=& \frac{1}{16\pi^2}\lambda^{Q H_u p} \gamma^{(1)U}_p + \frac{1}{16\pi^2}\lambda^{U H_u p} \gamma^{(1)Q}_p + \frac{1}{16\pi^2}\lambda^{Q U p} \gamma^{(1)H_u}_p\nonumber
\eea
where the index $p$ runs over all chiral superfields. The first factor $\lambda^{Q H_u p}\gamma^{(1)U}_p$ returns:
\bea
\lambda^{Q H_u p}\gamma^{(1)U}_p &=& -\frac12 \lambda_{p i j}\lambda^{U i j} - 2\delta^{U}_{p} \sum_a g_a^2 C_a(p)\nonumber
\eea
where here also, an index appearing once as a subscript and once as a superscript is summed. From the MSSM's superpotential, we deduce that the only non zero terms are:
\bea
\lambda^{Q H_u U}\gamma^{(1)U}_U &=& -\frac12 2\times \lambda^{Q H_u U}\lambda_{U H_u Q}\lambda^{U H_u Q} - 2 \big( g^2 \frac49 + g^2 \frac43 + g_s^2 \frac34 \big) \nonumber
\eea
A few remarks are in order here. First, the overall factor of two is added to take into account the permutations. Secondly one must pay attention to the index structure of this expression to know which are free and which are summed. Let $f_i$ be a set of flavor indices, $c_i$ color indices and $i_j$ a set of $SU(2)_L$ indices. We then re-write the expression with the indices explicit:
\bea 
\frac{d}{dt}\lambda^{Q_{c_1,f_1,i_1} H^{i_2}_u U^{c_2 f_2}} &=& \dots + \lambda^{Q_{c_1,f_1,i_1} H_u^{i_2} U^{c_3 f_3}}(\gamma^{(1)U}_U)^{c_2 f_2}_{c_3 f_3}\n
&=&\dots - \lambda^{Q_{c_1,f_1,i_1} H_u^{i_2} U^{c_3 f_3}} \lambda_{U_{c_3 f_3} H_u^{i_3} Q_{c_4 f_4} }\lambda^{U^{c_2 f_2} H_{u i_3} Q^{c_4 f_4} } \n
&& - 2 \big( g^2 \frac49 + g^2 \frac43 + g_s^2 \frac34 \big)\delta_{c_1}^{c_2} \epsilon_{i_1}^{i_2} \delta_{f_1}^{f_2} \n
\nonumber\eea
where dots stand for the other terms we did not include yet. The Yukawa couplings being diagonal in color space and $SU(2)$ products returning the totally antisymmetric tensor $\epsilon$ this formula reduces to:
\bea
\frac{d}{dt}y_u{}^{f_1}{}_{f_2} = -2 \times y_u{}^{f_1}{}_{f_3}y^\dagger_u{}^{f_3}{}_{f_4} y_u{}^{f_4}{}_{f_2} -2 \big( g^2 \frac49 + g^2 \frac43 + g_s^2 \frac34 \big) \nonumber 
\eea
where the new overall factor $2$ is due to the $\delta^{i_3}_{i_3}$ appearing from the product of the two last $\lambda$ couplings in the second line. Applying the same procedure for the other coefficients, we find the formula governing the evolution with respect to energy of the Yukawa coupling $y_u$:
\bea
\frac{d}{dt}y_u = \frac{1}{16\pi^2} y_u \Big[ 3 {\rm Tr}(y_u y_u^\dagger) + 3 y_uy_u^\dagger  + y_d y_d^\dagger - \frac{16}{3}g_s^2 - 3 g^2 - \frac{13}{15} g'^2 \Big] \nonumber
\eea
The trace appears from the term 
$$\frac{1}{16\pi^2}\lambda^{Q U p} \gamma^{(1)H_u}_p$$
because the superfield $H_u$ in $\gamma^{(1)H_u}_p$ does not carry any flavor index, in other terms the flavor index structure is:
$$ \lambda^{Q_{f_1} U^{f_2} H_u} (\gamma^{(1)H_u}_{H_u})_{f_3}^{f_3}.$$
\subsection{Status of the Minimal Supersymmetric Standard Model}
Other supersymmetry breaking mechanisms have been applied to the MSSM to explore their phenomenology. Be it gauge mediated (for a review see \cite{Ambrosanio:1997rv}) or anomaly mediated (see \cite{Huitu:2002fg}), the most important feature is that some unification mechanisms occur at high scale leading to a handful number of free parameters rendering these models highly predictive and easy to explore phenomenologically. The MSSM brought to particle physics other theoretical successes among which we can cite the fact that its $R$ conserving version provides a natural candidate for dark matter namely the lightest neutralino (admixture of the superpartners of the $W^3$-boson, the wino, and the Higgs bosons, the Higgsinos). The renormalization group equations also predict the unification of the gauge couplings at the one loop level and the hierarchy problem is solved provided that the breaking of supersymmetry happens around the TeV scale. In fact these successes, led to a very rich litterature about the MSSM and many predictions have been made. Unfortunately, however, this model, in addition to its inability to explain some theoretical issues, is more and more constrained on the experimental side.

\paragraph{Theoretical issues} Though a natural candidate for dark matter arises in the MSSM, this is done by supposing the discrete symmetry $R$-parity which may seem unnatural if one assumes the validity of the MSSM until the grand unification scale. This is due to the fact that the couplings $\lambda, \lambda' \et \lambda''$ are not forbidden by supersymmetry. Moreover, it is somehow difficult to take into account the smallness of neutrino masses in the framework of the minimal supersymmetric standard model and one is thus forced to imagine new mechanisms to generate these masses\cite{Hirsch:2004he}. Finally, in the process of ``supersymmetrization" of the Standard Model of particle physics, a $\mC\mP$ violation problem appears namely the supersymmetric $\mC\mP$ problem, in addition to the Strong $\mC\mP$ problem that is inherited from the Standard Model. These two issues pose a fine-tuning issue. \\
\begin{figure}
\hskip-1truecm
\subfigure[]{%
\label{fig:cms susy spectrum}   \includegraphics[scale=0.5]{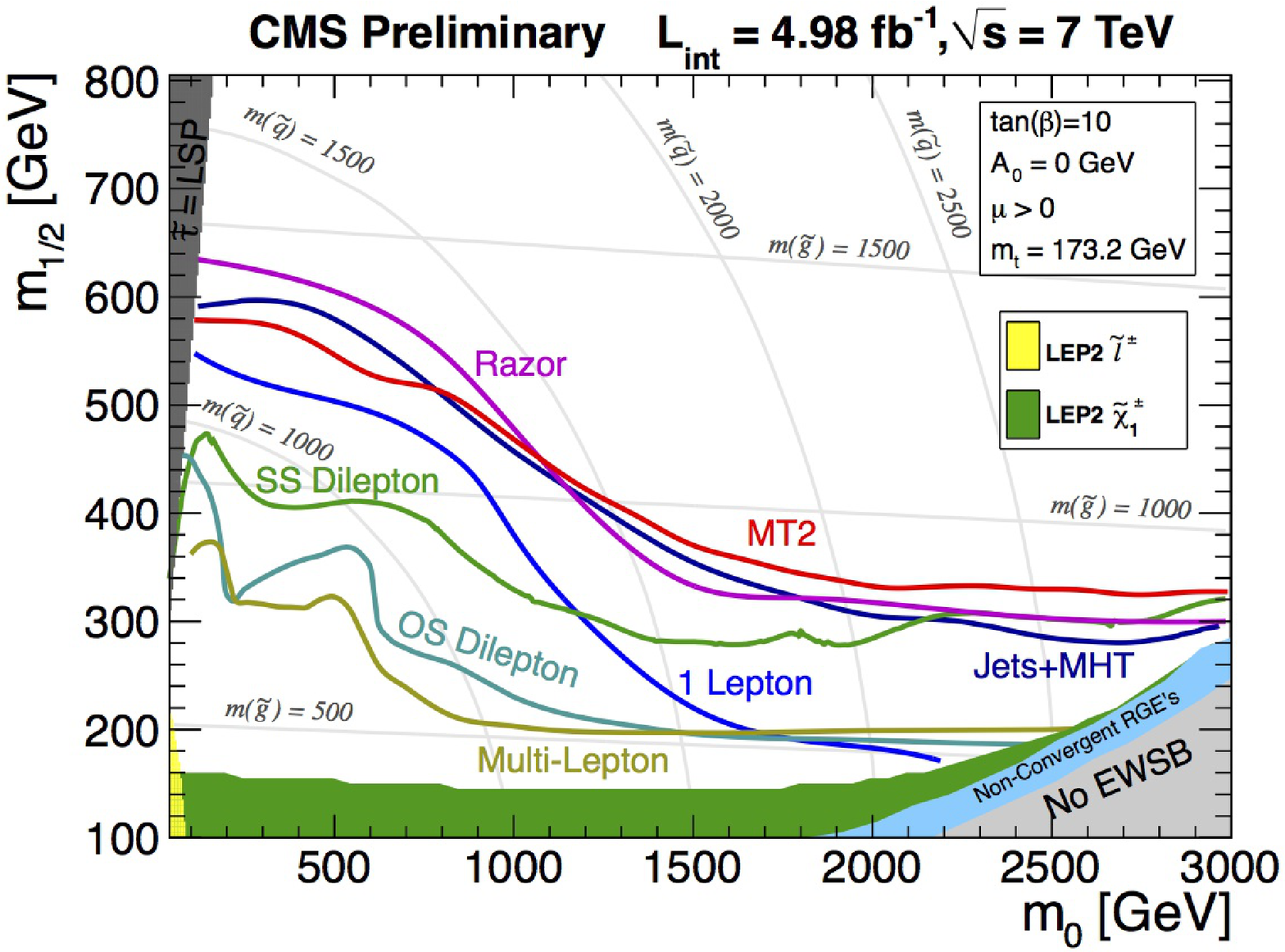} }%
\subfigure[]{%
\label{fig:atlas susy spectrum}   \includegraphics[scale=0.4]{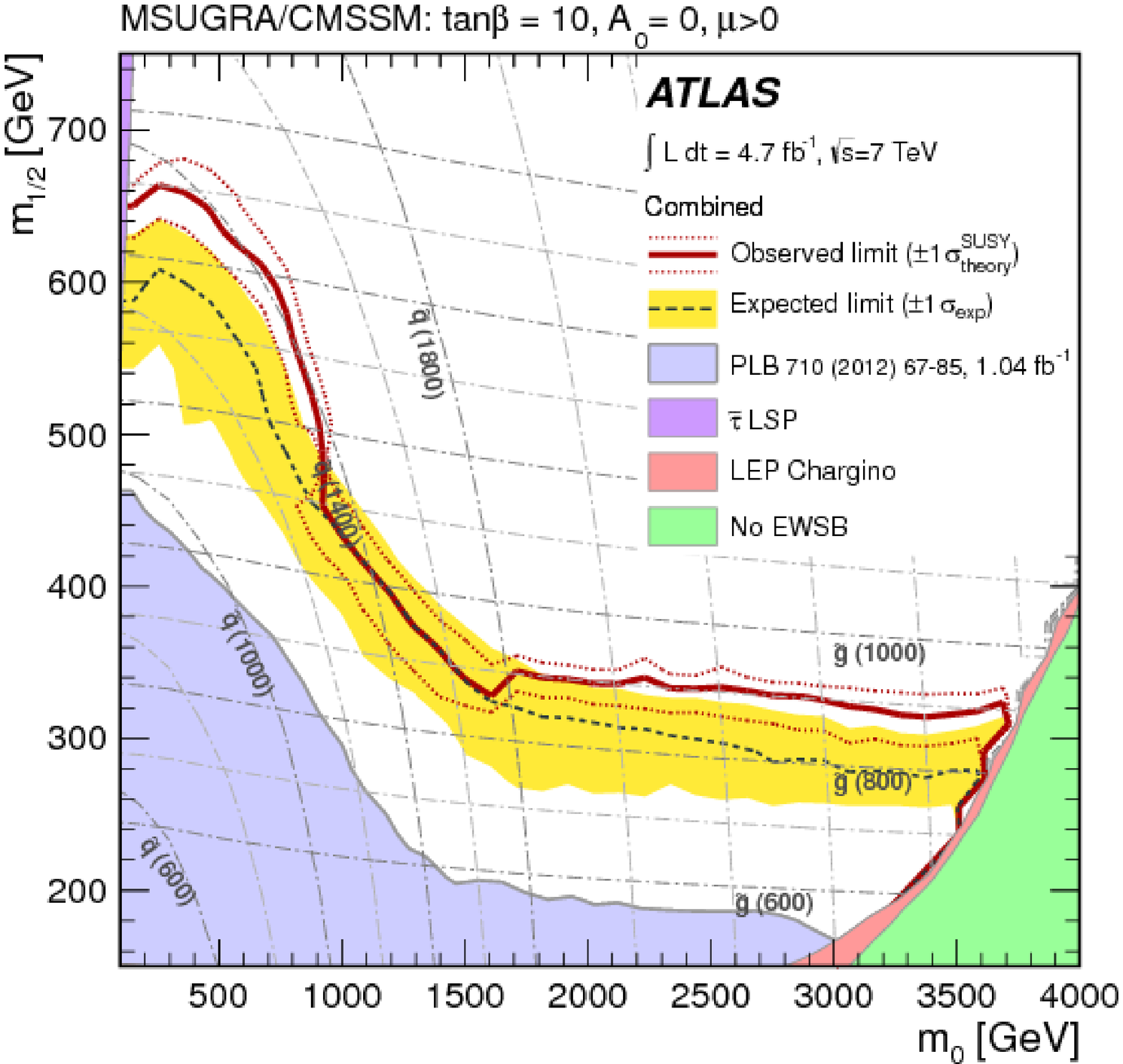}}\\ 

\caption{\footnotesize\label{susy limits} CMS (left panel) and ATLAS (right panel) 2011 exclusion limits at the $95\%$ CLs exclusion limits on the $(m_0,m_{\frac12})$ plane of cMSSM taken from. These exclusion limits (everything below the curves is excluded) are set for a center of mass energy $\sqrt{s}=7TeV$ and $4.98~fb^{-1}$ of data for CMS and $4.7~fb^{-1}$ for ATLAS}
\end{figure}

\paragraph{Experimental constraints} Though the Large Hadron Collider has been collecting data since early 2010, not a single superparticle (nor exotic) has been discovered yet. This led particle physicists to set limits on the mass of the superparticles, pushing the spectrum of the MSSM higher and higher in energy (see figure \ref{susy limits}). The problem is that we are expecting the supersymmetric particles to have masses lying in the range $100 ~-~ 1000~ {\rm GeV}$ so that the
stop (scalar partner of the top quark) and the Higgsinos (fermionic partners of the Higgses) do not have too large masses ($\lesssim 1~{\rm TeV}$) to avoid too much fine tuning in the Higgs sector. This is known as the naturalness problem or the little $\mu$ problem as it is directly linked to the value of the $\mu$ parameter in the superpotential \eqref{nrpv superpot}. The latter has to be of the order of $100~{\rm GeV}$ and the question is to know why this value is so far from the
$M_{Planck}$. This constraint put together with the recent discovery of a Higgs-like scalar \cite{:2012gu,:2012gk} at the LHC reduces significantly the allowed parameter space. Indeed, though the cMSSM predicts a Higgs boson mass lighter than around $ 135~{\rm GeV}$ \cite{Djouadi:2005gj,Akula:2011aa} setting the Higgs scalar mass at 125 GeV turns out to be complicated. Indeed, the latter receives important one-loop corrections from the top quark and its supersymmetric partner, the stop loops as can be seen from the approximate one-loop formula for the Higgs mass
$$ m_h^2 = m_z^2 \cos^2 2\beta + \frac{3m_t^4}{4\pi^2 v^2}\Bigg[ \log\Big(\frac{M_s^2}{m_t^2}\Big) + \frac{X_t^2}{M_s^2}\Big(q-\frac{X_t^2}{12M_s^2}\Big) \Bigg] $$
where $M_s = \sqrt{m_{\tilde{t}_1 \tilde{t}_2}}$ characterizes the scale of the two lightest scalar top quarks $\tilde{1}_1 \et \tilde{t}_2$ and $X_t = A_t - \mu \cot \beta$ characterizes the mixing in the stop sector with $A_t$ being the trilinear soft supersymmetry breaking parameter coupling the up-type squarks. This formula also indicates that to have a large Higgs mass one needs either a large mixing in the squark sector or high scalar masses. The first option implies that soft supersymmetry breaking terms are large but stop masses can be as low as few hunder GeV the second option implies automatically multi TeV stop masses. The problem with these two options is that either large $A_t$ or large stop masses imply large fine tuning, at least in the cMSSM \cite{Draper:2011aa,Hall:2011aa}.

\paragraph{Supersymmetry status} Since its theoretical construction, supersymmetry has been a very exciting  and promising field of research. Its ability to cure the hierarchy problem of the Standard Model as well as the facility to be embedded in more fundamental theories (e.g. supergravity) and to provide a candidate for dark matter have contributed to make of it one the most popular and studied alternatives to the Standard Model.\\

However, the experimental limits which are pushing the sparticles (i.e. supersymmetric particles) spectrum to high energies contribute to drastically narrow down the available parameter space for the constrained minimal supersymmetric standard model. Though this version of the MSSM is not completely dead yet \cite{Nath:2012nh,Allanach:2013cda}, one is more and more pushed to explore other alternatives to this particular model. Be it by choosing other supersymmetry breaking mechanisms like the gauge mediated one\cite{gmsb}, by extending the number of free parameters like in the so-called ``phenomenological MSSM" \cite{Arbey:2012dq} or by enriching the field content like in the so-called NMSSM \cite{Ellwanger:2009dp} and/or the gauge group, there is a huge number of alternatives one can study. However the attractiveness of these alternatives depends on both the solutions they bring to the open questions of the MSSM and the number of questions they rise but also and maybe more importantly on how complicated they are. Indeed, if the number of free parameters in the MSSM (at least at low energy) is already bigger than $100$, one might reasonably think that for more complicated models we get more free parameters. The calculational techniques get thus more complicated and equations impossible to solve analytically (except in some very specific cases like in ref. \cite{Mambrini:2001wt}). One is thus forced to turn towards automated tools to circumvent this complexity and do phenomenology.\\

In this scope, I present in the following section a phenomenological study of the left-right symmetric supersymmetric model, a model by far not minimal and which leads to interesting phenomenology. Besides highlighting the need of studying non minimal models and putting some limits on the possibility to discover it, I will also use this chapter to emphasize the importance of automated tools and the big help they bring to phenomenology. Almost as a consequence, the next chapter will be devoted to presenting an automated spectrum generator generator developped during my thesis and which might be of big help in such studies.

\chapter{The left-right symmetric supersymmetric model}\label{chap:lrsusy}

\section{Left-right symmetry in particle physics.}
\indent The first models exhibiting a symmetry between left-handed and right-handed fermions appeared in the early '70s. At that time, theorists were looking for a solution to $\mP$ and $\mC\mP$ violation and introducing this symmetry seemed to be a possible solution \cite{PhysRevD.11.2558,Pati:1974yy}. \\

Through the years, a rich literature has emerged proving how interesting and exciting these models could be. Indeed, assuming a symmetry between left-handed and right-handed fermions may lead to solving the strong $\mC\mP$ problem \cite{Guadagnoli:2011id}, explaining naturally the smallness of neutrinos masses through the tantalizing seesaw mechanism\footnote{This feature will be illustrated in the next sections.} \cite{Mohapatra:1980yp,Gu:2010zv,Chao:2007rm,FileviezPerez:2008sr} as well as providing a viable candidate for dark matter \cite{Nemevsek:2012cd, Guo:2011mi}. It was also found in ref.\cite{Shaban1992281} that this class of models could predict gauge couplings unification. However, at least one big problem was still there, compared to the Standard Model, namely the hierarchy problem. \\

Supersymmetrizing the left-right symmetric model offers a very elegant way to solve the hierarchy problem while keeping the non-supersymmetric version's advantages. Cveti\u{c} and Pati published in 1983 \cite{Cvetic:1983su} one of the first models ``linking $N=1$ supergravity with the minimal left-right symmetry". Since then, a rich literature emerged \cite{Ng:1983kd,Kuchimanchi:1993jg,Mohapatra:1996vg,Raidal:1998dz,Babu:2008ep,Borah:2012bb} and many nice features were brought to light. Some are common with the non supersymmetric version such as the ability to solve the strong $\mC\mP$ problem \cite{Babu:2001se,Mohapatra:1996vg,Babu:2008ep}, easy embedding of the seesaw mechanism \cite{Chakrabortty:2010rq,Borah:2012bb} leading to sizeable lepton flavor violation effects compared to the MSSM \cite{Esteves:2010si,Vicente:2010wa}. Some others are specific to the supersymmetric case like the solving of the susy $\mC\mP$ problem and the existence of a vacuum state conserving R-parity \cite{Babu:2008ep}.\\

From the experimental side this class of models is far from being excluded. The {\sc CMS} collaboration assuming exact left-right symmetry, {\ie} equality between $SU(2)_L \et SU(2)_R$ gauge coupling constants and a tiny mixing angle between $SU(2)_L$ and $SU(2)_R$ fields, used both runs of 2011 and 2012 to look for an excess in the events over the SM background due to the leptonic decays of the gauge boson $W_R$ associated the group $SU(2)_R$. No excess was found and a limit of 2.9 TeV on the mass of the latter was established. The {\sc ATLAS} collaboration following an effective-Lagrangian approach used 14.3 ${\rm fb}^{-1}$ of data acquired with a center-of-mass energy of 8 TeV in order to exclude the $W_R$ up to a mass of 1.84 TeV\cite{ATLAS-CONF-2013-050}.\\

In this chapter we propose to lead an analysis of such models starting from the construction of the model and pursuing the analysis until the Monte Carlo simulation. To this end, we write piece by piece all the necessary ingredients to build the Lagrangian of this model and derive the mass matrices associated to the neutralinos and charginos. The latter sector being very rich, we ocus in our Monte Carlo simulation on the production and the decay channels of these particles and
analyze the events leading to at least one charged light lepton in the final state. This choice being justified by the fact that the Standard Model background, overwhelming in the pure QCD regime becomes less and less dominant with an increasing number of charged light leptons in the final state to finally vanish for high multiplicities. The aim of this simulation will be twofold. First we will use the same selection criteria (large missing transverse energy, hard leptons) usually
present in analyses of supersymmetric models to show that they could be applied successfully in the case of left-right symmetric supersymmetric models. Secondly, we perform, with the same selection criteria, a comparison between our model and the Minimal Supersymmetric Standard Model. \\

This chapter is organized as follows. First we start in section \ref{sec:non susy} with a brief description of the non-supersymmetric version to illustrate some of its appealing properties that we shall meet also in supersymmetry and give some of the recent developments in this context. In section \ref{sec:lrsusy}, we present the details of the model building, derive the analytic expressions for the mass matrices of the charginos and neutralinos and give the minimization equations derived from the scalar potential. In section \ref{sec:lrsusy pheno}, we describe the phenomenological analysis we have performed and the results we have found. Finally, in section \ref{sec: lrsusy conclusion} we give our conclusions together with some ideas for some future work.\\

Finally we draw the reader's attention to the somehow similar study that was performed by Mohapatra, Setzer and Spinner in 2008 and published in \cite{Mohapatra:2008gz}. In the latter, the authors use a top-down\footnote{Here top-down means that they use the renormalization group equations to link high and low energy.} approach in the framework of a left-right symmetric supersymmetric where supersymmetry breaking is mediated via gauge anomalies and give some qualitative signatures that could arise, but do not perform any Monte Carlo simulation. Our study only focuses on the low energy phenomena but with a complete Monte Carlo simulation of the Standard Model Background.

\section{Non-supersymmetric left-right symmetric model} \label{sec:non susy}
\subsection{General considerations}
The gauge group of the left-right symmetric model can be taken to be 
\bea \label{gauge group} SU(3)_C \times SU(2)_L \times SU(2)_R \times U(1)_{B-L} \eea
where $B-L$ is the difference between baryon and lepton number\footnote{For an interpretation of the $U(1)$ see \cite{Marshak:1979fm}.}. The gauge sector is then very easily defined as it suffices to assign one vector multiplet for each direct factor of the gauge group {\ie} multiplets lying in the corresponding adjoint representation\\
\begin{center}\begin{tabular}{c c c c }
$SU(3)_C$ & $\rightarrow$ & $g^a = (\octet,\singlet,\singlet,0)$ \\
$SU(2)_L$ & $\rightarrow$ & $W_L = (\singlet,\triplet,\singlet,0)$ \\
$SU(2)_R$ & $\rightarrow$ & $W_R = (\singlet,\singlet,\triplet,0)$  \\
$U(1)_{B-L}$ & $\rightarrow$ & $B' = (\singlet,\singlet,\singlet,0)$ 
\end{tabular}\end{center}
where $g^a$ are the gluons. The electric charge is then defined using the Gell-Mann-Nishijima relation
\bea Q = T_{3L} + T_{3R} + \frac{B-L}{2}. \eea

The matter sector is a little more complicated as it depends on the definition we give to left-right symmetry. Indeed, while the known left-handed quarks and leptons remain the same as in the Standard Model\footnote{that is left-handed fermions belonging to the fundamental representation of $SU(2)_L$. } left-right symmetry can be defined in two ways. Originally (see for example ref.\cite{Pati:1974yy}), it was defined as the parity operation supplemented by the exchange of the left and right $SU(2)$ gauge groups. In this case, the field content is 
\bea
Q_L &=& \begin{pmatrix} u_L \\ d_L \end{pmatrix} = (\triplet, \doublet, \singlet, \frac13),~~~ Q_R = \begin{pmatrix} u_R \\ d_R \end{pmatrix} = (\triplet, \singlet, \doublet, \frac13), \n
L_L &=& \begin{pmatrix} \nu_L \\ e_L \end{pmatrix} = (\singlet,\doublet,\singlet,-1),~~~ L_R = \begin{pmatrix} \nu_R \\ l_R \end{pmatrix} = (\singlet,\singlet,\doublet,-1). \nonumber 
\eea
Another way to define the left-right symmetry is to use the charge conjugation. In this case, $SU(2)_R$ doublets become
\bea
 Q_R = \begin{pmatrix} u^c_R & d^c_R \end{pmatrix} = (\utilde{\mathbf{\bar{3}}}, \singlet, \utilde{\mathbf{2}}^*, -\frac13),~~~ L_R = \begin{pmatrix} \nu^c_R & l^c_R \end{pmatrix} = (\singlet,\singlet,\utilde{\mathbf{2}}^*,1) \nonumber 
\eea
where the superscript $c$ represents charge conjugation. The main difference between these approaches lies in the chirality of the fermions. Parity transformation induces $SU(2)_R$ fermions to be right-handed while charge conjugation leaves them left-handed. While from group theory arguments these two definitions may seem equivalent, on the phenomenology side, however, they lead to different constraints on the Yukawa couplings in the Lagrangian. Indeed, while parity induces
$$
\begin{array}{l l l}
	\mP: \left\{ \begin{array}{l} 
		Q_L \leftrightarrow Q_R \\
	    \Phi \leftrightarrow \Phi^\dagger \end{array}\right. &
\Rightarrow &
y = y^\dagger
\end{array}
$$
charge conjugation leads to
$$
\begin{array}{l l l}
	\mC: \left\{ \begin{array}{l} 
		Q_L \leftrightarrow (Q_R)^c \\
	    \Phi \leftrightarrow \Phi^t \end{array}\right. &
\Rightarrow &
y = y^t
\end{array}
$$
where we have noted $y$ a generic yukawa coupling and $\Phi$ a Higgs field. 
The authors of ref.\cite{Maiezza:2010ic} studied the phenomenological differences between these two definitions and showed that it was easier to obtain a $W_R$, the charged gauge boson associated to $SU(2)_R$, with a mass in the TeV range if one considered charge conjugation as the left-right symmetry rather than parity. In the following, we choose charge conjugation. \\
\subsection{Scalar potential}
The Higgs sector is dictated by the symmetry breaking pattern of this particular model. Indeed, left-right symmetry having not been observed yet, the only way to account for this experimental fact is to consider that it has been broken at some high scale\footnote{One could also say that left-right symmetry is not a symmetry of nature.} . The symmetry breaking pattern takes then the form\\
\vskip0.5truecm
\begin{tabular}{c c l}
$SU(3)_C \times SU(2)_L \times SU(2)_R \times U(1)_{B-L} \longrightarrow$ & $SU(3)_C \times SU(2)_L \times U(1)_{Y}\longrightarrow SU(3)_C \times U(1)_{em}$ \end{tabular}\\
\vskip0.5truecm
where $U(1)_Y$ represents the hypercharge group and $U(1)_{em}$ the electromagnetism group. The first step can be accomplished in two ways: either using a doublet under $SU(2)_R$ or a triplet. Then we end up with the standard electroweak group which breaking can be achieved using a bidoublet scalar field under $SU(2)_L \times SU(2)_R$. In the sequel, we set the Higgs sector to be\\
$$\delta_R = (\singlet,\singlet,\triplet,2) = \begin{pmatrix} \delta_{1R} \\ \delta_{2R} \\ \delta_{3R} \end{pmatrix},~~~ \delta_L = (\singlet,\triplet,\singlet,2) = \begin{pmatrix} \delta_{1L} \\ \delta_{2L} \\ \delta_{3L} \end{pmatrix},~~~ \Phi = (\singlet,\doublet,\doublet^*,0) = \begin{pmatrix} \phi^0 & \phi^+ \\ \phi'^- & \phi'^0 \end{pmatrix}.$$
\paragraph{Remarks} Two remarks are in order here. First, left-right symmetry makes ``mandatory" the addition of a triplet transforming non-trivially under $SU(2)_L$. Secondly, it is useful to recall here the matrix representation of triplet fields which is more convenient for building gauge invariant quantities in the lagrangian.  We define then
\bea \label{doubly charged} \Delta_{L,R} = \frac{1}{\sqrt2}\sigma_a \delta^a_{L,R} = \begin{pmatrix} \Delta^+_{L,R} & \Delta^{++}_{L,R}\\ \Delta^0_{L,R} & -\Delta^+_{L,R} \end{pmatrix} \eea
where $\sigma_{a_{=1\dots 3}}$ are the Pauli matrices and superscripts denote the electric charge of each field. In particular, $\Delta_{L,R}^{++}$ define doubly charged scalars\footnote{In chapter \ref{chap: doubly charged} is presented a study on these particles. One can also refer to the papers \cite{Rentala:2011mr,Demir:2008wt,Dutta:1998yy,Azuelos:2005uc}.}. The most general scalar potential describing the Higgs sector reads:
\bea 
\lag_{Higgs} &=& \mu_1^2 \Tr[\Phi^\dag \Phi] - \lambda_1 \Big(\Tr[\Phi^\dag \Phi]\Big)^2 - \lambda_2 \Tr[\Phi^\dag \Phi\Phi^\dag \Phi] - \frac12 \lambda_3 \Big( \Tr[\hat\Phi\Phi^t] + \Tr[\Phi^\dag \hat\Phi^{\dag t}]\Big)^2 \n
&-& \lambda_4 \Tr[\Phi^\dag \Phi \hat\Phi^t\hat\Phi^{\dag t}] - \frac12 \lambda_5 \Big( \Tr[\hat\Phi\Phi^t]  - \Tr[\Phi^\dag \hat\Phi^{\dag t}] \Big)^2 - \frac12 \lambda_6 \Big( \Tr[\Phi^\dag \hat\Phi^{\dag t} \Phi^\dag \hat\Phi^{\dag t}] +  \Tr[\hat\Phi^t \Phi \hat\Phi^t \Phi]\Big) \n 
&+& \mu_2^2 \Big(\Tr[\Delta_L^\dag \Delta_L] + \Tr[\Delta_R^\dag\Delta_R] \Big) - \rho_1 \Big( \Tr[\Delta_L^\dag\Delta_L]^2 + \Tr[\Delta_R^\dag\Delta_R]^2\Big)\n
&-& \rho_2 \Big( \Tr[\Delta_L^\dag \Delta_L\Delta_L^\dag \Delta_L] + \Tr[\Delta_R^\dag \Delta_R\Delta_R^\dag \Delta_R]\Big) - \rho_3 \Tr[\Delta_L^\dag \Delta_L] \Tr[\Delta_R^\dag \Delta_R]\n
&-& \alpha_1 \Tr[\Phi^\dag \Phi] \Big(\Tr[\Delta_L^\dag \Delta_L] + \Tr[\Delta_R^\dag \Delta_R]\Big) - \alpha_2 \Big(\Tr[\Delta_R^\dag \Phi^\dag \Phi \Delta_R] + \Tr[\Delta_L^\dag \Phi \Phi^\dag \Delta_L]\Big)\n
&-& \alpha_3 \Big(\Tr[\Delta_R^\dag \hat\Phi^t \hat\Phi^{\dag t}\Delta_R] + \Tr[\Delta_L^\dag \hat\Phi^{\dag t}\hat\Phi^t \Delta_L]\Big) . 
\eea
where we introduced the quartic Higgs couplings $\lambda_i, \rho_i \et \alpha_i$ and the bilinear couplings $\mu_1 \et \mu_2$. For gauge invariance, we also had to define \\
$$ \hat \Phi_i{}^{i'} = \epsilon_{ij}\epsilon^{i'j'} \Phi^j{}_{j'},~~~ \hat L_{Li} = \epsilon_{ij} L_L^j,~~~ \hat L_R^{i'} = \epsilon^{i'j'} L_{Rj'}.$$
At the minimum of the potential, the neutral scalars acquire a vev which we suppose real
\bea \langle \Delta_R \rangle =\frac{1}{\sqrt2} \begin{pmatrix} 0 & 0 \\ v_R & 0 \end{pmatrix},~~~ \langle \Phi \rangle = \frac{1}{\sqrt2}\begin{pmatrix} v_1 & 0 \\ 0 & v'_1 \end{pmatrix},~~~ \langle \Delta_L \rangle = \frac{1}{\sqrt2}\begin{pmatrix} 0 & 0 \\ v_L & 0 \end{pmatrix},
\eea
and the scalar potential reads at the minimum
\bea
\langle V \rangle &=&
- \frac{\lambda_{1}}{4}(v_1^4 + v'^4_1 + 2 v_1^2 v'^{2}_1) - \frac{\lambda_{2}}{4}(v_1^4 - v'^{4}_1) - 2\lambda_{3} v_1^2 v'^{2}_1- \frac12\lambda_5 v_1^2v'^2_1 - \frac12\lambda_{6} v_1^2 v'^{2}_1 \n
&-& \frac{\alpha_1}{4}(v_1^2 v_L^2 +v'^{2}_1 v_L^2 + v_R^2 v_1^2 +v_R^2 v'^{2}_1 ) - \frac{\alpha_2}{4} (v'^{2}_1v_L^2 +v_R^2 v'^{2}_1)- \frac{\alpha_3}{4} (v_1^2 v_L^2 + v_R^2 v_1^2 ) \n
&-& \frac{\rho_1}{4} (v_L^4 +v_R^4 )- \frac{\rho_2}{4} (v_L^4 +v_R^4 ) - \frac{\rho_3}{4}v_L^2v_R^2+ \frac{\mu_{1}^2}{2}(v_1^2 + v'^{2}_1)+ \frac{\mu_{2}^2}{2}(v_L^2 + v_R^2).
\eea
Minimizing the scalar potential gives the following values for $\mu_1 \et \mu_2$
\bea \label{minimisation pot scalaire}
\frac{\partial V}{\partial v_1} = 0&\Rightarrow&  \mu_1^2 =  \lambda_1 (v_1^2+ v'^2_1) + \lambda_2 v_1^2 
+ 4\lambda_3 v'^2_1 + \lambda_5v'^2_1 + \lambda_6v'^2_1 + \frac{\alpha_1 + \alpha_3}{2}(v_L^2+v_R^2),\n
\frac{\partial V}{\partial v'_1} = 0 &\Rightarrow& \mu_1^2= \lambda_1 (v_1^2 + v'^2_1)
+ \lambda_2v'^2_1 + 4\lambda_3v_1^2 + \lambda_5v_1^2 + \lambda_6v_1^2 + \frac{\alpha_1 + \alpha_2}{2}(v_L^2 + v_R^2),\n
\frac{\partial V}{\partial v_L} = 0 &\Rightarrow& \mu_2^2=  \frac{\alpha_1}{2}(v_1^2 +v'^2_1) + \frac{\alpha_2v'^2_1}{2} + \frac{\alpha_3v_1^2}{2} + \rho_1v_L^2 + \rho_2v_L^2 + \frac{\rho_3v_R^2}{2},\n
\frac{\partial V}{\partial v_R} = 0 &\Rightarrow&  \mu_2^2 = \frac{\alpha_1}{2}(v_1^2 + v'^2_1) + \frac{\alpha_3v_1^2}{2} + \frac{\alpha_2v'^2_1}{2} + \rho_1v_R^2 + \rho_2v_R^2 + \frac{\rho_3v_L^2}{2}.
\eea
One important remark here is that if one combined the third and fourth conditions, one would get
\bea 
v_L \times \frac{\partial V}{\partial v_L} - v_R \times \frac{\partial V}{\partial v_R} = v_Lv_R(2 \rho_1 + 2\rho_2 - \rho_3)(v_L^2 - v_R^2) = 0
\eea
which implies that either $v_L \mbox{ and/or } v_R$ is zero or $v_L = v_R \neq 0$ or the $\rho$ couplings satisfy the relation $2 \rho_1 + 2\rho_2 - \rho_3 =0$. \\
On the one hand, it is obvious that the solution $v_R = 0$ is ruled out since we need left-right symmetry breaking. On the other hand, only a carefull study of the Yukawa lagrangian can help to dismiss the other solutions.\\

\subsection{The seesaw mechanism}\label{sec:seesaw}
The most general Yukawa lagrangian preserving the symmetries of our model reads after having removed all the indices for clarity
\bea \lag_{\text{Yukawa}} &=& \bar{Q}^c_L {\bf y_Q^{(1)}} \hat\Phi Q_R +   \bar{L}^c_L {\bf y_\ell^{(1)}} \hat \Phi L_R + \bar{Q}^c_R {\bf y_Q^{(2)}}  \Phi^\dag Q_L +   \bar{L}^c_R {\bf y_\ell^{(2)}} \Phi^\dag L_L \n
   &+& \hat{\bar{L}}^c_L {\bf y_\ell^{(3)}} \Delta_L L_L + \hat{\bar{L}}_R {\bf y_\ell^{(4)}} \Delta_R L_R^c + {\rm h.c.} .
\eea
where the couplings $y_\ell^{(3)} \et y_\ell^{(4)}$ induce Majorana mass terms for neutrinos and the others, {\ie} ${\bf y_Q^{(1)}}, {\bf y_\ell^{(1)}}, {\bf y_Q^{(2)}}$,  induce Dirac mass terms for quarks and leptons.\\

To constrain the value of $v_L$, we need to compute the masses of the neutrinos. To do so, let us introduce gauge eigenstates 
$$ \begin{pmatrix} \nu \\ N \end{pmatrix} = \begin{pmatrix} \nu_{Le} \\ \nu_{L\mu} \\ \nu_{L\tau} \\ \nu^c_{Re} \\ \nu^c_{R\mu} \\ \nu^c_{R\tau}\end{pmatrix}$$
and the mass eigenstates 
$$ \begin{pmatrix} \nu_M \\ N_M \end{pmatrix} = \begin{pmatrix} \nu_1 \\ \nu_2 \\ \nu_3 \\ N_1 \\ N_2 \\ N_3 \end{pmatrix} $$
where the subscripts $1,2 \et 3$ stand for generation number. We also assume for simplicity that Yukawa couplings are diagonal. The mass matrix is finally given by
\bea 
\sqrt2\begin{pmatrix}y_\ell^{(3)}v_L & -\frac{v_1 y_\ell^{(2)}}{2} \\
-\frac{v_1 y_\ell^{(2)}}{2} & v_R y_\ell^{(4)} 
\end{pmatrix} 
\eea 
in the basis $(\nu , N)$
where all the parameters are supposed real for simplicity. The ratio $\frac{v_1}{v'_1}$ being constrained by both the ratio at tree level of the top quark mass over the bottom quark mass $\frac{m_t}{m_b}$ and the $K^0 - \bar{K}^0$ mixing, we consider simply that $v'_1 = 0$. The eigenvalues of this matrix read
\bea 
m_{1} = \frac{v_L y_\ell^{(3)} + v_R y_\ell^{(4)} - \sqrt{v_1^2 y_\ell^{(2)2} + (v_L y_\ell^{(3)} - v_R y_\ell^{(4)})^2}}{\sqrt2} ,\n
m_{2}= \frac{v_L y_\ell^{(3)} + v_R y_\ell^{(4)} + \sqrt{v_1^2 y_\ell^{(2)2} + (v_L y_\ell^{(3)} - v_R y_\ell^{(4)})^2}}{\sqrt2}. \nonumber
\eea
From these two eigenvalues, we do a Taylor expansion around $\frac{v_L}{v_R} \sim 0$ and find
\bea m_1 = \frac{v_L}{v_R} \frac{y_\ell^{(3)}}{y_\ell^{(4)}} m_2 \eea
which clearly indicates a seesaw mechanism leading to a mass hierarchy depending on the ratio $\frac{v_L}{v_R}$\footnote{Provided the ratio $\frac{y_\ell^{(3)}}{y_\ell^{(4)}}$ is not very big.}. When this ratio goes to zero, the mixing matrix becomes diagonal and no mixing rises between gauge eigenstates. To perform a phenomenological study of this model, one can thus assume the hierarchy
$$ v_{1R} \gg v_{1} \gg v_1' \simeq v_L =0 $$
and only keep the first and last equations from \eqref{minimisation pot scalaire}.

\subsection{Status of non-supersymmetric left-right symmetry}
As emphasized earlier, left-right symmetry has not been discovered at the LHC and one is pushed to put limits from the latest available experimental data. In the case of parity, it was shown in \cite{Xu:2009nt} that the neutron electric dipole moment imposes a huge constraint on left-right symmetry scale pushing the mass of the $W_R$ to be at $(10\pm 3)$ TeV thus out of reach of the LHC. In another paper published by Maiezza \textit{et al.} \cite{Maiezza:2010ic}, it was shown that one could release the constraint from the neutron dipole moment by fine tuning the QCD parameter $\bar\theta$ leading to a $W_R$ gauge boson with a mass as small as $4$ TeV. In the same paper, they also treat the case of charge conjugation and show that $m_{W_R}$ can be lowered to $2 - 3$ TeV which makes it reachable by direct searches at LHC. These bounds are in agreement with the experimental results presented in the introduction \cite{CMS:2012eza,:2012rz}.\\

\unitlength = 1mm
\begin{figure}
\begin{center}
\begin{fmffile}{dblbeta}
\begin{fmfgraph*}(40,55)
	\fmfstraight
\fmfleft{i0,i1,i2,i3,i4,i5}
\fmflabel{n}{i1}\fmflabel{n}{i4}
\fmfright{o0,o1,o2,o3,o4,o5}
\fmflabel{p}{o1}\fmflabel{$e_L^-$}{o2}\fmflabel{$e_L^-$}{o3}\fmflabel{p}{o4}
\fmf{fermion,tension=2.5}{i1,v5}
\fmf{fermion}{v5,o1}
\fmf{fermion,tension=2.5}{i4,v2}
\fmf{fermion}{v2,o4}
\fmffreeze
\fmf{fermion}{v3,o3}
\fmf{plain,label=$\nu_i$,tension=0.1}{v3,v4}
\fmf{fermion,tension=0.7}{v4,o2}
\fmf{photon,label=$W_L$,tension=2.0}{v2,v3}
\fmf{photon,label=$W_L$,tension=0.1}{v4,v5}
\fmf{phantom}{i2,v4}
\fmf{phantom}{i1,v3}
\end{fmfgraph*}
\end{fmffile}
\end{center}
\caption{\footnotesize\label{fig:double beta}Feynman diagrams contributing to neutrinoless double beta decay due to the exchange of a left-handed (light) neutrino. The same diagram exists for right-handed neutrinos by exchanging left-handed particles by right-handed ones.}
\end{figure}

In any case, if discovered, this class of models will have, due to the Majorana mass terms for neutrinos, spectacular signatures of lepton-number violation in the form of same-sign dilepton \cite{Maiezza:2010ic} and double-beta decay\cite{Barry:2013xxa} (see fig.\ref{fig:double beta}). Moreover, as seen in eq.\eqref{doubly charged}, these models may predict doubly charged scalars which would offer a very clean signature at the LHC (see ref.\cite{Azuelos:2005uc} and chapter \ref{chap: doubly charged}).

\section{Building a Left-right symmetric supersymmetric model} \label{sec:lrsusy}
\subsection{Some notations and conventions}
In the sequel we assume the gauge group to be that of eq.\eqref{gauge group}
\bea \label{gauge group 2}SU(3)_c \times SU(2)_L \times SU(2)_R \times U(1)_{B-L}. \eea
The coupling constants associated to this gauge group will be noted $g_s, g_L, g_R \mbox{ and } g_{B-L}$ respectively. The fundamental representations for the gauge groups $SU(3) \mbox{ and } SU(2)$ will be spanned by $T_a \mbox{ and } \frac{\sigma_k}{2}$.\\
The lagrangian will be split into six parts
\bea \label{lagr lrsusy} \lagr = \lagr_{gauge} + \lagr_{chiral} + \lagr_{int} + V_D + V_F + \lagr_{soft}, \eea
where we have introduced
\begin{itemize}
\item $\lagr_{gauge}$ the part of the lagrangian containing the kinetic terms for the gauge fields;
\item $\lagr_{chiral}$ the part containing the kinetic terms for chiral superfields as well as matter-gauge interactions;
\item $\lagr_{int}$ the lagrangian originating from the superpotential $W$ of the theory and describing thus chiral interactions;
\item $V_D \mbox{ and } V_F$ the $D$ and $F$ contributions to the scalar potential;
\item $\lagr_{soft}$ the soft supersymmetry breaking lagrangian.
\end{itemize}
We shall also refer to the appendix \ref{annex:lrsusy} for the conventions and notations we chose. We use the Gell-Mann-Nishijima relation to get the electric charge of every field
$$ Q = T_{3L} + T_{3R} + \frac{Y_{B-L}}{2} $$
where $T_{3L}, T_{3R} \mbox{ and } Y_{B-L}$ are the $SU(2)_L, SU(2)_R \mbox{ and } U(1)_{B-L}$ quantum numbers.

\subsection{Gauge sector}
The gauge sector is quiet easy to determine as to every direct factor we associate one vector supermultiplet lying in the corresponding adjoint representation
\bea \label{gauge content}
SU(3)_c \longrightarrow V_3 &=& ( \octet, \singlet, \singlet, 0 ) \equiv (\tilde{g}^a,g_\mu^a) , \n
SU(2)_L \longrightarrow V_{2L} &=& ( \singlet, \triplet, \singlet, 0 ) \equiv (\tilde{W}^k_L,W^k_{L\mu}) , \n
SU(2)_R \longrightarrow V_{2R} &=& ( \singlet, \singlet, \triplet, 0 ) \equiv (\tilde{W}^k_R,W^k_{R\mu}) , \n
U(1)_{B-L} \longrightarrow V_{1} &=& ( \singlet, \singlet, \singlet, 0 ) \equiv (\tilde{B},B_{\mu}). \n
\eea
The $~\tilde{}~$ symbol denotes the fermionic partner of the gauge boson, and the numbers between brackets denote the representations in which lie the various supermultiplets with respect to the gauge group \eqref{gauge group 2}.\\
Refering to the equation \eqref{general renormalizable lagrangian}, the gauge lagrangian reads
\bea \lagr_{gauge} = -\frac{1}{4} V_{k}^{\mu\nu}V^k_{\mu\nu} + \frac{i}{2}(\tilde{V}^k \sigma^\mu D_\mu \bar{\tilde{V}}_k - D_\mu \tilde{V}^k \sigma^\mu \bar{\tilde{V}}_k) \eea
where $V^k$ is one of the gauge bosons in \eqref{gauge content}, $\tilde{V}^k$ is its fermionic superpartner, and $V_{\mu\nu}^k$ and $D_\mu$ are the field strength and the covariant derivative respectively that we define as follows
\bea 
V_{\mu\nu}^k &=& \delm V_\nu^k - \deln V_\mu^k + g f_{ij}{}^k V_\mu^i V_\nu^j,\n
D_\mu\tilde{V}^k &=& \delm \tilde{V}^k + g f_{ij}{}^k V_\mu^i \tilde{V}^j .\eea
We have introduced the gauge coupling constant $g$ and the structure constant $f_{ij}{}^k$ associated with every gauge group.

\subsection{Chiral lagrangian}
The matter sector of the model is simply made of right-handed and left-handed quarks and leptons chiral superfields which we define as follows\footnote{see appendix \ref{annex:lrsusy} for conventions}
\bea
(Q_L)^{fmi} = \begin{pmatrix} u^{fm}_L \\ d_L^{fm} \end{pmatrix} = (\triplet,\doublet,\singlet,\frac13), && (Q_R)_{fmi'} = \begin{pmatrix} u_{Rfm}^c & d_{Rfm}^c\end{pmatrix} = (\utilde{\mathbf{\bar{3}}}, \singlet, \doublet^{*}, -\frac13),\n
(L_L)^{fi} = \begin{pmatrix} \nu_L^f \\ l_L^f \end{pmatrix} = (\singlet,\doublet,\singlet,-1), && (L_R)_{fi'} = \begin{pmatrix} \nu^c_{Rf} & l_{R_f}^c \end{pmatrix} = (\singlet,\singlet,\doublet^*,1),
\eea
where the superscript $c$ denotes charge conjugation, the index $f$ is a generation index and $m$ is a color index. \\
The Higgs sector, just like in the non-supersymmetric version, is a little more complicated than in the MSSM case and actually can be defined in two different ways following the desired effects. Indeed one can choose to break the left-right symmetry with either Higgs doublets or Higgs triplets with respect to $SU(2)$. In the first case we define the following fields
\bea 
h_{1L} = ( \singlet, \doublet, \singlet, 1) = \begin{pmatrix} h^+_{1L}\\ h^0_{1L} \end{pmatrix}, && h_{1R} = (\singlet, \singlet, \doublet, 1) = \begin{pmatrix} h^+_{1R} \\ h^0_{1R} \end{pmatrix}, \n
h_{2L} = ( \singlet, \doublet, \singlet, -1) = \begin{pmatrix} h^0_{2L}\\ h^-_{2L} \end{pmatrix}, && h_{2R} = (\singlet, \singlet, \doublet, -1) = \begin{pmatrix} h^0_{2R}\\ h^-_{2R} \end{pmatrix}. \nonumber
\eea
where the superscripts denote the electric charges. To induce the desired symmetry breaking, the field $h_{1R} \et h_{2R}$ acquire vacuum expectation values $v_{1R} \et v_{2R}$. Their ``left-handed" counterparts $h_{1L} \et h_{2L}$ are introduced to restore parity in the lagrangian.\\

In the second case, the left-right symmetry breaking is induced by Higgs triplets acquiring a vacuum expectation value. We introduce thus the four Higgs fields
\bea
\delta_{1L} = (\singlet, \triplet, \singlet, -2) = \begin{pmatrix} \delta^1_{1L} \\ \delta^2_{1L} \\ \delta^3_{1L}\end{pmatrix}, && \delta_{1R} = (\singlet, \singlet, \triplet, -2) = \begin{pmatrix} \delta^1_{1R} \\ \delta^2_{1R} \\ \delta^3_{1R}\end{pmatrix}, \n
\delta_{2L} = (\singlet, \triplet, \singlet, +2) = \begin{pmatrix} \delta^1_{2L} \\ \delta^2_{2L} \\ \delta^3_{2L}\end{pmatrix}, && \delta_{2R} = (\singlet, \singlet, \triplet, +2) = \begin{pmatrix} \delta^1_{2R} \\ \delta^2_{2R} \\ \delta^3_{2R}\end{pmatrix}.
\eea
We remind the reader of the matrix notation for these fields which is very convenient to build gauge invariant quantities as well as to exhibit the electric charges of the different fields
\bea \Delta_{1 \{L,R\}} = \begin{pmatrix} \frac{\Delta^{-}_{1\{L,R\}}}{\sqrt2} & \Delta^{0}_{1\{L,R\}} \\ \Delta^{--}_{1\{L,R\}} & -\frac{\Delta^{-}_{1\{L,R\}}}{\sqrt2} \end{pmatrix},~~~\Delta_{2 \{L,R\}} = \begin{pmatrix} \frac{\Delta^{+}_{2\{L,R\}}}{\sqrt2} & \Delta^{++}_{2\{L,R\}} \\ \Delta^{0}_{2\{L,R\}} & -\frac{\Delta^{+}_{2\{L,R\}}}{\sqrt2} \end{pmatrix} \eea
Finally, to induce electroweak symmetry breaking and give masses to both up-type and down-type fermions we need two Higgs fields transforming non-trivially under $SU(2)_L \times SU(2)_R$. We define thus
\bea \Phi_a = (\singlet, \doublet, \doublet, 0) = \begin{pmatrix} \phi^0_a & \phi^+_a \\ \phi^-_a & \phi^{'0}_a \end{pmatrix},~~a = 1,2. \eea

If one uses Higgs scalar fields lying in the adjoint representation of $SU(2)$ than the most general gauge invariant renormalizable superpotential preserves R-parity. This is a real advantage with respect to the MSSM in which R-parity, when present, is put by hand. We thus choose the triplet fields instead of the doublet fields. However, though explicit R-parity violation is forbidden in this case, it was shown in \cite{Kuchimanchi:1993jg} that such a setup induced automatically vacuum expectation values for the scalar partners of the neutrinos. To elude this problem, we follow the same path as in \cite{Babu:2008ep} and introduce a new singlet field $S$ in order to have stable R-parity conserving vacua. It has also been shown in ref. \cite{Babu:2008ep}, that such a setup can cure both strong and supersymmetric $\mC\mP$ problems
\bea S = ( \singlet, \singlet, \singlet, 0 ). \nonumber \eea

Following the lagrangian in eq.\eqref{general renormalizable lagrangian}, the kinetic and gauge interaction terms for chiral superfields read
\bea \lagr_{chiral} = D_\mu \phidagger D^\mu \phi + \frac{i}{2} ( \psi \sigma^\mu D_\mu \psibar - D_\mu \psi \sigma^\mu \psibar) + (i g\sqrt2 \bar{\tilde{V}}^k \cdot \psibar_i T_k \phi^i + {\rm h.c.} ) \eea
where a sum over all gauge groups is understood. The choice for the matrices $T_k$ generating the representations of the gauge group depends on every field. The covariant derivative for the sleptons reads
\bea
(D_\mu \tilde{L}_L)^{fi} &=& (\delm \tilde{L}_L)^{fi} - \frac{i}{2} g_L W^k_{L\mu} (\sigma_k \tilde{L}_L)^{fi} + \frac{i}{2} g_{B-L} B_\mu (\tilde{L}_L)^{fi}, \n
(D_\mu \tilde{L}_R)_{fi'} &=& (\delm \tilde{L}_R)_{fi'} + \frac{i}{2} g_L W^k_{R\mu} (\tilde{L}_R\sigma_k)_{fi'} - \frac{i}{2} g_{B-L} B_\mu (\tilde{L}_R)^{fi'}. \nonumber \eea
For the squarks
\bea 
(D_\mu \tQ_L)^{fmi} &=& (\delm \tQ_L)^{fmi} - i g_s g_\mu^a (T_a \tQ_L)^{fmi} - \frac{i}{2}g_L W^k_{L\mu}(\sigma_k \tQ_L)^{fmi} -\frac{i}{6} g_{B-L}B_\mu(\tQ_L)^{fmi}, \n
(D_\mu \tQ_R)_{fmi'} &=& (\delm \tQ_R)_{fmi'} + i g_s g_\mu^a (\tQ_R T_a)_{fmi'} + \frac{i}{2}g_R W^k_{R\mu}(\tQ_R\sigma_k)_{fmi'} +\frac{i}{6} g_{B-L}B_\mu(\tQ_R)_{fmi}. \n
\eea
For the Higgs bidoublets it is defined as follows
\bea \label{bidoublet covariant}
(D_\mu \Phi_a)^i{}_{i'} &=& (\delm \Phi_a)^i{}_{i'} - \frac{i}{2}g_L W_{L\mu}^k ( \sigma_k \Phi_a)^i{}_{i'} + \frac{i}{2} g_R W^k_{R\mu} (\Phi_a \sigma_k)^i{}_{i'}.
\eea
As for the triplets, one should differentiate between the triplet and the matrix representations
\bea \label{triplet covariant}
(D_\mu \delta_{\{1,2\}L})^i &=& \delm \delta_{\{1,2\}L}^i + g_L \epsilon_{jk}{}^i W_{L\mu}^j\delta_{\{1,2\}L}^k \pm 2 g_{B-L} B_\mu (\delta_{\{1,2\}L})^i , \n
(D_\mu \Delta_{\{1,2\}L})^i{}_j &=& (\delm \Delta_{\{1,2\}L})^i{}_j - \frac{i}{2}g_L W^k_{L\mu} (\sigma_k\Delta_{\{1,2\}L})^i{}_j\n && + \frac{i}{2}g_L W^k_{L\mu}(\Delta_{\{1,2\}L}\sigma_k)^i{}_j \pm 2 g_{B-L} B_\mu (\delta_{\{1,2\}1L})^i{}_j.
\eea
In these equations care must be taking in writing the actions of $SU(2)_L$ and $SU(2)_R$ operators. Indeed while the former act on the elements of the columns which defines thus a left action, $SU(2)_R$ operators act on the elements forming the lines which defines the right action. 
\subsection{Interactions}
With the chiral content described above, the most general gauge invariant superpotential inducing renormalizable interactions reads in terms of scalar fields (here $\Phi \et \Delta$ are scalar fields)
\bea
\label{eq:superpot lrsusy}
W &=& (\tQ_L)^{mi} y_Q^1 (\hPhi)_i{}^{i'} (\tQ_R)_{mi'} + (\tQ_L)^{mi}y_{Q}^2 (\hPhi_2)_i{}^{i'} (\tQ_R)_{mi'} + (\tL_L)^iy_L^1(\hPhi)_i{}^{i'}(\tL_R)_{i'} + (\tL_L)^iy_L^2 (\hPhi_2)_i{}^{i'} (\tL_R)_{i'} \n
  &+& (\hat{\tL}_L)_i y_L^3 (\Delta_{2L})^i{}_j(\tL_L)^j + (\hat{\tL}_R)_{i'} y_L^4 (\Delta_{1R})^{i'}{}_{j'} (\tL_R)^{j'} + (\mu_L + \lambda_L S) \Delta_{1L} \cdot \hDelta_{2L} \n
  &+& (\mu_R + \lambda_R S)\Delta_{1R}\cdot\hDelta_{2R} + (\mu_3 + \lambda_3 S) \Phi_1 \cdot \hPhi_2 + \frac13 \lambda_s S^3 + \mu_s S^2 + \xi_S S.
\eea
\paragraph{Remarks} Some remarks are in order here. First, to construct a gauge invariant Lagrangian,  we have introduced
\bea 
(\hat{\tL}_L)_i &=& \epsilon_{ij}(\tL_L)^j, ~~~ (\hat{\tL}_R)^{i'} = \epsilon^{i'j'}(\tL_R)_{j'}, \n
(\hDelta_{2L})_i{}^j &=& \epsilon_{ik}\epsilon^{jl} (\Delta_{2L})^k{}_l, ~~~ (\hDelta_{2R})_{i'}{}^{j'} = \epsilon_{i'k'}\epsilon^{j'l'} (\Delta_{2R})^{k'}{}_{l'}, \n
\Delta_{1L} \cdot \hDelta_{2L} &=& {\rm Tr}(\Delta^t_{1L} \hDelta_{2L}) = (\Delta_{1L})^i{}_j (\hDelta_{2L})_i{}^j, ~~~ \Delta_{1R} \cdot \hDelta_{2R} = {\rm Tr}(\Delta^t_{1R} \hDelta_{2R}) = (\Delta_{1R})^{i'}{}_{j'} (\hDelta_{2R})_{i'}{}^{j'}, \n
(\hPhi_{1,2})_i{}^{i'} &=& \epsilon^{i'j'}\epsilon_{ij}(\Phi_{1,2})^j{}_{j'},~~~ \Phi_1\cdot\hPhi_2 = {\rm Tr}(\Phi^t_1 \hPhi_2) = (\Phi_1)^i{}_{i'} (\hPhi_2)_i{}^{i'}.
\eea
We also wanted to keep the number of parameters as small as possible. This led us to assume that the bilinear terms as well as the linear term are equal to zero
\bea \mu_L = \mu_R = \mu_3 = \mu_s = \xi_s = 0. \eea
Motivated, for example, by a $\mathbb{Z}_3$-symmetry, this setup has the advantage to provide an explanation for the $\mu$ terms through the vacuum expectation value $v_s$ of the scalar component of $S$. A $\mathbb{Z}_3$ symmetry, when broken, has also the disadvantage to lead to the formation of domain walls \cite{Zeldovich:1974uw,Vilenkin:1984ib}. This issue can be eluded by supposing the existence of higher-dimensional, non-renormalizable, Planck-scale suppressed operators in the superpotential far beyond the electroweak scale. In our study, we set $v_s$ much higher than that of $\langle\Delta^0\rangle$ \footnote{$\Delta^0$ represents all the neutral components of the triplet fields.}.
\subsection{Soft supersymmetry breaking lagrangian}
The soft supersymmetry breaking lagrangian is the sum of mass terms for both the gauginos (superpartners of the gauge bosons) and the scalar fields and a part which is completely dictated by the superpotential. We have thus
\bea \label{eq:lsoft lrsusy}
\lagr_{soft} &=& -\frac12 \big[ M_1 \tilde{B}\cdot\tilde{B} + M_{2L} \tilde{W}^k_L \cdot \tilde{W}_{Lk} + M_{2R} \tilde{W}^k_R \cdot \tilde{W}_{Rk} + M_3 \tilde{g}^a \cdot \tilde{g}_a + {\rm h.c.} \big] \n
             &-& \Big[\tilde{Q}^\dagger m_{Q_L}^2 \tilde{Q}_L + \tilde{Q}_R m_{Q_R}^2 Q_R^\dagger + \tL_L^\dagger m_{L_L}^2 \tL_L + \tL_R m_{L_R}^2 \tL_R^\dagger - (m_\Phi^2)^{ff'} {\rm Tr}(\Phi^\dagger_f \Phi_{f'}) \n
&+& m^2_{\Delta_{1L}}{\rm Tr}(\Delta^\dagger_{1L} \Delta_{1L}) +  m^2_{\Delta_{2L}}{\rm Tr}(\Delta^\dagger_{2L} \Delta_{2L}) + m^2_{\Delta_{1R}}{\rm Tr}(\Delta^\dagger_{1R} \Delta_{1R}) +  m^2_{\Delta_{2R}}{\rm Tr}(\Delta^\dagger_{2R} \Delta_{2R}) + m^2_S S^\dagger S \Big]\n
&-& \Big[ \tQ_L T_Q^1 \hPhi_1 \tQ_R + \tQ_L T_Q^2 \hPhi_2 \tQ_R + \tL_L T_L^1\hPhi_1\tL_R + \tL_L T_L^2\hPhi_2\tL_R + \hat{\tL}_LT^3_L\Delta_{2L}\tL_L + \tL_R T^4_L \Delta_{1R} \hat{\tL}_R + {\rm h.c.} \Big]\n
&-& \Big[ T_L S \Delta_{1L} \cdot \hDelta_{2L} + T_R S \Delta_{1R} \cdot \hDelta_{2R} + T_{3}S \Phi_1 \cdot \hPhi_2 + {\rm h.c.} \Big]
\eea
where we omitted all the indices for more clarity.

\subsection{Hierarchies and simplifications}
The symmetry breaking pattern for this model is exactly the same than that of the non-supersymmetric left-right symmetric model and is given schematically by
\bea SU(3)_C \times SU(2)_L \times SU(2)_R \times U(1)_{B-L} \longrightarrow SU(3)_C \times SU(2)_L \times U(1)_{Y}\longrightarrow SU(3)_C \times U(1)_{em} \nonumber \eea
At the minimum of the scalar potential, the neutral components of the scalar Higgs fields acquier a vacuum expectation value:
\bea
\langle \Delta_{1R} \rangle &=& \begin{pmatrix} 0 & \frac{v_{1R}}{\sqrt2} \\ 0 & 0 \end{pmatrix},~~~ \langle\Delta_{2R} \rangle = \begin{pmatrix} 0 & 0 \\ \frac{v_{2R}}{\sqrt2}&0 \end{pmatrix},~~~\langle \Delta_{1L} \rangle =\begin{pmatrix} 0 & \frac{v_{1L}}{\sqrt2} \\ 0 & 0 \end{pmatrix},~~~ \langle\Delta_{2L} \rangle = \begin{pmatrix} 0 & 0 \\ \frac{v_{2L}}{\sqrt2}&0 \end{pmatrix},\n
\langle \Phi_1 \rangle  &=& \begin{pmatrix} \frac{v_1}{\sqrt2} & 0 \\ 0 & \frac{1}{\sqrt2} v'_1 e^{i\alpha_1}\end{pmatrix}, ~~~ \langle \Phi_2 \rangle = \begin{pmatrix} \frac{1}{\sqrt2}v_2'e^{i\alpha_2} & 0 \\ 0 & \frac{v_2}{\sqrt2} \end{pmatrix} \n
\langle S \rangle &=& \frac{1}{\sqrt2} v_S e^{i\alpha_s}.
\eea
As in section \ref{sec:seesaw}, we find that requiring the seesaw mechanism as an explanation to neutrinos masses and applying $K-\bar{K}^0$ mixing data constaints induces the same hierarchy in the vevs, that is
\bea \label{scalar vevs}  v_{\{1,2\}R} \gg v_{\{1,2\}} \gg v_{\{1,2\}L} \simeq v_{\{1,2\}}' \simeq 0. \eea
Also, as stated above, we take $v_s \gg v_{\{1,2\}R}$. Finally the vev of the singlet field can be rotated away by the mean of field redefinition; we hence consider the phase as equal to zero
$$ \alpha_s = 0 .$$
\subsection{Scalar potential}
The scalar potential of this model is given by the sum of the contributions of both the $F$- and $D$-term and the soft supersymmetry breaking lagrangian $\lagr_{soft}$. In a general setup where the simplifications and assumptions given above are ignored but keeping all the parameters real and shifting the neutral Higgs fields by their vevs, the scalar potential reads at the minimum
\bea
V &=& \frac{g_L^2}{32} \big[v_1^2 - v_2^2 + v_2'^2 - v_1'^2 + 2v_{1L}^2 - 2v_{2L}^2 \big]^2 + \frac{g_R^2}{32} \big[v_1^2 - v_2^2 + v_2'^2 - v_1'^2 - 2v_{1R}^2 + 2v_{2R}^2 \big]^2 \n
&+& \frac{g^2_{B-L}}{8} \big[v_{1R}^2 - v_{2R}^2 + v_{1L}^2 - v_{2L}^2\big]^2 + \frac12 \Big[ m^2_{\Delta_{1R}} v_{1R}^2 + m^2_{\Delta_{2R}} v_{2R}^2 + m^2_{\Delta_{1L}}v_{1L}^2 + m^2_{\Delta_{2L}} v_{2L}^2 + m_S v_S^2 \Big] \n
&+& \frac12 \Big[ (m^2_\Phi)^{11} (v_1^2 + v_1'^2) + (m^2_\Phi)^{22}(v_2^2 + v_2'^2) + 2\Re\big\{ (m_\Phi^2)^{21}(v_2v'_1e^{i\alpha_1} + v_2' e^{i\alpha_2} v_1) \big\}\Big] \n
&+& \frac12 \Bigg[ \Big| \mu_L + \frac{1}{\sqrt2} \lambda_L v_s \Big|^2(v_{1L}^2 + v_{2L}^2) + \Big| \mu_R + \frac{1}{\sqrt2} \lambda_R v_s \Big|^2(v_{1R}^2 + v_{2R}^2) +  \Big| \mu_3 + \frac{1}{\sqrt2}\lambda_3 v_s \Big|^2(v_{1}^2 + v_{2}^2 + v_1'^2 + v_2'^2) \Bigg]\n
&+& \Re\Bigg\{ \Big[B_L + \frac{T_Lv_s}{\sqrt2} \Big] v_{1L}v_{2L} + \Big[B_R + \frac{T_Rv_s}{\sqrt2}\Big] v_{1R}v_{2R} - \Big[B_3 + \frac{T_3v_s}{\sqrt2}\Big](v_1v_2 + v'_1 v'_2) B_S v_S^2 + \frac{T_3v_S^3}{3\sqrt2} + \sqrt2 v_S \xi_S \Bigg\}\n
&+& \frac14\Big[ \lambda_L v_{1L} v_{2L} + \lambda_R v_{1R} v_{2R} - \lambda_3 (v_1 v_2 + v_1'v_2') + \lambda_s v_s^2 + 2\sqrt2 \mu_s v_s + 2\xi_S \Big]^2.
\eea
The minimization equations resulting from this scalar potential are given in the appendix \ref{annex:lrsusy}.

\subsection{Mass matrices}
As our study focuses on the chargino-neutralino sector of this model, we give here the related mass matrices as well as the mass matrices for gauge bosons. For more compact expressions, we define 
\bea \label{compact} v_L^2 &=& v_{1L}^2 + v_{2L}^2, ~~~ v_R^2 = v_{1R}^2 + v_{2R}^2, ~~~ v^2 = v_1^2 + v_2^2, ~~~ v'^2 = v'^2_1 + v'^2_2, ~~~ vv' = v_1v_1' e^{i\alpha_1}+v_2v_2' e^{i\alpha_2}\n
 \tilde{\mu}_{L,R,3,S} &=& \mu_{L,R,3,S} + \frac{1}{\sqrt2} \lambda_{L,R,3,S}v_s e^{i\alpha_s} \eea
\paragraph{Gauge bosons}
Covariant derivatives for the Higgs fields defined in eq.\eqref{bidoublet covariant} and eq.\eqref{triplet covariant} give rise to mass terms for gauge bosons when neutral Higgses are shifted by their vevs. Consequently, we get the two mass matrices
\bea 
\begin{pmatrix} Z_\mu \\ A_\mu \\ Z'_\mu \end{pmatrix} &=& \begin{pmatrix} \frac14 g_L^2 (4v_L^2+v^2+v'^2) & -\frac14 g_Lg_R(v^2 + v'^2) & -g_{B-L}g_Lv_L^2 \\
-\frac14 g_Lg_R(v^2 + v'^2) &\frac14 g_R^2 (4v_R^2+v^2+v'^2) & -g_{B-L}g_Rv_R^2 \\
-g_{B-L}g_Lv_L^2 & -g_{B-L}g_Rv_R^2 & g_{B-L}^2(v_L^2+v_R^2) \end{pmatrix} \begin{pmatrix} W_{L\mu}^3 \\ W_{R\mu}^3 \\ B_\mu \end{pmatrix},\n
\begin{pmatrix} W^+ \\ W'^+ \end{pmatrix} &=& \begin{pmatrix} \frac14 g_L^2 (2v_L^2+v^2+v'^2) & -\frac12 g_Lg_R(vv')^* \\
-\frac12 g_Lg_R(vv') & \frac14 g_R^2 (2v_R^2+v^2+v'^2) \end{pmatrix} \begin{pmatrix} W_{L\mu}^+ \\ W_{R\mu}^+ \end{pmatrix},
\eea
with $W_{\{L,R\}\mu}^{\pm} = \frac{1}{\sqrt2}(W_{\{L,R\}\mu}^{1} \mp W_{\{L,R\}\mu}^{2})$. In the limit defined in equation \eqref{scalar vevs}, the first mass matrix gives rise to a massless state that we identify as being the photon and two neutral gauge bosons with their masses given by
\bea m_{Z} = v^2 \frac{g_L^2g_R^2 + g_{B-L}^2(g_L^2 + g_R^2)}{4(g_{B-L}^2 + g_R^2)},~~~  m_{Z'} = (g_{B-L}^2 + g_R^2)v_R^2 \nonumber\eea
The model breaking pattern is in two steps: first $SU(2)_R \times U(1)_{B-L} \rightarrow U(1)_Y$ induces the mixing of $W_{R\mu}^3 \mbox{ and } B_\mu$ into a massless state $B'_\mu$ indentifed as the hypercharge gauge boson and a massive $Z'$ and secondly $SU(2)_L\times U(1)_Y \rightarrow U(1)_{e.m}$ where the hypercharge field $B'_\mu$ mixes with the neutral $W_{L\mu}^3$ into the photon $A_\mu$ and a massive $Z$-boson. By consequence, we introduce two angles $\phi \mbox{ and } \theta$ which we define to be
\bea \label{angles}
\cos\phi = \frac{g_R}{\sqrt{g_R^2 + g_{B-L}^2}}&,& ~~~ \sin\phi = \frac{g_{B-L}}{\sqrt{g_R^2+g_{B-L}^2}}.\n
\cos\theta_w = \frac{g_L}{\sqrt{g_R^2\sin^2\phi + g_L^2}}&,&~~~ \sin\theta_w = \frac{g_R\sin\phi}{\sqrt{g_R^2\sin^2\phi + g_L^2}}.
\eea
We re-express the masses for the $Z \et Z'$
\bea m_{Z'} = \frac{g_R^2}{\cos^2\phi}v_R^2,~~~ m_{z} = v^2\frac{g_L^2}{4\cos^2\theta_w}. \eea

\paragraph{Neutralinos and charginos}
All the partners of the gauge and Higgs bosons with the same quantum numbers (electric charge and color representation) mix after breaking of electroweak symmetry to electromagnetism. The model we are considering contains twelve neutralinos whose mass matrix, expressed in the basis $(i\tilde{W}_L^3, i\tilde{W}_R^3, i\tilde{B}, \tilde{\Phi}'^0_2,\tilde{\Phi}_2^0,\tilde{\Phi}_1^0,\tilde{\Phi}_1'^0,\tilde{\Delta}_{2L}^0, \tilde{\Delta}_{2R}^0, \tilde{\Delta}_{1L}^0, \tilde{\Delta}_{1R}^0,\tilde{S} )$, reads
\begin{align}
M_{\chi^0}&=\n
&\footnotesize\hskip-1truecm\begin{pmatrix}
 -M_{2L} & 0 & 0 & \frac{g_L \tilde v_2^\prime}{2} & -\frac{g_L v_2 }{2} & \frac{g_L v_1 }{2} & -\frac{g_L \tilde v_1^\prime }{2} & -g_L v_{2L} & 0 & g_L v_{1L} & 0 & 0\\
      0 & -M_{2R} & 0 & -\frac{g_R \tilde v_2^\prime}{2}  & \frac{g_R v_2}{2} & 
        -\frac{g_R  v_1 }{2}& \frac{g_R \tilde v_1^\prime}{2}  & 0 & -g_R v_{2R} &  
        0  & g_R v_{1R} & 0\\
      0 & 0 &- M_1 & 0 & 0 & 0 & 0 & g_{B-L} v_{2L} & g_{B-L} v_{2R} & 
        -g_{B-L} v_{1L}& -g_{B-L} v_{1R} &0 \\
      \frac{g_L \tilde v_2^\prime}{2}  & -\frac{g_R \tilde v_2^\prime }{2} & 0 & 0
        & 0 & 0 & -\tilde\mu_3 & 0 & 0 & 0 & 0 & -\frac{\lambda_3 \tilde
        v_1^\prime}{\sqrt{2}} \\ 
      -\frac{g_L  v_2 }{2}&  \frac{g_R v_2}{2}  & 0 & 0 & 0 & 
         -\tilde\mu_3 & 0 & 0 & 0 &0 &0&
     -\frac{\lambda_3 v_1}{\sqrt{2}}\\
      \frac{g_L v_1}{2}  & -\frac{g_R  v_1}{2} & 0 & 0 &-\tilde\mu_3&0&0&0&0&0&0
& -\frac{\lambda_3 v_2}{\sqrt{2}}\\
      -\frac{g_L  \tilde{v}_1^\prime}{2}  & \frac{g_R \tilde v^\prime_1}{2}  & 0 &
      -\tilde\mu_3  &0& 0&0&0&0&0&0 &  -\frac{\lambda_3 \tilde v_2^\prime}{\sqrt{2}} \\
      -g_L v_{2L} & 0 & g_{B-L} v_{2L}&0&0&0&0&0&0&\tilde\mu_L&0 & \frac{\lambda_L v_{1L}}{\sqrt{2}}
\\
      0&-g_R v_{2R} & g_{B-L} v_{2R}&0&0&0&0&0&0&0&\tilde\mu_R &\frac{\lambda_R v_{1R}}{\sqrt{2}}  \\
      g_L v_{1L} & 0 & -g_{B-L} v_{1L}&0&0&0&0&\tilde\mu_L&0&0&0&  \frac{\lambda_L v_{2L}}{\sqrt{2}}
\\
      0&g_R v_{1R} & -g_{B-L} v_{1R}&0&0&0&0&0&\tilde \mu_R &0&0& \frac{\lambda_R v_{2R}}{\sqrt{2}} \\
      0& 0 & 0 & -\frac{\lambda_3 \tilde v_1^\prime}{\sqrt{2}} &
     -\frac{\lambda_3 v_1}{\sqrt{2}}& -\frac{\lambda_3 v_2}{\sqrt{2}}&
     -\frac{\lambda_3 \tilde v_2^\prime}{\sqrt{2}} & \frac{\lambda_L v_{1L}}{\sqrt{2}}
     & \frac{\lambda_R v_{1R}}{\sqrt{2}} & \frac{\lambda_L v_{2L}}{\sqrt{2}}
    & \frac{\lambda_R v_{2R}}{\sqrt{2}} & 2\tilde \mu_s
\end{pmatrix}\n
\end{align}
It also contains six singly-charged charginos whose mass matrix expressed in the $(i\tilde{W}_L^{+}, i \tilde{W}_R^+, \tilde{\Phi}_2^+, \tilde{\Phi}_1^+, \tilde{\Delta}_{2L}^+, \tilde{\Delta}_{2R}^+)$ and $ (i\tilde{W}_L^-, i \tilde{W}_R^-, \tilde{\Phi}_2^-, \tilde{\Phi}_1^-, \tilde{\Delta}_{1L}^-, \tilde{\Delta}_{2R}^-)$ bases reads
\bea \label{mat: charginos}
\begin{pmatrix}
M_{2L} & 0 & \frac{g_L}{\sqrt2}\tilde{v}_2' & \frac{g_L}{\sqrt2}v_1 & -g_L v_{1L} & 0 \\
0 & M_{2R} & -\frac{g_R}{\sqrt2}v_2 & -\frac{g_R}{\sqrt2}\tilde{v}_1' & 0 & -g_R v_{1R} \\
\frac{g_L}{\sqrt2}v_2 & -\frac{g_R}{\sqrt2}\tilde{v}_2' & 0 & \tilde{\mu_3} & 0 & 0 \\
\frac{g_L}{\sqrt2}\tilde{v}_1' & -\frac{g_R}{\sqrt2}v_1 & \tilde{\mu}_3 & 0 & 0 & 0 \\
g_L v_{2L} & 0 & 0 & 0 & \tilde{\mu}_L & 0 \\
0 & g_R v_{2R} & 0 & 0 & 0 & \tilde{\mu}_R
\end{pmatrix}
\eea
Finally, as seen earlier, using $SU(2)$ triplets induces the apparition of doubly charged particles in the theory. In the setup we chose, these exotic particles have the following mass matrix
\bea
\begin{pmatrix}
\tilde{\mu}_L & 0 \\
0 & \tilde{\mu}_R
\end{pmatrix}
\eea

\section{Monte Carlo analysis} \label{sec:lrsusy pheno}

\subsection[Automated tools and Monte Carlo simulation]{Automated tools and Monte Carlo simulation} As indicated in the introduction, in this analysis we want to use a Monte Carlo simulation in order to extract the signatures that could be linked to processes from a left-right symmetric supersymmetric model. To this end, we have implemented our model in {\sc FeynRules}\cite{Christensen:2008py,Christensen:2009jx,Alloul:2013yy,Alloul:2013fw} and taking advantage of both the supersymmetry package \cite{Duhr:2011se} and the {\sc UFO} interface \cite{Degrande:2011ua} we have generated the necessary files for the Monte Carlo event generator {\sc MadGraph 5}\cite{Maltoni:2002qb,Alwall:2007st,Alwall:2008pm,deAquino:2011ub} to work. The latter is then used to calculate the decay widths for all the particles relevant to our study (charginos, neutralinos and sleptons) but also to generate the Monte Carlo events for the processes leading to the production of two gauginos. For the latter step, matrix elements are convolved with the parton density set CTEQ6L1\cite{Pumplin:2002vw} and set the center-of-mass energy to 8 TeV.\\

The parton-level events that we have generated only contain supersymmetric particles in the final states. It is then necessary to process them in order for these unstable particles to be decayed but also for both the parton showering and hadronization to be performed properly. We choose the {\sc C++} package {\sc Pythia 8}\cite{Ilten:2012zb,Sjostrand:2007gs} to fulfill this task. \\

The last step before being able to analyze the generated signal events consists in reconstructing the jet objects. To this end we use the package {\sc FastJet}\cite{Cacciari:2011ma} through the interface provided by {\sc MadAnalysis~5}\cite{Conte:2012fm} and assume perfect reconstruction for both electrons and muons but allow for a b-tagging efficiency of 0.6, c-jets mistagging of 0.1 and light-jets mistagging of $0.01$. The cross sections associated to the signal processes are then reweighted with a $K$-factor of 1.2. The latter is defined as 
$$ K = \frac{\sigma_{\rm NLO}}{\sigma_{\rm LO}} $$
where $\sigma_{\rm LO} \et \sigma_{\rm NLO} $ are the cross sections calculated at the leading and the next-to-leading order, respectively. This factor, when considering the production of the neutralino and chargino states in the context of the MSSM is known to be equal to 1.2-1.25 after combining NLO calculations \cite{Beenakker:1999xh,Debove:2010kf,Debove:2009ia,Debove:2011xj} together with the resummation of the leading and next-to-leading logarithms to all orders\cite{Debove:2010kf,Debove:2009ia,Debove:2011xj}. Such calculations do not exist in the context of the left-right symmetric supersymmetric model and, seen the masses involved in our models, we take the value of 1.2 as the structure of the calculation is very similar in both models. \\

Finally we use the facilities provided by the {\sc MadAnalysis 5} package to analyze the events, assuming an integrated luminosity of 20 $fb^{-1}$ which corresponds to the amount of data acquired by the LHC during the 2012 run.

\subsection{Background simulation} To be able to present quantitative results, we have also simulated the main processes in the Standard Model leading to final states with at least one lepton. When possible the cross sections have been normalized to the next-to-leading order (NLO) and even to the next-to-next-to-leading order (NNLO) accuracy and were convolved with the CT10 parton densities\cite{Guzzi:2011sv}. We classify these background events into six categories:
\begin{itemize}
    \item Single top events, including the $t-$, $tW-$ and $s-$ channel topologies. These processes which correspond to the production of a single top quark in conjunction with either other lighter quarks or a $W$ gauge boson may contribute to both the single-lepton and the dilepton channel when the $W$-boson and/or the top quark decay leptonically. These events have been reweighted to an approximate NNLO accuracy leading to the cross sections 87.2 pb, 22.2 pb and 5.5 pb for the $t-$, $tW-$ and $s-$ channels respectively\cite{Kidonakis:2012db}.

    \item Di-boson events include the production of a pair of $W$ gauge bosons, a pair of $Z$ bosons and the associated production of a $W$ and a $Z$ boson. These events have been normalized to the NLO cross sections 4.5 pb, 11.8 pb and 30.2 pb respectively as obtained using the package {\sc Mcfm}\cite{Campbell:2010ff}. In this case, we did not simulate the fully hadronic decays of the gauge bosons as they are irrelevant for our analysis.

    \item The production of a pair of top anti-top quarks can lead to final states with up to two leptons and are thus to be taken into account. The events simulated have been normalized to the cross section 255.8 pb which includes full NLO predictions and genuine NNLO contributions as derived by the package {\sc HATHOR}\footnote{NNNLO results have been made public recently \cite{Czakon:2013goa} but we did not take them into account.}.
    
    \item $W$ + jets regroups the processes where a $W$ boson is produced together with jets. It contributes to the background in the single lepton channel when the weak boson decays leptonically and its cross section is taken to be 35678 pb as predicted by NNLO simulations done with the package {\sc FEWZ}\cite{Li:2012wna}. Here also, we ignore the full hadronic decays of the gauge boson.

    \item $Z$ + jets are also of interest to us as they lead to final states with up to two light charged leptons in the final state. Ignoring the full hadronic decays of the $Z$ boson, the associated cross section as given by the {\sc FEWZ} package is 10319 pb and our simulated events are reweighted consistently.

    \item Rare processes are those whose cross section is small compared to the processes cited above. This background includes the production of a pair of top anti-top quarks in association with one or two gauge bosons, a top quark with a $Z$-boson and another jet and processes where two pairs of top anti-top quarks are produced. We use the {\sc Mcfm} package to determine both values 0.25 pb and 0.21 pb for $t~t~W$ and $t~t~Z$ events, respectively. The other processes, {\ie} $t~Z~j$, $t~t~W~W$ and $t~t~t~t$ are associated to the cross sections 46 fb, 13 fb and 0.7 fb respectively as predicted by {\sc MadGraph 5}.
\end{itemize}
Finally, we do not simulate multijet events because they require data driven methods but we are confident that this background is under good control as in our analysis we only select events with large missing transverse energy (MET) and hard leptons\cite{Chatrchyan:2012bd,Aad:2012wm,Aad:2012twa}.

\subsection{Setup for the analysis}
Before moving to the analysis, let us summarize our setup. We have the following hierarchy amongst the vacuum expectation values
$$ v_s \gg v_{\{1,2\}R}\gg v_{1,2} \gg v'_{1,2}\simeq v_{\{1,2\}L} = 0. $$
We also suppose all the physical phases equal to zero
$$ \alpha_1 = \alpha_2 = 0 .$$ 
This setting decouples most of the neutralinos and charginos from the low energy theory and allows us to only focus on the three lightest neutralinos and two lightest charginos. To reduce a little more the number of free parameters characterizing the higgsino sector, we assume exact left-right symmetry which implies the equality between $SU(2)_L$ and $SU(2)_R$ gauge coupling constants
$$ g_L = g_R. $$
The latter equality should hold true at all energies as the renormalization group equations associated to these parameters are exactly the same (see section \ref{sec:rges}). We also set generically all Yukawa couplings to a value of 0.1, the ratio of the vevs of the bidoublets to one while that of $SU(2)_R$ triplets allows us to define two scenarios:
$$ \tan\beta = 1, ~~~~\tan\tilde{\beta} = \frac{v_{2R}}{v_{1R}} = 1 \mbox{ or } 1.05 .$$
Finally, only remain the gaugino soft breaking masses $M_1, M_{2L}, M_{2R}$ as free paremeters. To set their values, we perform a scan over the paremeter space they span and only keep those possibilities where the lightest supersymmetric particle is a neutralino ensuring thus a candidate for dark matter\footnote{Though not excluded, we will not consider in this work the case where the right handed sneutrino is the dark matter candidate. See for example \cite{Borah:2012bb}.}. This scan allowed us to split the possible scenarios into two categories, ``pure" scenarios being those where mixing matrices diagonalizing chargino and neutralino mass matrices are almost diagonal and ``mixed" scenarios where the mixing between gauge eigenstates is larger. For each category we choose two representative scenarios each exhibiting a different hierarchy between the particles. All the numerical values are gathered in table \ref{4 scenarios}.

\begin{table}
\begin{center} 
\begin{tabular}{l | r r r r}
  \hline\hline
  Parameter &  Scenario I.1 & Scenario I.2 & Scenario II & Scenario III\\
  \hline
  \hline
  $M_1$ [GeV]    & 250  & 250  & 100 & 359\\
  $M_{2L}$ [GeV] & 500  & 750  & 250 & 320\\
  $M_{2R}$ [GeV] & 750  & 500  & 150 & 270\\
  \hline
  $m_Z$ [GeV]    & 91.1876 & 91.1876 & 91.1876 & 91.1876\\
  $m_W$ [GeV]    & 80.399  & 80.399  & 80.399  & 80.399 \\
  $\alpha(m_Z)^{-1}$ & 127.9 & 127.9 & 127.9 & 127.9 \\
  \hline
  $v_R$ [GeV]       & 1000   & 1000   & 1300 & 1300\\
  $v_s$ [GeV]       & $10^5$ & $10^5$ & $10^5$ & $10^5$\\
  $\tan\beta$       & 10 & 10 & 10 & 10\\
  $\tan\tilde\beta$ & 1  & 1  & 1.05 & 1.05\\
\hline
  $\lambda_L$ & 0.1 & 0.1 & 0.1 & 0.1\\
  $\lambda_R$ & 0.1 & 0.1 & 0.1 & 0.1\\
  $\lambda_s$ & 0.1 & 0.1 & 0.1 & 0.1\\
  $\lambda_3$ & 0.1 & 0.1 & 0.1 & 0.1\\
 \hline\hline
\end{tabular}
\caption{\footnotesize \label{4 scenarios} Summary of the numerical values we have used to construct our benchmark scenarios. Slepton masses are set to 400 GeV for all the scenarios but for scenario II where we also consider the case where sleptons have a universal mass of 200 GeV.}
\end{center}
\end{table}

\begin{figure}
\begin{center}
	\includegraphics[width=.49\columnwidth]{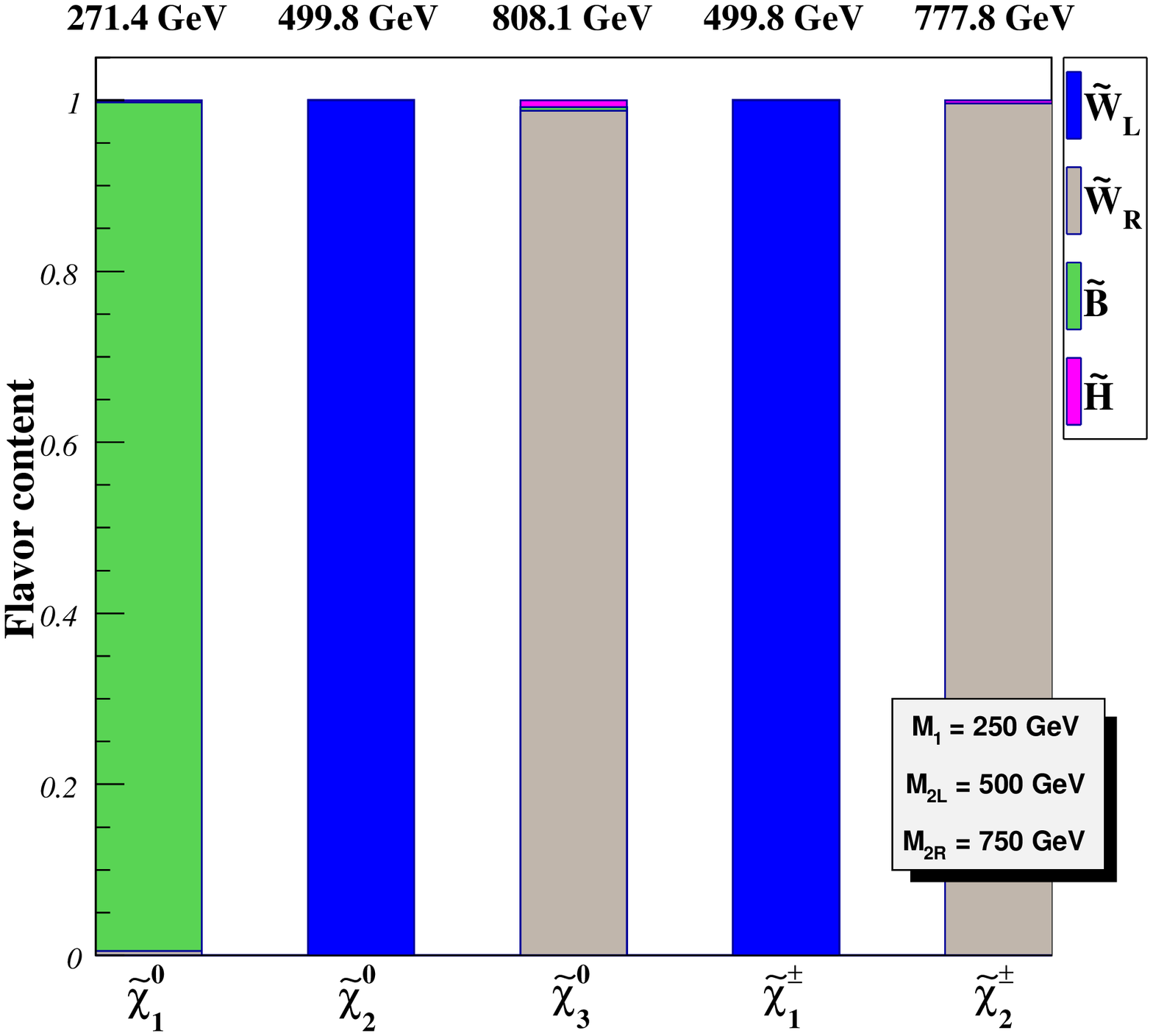} 
	\includegraphics[width=.49\columnwidth]{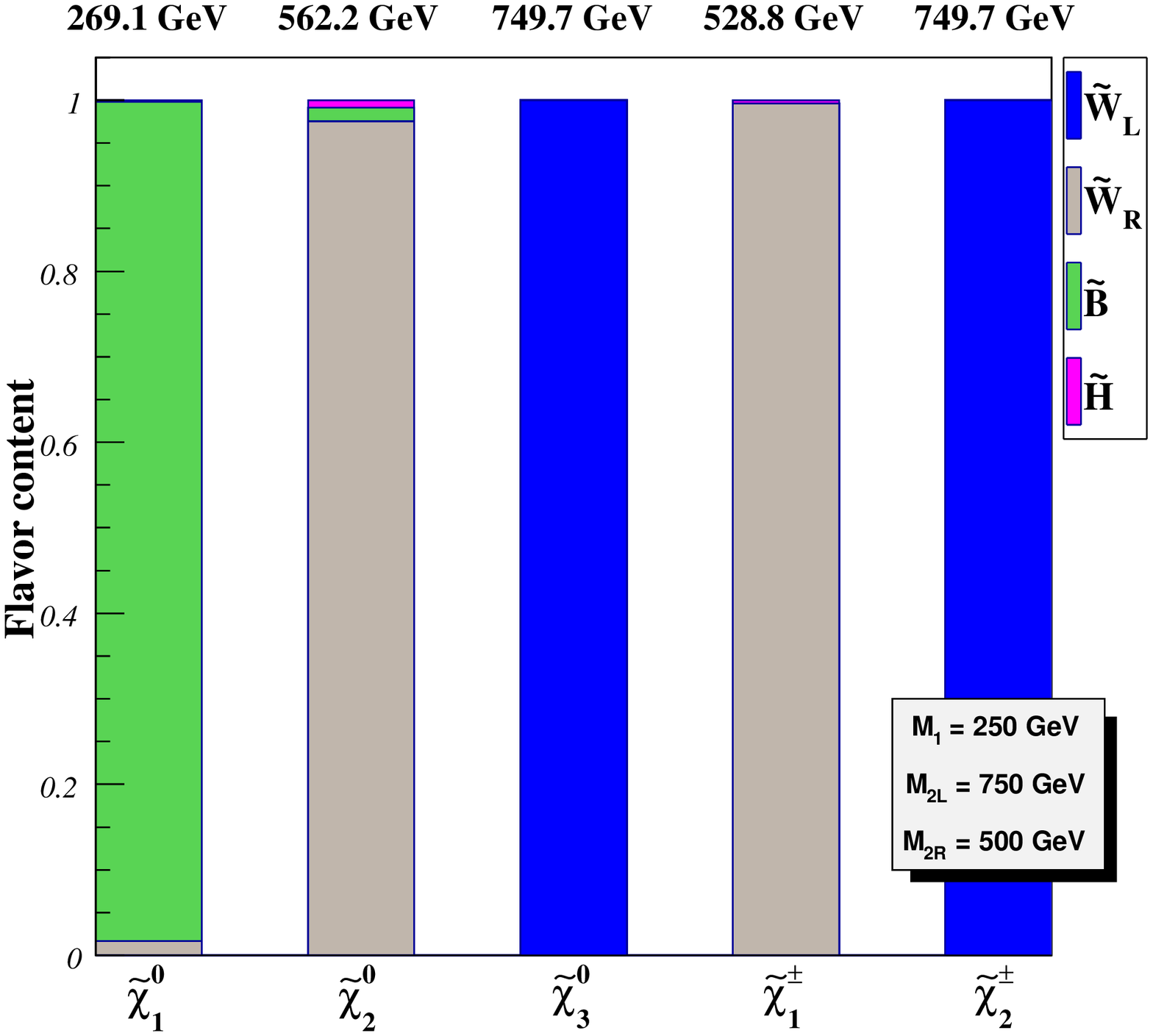}
\end{center}
\caption{\footnotesize Flavor decomposition and masses of the lightest neutralino and chargino
states for our benchmark scenarios I.1 (left panel) and I.2 (right panel) as
defined in Table \ref{4 scenarios}. The bino, $SU(2)_L$ and $SU(2)_R$ wino
component are presented in green, blue and gray, respectively, whilst the
Higgsino component is shown in pink.}
\label{fig:pure}
\end{figure}
\begin{figure}
\begin{center}
	\includegraphics[width=.49\columnwidth]{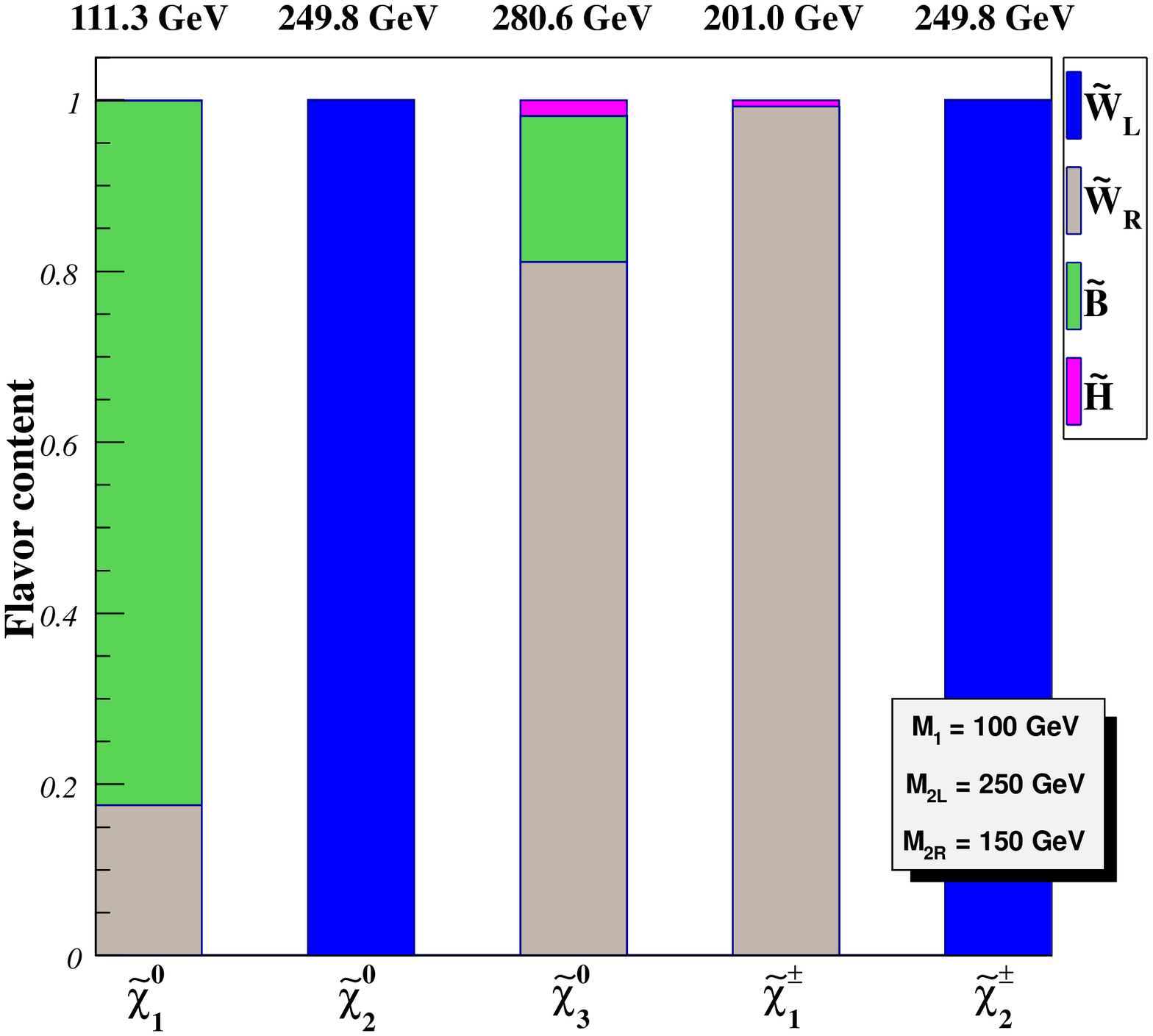} 
	\includegraphics[width=.49\columnwidth]{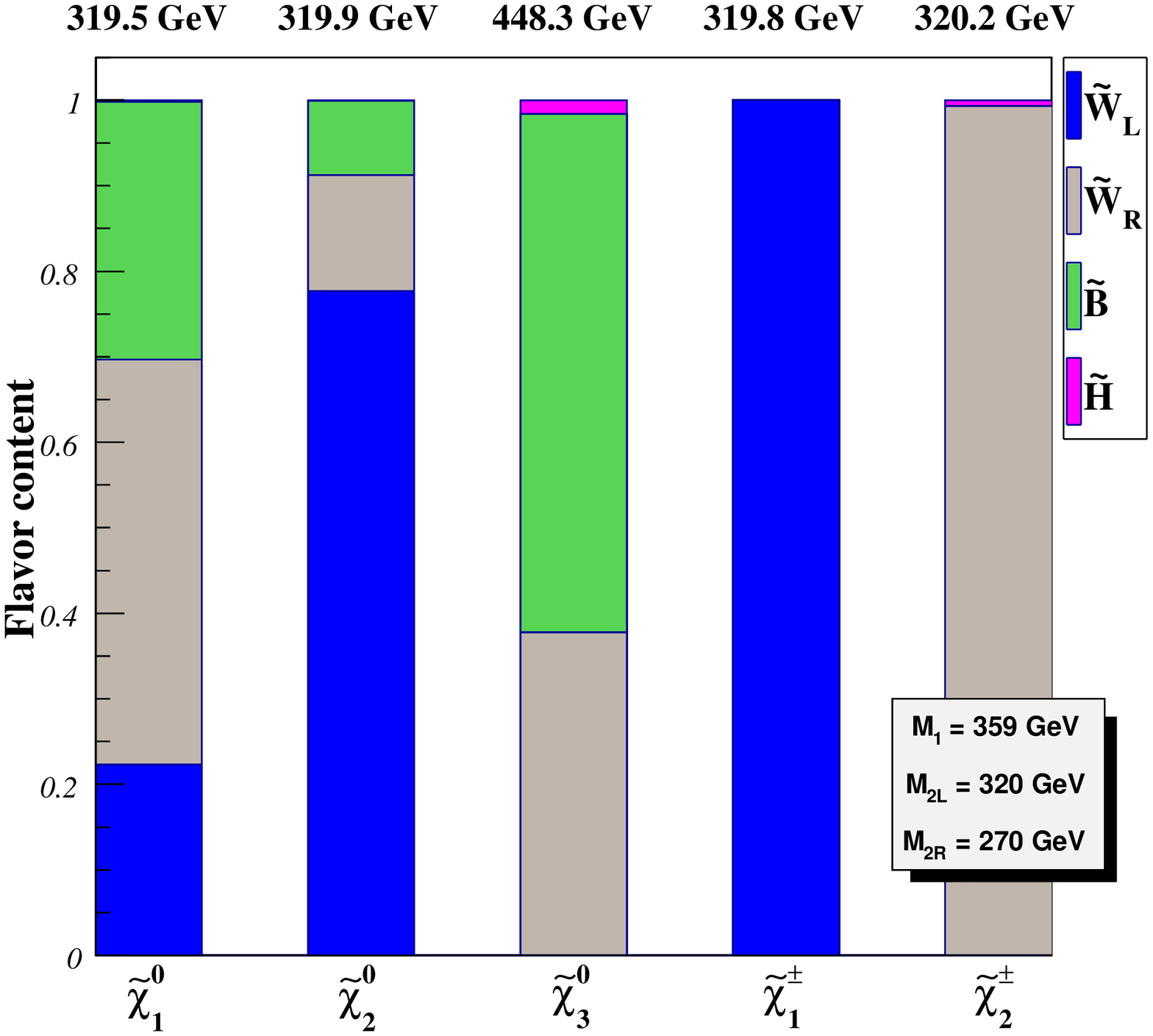}
\end{center}
\caption{\footnotesize Same as in Figure \ref{fig:pure} but for our benchmark scenarios II
(left panel) and III (right panel).}
\label{fig:mix}
\end{figure}
\paragraph{Two ``pure" scenarios}The first two scenarios presented schematically in figure \ref{fig:pure} exhibit very few mixings. In these two ``pure" scenarios, dubbed SI.1 and SI.2, the lightest neutralino in both cases only has a bino component while the decomposition of the second and third neutralino/chargino states depends on the hierarchy between $M_{2L} \et M_{2R}$. \\

Some general features can already be deduced from the spectrum we obtain. The unstable mass eigenstates being pure ($\tchi_2^0 \et \tchi_1^\pm$ are pure left-handed winos and $\tchi_3^0 \et \tchi_2^\pm$ are pure right-handed winos) their masses are equal to those of the winos. In scenario I.1, this degeneracy in mass imposes to the particles $\tchi_2^0 \et \tchi_1^\pm$ to only decay to the lightest neutralino and leptons while the two other ``inos" $\tchi_3^0 \et \tchi_2^\pm$ having a mass larger than that of sleptons have longer cascade decays where sleptons are expected to be produced and to decay into the bino state and charged leptons. In scenario I.2, where the hierarchy between the winos is inverted, we expect all the heavy mass eigenstates (in contrast with the lightest neutralino) to have long cascade decays involving sleptons and lighter neutralinos/charginos. However, one must keep in mind that states in these two scenarios are ``pure" and thus, due to the different helicities, no decay should be expected from a pure left- (right-) wino to a right- (left-) wino. 

\paragraph{Two ``mixed" scenarios} The two other scenarios we constructed, dubbed SII and SIII (see fig.\ref{fig:mix} for flavor decomposition) allow for more mixings in the mass eigenstates. Scenario II, where the mixing is moderate, has two pure left wino states (both second neutralino and chargino) the third lightest neutralino has a non-negligible bino part while the lightest chargino is a pure $SU(2)_R$ wino. The last scenario, SIII, allows the winos to appear in the flavor decomposition of all the neutralinos but the charginos which remains pure.\\

In these two scenarios, due to the presence of bino parts in the heavy states, we expect more envolved cascade decays where a heavy neutralino decays into a lighter chargino and a charged light lepton
$$ \tchi_0^3 \to \tchi_1^+ ~ \tilde{l}^- ,$$
for example. However, as one can see from figure \ref{fig:mix}, the splitting in mass between the various mass eigenstates is very small and one expects rather small branching ratios where a neutralino (chargino) decays into another neutralino or chargino. In scenario SIII, due to R-parity conservation, we even expect the particles $\tchi_2^0, \tchi_1^\pm \et \tchi_2^\pm$ to have vanishing two-body decays and tiny three-body decays.

\paragraph{Other masses} We decouple all squarks and gluinos from the low energy theory as we are only interested in leptonic final states. This choice can be justified by the fact that renormalization group equations evolution should push the spectrum of the latter particles to higher values due to the strong corrections. Slepton masses are set to a universal value of 400 GeV for all scenarios. However, in the case of the scenario SII, mass eigenstates being no heavier than 300 GeV cascade decays involving sleptons are thus kinematically impossible. We hence construct out of the latter scenario two benchmarks, one with slepton masses set to 200 GeV and one where they are set to 400 GeV.\\

Now all the numerical values are set, we can compute all the decay widths and cross sections associated to the production and cascade decays of the gauginos. In tables \ref{tab: lrsusy pure} are gathered together the branching ratios (left table) of the chargino and neutralino states to zero, one and two leptons and the cross sections (right panel) expressed in femtobarns (fb) associated to the production of these new heavy states. Tables \ref{tab: lrsusy mix} present the same
results but for the ``mixed" scenarios. the main conclusion to draw from these tables is that multilepton final states, {\ie} those with more than two charged light leptons, are quiet difficult to obtain, especially in the case of scenario III where particles decay preferably to jets and missing transverse energy. \\
\newpage

\begin{sidewaystable}
\centering
\begin{tabular}{c | c | c | c | c }
\hline
\hline
 & Process & n = 0 & n = 1 & n = 2 \\
\hline
\hline
\multirow{4}{*}{SI.1} & $ \tchi_2^0   \to \tchi_1^0 + n~\ell+X\quad$  & 0.57 & 0.08 & 0.35 \\
& $ \tchi_3^0   \to \tchi_1^0 + n~\ell+X\quad$ & 0.14 & 0.15 & 0.71 \\
& $ \tchi_1^\pm \to \tchi_1^0 + n~\ell+X\quad$ & 0.22 & 0.78 & 0\\
& $ \tchi_2^\pm \to \tchi_1^0 + n~\ell+X\quad$ & 0.22 & 0.78 & 0\\
\hline
\multirow{4}{*}{SI.2} & $ \tchi_2^0   \to \tchi_1^0 + n~\ell+X\quad$ & 0.15 & 0.15 & 0.70 \\
& $ \tchi_3^0   \to \tchi_1^0 + n~\ell+X\quad$ & 0.57 & 0.08 & 0.35 \\
& $ \tchi_1^\pm \to \tchi_1^0 + n~\ell+X\quad$ & 0.22 & 0.78 & 0\\
& $ \tchi_2^\pm \to \tchi_1^0 + n~\ell+X\quad$ & 0.22 & 0.78 & 0 \\
\hline
\end{tabular}$\quad$
\begin{tabular}{c | c | c | c | c }
\hline
\hline
 & Process & 7 TeV & 8 TeV & 14 TeV \\
\hline
\hline
\multirow{4}{*}{SI.1} & $p~p \to \tchi_2^0 \tchi_1^\pm\quad$  & $\quad 13.2\quad$ & $\quad 20.6\quad$ &$\quad  89.6\quad$ \\
                      & $p~p \to \tchi_3^0 \tchi_2^\pm\quad$  & $\quad 0.71\quad$ & $\quad 1.40\quad$ &$\quad  11.4\quad$ \\
                      & $p~p \to \tchi_1^0 \tchi_2^\pm\quad$  & $\quad < 0.1\quad$ & $\quad< 0.1\quad$ &$\quad  0.39\quad$ \\
                      & $p~p \to \tchi_1^+ \tchi_1^-\quad$    & $\quad 2.90\quad$ & $\quad 4.61\quad $ & $\quad 21.2\quad $ \\
                      & $p~p \to \tchi_2^+ \tchi_2^-\quad$    & $\quad 0.21\quad$ & $\quad 0.42\quad $ & $\quad 3.41 \quad$\\

\hline
\multirow{4}{*}{SI.2} & $p~p \to \tchi_2^0 \tchi_1^\pm\quad$  & $\quad 10.2 \quad$ & $\quad 16.3 \quad$ &$\quad 76.0 \quad$ \\
                      & $p~p \to \tchi_3^0 \tchi_2^\pm\quad$  & $\quad 0.98 \quad$ & $\quad 1.86\quad$ &$\quad 13.8 \quad$ \\
                      & $p~p \to \tchi_1^0 \tchi_1^\pm\quad$  & $\quad 1.16 \quad$ & $\quad 1.67\quad$ &$\quad 5.88 \quad$ \\
                      & $p~p \to \tchi_1^+ \tchi_1^-\quad$  & $\quad 4.49 \quad$ & $\quad 7.13 \quad $ &$\quad 32.9 \quad $ \\
                      & $p~p \to \tchi_2^+ \tchi_2^-\quad$  & $\quad 0.21 \quad$ & $\quad 0.40 \quad $ & $\quad 3.14 \quad$\\
\hline
\end{tabular}$\quad$

\caption{\label{tab: lrsusy pure} Tables summarizing the branching ratios of the various neutralinos and charginos states (left) and the cross sections (right) associated to their production at the LHC running at a center-of-mass energy of 7, 8 and 14 TeV. Cross sections are expressed in femtobarns {fb}. }
\end{sidewaystable}
\begin{sidewaystable}
\centering
\begin{tabular}{c | c | c | c | c }
\hline
\hline
 & Process & n = 0 & n = 1 & n = 2 \\
\hline
\hline
\multirow{4}{*}{SII.200} & $ \tchi_2^0   \to \tchi_1^0 + n~\ell + X\quad$ & 0.57 & 0.08 & 0.35 \\
&$ \tchi_3^0   \to \tchi_1^0 + n~\ell + X\quad$& 0.26 & 0.13 & 0.61 \\
&$ \tchi_1^\pm \to \tchi_1^0 + n~\ell + X\quad $ & 1 & 0 & 0\\
&$ \tchi_2^\pm \to \tchi_1^0 + n~\ell + X\quad $ & 0.22 & 0.78 & 0 \\
\hline
\multirow{5}{*}{SII.400} & $ \tchi_2^0 \to \tchi_1^0 + n~\ell + X\quad $ & 0.57 & 0.08 & 0.35 \\
& $ \tchi_3^0 \to \tchi_1^0 + n~\ell + X \quad $ & 0.12 & 0.09 & 0.43 \\
& $ \tchi_3^0 \to \tchi_1^\pm  + n~\ell + X\quad $ & 0.35 & 0 & 0\\
& $ \tchi_1^\pm \to \tchi_1^0 + n~\ell + X\quad  $ & 1 & 0 & 0\\
& $ \tchi_2^\pm \to \tchi_1^0 + n~\ell + X\quad  $ & 0.22 & 0.78 & 0 \\
\hline
\multirow{8}{*}{SIII} & $ \chi_2^0 \to \chi_1^0 + n~\ell   +X \quad $ & 0.22 & 0 & 0.16 \\
& $ \chi_2^0 \to \chi_1^\pm+ n~\ell  +X \quad $ & 0.52 & 0.10 &0 \\
& $ \chi_3^0 \to \chi_1^0 + n~\ell   +X \quad $ & 0.10 & 0 & 0 \\
& $ \chi_3^0 \to \chi_2^0 + n~\ell   +X \quad $ & 0.10 & 0.03 &  0.13 \\
& $ \chi_3^0 \to \chi_1^\pm  + n~\ell+X \quad $ & 0.14 & 0.51 &  0\\
& $ \chi_1^\pm \to \chi_1^0 + n~\ell +X \quad $ & 0.84 & 0.16&0  \\
& $ \chi_2^\pm \to \chi_1^0 + n~\ell +X \quad $  & 0.996 & 0 & 0 \\
& $ \chi_2^\pm \to \chi_2^0 + n~\ell +X \quad $  & 0.004 & 0 & 0 \\
\hline\hline
\end{tabular}
\begin{tabular}{c | c | c | c | c }
\hline
\hline
 & Process & 7 TeV & 8 TeV & 14 TeV \\
\hline
\hline
\multirow{7}{*}{SII.200} & $p~p \to \tchi_3^0 \tchi_1^\pm\quad$  & $\quad 4999 \quad$ & $\quad 6530 \quad$ &$\quad 17490 \quad$ \\
                         & $p~p \to \tchi_1^0 \tchi_1^\pm\quad$  & $\quad 3139 \quad$ & $\quad 4085\quad$ &$\quad 10830 \quad$ \\
                         & $p~p \to \tchi_2^0 \tchi_2^\pm\quad$  & $\quad 387 \quad$ & $\quad 514 \quad$ &$\quad 1452 \quad$ \\
                         & $p~p \to \tchi_2^0 \tchi_1^\pm\quad$  & $\quad 0.83 \quad$ & $\quad 1.09\quad$ &$\quad 2.88\quad$ \\
                         & $p~p \to \tchi_3^0 \tchi_2^\pm\quad$  & $\quad < 0.1 \quad$ & $\quad < 0.1\quad$ &$\quad 0.11 \quad$ \\
                         & $p~p \to \tchi_1^+ \tchi_1^-\quad$  & $\quad 532 \quad$ & $\quad 780\quad $ & $\quad 2851\quad $ \\
                         & $p~p \to \tchi_2^+ \tchi_2^-\quad$  & $\quad 92.2 \quad$ & $\quad 123 \quad $ & $\quad 355.9 \quad$\\
\hline
\multirow{7}{*}{SII.400} & $p~p \to \tchi_3^0 \tchi_1^\pm\quad$  & $\quad 5188 \quad$ & $\quad 6776 \quad$ &$\quad 18140 \quad$ \\
                         & $p~p \to \tchi_1^0 \tchi_1^\pm\quad$  & $\quad 3255 \quad$ & $\quad 4236\quad$ &$\quad 11230 \quad$ \\
                         & $p~p \to \tchi_2^0 \tchi_2^\pm\quad$  & $\quad 387 \quad$ & $\quad 514 \quad$  &$\quad 1451 \quad$ \\
                         & $p~p \to \tchi_2^0 \tchi_1^\pm\quad$  & $\quad 0.86 \quad$ & $\quad 1.13\quad$ &$\quad 3 \quad$ \\
                         & $p~p \to \tchi_3^0 \tchi_2^\pm\quad$  & $\quad < 0.1 \quad$ & $\quad < 0.1\quad$ &$\quad 0.11 \quad$ \\
                         & $p~p \to \tchi_1^+ \tchi_1^-\quad$  & $\quad 572 \quad$ & $\quad 838\quad $ & $\quad 3059\quad $ \\
                         & $p~p \to \tchi_2^+ \tchi_2^-\quad$  & $\quad 92.2 \quad$ & $\quad 123 \quad $ & $\quad 356 \quad$\\
\hline
\multirow{7}{*}{SIII} & $p~p \to \tchi_2^0 \tchi_1^\pm\quad$  & $\quad 99.6 \quad$ & $\quad 137\quad$ &$\quad 433 \quad$ \\
                      & $p~p \to \tchi_1^0 \tchi_2^\pm\quad$  & $\quad 93.6 \quad$ & $\quad 128\quad$ &$\quad 393 \quad$ \\
                      & $p~p \to \tchi_1^0 \tchi_1^\pm\quad$  & $\quad 28.5 \quad$ & $\quad 39.3 \quad$ &$\quad 125 \quad$ \\
                      & $p~p \to \tchi_2^0 \tchi_2^\pm\quad$  & $\quad 26.7 \quad$ & $\quad 36.6 \quad$ &$\quad 113 \quad$ \\
                      & $p~p \to \tchi_3^0 \tchi_2^\pm\quad$  & $\quad 13.0 \quad$ & $\quad 19.0\quad$ &$\quad 69.2 \quad$ \\
                      & $p~p \to \tchi_2^+ \tchi_2^-\quad$  & $\quad 537 \quad$ & $\quad 788 \quad $ & $\quad 2887\quad$\\
                      & $p~p \to \tchi_1^+ \tchi_1^-\quad$  & $\quad 29.8 \quad$ & $\quad 41.7\quad $ & $\quad 137 \quad $ \\
\hline\hline
\end{tabular}
\caption{\label{tab: lrsusy mix} Same as in table \ref{tab: lrsusy pure} but for scenarios II and III. Here symbols ``SII.200" and ``SII.400" refere to the benchmarks constructed out of the scenario II where slepton masses are set to 200 and 400 GeV respectively. Cross sections are expressed in femtobarns.}
\end{sidewaystable}
\clearpage

\subsection{Analysis and results}
We now turn to the Monte Carlo analysis. Requiring only events where the final state contains at least one charged lepton, we decided to subdivide our analysis following the number of charged leptons in the final state. We hence present the results in the single-lepton, the dilepton and the multilepton\footnote{From now on, multilepton will refer to a number of leptons strictly larger than two.} channels. The strategy we shall follow to enhance the sensitivity of our analysis to the signal, which is defined as the ratio
$$ \frac{\rm S}{\sqrt{\rm S+B}} $$
where S and B are the number of signal and background events, respectively; will be based on constraining the values of some observables. Before proceeding, however, one must make sure that leptons and jets can be detected by both the {\sc CMS} and the {\sc ATLAS} detectors. Consequently, for each of the following analyses, we will require the leptons to have a transverse momentum ($p_T$) strictly larger than 10 GeV and the jets to have a transverse momentum strictly larger than 20
GeV. Furthermore, we apply two isolation criteria to make sure, first, that no jet object lies within a cone around an electron at an angular distance of 0.1 and then no lepton remains within a cone centered on the surviving jet objects and within a radius of 0.4. Finally, we reject all events where a b-jet was tagged in order to reduce the background events coming from the processes leading to the production of a top-quark. Signal events\footnote{Signal events refers to events coming from the decay of the charginos and the neutralinos.} where we expect only lighter quarks to be produced are also affected by this constraint as we allow for a non-vanishing mis-tagging rate for the latter jet objects, however the loss is rather tiny. Finally, table \ref{tab: events before cut} summarizes the number of events we get after all the above selection criteria were applied and after selecting the events where the number of leptons is equal to one, two and more.

\begin{table}
\centering
\begin{tabular}{c|c||c|c|c}
\cline{2-5}
\cline{2-5}
& \multirow{2}{*}{Process} &\multicolumn{3}{|c}{Number of events in each channel}\\
\cline{3-5}
&  & One lepton & Two leptons & Three or more leptons\\
\hline
\hline
\multirow{5}{*}{LRSUSY scenarios}& SI.1 & $221.8\pm 11.4$ & $118.14 \pm 9.61$ & $99.85 \pm 9.03 $    \\
\cline{2-5}
& SI.2 & $ 156.6 \pm 10.6 $ & $173.5 \pm 10.9$ & $139.4 \pm 10.2$  \\
\cline{2-5}
& SII.200 & $ 30095 \pm 162 $ & $41498 \pm 186 $ & $1763.0 \pm 41.8$ \\
\cline{2-5}
& SII & $24550 \pm 148 $ & $ 66329 \pm 220 $ & $1987.0 \pm 44.4 $\\
\cline{2-5}
& SIII & $169.9 \pm 13.0$ & $42.72 \pm 6.53$ & $\approx 0 $ \\
\hline
\hline
\multirow{6}{*}{Background processes} & Single top & $135590.2\pm 542.7$ & $5543.6 \pm 74.0$ & $ \approx 0 $\\
\cline{2-5}
& Diboson & $373640.79\pm 1033.88$ & $67780.4\pm 451.27$ & $ 5394.0\pm 83 $ \\
\cline{2-5}
& $t~\bar{t}$ & $248763\pm 628$ & $ 35096 \pm 181$ & $\approx 0$ \\
\cline{2-5}
& $W$+ jets & $344732715\pm 13348 $ & $ \approx 0 $ & $ \approx 0$ \\
\cline{2-5}
& $Z$+ jets & $21087970 \pm 4351 $ & $22979668 \pm 4518$ & $\approx 0$ \\
\cline{2-5}
& Rare &  $479.95\pm 37.7$ & $133.36\pm 20.40$ & $25.44\pm 9.25 $\\
\hline
\hline
\end{tabular}
\caption{\footnotesize \label{tab: events before cut}Number of events surviving and the related uncertaintities in each channel after having applied all the selection criteria to make sure our events are detected.}
\end{table}

\paragraph{One lepton channel} We now try to extract the signal from the background when the final state contains only one light charged lepton. From table \ref{tab: events before cut} we see that in this channel the situation is rather complicated as the scenarios that we have simulated induce a low number of events compared to the Standard Model predictions. For the latter, most events (94$\%$) come from the processes producing a $W$ boson and jet objects, followed by $Z$+jets events which only represent 5$\%$. In order to reduce this background we first concentrate on the transverse mass of the pair comprised of the charged lepton and the missing transverse energy which is defined as 
$$ M_T^2 =  \sqrt{ 2p_T^l \slashed{E}_T [ 1 - \cos\Delta\phi_{l,\slashed{E}_T}]} $$ 
where 
$\slashed{E}_T$ is the missing transverse energy, $\Delta\phi_{l,\slashed{E}_T}$ is the angular distance, in the azimuthal direction with respect to the beam and $p_T^l$ is the transverse momentum of the lepton. In the case where the lepton and the missing energy originate both from a $W$ gauge boson, the transverse mass is identified to that of the latter and is hence small compared to the case where both the lepton and the missing transverse energy originate from a supersymmetric particle. Figure \ref{fig: 1l mtl1} illustrates this fact and motivates the choice to keep only events where this variable is larger than 200 GeV. This first requirement allows to divide the background by 300 but does not allow our signal to rise yet and one has to turn now towards other kinematical variables. \\
\begin{figure}
\centering
\includegraphics[scale=0.5]{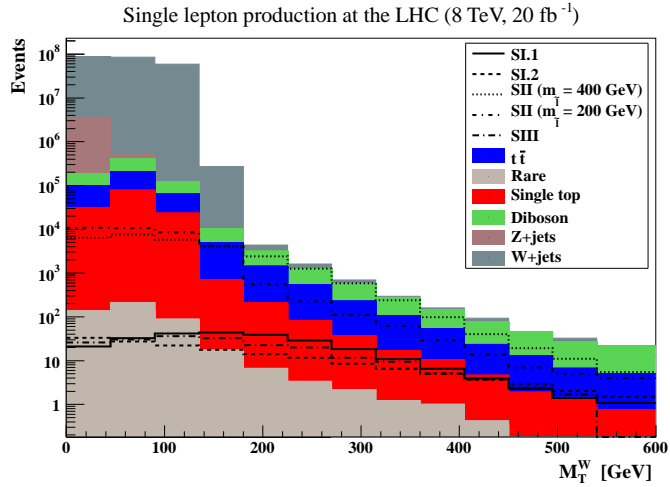}
\caption{\footnotesize \label{fig: 1l mtl1} Distribution of the transverse mass of the pair lepton - MET before requiring the latter to be at least equal to 200 GeV. Here the signal was superimposed to the distribution of the background.}
\end{figure}

The next kinematical variable which we find useful to enhance the visibility of the signal is the transverse missing energy carried by the undetected particles. From figure \ref{fig: 1l met} where we plot the distribution related to this observable (after the requirement on the transverse mass was applied) we see that background events peak around small values while the signal peaks at higher values. Requiring thus a missing transverse energy at least equal to 100 GeV allows to furthermore reduce the background from the Standard Model by a factor $\sim 2$. \\

Continuing our analysis, we find one more key observable we can constrain in order to reduce the overwhelming background, namely the lepton transverse momentum. From the distribution depicted in figure \ref{fig: 1l ptl1}, we see that, as expected, leptons originating from our benchmarks are harder than those produced by SM processes. We hence proceed to reject all the events where the lepton is softer than 80 GeV allowing us to further reduce the background events by a factor larger than 3. \\

All in all, with these simple requirements, though we succeeded in reducing the number of background events by a factor of almost 1830, their number is still very high with respect to our benchmark scenarios predictions. Moreover, it turns out that any additional requirement would decrease significantly the number of signal events. However, scenario II in both benchmarks turns out to be rather interesting as the significance of the signals compared to the Standard Model expectations is of 5.29 and 10.91 when the universal slepton mass is set to 200 and 400 GeV respectively. In table \ref{tab: 1l summary} we give the number of events expected for each scenario and the deviation it represents with respect to the Standard Model background.\\

\begin{figure}
\subfigure[]{%
\label{fig: 1l met}\includegraphics[scale=0.4]{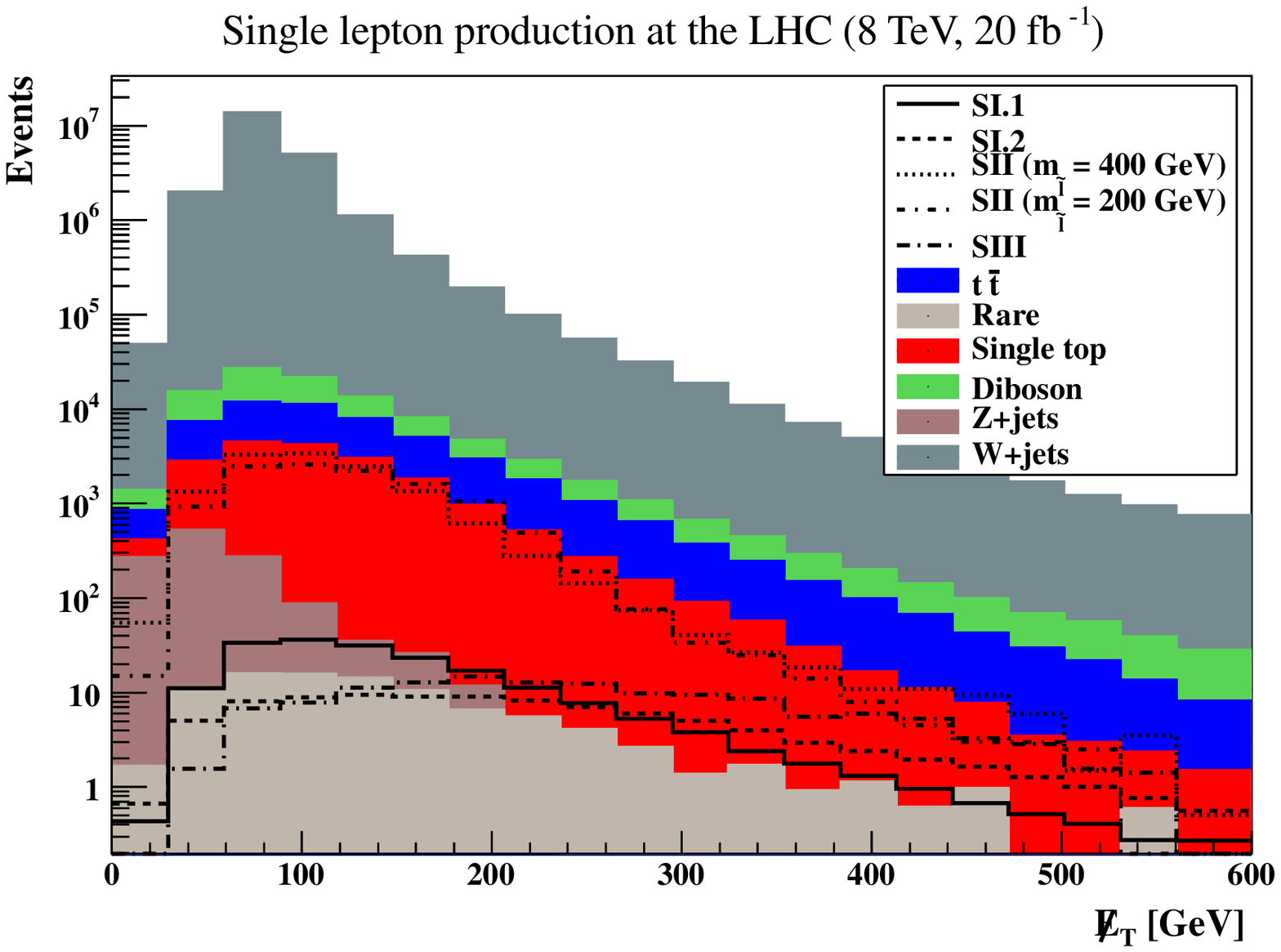}}
\subfigure[]{%
\label{fig: 1l ptl1}\includegraphics[scale=0.4]{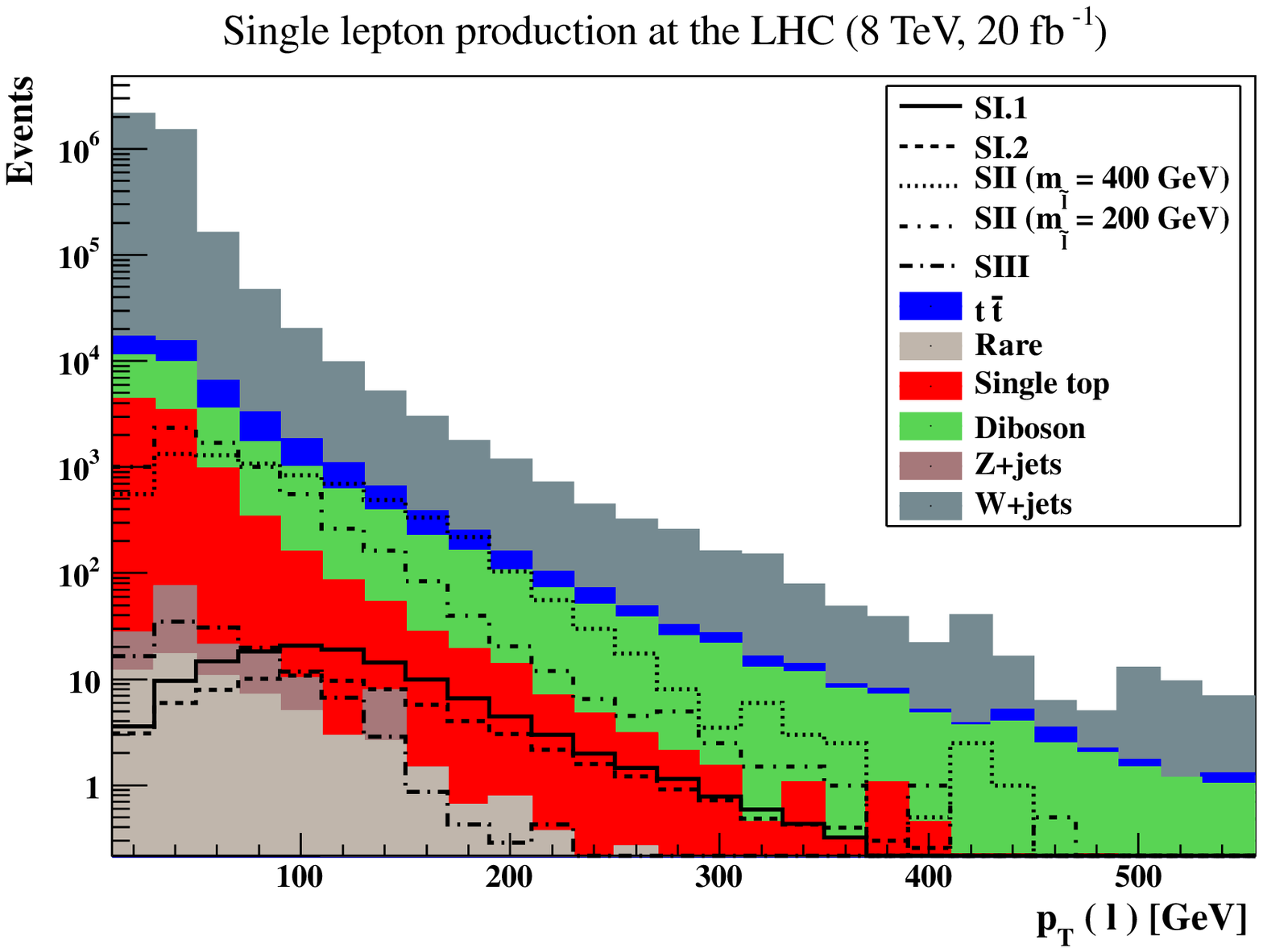}}
\caption{\footnotesize Distribution of the missing transverse energy (left panel) before we required it to be at least equal to 100 GeV and distribution of the transverse momentum of the unique charged lepton just before the cut selecting only events with leptons harder than 80 GeV.}
\end{figure}

\begin{table}[h]
\begin{center}
\begin{tabular}{|c|c|c|c|}
\hline
& Signal & Background & $\frac{\rm S}{\sqrt{\rm S+B}}$ \\
\hline
Scenario I.1 & $ 79.07 \pm 8.22 $ & \multirow{5}{*}{$ 61447 \pm 247 $} & $ 0.32\pm 0.08$ \\
\cline{1-2}\cline{4-4}
Scenario I.2 & $ 46.76 \pm 6.54 $ &  & $ 0.18 \pm 0.06 $  \\
\cline{1-2}\cline{4-4}
Scenario II, $m_{\tilde{l}}=200{\rm  GeV}$ & $ 1328.2 \pm 36.3 $ & & $ 5.29 \pm 0.34$\\
\cline{1-2}\cline{4-4}
Scenario II, $m_{\tilde{l}}=400{\rm  GeV}$ & $ 2778.3 \pm 52.4 $ &  &$ 10.91 \pm 0.48$\\
\cline{1-2}\cline{4-4}
Scenario III & $ 26.54 \pm 5.15 $ & & $0.10\pm 0.05$  \\
\hline
\end{tabular}
\end{center}
\caption{\footnotesize \label{tab: 1l summary} Table summarizing the number of events we find after all requirements were applied in the one lepton channel. It is clear from these results that scenario SII is the most promising one compared to the other benchmark points we have simulated.}
\end{table}

\newpage
\paragraph{Two leptons channel} After selecting only events where the number of charged leptons is exactly equal to two, we find that the situation is better for both benchmarks constructed out of scenario II and for the scenario I.2 where the number of events increases with respect to one-lepton final states. For the other scenarios, however, the number of events decreases. As to the background events, $Z$+jets events are here dominant as they represent more than 99$\%$ of the background events.\\

This is a good news because these events are known to lead too few missing transverse energy in the final states as can be seen on figure \ref{fig: 2l met}. We seize this opportunity and impose the constraint to only keep events where the missing transverse energy is higher than 80 GeV which allows us to also reduce the other background sources, {\ie} $t~\bar{t}$ and diboson events. This constraint reduces the total background events by a factor larger than 500. \\

After the latter requirement is applied, the situation is rather lukewarm. On the one hand, both benchmark scenarios constructed out of scenario II are expected to produce a number of events very close to Standard Model predictions reaching a significance at least equal to 90; on the other hand scenarios I.1, I.2 and III predict less events. Let us thus try to enhance the sensitivity of our analysis by considering other observables.\\

Considering the leading lepton transverse momentum, we remark from figure \ref{fig: 2l ptl1} that its value in the case of Standard Model processes is rather low compared to LRSUSY processes and one may thus safely impose a minimum value of 80 GeV.\\

Finally we also consider the distribution of the second lepton present in this channel. From figure \ref{fig: 2l ptl2}, one can see that the distributions induced by Standard Model processes are characterized by low transverse momenta, contrary to those from LRSUSY. One exception for our supersymmetric signal is the scenario III which predicts rather low values for this observable but the number of events being already very low, we do not take this scenario into account and choose to require a next-to-leading lepton with a minimum $p_T$ of 70 GeV. \\

\begin{figure}
\centering
\includegraphics[scale=0.5]{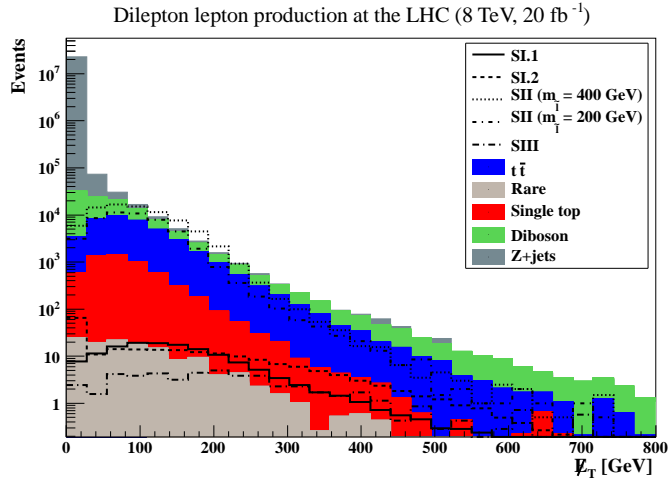}
\caption{\footnotesize \label{fig: 2l met} Distribution of the missing transverse energy after having selected events containing exactly two charged light leptons. One can see that scenario SII already displays a high number of events by contrast to scenario SIII which, here also, seems to be disfavoured.}
\end{figure}

\begin{figure}
\subfigure[]{\label{fig: 2l ptl1} \includegraphics[width=.5\columnwidth]{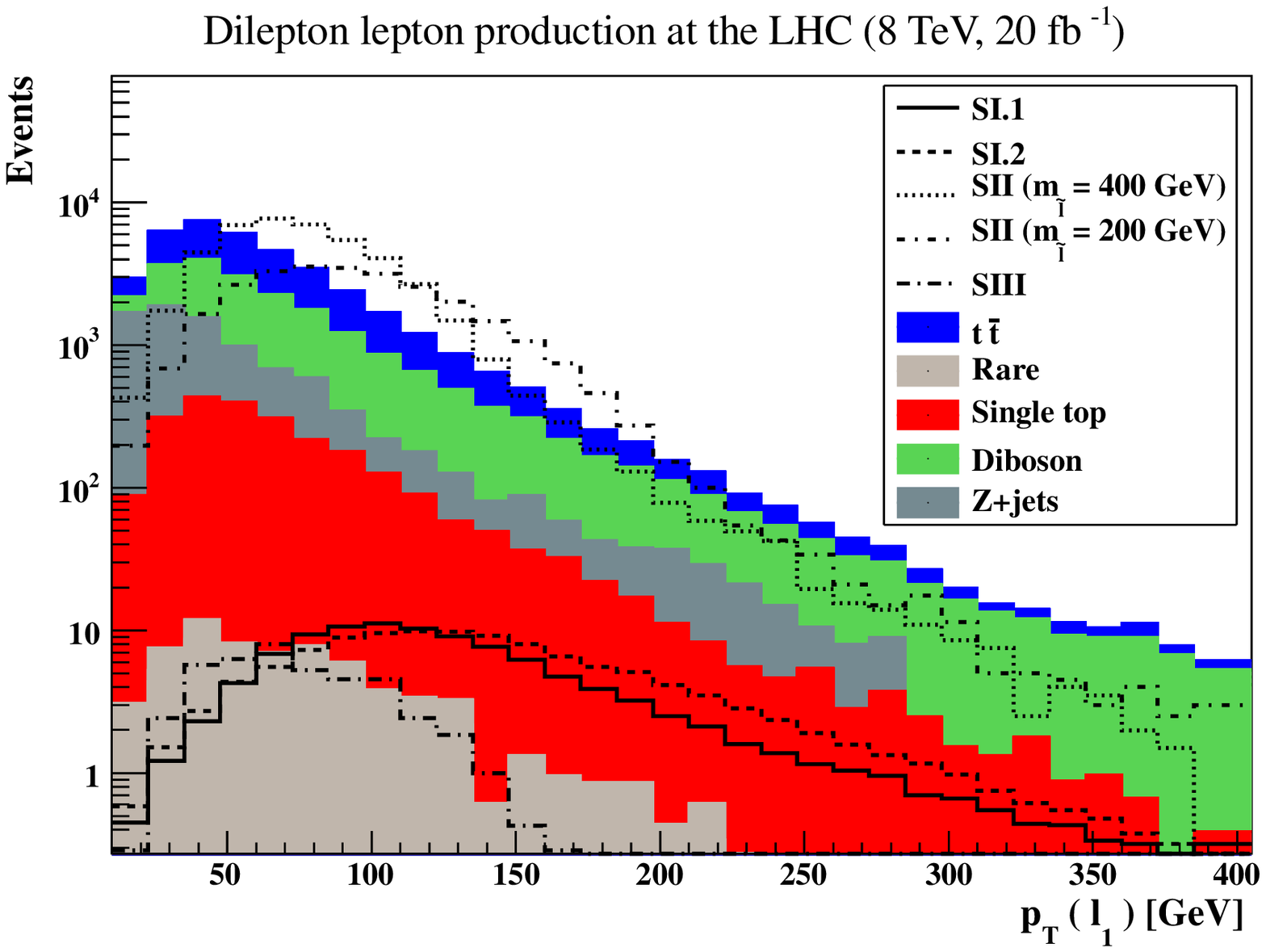}}
\subfigure[]{\label{fig: 2l ptl2} \includegraphics[width=.5\columnwidth]{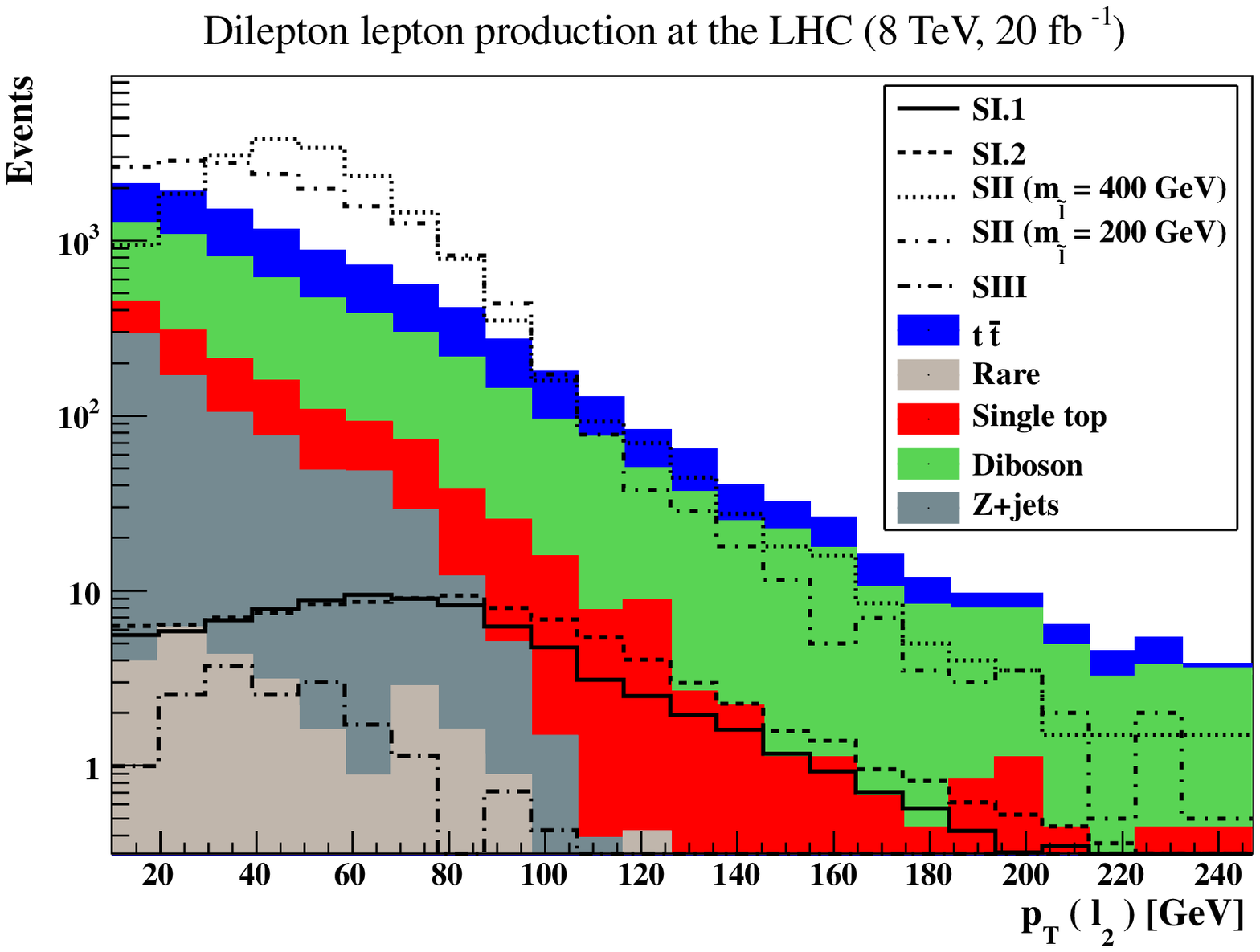}}
\caption{\footnotesize Distribution of the transverse momenta of both the leading lepton (left panel) and the next-to-leading lepton (right panel) in the two lepton channel. Leading lepton $p_T$ was drawn after the constrain on the value of the missing transverse energy while the $p_T$ of the next-to-leading lepton was drawn after requiring a leading lepton with a $p_T$ at least equal to 80 GeV. }
\end{figure}
The combined effects of these two constraints allows us to reduce the number of background events by a factor larger than 20 but this however is not enough to improve the sensitivity of the analysis to scenarios I.1, I.2 and III. In the latter case the number of events is even insignificant. The events from scenario II, though reduced by an order of magnitude by the requirements on the lepton transverse momenta, now predict a higher event rate with respect to the Standard Model. In table \ref{tab: 2l summary} we summarize these results.\\
\begin{table}
\begin{center}
\begin{tabular}{|c|c|c|c|}
\hline
& Signal & Background & $\frac{\rm S}{\sqrt{\rm S+B}}$ \\
\hline
Scenario I.1 & $ 41.2 \pm 6.8$ & \multirow{5}{*}{$1748.3 \pm 41.7$} & $ 0.97 \pm 0.32 $ \\
\cline{1-2}\cline{4-4}
Scenario I.2 & $ 53.9 \pm 7.7$ & & $ 1.27 \pm 0.36 $ \\
\cline{1-2}\cline{4-4}
Scenario II, $m_{\tilde{l}}=200{\rm  GeV}$ & $ 2610 \pm 56 $ & & $ 39.5 \pm 1.2 $ \\
\cline{1-2}\cline{4-4}
Scenario II, $m_{\tilde{l}}=400{\rm  GeV}$ & $ 2686 \pm 57 $ & & $ 40.3 \pm 1.2 $ \\
\cline{1-2}\cline{4-4}
Scenario III & $ 2.6 \pm 1.8 $ & & $ 0.06 \pm 0.08 $\\
\hline
\end{tabular}
\end{center}
\caption{\footnotesize \label{tab: 2l summary} Number of events we expect per benchmark point and the sensitivity of the LHC running at 8 TeV and with 20 $fb^{-1}$ of data if we required exactly two charged light leptons whose transverse momenta are larger than 80 GeV and 70 GeV and with at least 80 GeV of missing transverse energy. Scenario SIII remains invisible even in this channel.}
\end{table}

From the above analyses, we can already draw some preliminary conclusions. Scenarios exhibiting a moderate mixing in the gauge eigenstates seem to be the most favoured, leading to very high signal efficiencies. This result can be explained by the interplay of both the relatively large mass splitting between mass eigenstates but also and more importantly by the low masses of the gauginos which lead to high production cross sections. \\

Scenario SIII suffers from both a heavy spectrum leading to small cross sections and tiny mass splitting which reduces furthermore the branching ratios into more than one lepton. Such a compressed spectrum also imposes the decay products to carry small transverse momenta. \\

\paragraph{Multilepton channel} 
In scenario SI.1 and SI.2, the branching ratios to one and two charged leptons of some of the neutralinos and charginos ( $\tchi_3^0, \tchi_{\{1,2\}}^\pm$ for scenario SI.1 and  $\tchi_2^0, \tchi_{\{1,2\}}^\pm$ for scenario SI.2) are dominant which makes the multilepton channel promising. We thus consider now multilepton final states.\\

As expected, the simple fact of rejecting all events with less than three charged light leptons reduces significantly the number of background events leaving only diboson ($99\%$ of the background events) and a tiny fraction from rare processes. More satisfactory, all scenarios but SIII are now in good shape as SI.1 already has a significance of 1.3, SI.2 a signficance of 1.86, SII with low slepton masses has a significance of 20 and finally benchmark SII with slepton masses set to 400 GeV has a significance of 23. Obviously, we cannot state that scenarios SI.1 and SI.2 are visible yet but these first results are promising.\\

We thus proceed with the kinematical constraints. Following the same pattern as in the two lepton channel, we start by the missing transverse energy. From figure \ref{fig: 3l met}, we can see that requiring this observable to be higher than about 70 - 100 GeV should allow us to reduce significantly the background while conserving a good amount of signal events. We choose the value of 100 GeV and reduce by consequence the background by a factor of 7 enhancing the sensitivity of our analysis to scenario SI.1, \textit{e.g.}, by a factor of 2, which is good but not enough. \\

Moving to the transverse momenta of both leading leptons, we find from figure \ref{fig: 3l ptl1} that the leading lepton arising from diboson processes has a mean value around 100 GeV while those from LRSUSY benchmarks range from 110 for both scenarios constructed out of SII to 170 GeV in the case of scenarios SI.1 and SI.2 . Choosing a rather conservative option with respect to both scenarios SII, we require this observable to be at least equal to 80 GeV. Finally, we also consider the second (ordered by decreasing $p_T$) lepton transverse momentum and deduce from figure \ref{fig: 3l ptl2} a minimum value of 70 GeV. \\

With all these selection criteria, we succeed in achieving a sensitivity of 4.64 for scenario SI.1 and 6.98 for scenario SI.2. As to the benchmark points constructed from scenario SII, we find that the number of events surviving all the selection criteria and requirements is higher than those from the Standard Model background. Finally, scenario SIII did not pass the selection criteria asking for at least three charged leptons which is certainly due to its compressed spectrum. In table \ref{tab: 3l summary}, we summarize these results.

\begin{figure}
\centering
\includegraphics[scale=0.4]{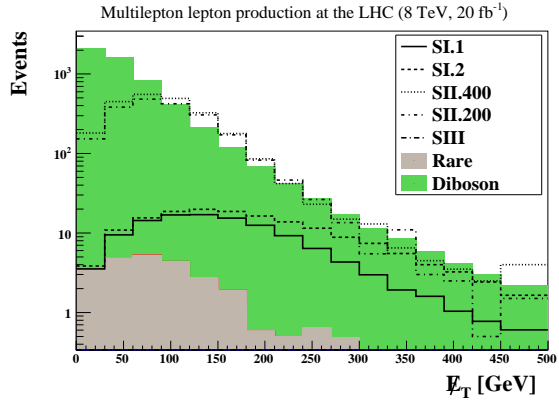}
\caption{\footnotesize \label{fig: 3l met} Distribution of the transverse missing energy when the selected events contain at least three charged light leptons. We already see that scenario II is still the most promising but with a cut selecting only events with MET higher than 100 GeV, one should be able to improve the sensitivity of our analysis to the other scenarios.}
\end{figure}
\begin{figure}
\subfigure[]{%
\label{fig: 3l ptl1}\includegraphics[width=.49\columnwidth]{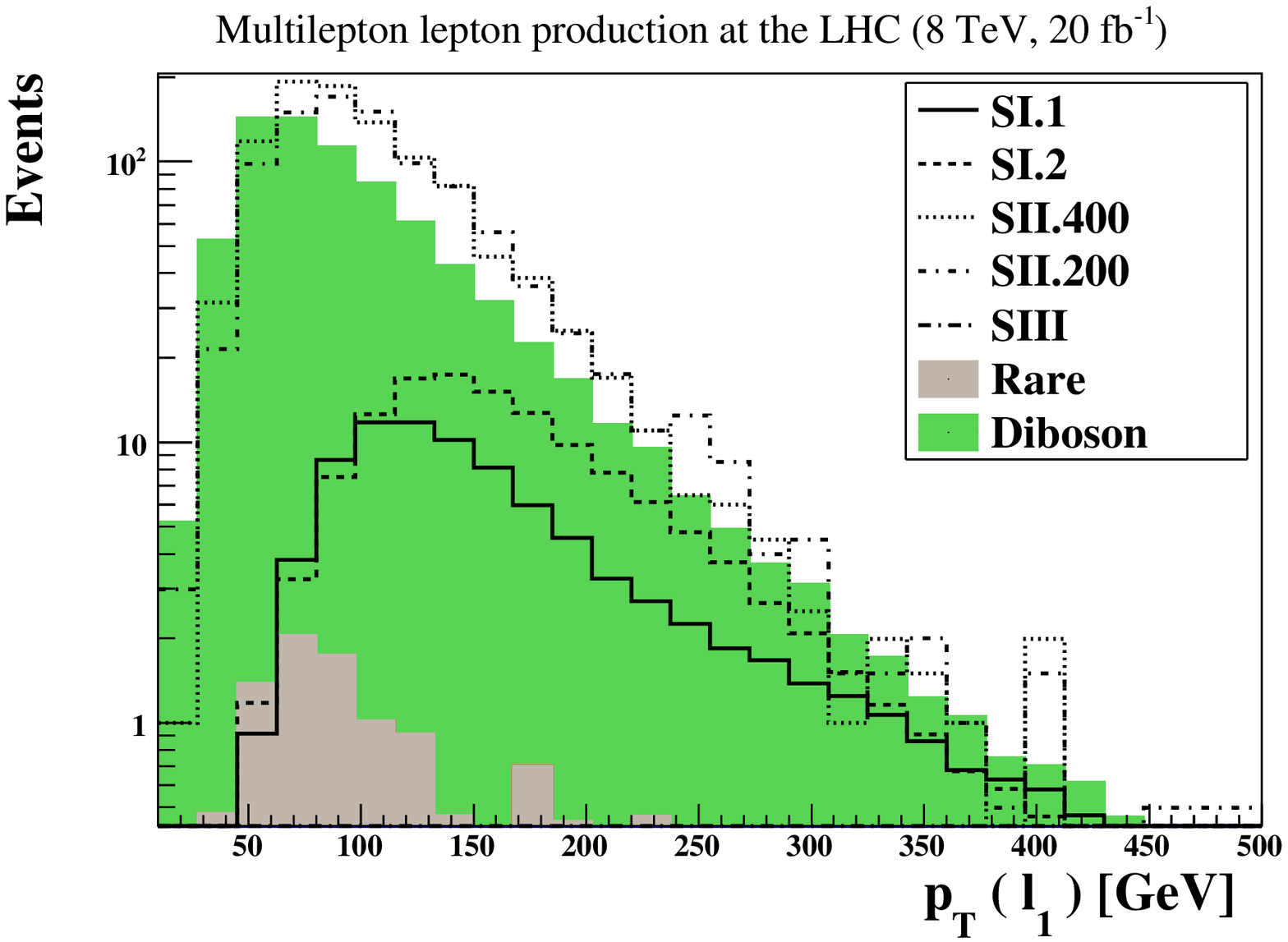}}
\subfigure[]{%
\label{fig: 3l ptl2}\includegraphics[width=.49\columnwidth]{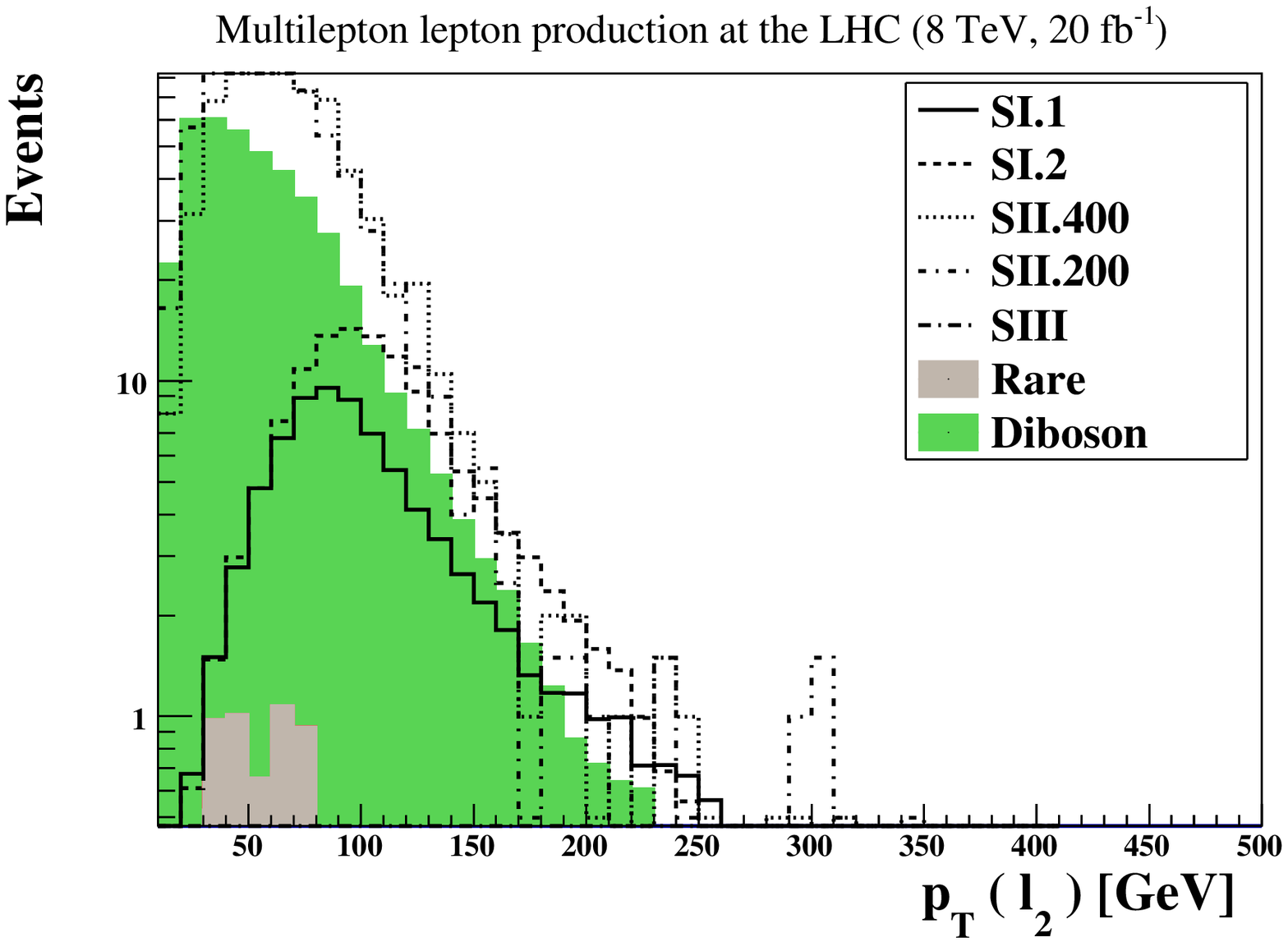}}
\caption{\footnotesize Distributions of the transverse momenta of the leading and next-to-leading lepton. Histogram in the left panel is drawn after the cut on the MET and that in the right panel is drawn after the cut on the leading lepton's $p_T$.}
\end{figure}
\begin{table}[!h]
\begin{center}
\begin{tabular}{|c|c|c|c|}
\hline
& Signal & Background & $\frac{\rm S}{\sqrt{\rm S+B}}$ \\
\hline
Scenario I.1 & $ 65.4 \pm 8.4 $ & \multirow{5}{*}{$133.4 \pm 11.5$} & $ 4.64 \pm 1.03 $ \\ 
\cline{1-2}\cline{4-4}
Scenario I.2 & $ 108 \pm 10 $ & & $ 6.98 \pm 1.09 $ \\
\cline{1-2}\cline{4-4}
Scenario II, $m_{\tilde{l}}=200{\rm  GeV}$ & $ 259 \pm 18 $ & & $ 13.1 \pm 1.3 $ \\
\cline{1-2}\cline{4-4}
Scenario II, $m_{\tilde{l}}=400{\rm  GeV}$ & $ 289 \pm 19 $ & & $ 14.1 \pm 1.3 $ \\
\cline{1-2}\cline{4-4}
Scenario III & $ 0 $ & & $ \approx 0 $ \\
\hline
\end{tabular}
\end{center}
\caption{\footnotesize \label{tab: 3l summary} Number of events we expect per benchmark point and the sensitivity of the LHC running at 8 TeV and with 20 $fb^{-1}$ of data if we required three charged light leptons or more in the final state whose transverse momenta are larger than 80 GeV and 70 GeV (for the leading and next-to-leading lepton, resp.) and with at least 100 GeV of missing transverse energy. Here, just like in the previous channels, scenario III, due to the tiny cross sections and the compressed spectrum does not yield a large number of events.}
\end{table}

\section{Discussion of the results} As expected before starting the analysis, the richness of the gaugino sector of left-right symmetric supersymmetric models does induce an enhancement in the number of events in the leptonic channels with respect to the predictions from the Standard Model. \\

In our Monte Carlo simulation, we have built five benchmark scenarios where both mixing patterns and hierarchies amongst the mass eigenstates were different in order to try draw some general features of this class of models. \\

We start by the least promising one, that is scenario SIII. The latter was built in order to achieve a large mixing in the mass eigenstates and thus have long cascade decays. We find it difficult, within our setup, to produce hard leptons or multilepton states in the cascade decays of the charginos because of the low cross sections due to the heaviness of the gaugino states but also to the compressed spectrum in the mass eigenstates. However, the third neutralino is the heaviest and this is explained by the small part of Higgsino present in its flavor decomposition as can be seen in figure \ref{fig:pure}. Allowing for larger parts of the Higgsino in the flavor decomposition of both the neutralinos and charginos could therefore enhance the cascade decays and the cross sections. \\

Scenarios I.1 and I.2 were constructed to illustrate scenarios where $SU(2)_L$ and $SU(2)_R$ gauge eigenstates do not mix at all. The main difference between these scenarios is in the hierarchy amongst the mass eigenstates which we find to depend on the hierarchy between the gaugino masses $M_{1L} \et M_{2L}$. The mass spectrum being rather high here also, we find that in order to achieve a high sensitivity for both scenarios one needs to focus on the multilepton channels and demand large missing transverse energy together with two hard leptons.\\

Finally, the most promising type of scenarios is that exhibiting both moderate mixing between the gauge eigenstates and low masses. The combined effects of these spcificities induce then high cross sections and allow both charginos and neutralinos to decay more often into one or two light charged leptons. For both masses of sleptons, 200 and 400 GeV, we find that this scenario yields a rather high number of events in all final states we have considered and a high sensitivity.\\

It is also noteworthy that both the selection criteria and constraints applied in the above analyses are usually those used in the searches for supersymmetric models like the MSSM. Therefore it is interesting to compare the MSSM to our benchmark scenarios in order to find the key observables one should use to distinguish between these two scenarios. To this end, we have constructed three scenarios in the MSSM (MSSM-I.1, MSSM-I.2 and MSSM-II) where the spectrum is similar to SI.1, SI.2 and SII with slepton masses set to 400 GeV. We also include a comparison between the MSSM and LRSUSY in the case of scenario II with slepton masses set to 200 GeV.  We do not however compare the MSSM against scenario SIII as the latter does not lead to any sizeable effect in leptonic final states. \\

\begin{table}
\begin{center}
\begin{tabular}{| c | c | c | c |}
\hline
Scenario & ($M_1$, $M_2$) [GeV] & ($m_{\chi^0_1}, m_{\chi^0_2}, m_{\chi_1^+}$) [GeV] & LRSUSY (GeV)\\
\hline
MSSM I.1 &  (270, 506) & (270, 500, 500) & (271.4, 299.8, 499.8)\\
\hline
MSSM I.2 & (270, 760) & (269, 747, 747) & (269.1, 749.7 , 749.7)\\
\hline 
MSSM II &  (112, 254) &(111, 250, 250) & (111.3, 249.8, 249.8)\\
\hline
\end{tabular}
\end{center}
\caption{\footnotesize \label{tab: tableau mssm} Summary of the values for the gaugino masses we choose so that the spectrum of the two lightest neutralinos and the lightest chargino be similar to those of the LRSUSY.}
\end{table}

\begin{figure}
\subfigure[]{%
\label{fig: mssm 2l met}\includegraphics[width=.5\columnwidth]{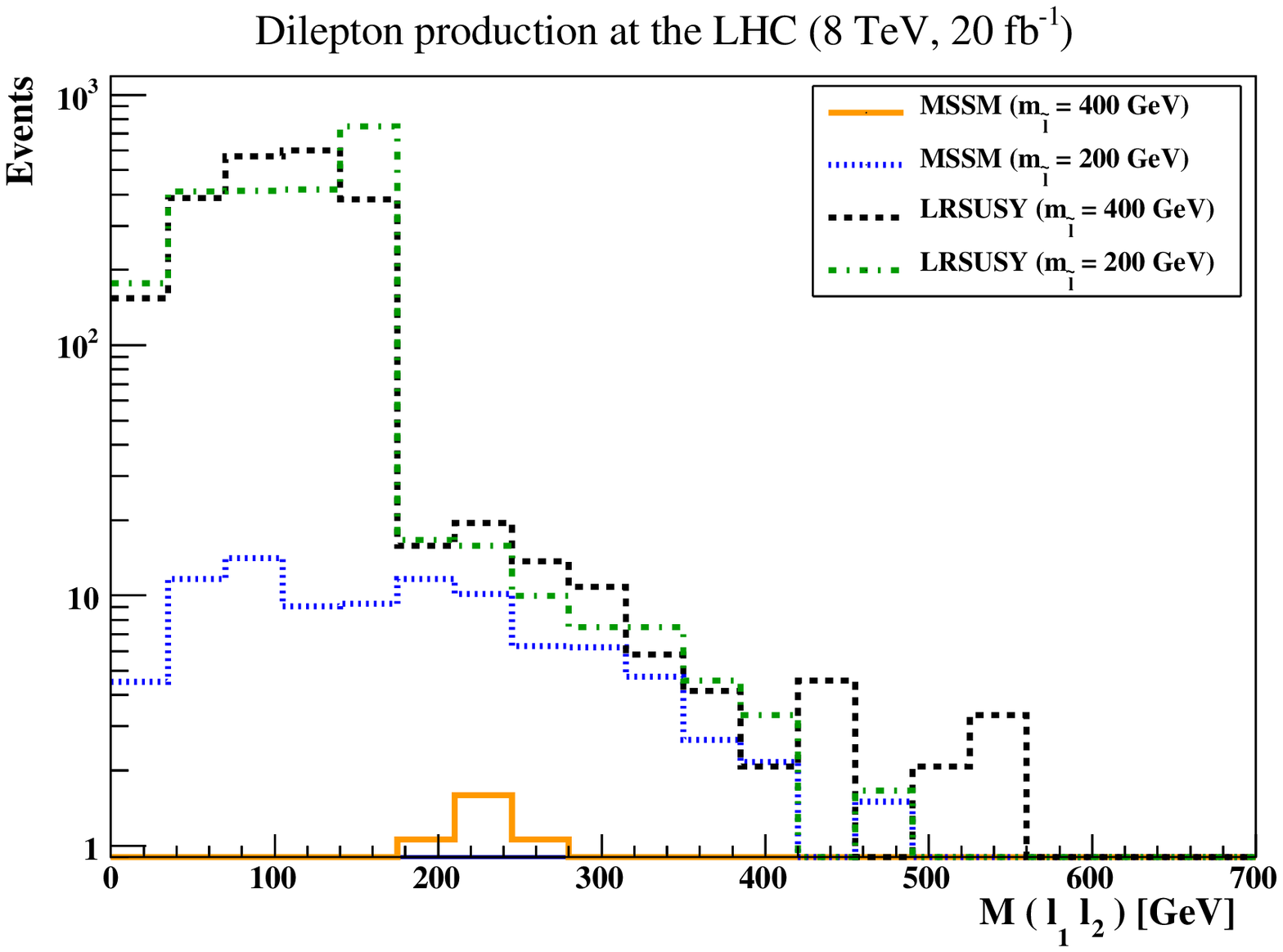}}
\subfigure[]{%
\label{fig: mssm 2l ptl1}\includegraphics[width=.5\columnwidth]{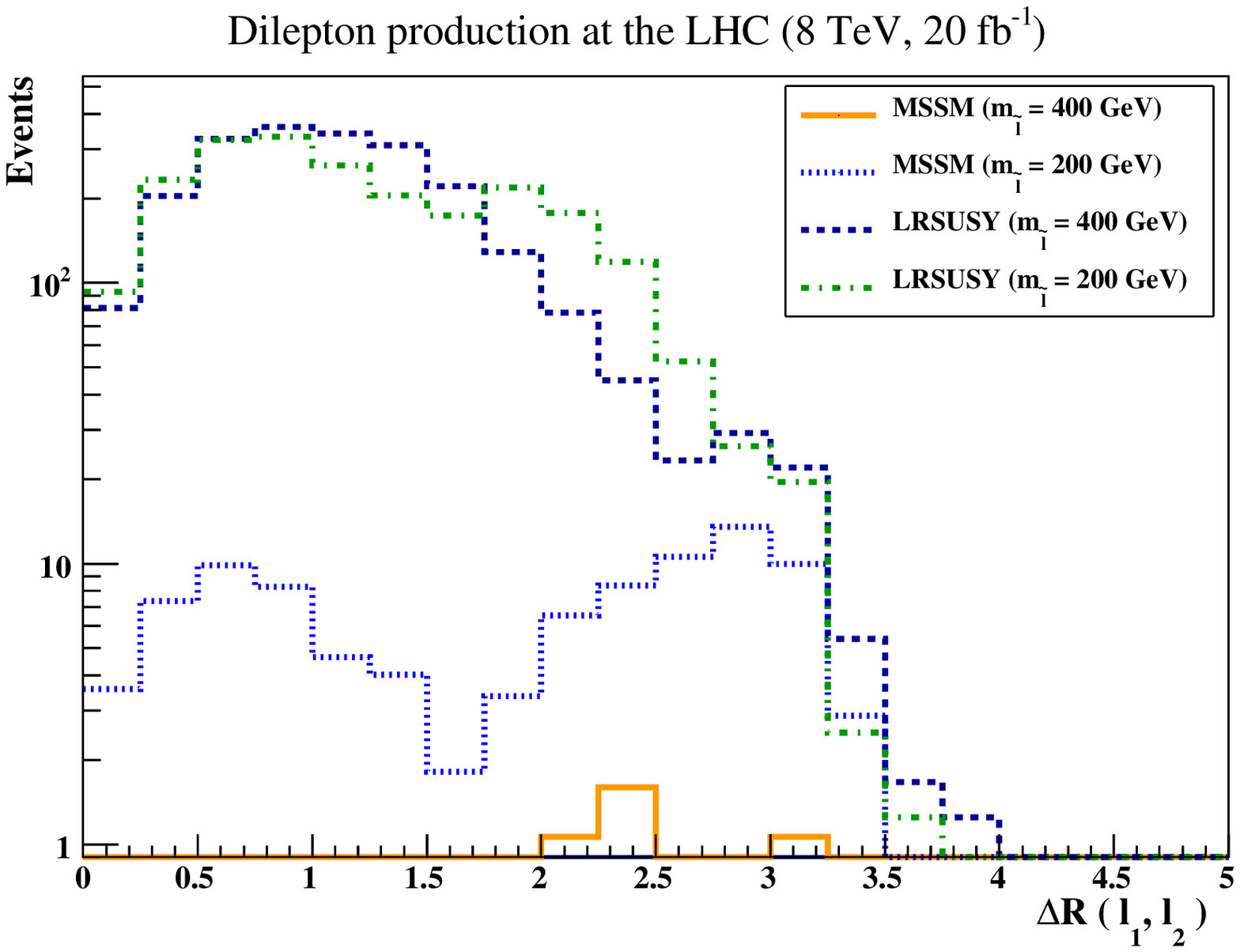}}\\
\caption{\footnotesize \label{fig: mssm 2l} Comparing the invariant mass (left) and the relative distance between the two leptons (right) in both scenarios MSSM-II and SII. These two variables lead to clean differences between these two models.}
\end{figure}

To build the MSSM scenarios we start with the values of the parameters in snowmass point slope 1a (sps1a \cite{Allanach:2002nj}) and change the value of the soft breaking mass of the Higgs $H_u$ in order to decouple the Higgsinos. This results in decoupling one chargino and two neutralinos. We then use both mass matrices and scalar potential minimization equations to set the values of the gaugino soft breaking masses  $M_1 \et M_2$ and the bilinear couplings $\mu \et b$. Just like we did in our benchmarks, we decouple the squarks and gluinos and set the slepton masses to the value of 400 GeV. We then use the package {\sc ASperGe}\cite{Alloul:2013fw} to generate the parameter card and use the same set of tools to obtain the Monte Carlo events. Table \ref{tab: tableau mssm} summarizes the setting for the MSSM.\\

Note that for scenario MSSM-I.2, the first chargino and the second neutralino are taken to have the same mass as the third neutralino and second chargino in scenario SI.2 from LRSUSY. This is due to the fact that the second neutralino and the lightest chargino in the latter scenario are pure $SU(2)_R$ winos which makes it difficult to accomodate their masses together with the constraint of decoupling. The same constraint appear for scenario II. Furthermore, scenarios MSSM-I.1 and MSSM-I.2 are compared to their LRSUSY equivalents SI.1 and SI.2 only in the three-or-more leptons channel.\\

Our events generated, we find that cross sections in the supersymmetric version of the Standard Model are systematically lower than those involved in LRSUSY which is mainly due to the fact that the higgsino and gaugino sector of left-right symmetric supersymmetric models is much richer. We also find that applying the same requirements than in the multilepton case, {\ie}
\begin{itemize}
\item reject any event with less than 80 GeV missing transverse energy;
\item reject any event where the leading lepton has a transverse momentum smaller than 80 GeV;
\item reject any event whose sub-leading lepton has a transverse momentum smaller than 70 GeV;
\end{itemize}
kills both benchmarks MSSM-II and MSSM-I.2 and we are only left with events coming from MSSM-I1.\\

In the histograms drawn in figure \ref{fig: mssm 2l} is presented the comparison between both scenarios MSSM-II and SII when the events contain exactly two charged light leptons. In the left corner, the distribution corresponds to that of the invariant mass of the pair of charged leptons in the final state, while in the right panel is presented the angular distance between the same pair of charged leptons.\\

The first remark one can make is that the number of events predicted by the MSSM in this setup is much smaller than that of SII. Moreover, the shapes of the distributions are very different, the tails in the case of the LRSUSY benchmarks extending to much higher values. \\

Turning on to the three lepton case, we find that after applying the same requirements than above, {\ie}
\begin{itemize}
\item we reject any event with less than 100 GeV of missing transverse momentum;
\item we reject any event where the leading lepton has a transverse momentum smaller than 80 GeV;
\item we reject any event whose sub-leading lepton has a transverse momentum smaller than 70 GeV;
\end{itemize}
only remain the events from scenario MSSM-I.1. In figure \ref{fig: mssm 3l}, the same distributions than in the two-lepton case are presented, that is, the invariant mass and the angular distance between the two leading charged leptons in the final state. Here, as in the two lepton case, clear differences in both shapes and number of events appear between the MSSM and LRSUSY model SI.1. 
 
\begin{figure}
\subfigure[]{%
\label{fig: mssm 3l met}\includegraphics[width=.5\columnwidth]{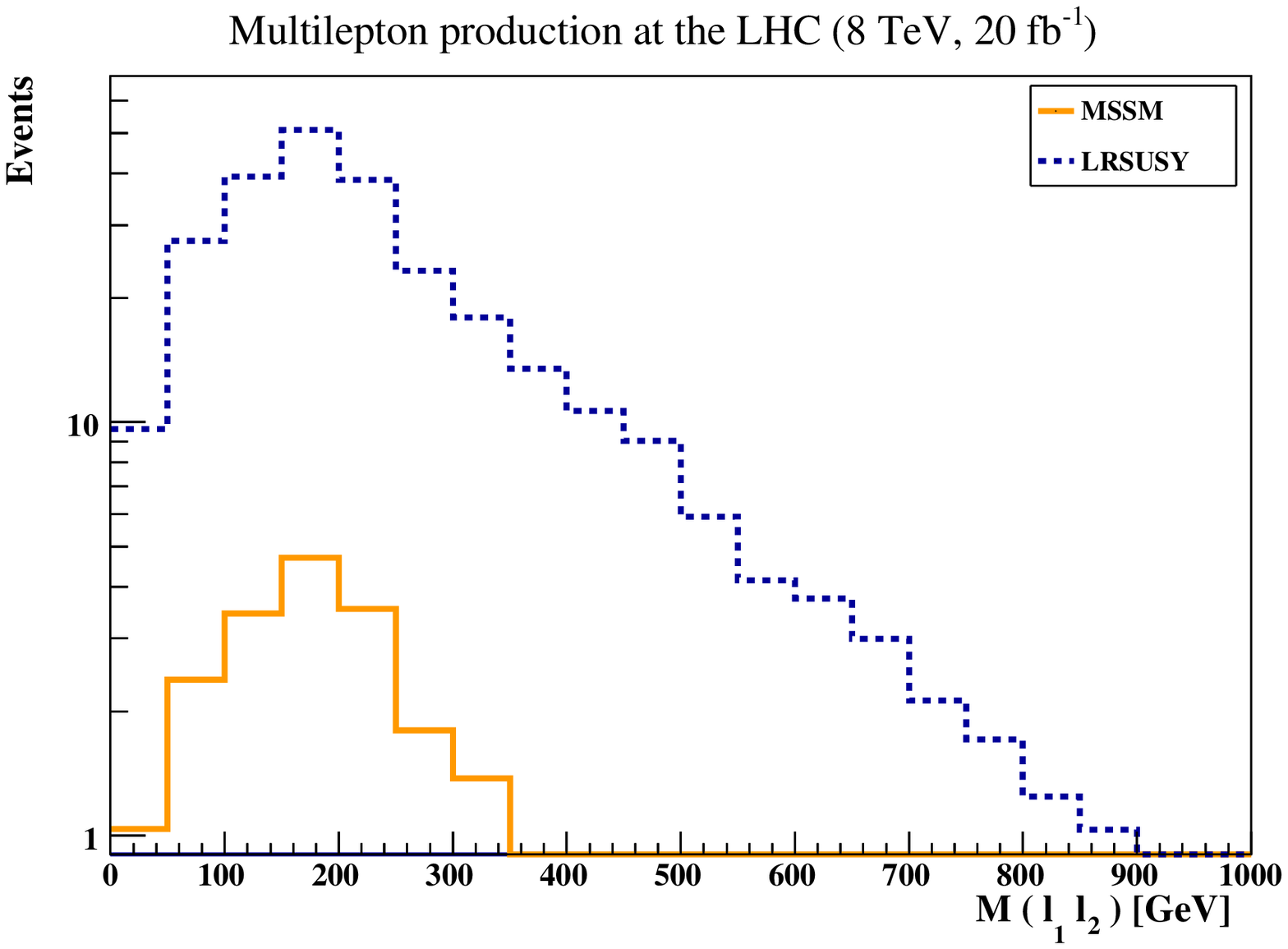}}
\subfigure[]{%
\label{fig: mssm 3l ptl1}\includegraphics[width=.5\columnwidth]{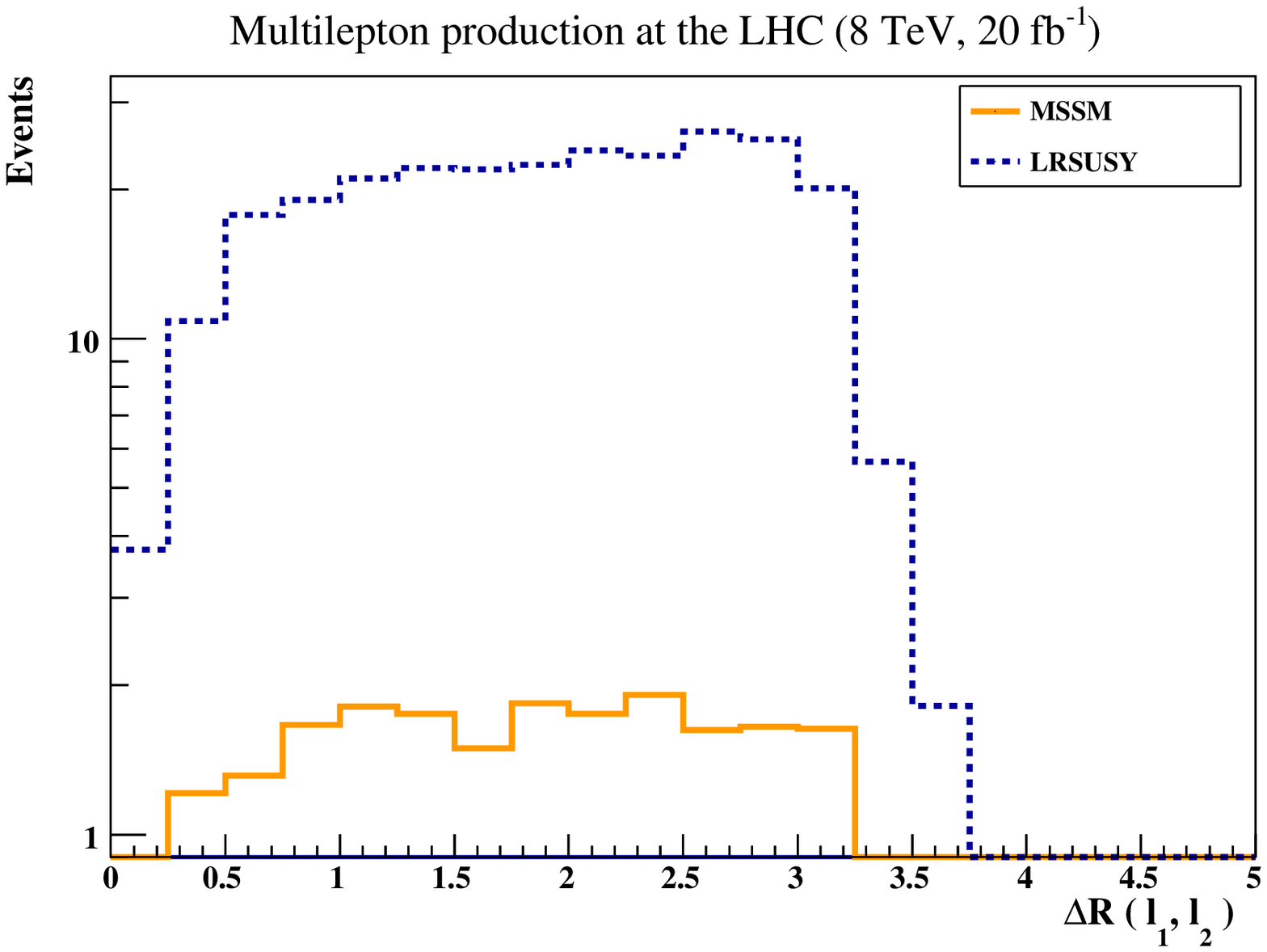}}
\caption{\footnotesize \label{fig: mssm 3l} Distribution of the invariant mass (left) and of the angular distance between the leading and the next-to-leading lepton (right) after selecting events with at least three charged leptons in the final state and having applied all constraints described in the multilepton analysis. We see clear differences between both models compared here.}
\end{figure}

\newpage
\section{Conclusion}\label{sec: lrsusy conclusion}
In this chapter, we have seen how one can achieve a successful building of a left-right symmetric supersymmetric model and deduce some general features characterizing the signatures induced by the production of charginos and neutralinos from this model.\\

In the first part of this job, we have written every quantity entering the Lagrangian explicitely taking care to properly write gauge invariant products. In particular, we have been careful in the definition of the covariant derivatives where one should differentiate between the left action of $SU(2)_L$ operators and the right action of $SU(2)_R$ operators. From this Lagrangian, we have calculated both mass and mixing matrices of the chargino and neutralino fields and made some legitimate assumptions on the vevs hierarchy to reduce the number of free parameters. Finally, we have set up four benchmark scenarios to find the general behaviour of scenarios where the mixing in the gaugino sector is either maximal or not and where the splitting in mass is more or less large. \\

In the second part of the analysis, we have used a Monte Carlo simulation to extract the phenomenological features of the model we considered. Using only general considerations based on the behaviour of the Standard Model processes, we have determined a set of constraints for the final states with one, two or more leptons that enhance the potential of discovery for such left-right supersymmetric models.\\

These constraints being generic in the sense that they are not specific to left-right symmetric supersymmetric models, we have performed a comparison between the latter model and the MSSM. Constructing benchmark scenarios in the latter corresponding to those we have analyzed, we find that the MSSM predicts systematically a much lower number of events when the final state contains at least two charged light leptons. We also find that the transverse momenta of the leptons, the missing transverse energy, the relative angular distance $\Delta_R$ and the transverse mass of any pair of leptons could lead to significant differences between both models allowing thus for easy distinction.\\

Of course our quantitative results hold in the limit where our hypotheses are satisfied, in particular when the higgsinos are decoupled from the low energy theory. When the latter fields are present in the flavor decomposition of the charginos and neutralinos, they lift the degeneracy between the mass eigenstates and relaxing this hypothesis may lead thus to a less compressed spectrum especially in the case of scenario SIII. This would enhance the decay rates and maybe lead to even richer multileptonic final states where one can imagine cascade decays like
$$ p ~ p \to (\tchi_3^0 \dashrightarrow l^\pm ~ l^\mp ~ l'^\pm ~ l'^\mp ~ \tchi_1^0 ) ~ (\tchi_3^0 \dashrightarrow l''^\pm ~ l''^\mp ~ l'''^\pm ~ l'''^\mp ~ \tchi_1^0 )$$
where $\dashrightarrow$ stands for the potential production and the decay of any intermediate unstable particle. However, the qualitative results describing the general behaviour of the model should hold especially for final states with high lepton multiplicity.

\chapter{ Spectrum generator}\label{chap: spec gen}

\section{Automated tools in particle physics}\label{sec: motivate automated tools}
As already stated in the previous chapters, we are nowadays at a turning point in particle physics history. The Large Hadron Collider, though now in an upgrade phase and thus not acquiring any data, has accumulated a huge amount of collisions (20 $fb^{-1}$ for 2012 run) that still needs to be processed in order to find new signatures not predicted by the Standard Model of particle physics. These excesses over the Standard Model predictions will probably not be easy to detect and only reliable and complete Monte Carlo simulations of both Standard Model and other models processes will allow us to extract the new physics signal from the experimental data. \\

Actually, to be complete, one needs to combine several tools in order to be able to have reliable simulations that can be compared to the experimental results. In a few words, one starts with general purpose event generators that automate the generation of parton level Monte Carlo events. {\sc MadGraph\cite{Stelzer:1994ta,Maltoni:2002qb,Alwall:2007st,Alwall:2008pm,Alwall:2011uj,deAquino:2011ub}, CalcHep/CompHep\cite{Pukhov:2004ca,Belyaev:2012qa}, Herwig/Herwig++\cite{Bahr:2008pv}, Pythia\cite{Ilten:2012zb}, Sherpa\cite{Gleisberg:2003xi,Gleisberg:2008ta}, Whizard\cite{Kilian:2007gr}} are some of the public tools able to perform this task and store the parton-level events in the so-called LHE format \cite{Alwall:2006yp}, a file format based on a XML-like structure. Proceeding with the simulation, one needs now to turn towards tools able to handle properly the integration of the matrix-element hard processes. Tools like {\sc Pythia}\cite{Ilten:2012zb}, Herwig/Herwig++\cite{Bahr:2008pv} or {\sc Sherpa}\cite{Gleisberg:2003xi,Gleisberg:2008ta} provide the necessary algorithms for parton showering and hadronization and usually store the events in either {\sc STDHEP}\cite{Stdhep} or {\sc HepMC}\cite{Dobbs:2001ck} formats. Finally, jet clustering algorithms like those provided in {\sc FastJet}\cite{Cacciari:2011ma} and detector response simulation with {\sc PGS}\cite{pgs},  or {\sc Delphes\cite{Ovyn:2009tx}} must be carried out. Detector level events are then stored in the {\sc LHCO} format \cite{lhco} and one can use a tool like {\sc MadAnalysis 5}\cite{Conte:2012fm} to carry his phenomenological analysis\footnote{Actually, to be more precise, the {\sc MadAnalysis 5} package can be used at any level of the analysis.}.\\

In this quite complete scheme (see fig.\ref{fig:scheme tools} for a schematic description), there is however a very important step that we did not mention yet, namely, the implementation of the model we want to simulate in the Monte Carlo generators. In practice, this means that one needs to define, in a format that is usually specific to the Monte Carlo generator one wants to use, all the Feynman rules associated to all the vertices of the model together with the numerical values of every parameter. This also means that if one wanted to use several Monte Carlo tools at the same time in order for example to compare the results or overcome the limitations of either one or the other tool, one would need to go through this task several times. This is clearly not efficient and moreover error-prone and painstaking. Obviously, several models have already been implemented in these Monte Carlo tools to avoid users all these steps but this generally concerns only most studied models. For example, the Left-Right symmetric model as defined in the previous chapter was not included in any of the public tools. To implement it in {\sc MadGraph 5} (the tool we have chosen to carry out the simulation), we have made use of {\sc FeynRules}\cite{Christensen:2008py,Christensen:2009jx,Duhr:2011se,Alloul:2013yy}.\\

{\sc FeynRules}, {\sc LanHep\cite{Semenov:2010qt}} and {\sc Sarah}\cite{Staub:2008uz,Staub:2012pb} can be qualified as Lagrangian based tools, that is they only need the basic information about the model under consideration such as the gauge group, the field content and the Lagrangian. From these information they are able to compute automatically all the Feynman diagrams and the associated rules and then, through dedicated interfaces, generate automatically the necessary files for the general purpose Monte Carlo tools to work. \\

During my thesis, I have been envolved in the development of two modules in {\sc FeynRules}. First of them, takes full advantage of the superspace module included in {\sc FeynRules}\cite{Duhr:2011se} in order to generate the renormalization group equations at the two-loop level for any renormalizable supersymmetric theory. The other one is a spectrum generator generator, that is a module able to extract the mass matrices automatically from a Lagrangian, diagonalize them and return both mixing matrices and masses of the mass eigenstates in an {\sc SLHA}-like format \cite{Skands:2003cj,Allanach:2008qq}.\\

This chapter is intended to give a description of these two modules, published in the proceedings of the 2011 Les Houches conference \cite{Brooijmans:2012yi} and in the paper\cite{Alloul:2013fw}. To this end, the first section will be dedicated to a brief introduction to {\sc FeynRules}; the second section will be devoted to the RGEs module and the third section to how one can generate automatically the spectrum of his model. \\

Before concluding this introduction, I would like to draw the reader attention to the fact that a great progress in automating NLO calculations has been achieved last years leading to including such precision in Monte Carlo generators \cite{Bern:2013gka,Cullen:2011xs,Pittau:2012fn,Frixione:2002ik}. Finally, for a more detailed review of automated tools and for examples of their use, the reader can refer to the papers \cite{Christensen:2009jx,Fuks:2012im,PerretGallix:2013bg}.

\begin{figure}
\hskip-0.5truecm
\scalebox{1} 
{
\begin{pspicture}(0,-5.72)(18.462149,5.72)
\psellipse[linewidth=0.04,dimen=outer](6.95,4.91)(3.03,0.81)
\usefont{T1}{ptm}{m}{n}
\rput(7.011836,4.945){Idea of a model}
\psframe[linewidth=0.04,dimen=outer](11.58,2.54)(2.8,0.86)
\usefont{T1}{ptm}{m}{n}
\rput(4.4795556,1.805){\small Spectrum generators}
\usefont{T1}{ptm}{m}{n}
\rput(9.717442,1.985){\small Feynman Rules}
\psline[linewidth=0.04cm,linestyle=dashed,dash=0.16cm 0.16cm](7.16,2.48)(7.18,0.86)
\usefont{T1}{ptm}{m}{n}
\rput(9.694038,1.485){\small Calculators}
\psline[linewidth=0.04cm,arrowsize=0.05291667cm 2.0,arrowlength=1.4,arrowinset=0.4]{->}(6.62,2.08)(7.66,2.08)
\psline[linewidth=0.04cm,arrowsize=0.05291667cm 2.0,arrowlength=1.4,arrowinset=0.4]{->}(7.56,1.44)(6.54,1.42)
\psframe[linewidth=0.04,dimen=outer](5.98,-0.72)(2.7934065,-1.7558566)
\psframe[linewidth=0.04,dimen=outer](5.98,-2.6921115)(2.74,-3.7279682)
\psframe[linewidth=0.04,dimen=outer](5.98,-4.6841435)(2.7578022,-5.72)
\psframe[linewidth=0.04,dimen=outer](11.96,-0.68)(9.24,-5.72)
\psline[linewidth=0.268cm,arrowsize=0.05291667cm 3.69,arrowlength=1.4,arrowinset=0.4]{->}(12.28,-3.08)(14.46,-3.08)
\usefont{T1}{ptm}{m}{n}
\rput(16.533554,-3.06){\Huge RESULTS}
\usefont{T1}{ptm}{m}{n}
\rput(4.40333,-1.155){Parton Level}
\usefont{T1}{ptm}{m}{n}
\rput(4.4073634,-4.995){Reconstructed}
\usefont{T1}{ptm}{m}{n}
\rput(4.313242,-3.115){Hadron Level}
\usefont{T1}{ptm}{m}{n}
\rput(10.592939,-3.11){\large Analyzer}
\psarc[linewidth=0.04,arrowsize=0.05291667cm 2.0,arrowlength=1.4,arrowinset=0.4]{->}(2.63,3.41){1.49}{99.34467}{259.9025}
\psarc[linewidth=0.04,arrowsize=0.05291667cm 2.0,arrowlength=1.4,arrowinset=0.4]{->}(2.49,0.11){1.19}{101.976135}{262.7757}
\psarc[linewidth=0.04,arrowsize=0.05291667cm 2.0,arrowlength=1.4,arrowinset=0.4]{->}(2.42,-2.16){0.82}{101.976135}{262.7757}
\psarc[linewidth=0.04,arrowsize=0.05291667cm 2.0,arrowlength=1.4,arrowinset=0.4]{->}(2.36,-4.14){0.82}{101.976135}{262.7757}
\psline[linewidth=0.111999996cm,arrowsize=0.05291667cm 2.0,arrowlength=1.4,arrowinset=0.4]{->}(6.12,-1.2)(9.18,-1.24)
\psline[linewidth=0.111999996cm,arrowsize=0.05291667cm 2.0,arrowlength=1.4,arrowinset=0.4]{->}(6.08,-3.16)(9.24,-3.18)
\psline[linewidth=0.111999996cm,arrowsize=0.05291667cm 2.0,arrowlength=1.4,arrowinset=0.4]{->}(6.1,-5.14)(9.18,-5.14)
\psframe[linewidth=0.04,dimen=outer,fillstyle=solid](2.38,3.78)(0.0,3.2)
\usefont{T1}{ptm}{m}{n}
\rput(1.1518066,3.505){Model File}
\psframe[linewidth=0.04,dimen=outer,fillstyle=solid](2.52,0.28)(0.14,-0.3)
\usefont{T1}{ptm}{m}{n}
\rput(1.2918067,0.0050){Model File}
\psframe[linewidth=0.04,dimen=outer,fillstyle=solid](2.82,-1.92)(0.44,-2.5)
\usefont{T1}{ptm}{m}{n}
\rput(1.5822167,-2.195){LHE}
\psframe[linewidth=0.04,dimen=outer,fillstyle=solid](2.78,-3.94)(0.04,-4.52)
\usefont{T1}{ptm}{m}{n}
\rput(1.410625,-4.24){\footnotesize HEPMC/STDHEP}
\psframe[linewidth=0.0020,dimen=outer,fillstyle=solid](8.0,1.98)(6.22,1.5)
\usefont{T1}{ptm}{m}{n}
\rput(7.0800147,1.765){\small SLHA-like}
\usefont{T1}{ptm}{m}{n}
\rput(4.367041,-5.355){events}
\end{pspicture} 
}
\label{fig:scheme tools}\caption{\footnotesize Automated tools in high energy particle physics form a complete scheme connecting ideas to predictions.}	
\end{figure}
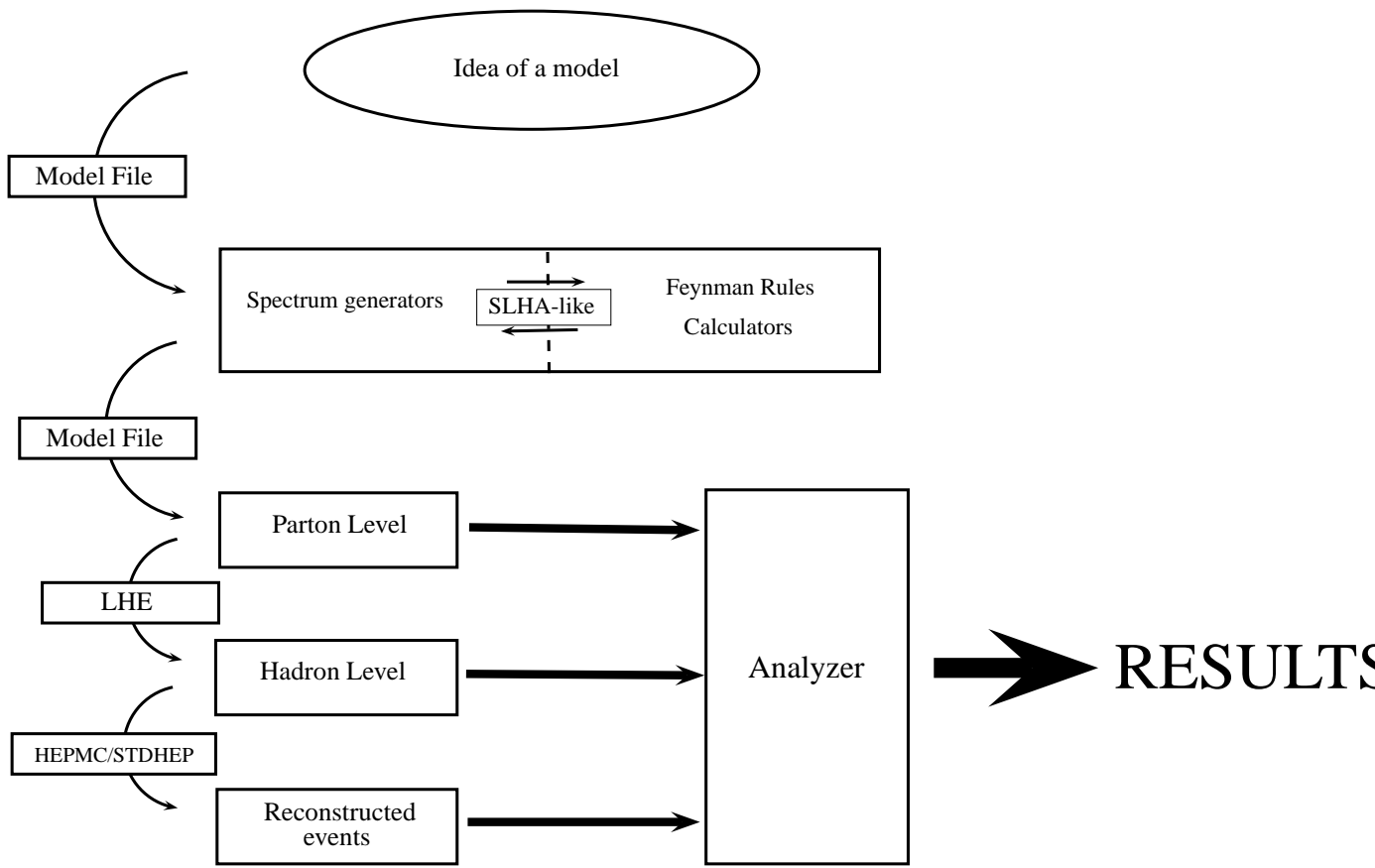


\section{Introduction to {\sc FeynRules}}\label{sec:feynrules}
The idea behind the {\mk} package {\fr} is to facilitate the implementation of any new quantum field theory into various Monte Carlo (MC) tools provided a Lagrangian and some basic inputs have been defined. The package comes along with a set of various routines able to calculate some characteristic quantities like the analytic formulas for the decays of particles, the mass matrices associated to the model or the renormalization group equations in the case of a supersymmetric model.\\

In this section, we shall describe some general features of {\fr} starting from the implementation of a model to the generation of model files for MC tools. For a detailed description of the package, the reader can refer to \cite{Christensen:2008py,Christensen:2009jx,Duhr:2011se,Alloul:2013yy}.

\subsection{Implementing a model in {\fr}}
Implementing a new model in {\fr} is the very first step to achieve before starting using it. It goes by the definition of the gauge group, the parameters and the fields following a structure based on lists and rules
\begin{verbatim}
List1 = {
  parameter1 == { option1 -> value1, option2 -> value2, ....},
  ...
}
\end{verbatim}
In consequence, all the subgroups of the gauge group of the model are stored as elements of the list \verb?M$GaugeGroups?. For each entry of this list, a set of options has to be provided together with the corresponding values. For example, the following declaration
\begin{verbatim}
M$GaugeGroups = {
  SU2L == {
   Abelian           -> False, 
   CouplingConstant  -> gw, 
   GaugeBoson        -> Wi, 
   StructureConstant -> Eps, 
   Representations   -> {Ta,SU2D}, 
   Definitions       -> {Ta[a_,b_,c_]->PauliSigma[a,b,c]/2}
  }
}
\end{verbatim}
can be used for the non-abelian gauge group $SU(2)_L$. The options inform {\fr} on the abelian nature, the coupling constant, the gauge boson, the structure constant and the representations associated with this gauge group. The latter information comes along with the associated index, here \verb?SU2D?. Finally the option \verb?Definitions? is used to set the generator of the fundamental representation to be \verb?PauliSigma[a,b,c]/2? {\ie} the element $(b,c)$ of the Pauli matrix $\sigma^a$ divided by two. The generator in the adjoint representation, that is $i \epsilon^{ijk}$ is generated automatically from the structure constants.\\

The indices have to be declared separately using the {\fr} dedicated functions \verb?IndexRange? and \verb?IndexStyle?. In the case of the gauge group $SU(2)_L$ declared above one needs the following two lines
\begin{verbatim}
IndexRange[Index[ SU2W ]] = Unfold[Range[3]]; IndexStyle[SU2W, j];
IndexRange[Index[ SU2D ]] = Unfold[Range[2]]; IndexStyle[SU2D, k];
\end{verbatim}
where the index \verb?SU2W? (resp. \verb?SU2D?) is set to run from 1 to 3 (resp 1 to 2).\\

Proceeding with the implementation, we now turn to the case of the parameters which have to be gathered in the list \verb?M$Parameters?. For example, the coupling constant \verb?gw? used above can be declared as follows
\begin{verbatim}
M$Parameters = {
 gw == { 
   ParameterType    -> External, 
   BlockName        -> Gauge,
   OrderBlock       -> 2,
   Value            -> 0.663922,   
   Description      -> "Weak coupling constant at the Z pole"
 }
}
\end{verbatim}
Here some important remarks are in order
\begin{itemize}
\item The option \verb?ParameterType? specifies whether the parameter is \verb?External? or \verb?Internal?. In the first case, it means that the parameter is considered as an input in the SLHA-format and consequently the flags \verb?BlockName? and \verb?OrderBlock? corresponding to the block and the index, respectively, associated to the parameter have to be provided. If not present, the block name is set to \verb?FRBlock? and the index set to a unique integer. In the case where the parameter is set to \verb?Internal?, the block name and the order block do not need to be defined anymore.
\item Following from the separation between external and internal parameters, a very important rule to follow is to first define all the external parameters and then the internal ones. Moreover, the internal parameters have to be declared in such a way that a parameter cannot be defined in function of another one appearing below in the list.
\item The option \verb?Value? is not mandatory. When provided, it can be a formulae written in terms of other parameters or just a numerical value. If not provided, {\fr} assigns a default value of 1 to the parameter. 
\item The option \verb?Description? points to a description of the parameter.
\end{itemize}

We now turn to the declaration of the fields for which two main classes exist depending on their nature. If one is dealing with a supersymmetric theory and wants to use superfields than both lists \verb?M$Superfields? and \verb?M$ClassesDescription? need to be defined. In any other case, only \verb?M$ClassesDescription? is required.  In the latter, a unique class exists for every type of field:
\begin{itemize}
\item Dirac fermions \verb? F[_] == { option -> value, ... }?
\item Weyl fermions \verb?  W[_] == { option -> value, ... }?
\item Scalars \verb?        S[_] == { option -> value, ... }?
\item Vector bosons \verb?  V[_] == { option -> value, ...}?
\end{itemize}
As to the superfields case, one has to differentiate between chiral and vector superfields
\begin{itemize}
\item Chiral superfields \verb?CSF[_] == { option -> value, ... }?
\item Vector superfields \verb?VSF[_] == { option -> value, ... }?
\end{itemize}
where the underscore stands for an integer. \\
Several values for \verb?option? exist to define as precisely as possible the properties of the fields. Some of them are given in table \ref{table:field options}. A noteworthy point is that when dealing with non physical states, that is gauge eigenstates, one should provide the mixing relations. For example, in the Standard Model, the gauge bosons $B_\mu \et W_\mu$ mix to give rise to the photon, the neutral Z-boson and the charged W-boson. This information can be entered in {\fr} as follows
\begin{verbatim}
 V[1] == { 
   ClassName     -> B, 
   Unphysical    -> True, 
   SelfConjugate -> True, 
   Definitions   -> { B[mu_] -> -sw Z[mu]+cw A[mu]} 
 },
 V[2] == { 
   ClassName     -> Wi,
   Unphysical    -> True,
   SelfConjugate -> True, 
   Indices       -> {Index[SU2W]},
   FlavorIndex   -> SU2W,
   Definitions   -> { Wi[mu_,1] -> (Wbar[mu]+W[mu])/Sqrt[2], 
                      Wi[mu_,2] -> (Wbar[mu]-W[mu])/(I*Sqrt[2]), 
                      Wi[mu_,3] -> cw Z[mu] + sw A[mu]}
  }
\end{verbatim}
where \verb?B? and \verb?Wi? are the gauge eigenstates and \verb?A, Z, W? and \verb?Wbar? are the mass eigenstates corresponding to the photon, the $Z$ boson, $W^-$ and $W^+$ respectively. In section \ref{sec:asperge} we will see how this can be changed so that {\fr} calculates by itself the mass matrices and thus the values for $\cos\theta_w$ (\verb?cw? in the example above) et $\sin\theta_w$ (\verb?sw? in the example above).\\
\begin{table}[!h]
\begin{center}
\shadowbox{ 
\begin{tabular}{c| p{2.8cm} | p{4cm}| p{4.5cm}}
\mbox{Option} \newline & Applicable to & Description & Example: Down-type \newline left-handed leptons \\
\hline
\verb?ClassName? & All fields \newline Mandatory & Gives a symbol by which a field or a class of fields is represented & \verb?ClassName->uq?\\
\hline
\verb?ClassMembers? & All fields & List containing all the members of a class. If only one member, by default it will be set to the \verb?ClassName? & \verb?ClassMembers->{e,mu,ta}? \\
\hline
\verb?Unphysical? & All fields & Boolean determining if gauge eigenstate or not & \verb?Unphysical -> False?\\
\hline
\verb?Indices? & All fields & The list of indices carried by the fields & \verb?{Index[Generation]}? \\
\hline 
\verb?QuantumNumbers? & All fields & List of quantum numbers carried by fields & ~\verb?QuantumNumbers->{Q -> -1}?\\
\hline
\verb?PDG? & Physical fields & List of the {\sc PDG}-Ids for all \verb?ClassMembers? & \verb?PDG -> {11, 13, 15}?\\
\hline 
\verb?Chirality? & Weyl fermions \&\newline chiral sueprfields & Sets the chirality of the field & \verb?Chirality -> Left?\\
\hline 
\end{tabular}
}
\end{center}
\caption{\label{table:field options}{\footnotesize Some of the various options that exist in {\fr} to declare fields and set their properties properly}}
\end{table}

Finally, only the lagrangian remains to be defined. To facilitate this task, it is of course possible to use {\mk} built-in functions as well as a whole set of functions that has been made available in {\fr} allowing to calculate the covariant derivatives, to expand contracted indices, to expand in components a superpotential written in superfield language \dots etc. The example below shows how to write the kinetic terms for the quarks and the up-quark-Higgs Yukawa interaction  in the case of the Standard Model\\
\begin{verbatim}
Lkinetic = I*QLbar.Ga[mu].DC[QL, mu];
LYukawa = yu[ff1, ff2] QLbar[sp, ii, ff1, cc].uR [sp, ff2, cc] Phibar[jj] Eps[ii, jj];
\end{verbatim}
where \verb?DC? is the symbol for the covariant derivative, \verb?Ga[mu]? is the Dirac matrix $\gamma^\mu$ and \verb?Eps[ii,jj]? the totally antisymmetric tensor of order 2. In this example, \texttt{QLbar} is the symbol for the Dirac conjugate of the left-handed quark $Q$, \texttt{Phibar} the complex conjugate of the Higgs field $\phi$ and \texttt{ur} stands for the right-handed up-type quark. Note that, in {\sc FeynRules} the declaration of the anti-fields is done automatically {\ie} if the left-handed quark \texttt{QL} is declared then automatically the symbol \texttt{QLbar} is created for the antiquark.\\

\subsection{Available functionalities in {\fr}}
In addition to {\mk} built-in functions, {\fr} is provided with a whole set of routines allowing the user to calculate some characteristic quantities associated to his model. To make them available the user has to load both the package and his model by issuing in a {\mk} session 
\begin{alltt}
 <<FeynRules\(\grave{}\) ;
 LoadModel["Model.fr"]
\end{alltt}
The first task of {\fr} being the calculation of Feynman rules, one cannot skip describing the function \verb?FeynmanRules? allowing to extract them. Thus, typing in the {\mk} session
\begin{verbatim} feynmanrules = FeynmanRules[lag]\end{verbatim}
will result in the calculation of the Feynman rules associated to the lagrangian \verb?lag?. If the latter is big, the output generated by the command \verb?FeynmanRules[lag]? might be too big and thus providing the option \verb?ScreenOutput -> False? may reveal helpful. More generally, if one wants to know the available options associated to a certain function, it suffices to type in the {\mk} session
\begin{verbatim} Options[function]\end{verbatim}

Another function one could use is \verb?SelectVertices? which takes as input the output from \linebreak\texttt{FeynmanRules[lag]}. This routine allows to select vertices containing a certain type of particles. For example, if the user wants all the vertices in which the Z-boson represented by the symbol \verb?Z? is involved, he might type
\begin{verbatim} SelectVertices[feynmanrules, Contains->{Z}]\end{verbatim}

The supersymmetry package included in {\fr} supports fully the superspace formalism and provides thus routines to perform calculations in this particular formalism. For example, if we define \verb?QL? to be the chiral superfield associated to the left-handed quark in the MSSM, we can access 
\begin{itemize}
\item all its components with the command \verb?SF2Components[QL]?,
\item its $\theta^2$ component by typing \verb? Theta2Component[QL]?,
\item the $\theta^2\thetabar^2$ component of the kinetic terms simply by typing \\ \verb? Theta2Thetabar2Component[CSFKineticTerms[QL]]?
\end{itemize}
\verb?SolveEqMotionD, SolveEqMotionF? are two other routines that reveal very useful in building supersymmetric lagrangians as they solve the equations of motion for the D- and F-terms\ref{sec:susy}.\\

Since version 1.8, new functionnalities have been added in {\fr}. It is now possible to calculate all the $1 \to 2$ decay widths of the various fields in the model with the command 
\begin{verbatim}CalculateM2Decays[lag, ExtractFeynmanRules -> True]\end{verbatim}
spin-$\frac32$ fields are fully supported and a new NLO module is under development. For a detailed description of these functions and some examples of their use, the reader can refer to the manual associated to this version \cite{Alloul:2013yy}.

\subsection{Interfaces}
The {\fr} package is intended to facilitate the transition between the formal definition of a model and its implementation in Monte Carlo generators. To this end, interfaces with {\sc CalcHep/CompHep \cite{Christensen:2009jx}, FeynArts, MadGraph5, GoSam, Sherpa} and {\sc Whizard \cite{Christensen:2010wz}} have been developed. All these interfaces can be invoked in a similar procedure
\begin{verbatim} WriteXXX[lag]\end{verbatim}
where the sequence of letters \verb?XXX? takes one of the values \verb?CHoutput? ({\sc CalcHep}), \verb?FeynArtsOutput? ({\sc FeynArts}), \verb?SHOutput? ({\sc Sherpa}), \verb?UFO? ({\sc MadGraph5} or any other MC tool compatible with the {\sc UFO} format\cite{Degrande:2011ua}) or \verb?WOOutput? ({\sc Whizard}).\\

The {\sc UFO} interface is a {\sc Python} library that has been designed for flexibility. In a few words, this format consists in storing the information on the particles, parameters and vertices of the model as a set of Python objects, each of them being associated with a list of attributes related to their properties. The power of such an object-oriented format is clear as there is no need anymore to constrain the Lorentz and/or the color structure of the vertices.\\

Finally, to speed up the generation of model files one can use the routines \verb?WriteRestrictionFile[]? and \verb?LoadRestrictionFile[]?. The first one, goes through all the parameters in the model looking for those whose value is 0 and creates a file with these parameters. For example, if the CKM matrix in the quarks sector is diagonal than the file generated by \verb?WriteRestrictionFile[]? would look like\\
\begin{verbatim}
     M$Restrictions = {
         CKM[1, 2] -> 0,
         CKM[1, 3] -> 0,
         CKM[2, 1] -> 0,
         CKM[2, 3] -> 0,
         CKM[3, 1] -> 0,
         CKM[3, 2] -> 0
     }
\end{verbatim}
The \verb?LoadRestrictionFile[]? is then used to load these restrictions and delete the zero values from the lagrangian. \\

\section{Renormalization group equations in {\fr}}\label{sec:rges}
As seen in section \ref{sec:rge-formulas}, renormalization group equations (RGEs) that govern the evolution of the parameters of any supersymmetric renormlizable lagrangian can be derived from a set of general equations. These equations are known to the second order and can be written under the form
\bea \frac{\partial p}{\partial t} = \frac{1}{16\pi^2} \beta_1 + \frac{1}{(16\pi^2)^2} \beta_2 \eea
where $p$ is the parameter to run between two energies $E_1 \et E_2$, $t=\ln E$ and $\beta_1 \et \beta_2$ are, respectively, the 1-loop and the 2-loop coefficients of the beta function whose expressions are known and can be found easily in the litterature \cite{Falck:1985aa,Martin:1997ns,Derendinger:1990tj}. This property of the RGEs has led several authors to writing programs able to extract them and run them between different scales to get the full spectrum of the theory. Some softwares like {\sc SuSpect} or {\sc SoftSusy} only focus on the MSSM case but since a few years now the {\mk} package {\sc SARAH} is able to extract these equations at two loop level for any renormalizable supersymmetric theory.\\

At the beginning of my thesis, we decided to develop our own routine able to extract automatically the renormalization group equations at two-loop level for any supersymmetric theory implemented in {\fr} together with the 1-loop mass matrices. We would then have a full spectrum-generator generator that, coupled to {\fr}, would reveal very powerful in the study of supersymmetric theories either minimal or not. This work led us to the creation of two {\fr} modules {\insurge}\footnote{{\insurge} stands for Independant Supersymmetry Renormalization Group Equations.} for the extraction of the RGEs and {\asperge}\footnote{{\asperge} stands for Automated Spectrum Generation.} for the mass matrices. In the following, we will first show how {\insurge} works then we will focus on {\asperge}.\\

\subsection{Modifying the {\fr} model file}
To allow {\insurge} to extract efficiently the RGEs, it is asked to the user to implement his model using the superspace formalism\footnote{A detailed manual can be found in \cite{Duhr:2011se}.} and to follow a slightely different way of declaring the interactions in the superpotential and their counterpart in the soft supersymmetry breaking lagrangian. We also added a new option in the declaration of gauge groups.\\
\paragraph{Abelian groups} Charges associated to abelian groups have to be normalized in function of the underlying Grand Unified Theory. This choice depending on the user, we have added the option \verb?GUTNormalization? that takes as input the square of the normalization factor, to allow for a proper declaration. In the case of the MSSM where the gauge group of the unified theory is $SU(5)$, the normalization for the abelian group $U(1)_Y$ is equal to $\sqrt{\frac35}$ which can be entered in the model file as follows
\begin{verbatim} U1Y == { ... , GUTNormalization -> 3/5}\end{verbatim}
where the dots represent any other option relevant for the declaration of the subgroup $U(1)_Y$.
\paragraph{Interactions} As seen in section \ref{sec:rge-formulas}, the derivation of the coefficients of the beta-functions requires the computation of the multiplicity of each superfield. In order to allow {\fr} to perform this task efficiently, both the superpotential and the associated soft supersymmetry breaking lagrangian must be implemented using the function \verb?SUDot? for $SU(N)$-invariant product of $N$ superfields. Equivalently, one can expand the $SU(N)$-invariant product by introducing explicitely an epsilon tensor which is represented by the {\sc FeynRules} symbol \verb?SUEps?. Hence, taking the example of the MSSM, the $\mu$-term of the superpotential can be implemented in two ways
\begin{verbatim} MUH SUDOT[HU[aa], HD[aa], aa]
 MUH SUEps[aa,bb] HU[aa]HD[bb]\end{verbatim}
where \verb?MUH? is the symbol associated to the $\mu$-parameter and \verb?HU? and \verb?HD? the symbols associated to the Higgs superfields $H_u \et H_d$, respectively. As clear from the example, the \verb?SUDot? function receives as arguments the sequence of the contracted superfields with all indices explicit. The contracted $SU(N)$ indices are set to a single given value for all superfields, the latter being repeated as the last argument of the \verb?SUDot? function. In our example, it is labeld \verb?aa?. The second line in the example, shows how one can re-write the $SU(2)$ product by introducing explicitely the antisymmetric tensor \verb?SUEps?.  As another example, the up-type squark trilinear soft supersymmetry breaking interaction terms would be implemented as 
\begin{verbatim} -tu[ff1,ff2] URs[ff1,cc1] SUDot[QLs[aa,ff2,cc1], hus[aa], aa]\end{verbatim}
where \verb?URs? and \verb?QLs? are the symbols associated respectively to the left-handed anti up-squark and the left-handed squark.

\subsection{Generating the renormalization group equations}
Once the model file adapted for the RGE-extracting routine, one has to load both {\fr} and his model as usual. One has then access to several functions depending on the RGEs one wants calculated. The main call to the routines can be done through the function
\begin{verbatim}
 RGE[LSoft, SuperW, NLoop -> n]
\end{verbatim}
where \verb?LSoft? and \verb?SuperW? are the {\mk} symbols associated to the soft supersymmetry breaking lagrangian and the superpotential, respectively. The option \verb?NLoop? tells {\insurge} at which level the user wants the RGEs. Hence, the symbol \verb?n? can take either values \verb?1? or \verb?2?. The other available functions are listed in table \ref{table:rges functions}

\begin{table}
\begin{center}
\shadowbox{ 
\begin{tabular}{p{8cm} p{7cm}}
Function & Description\\
\texttt{GaugeCouplingsRGE[LSuperW,NLoop-> n]} & Function to obtain the renormalization group equations for the gauge coupling constants  \\
\\
\texttt{GauginoMassesRGE[LSuperW,LSoft, NLoop-> n]} & Function to obtain the renormalization group equations for the gaugino soft breaking masses \\
\\
\texttt{SuperpotentialRGE[LSuperW,NLoop-> n]} & Function to obtain the renormalization group equations for the superpotential parameters \\
\\
\texttt{ScaSoftRGE[LSoft,LSuperW, NLoop-> n]} & Function to obtain the renormalization group equations for soft supersymmetry breaking terms but gaugino masses\\
\\
\texttt{RGE[LSoft,LSuperW,NLoop-> n]} & Calls all the above functions and returns the RGEs for all the parameters of the lagrangian.
\end{tabular}
}
\end{center}
\caption{\label{table:rges functions}List of all available functions included in the module {\insurge}. Symbols \texttt{LSuperW} and \texttt{LSoft} stand for the superpotential and the soft supersymmetry breaking lagrangian, respectively. The symbol \texttt{n} defines the order of the RGEs, hence its values are either \texttt{1} or \texttt{2}.}
\end{table}

\subsection{Example of use: the left-right symmetric supersymmetric model}
In chapter \ref{chap:lrsusy}, we have introduced a \textit{top-down} approach to the study of a left-right symmetric supersymmetric model. For completeness and to illustrate the way {\insurge} works, we give here an example of how to implement this model and calculate its renormalization group equations. For clarity purposes, we only give in this subsection the 2-loop beta functions for the gauge coupling constants. \\

The superpotential of this model given in eq.\eqref{eq:superpot lrsusy}, is
\bea
W &=& (\tQ_L)^{mi} y_Q^1 (\hPhi)_i{}^{i'} (\tQ_R)_{mi'} + (\tQ_L)^{mi}y_{Q}^2 (\hPhi_2)_i{}^{i'} (\tQ_R)_{mi'} + (\tL_L)^iy_L^1(\hPhi)_i{}^{i'}(\tL_R)_{i'} + (\tL_L)^iy_L^2 (\hPhi_2)_i{}^{i'} (\tL_R)_{i'} \n
  &+& (\hat{\tL}_L)_i y_L^3 (\Delta_{2L})^i{}_j(\tL_L)^j + (\hat{\tL}_R)_{i'} y_L^4 (\Delta_{1R})^{i'}{}_{j'} (\tL_R)^{j'} + (\mu_L + \lambda_L S) \Delta_{1L} \cdot \hDelta_{2L} \n
  &+& (\mu_R + \lambda_R S)\Delta_{1R}\cdot\hDelta_{2R} + (\mu_3 + \lambda_3 S) \Phi_1 \cdot \hPhi_2 + \frac13 \lambda_s S^3 + \mu_s S^2 + \xi_S S.
\eea
It can be entered in the model file as follows
\begin{verbatim}
SPLR = yq1[ff1,ff2] SUEps[iip,jjp] SUEps[ii,jj] QL[ii,ff1,cc1] H1[jj,jjp] QR[iip,ff2,cc1] +
   yq2[ff1,ff2] SUEps[iip,jjp] SUEps[ii,jj] QL[ii,ff1,cc1] H2[jj,jjp] QR[iip,ff2,cc1] +
   yl1[ff1,ff2] SUEps[iip,jjp] SUEps[ii,jj] LL[ii,ff1] H1[jj,jjp] LR[iip,ff2] +
   yl2[ff1,ff2] SUEps[iip,jjp] SUEps[ii,jj] LL[ii,ff1] H2[jj,jjp] LR[iip,ff2] +
   yl3[ff1,ff2] PauliSigma[aa,ii,kk]/Sqrt[2] SUEps[ii,jj] LL[jj,ff1] H2L[aa] LL[kk,ff2] +
   yl4[ff1,ff2] PauliSigma[aa,ii,kk]/Sqrt[2] SUEps[ii,jj] LR[jj,ff1] H1R[aa] LR[kk,ff2] +
   muL PauliSigma[aa,ii,jj] H1L[aa] SUEps[ii,kk] SUEps[jj,jjp] PauliSigma[bb,kk,jjp] H2L[bb] +
   muR PauliSigma[aa,ii,jj] H1R[aa] SUEps[ii,kk] SUEps[jj,jjp] PauliSigma[bb,kk,jjp] H2R[bb] +
   mu3 H1[ii,iip] SUEps[iip,jjp] SUEps[ii,jj] H2[jj,jjp] + 
   lamL SPF PauliSigma[aa,ii,jj] H1L[aa] SUEps[ii,kk] SUEps[jj,jjp] PauliSigma[bb,kk,jjp] H2L[bb] +
   lamR SPF PauliSigma[aa,ii,jj] H1R[aa] SUEps[ii,kk] SUEps[jj,jjp] PauliSigma[bb,kk,jjp] H2R[bb] +
   lam3 SPF H1[ii,iip] SUEps[iip,jjp] SUEps[ii,jj] H2[jj,jjp] +
   1/3 lamS SPF^3 + muS SPF^3 + xiF SPF;
\end{verbatim}
where 
\begin{itemize} 
\item symbols \verb?QL, QR, H1, H2, H2L, H1R, SPF? correspond to the superfields $Q_L, Q_R, \Phi_1, \Phi_2, \delta_{2L}, \delta_{1R}, S$ respectively, 
\item \verb?PauliSigma? represents the Pauli matrices $\sigma$
\item \verb?yq1, yq2, yl1, yl2, yl3, yl4, lamL, lamR, lamS, muL, muR, mu3, muS, xiF? are the couplings $y_Q^1$, $y_Q^2$, $y_L^1$, $y_L^2$, $y_L^3$, $y_L^4$, $\lambda_L$, $\lambda_R$, $\lambda_S$, $\mu_L$, $\mu_R$, $\mu_3$, $\mu_S$, $\xi_S$, respectively.
\end{itemize}

The soft-supersymmetry breaking lagrangian given in eq.\eqref{eq:lsoft lrsusy} is
\bea 
\lagr_{soft} &=& -\frac12 \big[ M_1 \tilde{B}\cdot\tilde{B} + M_{2L} \tilde{W}^k_L \cdot \tilde{W}_{Lk} + M_{2R} \tilde{W}^k_R \cdot \tilde{W}_{Rk} + M_3 \tilde{g}^a \cdot \tilde{g}_a + {\rm h.c.} \big] \n
             &-& \Big[\tilde{Q}^\dagger m_{Q_L}^2 \tilde{Q}_L + \tilde{Q}_R m_{Q_R}^2 Q_R^\dagger + \tL_L^\dagger m_{L_L}^2 \tL_L + \tL_R m_{L_R}^2 \tL_R^\dagger - (m_\Phi^2)^{ff'} {\rm Tr}(\Phi^\dagger_f \Phi_{f'}) \n
&+& m^2_{\Delta_{1L}}{\rm Tr}(\Delta^\dagger_{1L} \Delta_{1L}) +  m^2_{\Delta_{2L}}{\rm Tr}(\Delta^\dagger_{2L} \Delta_{2L}) + m^2_{\Delta_{1R}}{\rm Tr}(\Delta^\dagger_{1R} \Delta_{1R}) +  m^2_{\Delta_{2R}}{\rm Tr}(\Delta^\dagger_{2R} \Delta_{2R}) + m^2_S S^\dagger S \Big]\n
&-& \Big[ \tQ_L T_Q^1 \hPhi_1 \tQ_R + \tQ_L T_Q^2 \hPhi_2 \tQ_R + \tL_L T_L^1\hPhi_1\tL_R + \tL_L T_L^2\hPhi_2\tL_R + \hat{\tL}_LT^3_L\Delta_{2L}\tL_L + \tL_R T^4_L \Delta_{1R} \hat{\tL}_R + {\rm h.c.} \Big]\n
&-& \Big[ T_L S \Delta_{1L} \cdot \hDelta_{2L} + T_R S \Delta_{1R} \cdot \hDelta{2R} + T_{3}S \Phi_1 \cdot \hPhi_2 + {\rm h.c.} \Big]
\eea
which can be implemented as follows
\begin{verbatim}
LSoft := Module[{(*Mino,MSca,TriBil*)},
  (*Gaugino mass terms*)
  Mino = - Mx1*bblow[sp].bblow[sp] - Mx2L*wlow[s, gl].wlow[s, gl] - Mx2R*wrow[s, gl].wrow[s, gl]
         - Mx3*gow[s, gl].gow[s, gl];  

  (*Scalar mass terms*)  
  MSca = -mH211*HC[h1s[ii, jj]]*h1s[ii, jj] - mH212*HC[h1s[ii, jj]]*h2s[ii, jj] 
   - mH221*HC[h2s[ii, jj]]*h1s[ii, jj] - mH222*HC[h2s[ii, jj]]*h2s[ii, jj]
   - mH12*HC[h1Ls[ii]]*h1Ls[ii] - mH22*HC[h2Ls[ii]]*h2Ls[ii] - mH32*HC[h1Rs[ii]]*h1Rs[ii] 
   - mH42*HC[h2Rs[ii]]*h2Rs[ii] - ms2*HC[SPs]*SPs - mLL2[ff1, ff2]*HC[LLs[ii, ff1]]*LLs[ii, ff2] 
   - mLR2[ff1, ff2]*HC[LRs[ii, ff1]]*LRs[ii, ff2] - mQL2[ff1, ff2]*HC[QLs[ii, ff1, cc1]]*QLs[ii, ff2, cc1] 
   - mQR2[ff1, ff2]*HC[QRs[ii, ff1, cc1]]*QRs[ii, ff2, cc1]];
  
  (*Trilinear and bilinear couplings*)
  TriBil = TQ1[ff1,ff2] SUEps[iip,jjp] SUEps[ii,jj] QLs[ii,ff1,cc1]* h1s[jj,jjp]*QRs[iip,ff2,cc1] +
      TQ2[ff1,ff2] SUEps[iip,jjp] SUEps[ii,jj] QLs[ii,ff1,cc1]* h2s[jj,jjp]*QRs[iip,ff2,cc1] +
      TL1[ff1,ff2] SUEps[iip,jjp] SUEps[ii,jj] LLs[ii,ff1]* h1s[jj,jjp]*LRs[iip,ff2] +
      TL2[ff1,ff2] SUEps[iip,jjp] SUEps[ii,jj] LLs[ii,ff1]* h2s[jj,jjp]*LRs[iip,ff2] + 
      TL3[ff1,ff2] PauliSigma[aa,ii,kk]/Sqrt[2] SUEps[ii,jj] LLs[jj,ff1] h2Ls[aa] LLs[kk,ff2] +
      TL4[ff1,ff2] PauliSigma[aa,ii,kk]/Sqrt[2] SUEps[ii,jj] LRs[jj,ff1] h1Rs[aa] LRs[kk,ff2] +
      BL PauliSigma[aa,ii,jj] h1Ls[aa] SUEps[ii,kk] SUEps[jj,jjp] PauliSigma[bb,kk,jjp] h2Ls[bb] +
      BR PauliSigma[aa,ii,jj] h1Rs[aa] SUEps[ii,kk] SUEps[jj,jjp] PauliSigma[bb,kk,jjp] h2Rs[bb] +
      B3 h1s[ii,iip] SUEps[iip,jjp] SUEps[ii,jj] h2s[jj,jjp] + 
      TL5 SPs PauliSigma[aa,ii,jj] h1Ls[aa] SUEps[ii,kk] SUEps[jj,jjp] PauliSigma[bb,kk,jjp] h2Ls[bb] +
      TR5 SPs PauliSigma[aa,ii,jj] h1Rs[aa] SUEps[ii,kk] SUEps[jj,jjp] PauliSigma[bb,kk,jjp] h2Rs[bb] +
      T3 SPs h1s[ii,iip] SUEps[iip,jjp] SUEps[ii,jj] h2s[jj,jjp] + 1/3 TS  SPs^3 + BS SPs^3 + xiS SPs;
    Return[MSca + Mino + TriBil + HC[Mino + TriBil]]
    ];
\end{verbatim}

Finally, the normalization constant associated to the abelian group $U(1)_{B-L}$ is $\sqrt{\frac{3}{16}}$

The renoramlization group equations for the gauge coupling constants at the 2-loop level are:
\begin{align*}
\frac{\partial g_{B-L}}{\partial t} &= \frac{115 g_{B-L}^5}{16384 \pi^4} + \frac{81 g_{B-L}^3 g_{L}^2}{2048 \pi^4} + \frac{81 g_{B-L}^3 g_{R}^2}{2048 \pi^4} + \frac{g_{B-L}^3 gs^2}{256 \pi^4} + \frac{3 g_{B-L}^3}{16 \pi^2} - \frac{27 g_{B-L}^3 \lambda_L \lambda_L^\dagger}{256 \pi^4} - \frac{27 g_{B-L}^3 \lambda_R \lambda_R^\dagger}{256 \pi^4} \\
& - \frac{3 g_{B-L}^3 {\rm Tr}(y_L^1)^2}{1024 \pi^4} - \frac{3 g_{B-L}^3 {\rm Tr}(y_L^2)^2}{1024 \pi^4} - \frac{27 g_{B-L}^3 {\rm Tr}(y_L^3)^2}{8192 \pi^4} - \frac{27 g_{B-L}^3 {\rm Tr}(y_L^4)^2}{8192 \pi^4} - \frac{g_{B-L}^3 {\rm Tr}(y_Q^1)^2}{1024 \pi^4} - \frac{g_{B-L}^3 {\rm Tr}(y_Q^2)^2}{1024 \pi^4} ;\\
\frac{\partial g_{L}}{\partial t} & = \frac{27 g_{B-L}^2 g_{L}^3}{2048 \pi^4} + \frac{5 g_{L}^5}{16 \pi^4} + \frac{3 g_{L}^3 g_{R}^2}{128 \pi^4} + \frac{3 g_{L}^3 g_s^2}{32 \pi^4} + \frac{3 g_{L}^3}{8 \pi^2} - \frac{g_{L}^3 \lambda_3 \lambda_3^\dagger}{64 \pi^4} - \frac{3 g_{L}^3 \lambda_L \lambda_L^\dagger}{8 \pi^4} - \frac{g_{L}^3 {\rm Tr}(y_L^1)^2}{64 \pi^4}\\
& - \frac{g_{L}^3 {\rm Tr}(y_L^2)^2}{64 \pi^4} - \frac{7 g_{L}^3 {\rm Tr}(y_L^3)^2}{512 \pi^4} - \frac{3 g_{L}^3 {\rm Tr}(y_Q^1)^2}{64 \pi^4} - \frac{3 g_{L}^3 {\rm Tr}(y_Q^2)^2}{64 \pi^4}; \\
\frac{\partial g_{R}}{\partial t} & = \frac{27 g_{B-L}^2 g_{R}^3}{2048 \pi^4} + \frac{3 g_{L}^2 g_{R}^3}{128 \pi^4} + \frac{5 g_{R}^5}{16 \pi^4} + \frac{3 g_{R}^3 g_s^2}{32 \pi^4} + \frac{3 g_{R}^3}{8 \pi^2} -  \frac{g_{R}^3 \lambda_3 \lambda_3^\dagger}{64 \pi^4} - \frac{3 g_{R}^3 \lambda_R \lambda_R^\dagger}{8 \pi^4} - \frac{g_{R}^3 {\rm Tr}(y_L^1)^2}{64 \pi^4} \\
& - \frac{g_{R}^3 {\rm Tr}(y_L^2)^2}{64 \pi^4} -  \frac{7 g_{R}^3 {\rm Tr}(y_L^4)^2}{512 \pi^4} - \frac{3 g_{R}^3 {\rm Tr}(y_Q^1)^2}{64 \pi^4} - \frac{3 g_{R}^3 {\rm Tr}(y_Q^2)^2}{64 \pi^4};\\
\frac{\partial g_s}{\partial t} & = \frac{g_{B-L}^2 g_s^3}{2048 \pi^4} + \frac{9 g_{L}^2 g_s^3}{256 \pi^4} + \frac{9 g_{R}^2 g_s^3}{256 \pi^4} + \frac{7 g_s^5}{128 \pi^4} - \frac{3 g_s^3}{16 \pi^2} - \frac{g_s^3 {\rm Tr}(y_Q^1)^2}{32 \pi^4} - \frac{g_s^3 {\rm Tr}(y_Q^2)^2}{32 \pi^4}
\end{align*}


\section{Automated spectrum generation}\label{sec:asperge}
Once the renormalization group equations generated automatically, the next step to achieve a full spectrum generator generator is the creation of a routine able to extract and diagonalize the mass matrices automatically. Though this spectrum generator can only work with supersymmetric theories, generalizing the mass matrices extraction routine to any lagrangian-based quantum field theory did not imply so much effort. As a result, we have implemented in {\fr} a module able to extract analytically the tree-level mass matrices for any quantum field theory. We also developped a {\sc C++} routine able to diagonalize these mass matrices and to generate an SLHA-compliant file containing the results. This work was published in january 2013 \cite{Alloul:2013fw} and this section is dedicated to the description of the modifications we brought to the {\fr} model file as well as how one can get the mass matrices and the numerical spectrum for his theory.

\subsection{Modifying the {\fr} model file}
As seen in section \ref{sec:feynrules}, the only way to declare in {\fr} how fields mix is to use the option \verb?Definitions? for the gauge eigenstates and to write manually the value of every mixing matrix. To illustrate this we gave the examples of the declaration of the gauge bosons $W_{L\mu} \et B_\mu$ 
\begin{verbatim}
 V[1] == { 
   ClassName     -> B, 
   Unphysical    -> True, 
   SelfConjugate -> True, 
   Definitions   -> { B[mu_] -> -sw Z[mu]+cw A[mu]} 
 },
 V[2] == { 
   ClassName     -> Wi,
   Unphysical    -> True,
   SelfConjugate -> True, 
   Indices       -> {Index[SU2W]},
   FlavorIndex   -> SU2W,
   Definitions   -> { Wi[mu_,1] -> (Wbar[mu]+W[mu])/Sqrt[2], 
                      Wi[mu_,2] -> (Wbar[mu]-W[mu])/(I*Sqrt[2]), 
                      Wi[mu_,3] -> cw Z[mu] + sw A[mu]}
  }
\end{verbatim}
where the mixing relations are given explicitely in function of the physical\footnote{In this section, physical (unphysical) will refer to mass (gauge, resp.) eigenstate.} gauge bosons $Z_\mu \text{ and the photon }  A_\mu$. The latter are declared as follows
\begin{verbatim}

  V[3] == {                               V[4] == { 
    ClassName       -> A,                     ClassName       -> Z, 
    SelfConjugate   -> True,                  SelfConjugate   -> True,
    Mass            -> 0,                     Mass            -> {MZ, 91.1876},
    Width           -> 0,                     Width           -> {WZ, 2.4952},    
    ParticleName    -> "a",                   ParticleName    -> "Z",  
    PDG             -> 22,                    PDG             -> 23,     
    PropagatorLabel -> "a",                   PropagatorLabel -> "Z",
    PropagatorType  -> W,                     PropagatorType  -> Sine,
    PropagatorArrow -> None,                  PropagatorArrow -> None,
    FullName        -> "Photon"},             FullName        -> "Z"}

\end{verbatim}
Here the mixing relations are easy and well determined however they may become very complicated when considering more complex theories where mass matrices are not known at all. To facilitate the implementation of mixing relations, we decided to add a new class to the {\fr} model file dedicated to particle mixings. Consequently, all mixing relations among the states can be declared on the same spirit as particles, gauge groups and parameters after having been gathered into a list dubbed \verb?M$MixingsDescription?
\begin{verbatim}
M$MixingsDescription == {
  Mix["l1"] == { option1 -> value1, ...}
  …
 }
\end{verbatim}
Each element of this list consists of an equality dedicated to one specific mixing relation. It associates a label, given as a string (here \verb?"l1"?) with a set of {\mk} replacement rules defining the mixing properties (\verb?option1 -> value1?). As a first example, we can examine the case of $W_{L}^1 \et W_L^2$ which mix to give rise to the charged gauge bosons $W^+ \et W^-$ through the relation
\bea W^{\pm} = \frac{W_L^1 \mp i W_L^2}{\sqrt2}. \nonumber \eea
This mixing can be simply declared in {\fr} as follows
\begin{verbatim}
   Mix["Wmix"] == {
     MassBasis  -> {W, Wbar}, 
     GaugeBasis -> {Wi[1], Wi[2]}, 
     Value      -> {{1/Sqrt[2], -I/Sqrt[2]}, {1/Sqrt[2], I/Sqrt[2]}}
   }, 
\end{verbatim}
where \verb?W? (\verb?Wbar?) is the {\mk} symbol for $W^-$ (resp. $W^+$). From this example it is clear that the option \verb?GaugeBasis? is used to declare the gauge eigenstates. The \verb?MassBasis? option while used in this example to declare mass eigenstates may take as attribute mass and/or gauge eigenstates. This possibility allows the user to declare his mixings in two steps or more as will be illustrated in subsection \ref{subsec:chain mixing}. The value of the mixing matrix being known, one can simply use the option \verb?Value? to enter it. Finally, with this declaration we have entered the relation
\begin{verbatim}
MassBasis = Value . GaugeBasis
\end{verbatim}
where the dot stands for the usual matrix product.\\

Let us now consider the case of $(W_{L}^3, B_\mu)$ mixing which we suppose unknown for illustration purposes. The declaration is as follows
\begin{verbatim}
   Mix["weakmix"] == {
     MassBasis    -> {A, Z}, 
     GaugeBasis   -> {B, Wi[3]}, 
     MixingMatrix -> UG, 
     BlockName    -> WEAKMIX}, 
\end{verbatim}
In this example where we supposed the mixing unknown, we introduced two new options namely \verb?MixingMatrix? and \verb?BlockName?. 
\begin{itemize}
\item The first one, \verb?MixingMatrix?, is the symbol associated to the mixing matrix. It will be treated as a {\fr} variable but the user will not have to declare it in the list \verb?M$Parameters?. This task will be done automatically by {\fr} which assumes that all mixing matrices are complex and will thus split them into a real and a complex part. In our example, {\fr} will thus create an entry for \verb?UG?, \verb?RUG? and \verb?IUG? where \verb?RUG? (\verb?IUG?) is the real (resp. complex) part of \verb?UG?\footnote{The prefixes \texttt{R} and \texttt{I} are added automatically by {\fr}.}:
\begin{verbatim} UG = RUG + I*IUG. \end{verbatim}
One exception to this rule is the case where the mixing matrix appears explicitely in the Lagrangian. In this case, it is asked to the user to declare this mixing matrix as he would do for any other parameter of the theory. As an example, one can cite the CKM matrix which relates the left-handed down quark gauge eigenstates $d_L^0$ to the mass eigenstates $d_L$ following the relation
$$ d_L^0 = V_{\text{CKM}} . d_L$$
and which appears explicitely in the Yukawa couplings of the Standard Model Lagrangian for example.
\item The option \verb?BlockName? is used to declare the {\sc SLHA}-block associated with the mixing matrix. The latter being assumed complex, {\fr} will create automatically a block for the complex part. In our example, the block \verb?WEAKMIX? will be associated to the real part of \verb?UG? that is \verb?RUG? while the automatically created \verb?IMWEAKMIX? will be associated to the complex part \verb?IUG?. It is important to remark that this option is mandatory when \verb?MixingMatrix? has been provided.
\end{itemize}

Before proceeding to the more specific cases of scalars and fermions, the attention of the reader is drawn to the fact that the Lorentz indices have not been used in the declaration of the mixings. This is a simplification that we have introduced, one is not obliged to provide Lorentz and spin indices as they are the same for all fields involved in the mixing declaration. More generally, if some indices are irrelevant {\ie} if they are identical for all the involved fields, underscores can be employed to simplify the mixing declaration. {\fr} will then refer to the indices declared in the fields declaration to restore them. \\

It is also to be noted that in the declaration of the parameters, for consistency reasons, the user should not use the masses of particles he wants calculated. Hence, a check over the definitions of the parameters is realized when generating the output for the {\asperge} package to avoid such problems.\\

\paragraph{Scalars} When neutral scalars are mixing, the gauge eigenstates might split into their real degrees of freedom so that one scalar and one pseudoscalar mass basis are required. Consequently, to allow {\fr} to handle this case efficiently a set of two gauge bases is needed as well as two entries for each of the options \verb?MixingMatrix, BlockName? and \verb?Value?. The convention in this case is that the first element of those lists will always refer to the scalar fields while the second one is related to the pseudoscalar fields. As an illustration, let us assume the complex scalar gauge eigenstates \verb?phi1? and \verb?phi2? which split into two real scalars \verb?h1? and \verb?h2? and two pseudoscalars \verb?a1? and \verb?a2?. Let us, moreover assume that the value for the mixing matrix of the pseudoscalars is known while that for the scalars in unknown. The declaration for this mixing reads
\begin{verbatim}
  Mix["scalarmix"] == {
    GaugeBasis   -> {ph1, phi2},
    MassBasis    -> {{h1,h2}, {a1,a2}},
    MixingMatrix -> {US, _},
    Value        -> {_, ...}
  }
\end{verbatim}
The dots in the attributes for the option \verb?Value? stand for explicit numerical value of the pseudoscalar mixing matrix. \\

In this declaration, we illustrate another use of the underscores. Indeed, as assumed in the example, only the mixing matrix for the pseudoscalars is known hence the use of the underscore in the attributes of the option \verb?Value?. It is there to tell {\fr} that the value of the scalar mixing matrix is not known. The attributes for the option \verb?MixingMatrix? also include an underscore but this time it is to tell {\fr} that we do not want the mixing matrix for the pseudoscalars to be computed. More generally, an underscore is used as an attribute of an option whenever we do not want that information to be computed.
\paragraph{Weyl fermions} Declaring electrically neutral Weyl fermions is quiet straightforward as it resembles that of gauge bosons: one attribute per option. \\
In the case of electrically charged Weyl fermions, the user may want to allow for two entries for all the options \verb?GaugeBasis, MassBasis, MixingMatrix, Value? and \verb?BlockName?. As an example, one can cite the case of charginos mixing in the left-right supersymmetric model studied in chapter \ref{chap:lrsusy}. Let us note \verb?chmw[i]? ( \verb?chpw[i]? ) the $i^{th}$ chargino with negative (resp. positive) charge, \verb?wowlp? and \verb?wowrp? (\verb?wowlm? and \verb?wowrm?) would be the charged gauginos $\tilde{W}_L^+ \et \tilde{W}_R^+$ (resp. $\tilde{W}_L^- \et \tilde{W}_R^-$). Finally the symbols \verb?h1w, h2w, h1Lw, h2Lw, h1Rw? and \verb?h2Rw? stand for the higgsinos $\tilde{H}_1, \tilde{H}_2, \tilde{\delta}_{1L}, \tilde{\delta}_{2L}, \tilde{\delta}_{1R} \et \tilde{\delta}_{2R}$ respectively. This mixing can be declared as follows
\begin{verbatim}
Mix["char"] == { MassBasis   -> {{chmw[1],chmw[2],chmw[3],chmw[4],chmw[5],chmw[6]}, 
                                 {chpw[1],chpw[2],chpw[3],chpw[4],chpw[5],chpw[6]}}, 
                 GaugeBasis   -> {{wowlp,wowrp,h2w[1,2],h1w[1,2],h2Lw[3], h2Rw[3]}, 
                                  {wowlm,wowrm,h2w[2,1],h1w[2,1],h1Lw[3], h1Rw[3]}}
                 BlockName    -> {UMIX, VMIX}, 
                 MixingMatrix -> {UU, VV}
                },
\end{verbatim}
which corresponds to
\bea
\begin{pmatrix} \tilde{W}_L^- & \tilde{W}_R^- & \tilde{H}_2^- & \tilde{H}^-_1 & \tilde{\delta}_{1L}^- & \tilde{\delta}_{1R}^- \end{pmatrix} M \begin{pmatrix} \tilde{W}_L^+ \\ \tilde{W}_R^+ \\ \tilde{H}_2^+ \\ \tilde{H}^+_1 \\ \tilde{\delta}_{2L}^+ \\ \tilde{\delta}_{2R}^+ \end{pmatrix} .
\eea
Symbols \verb?UMIX? (\verb?VMIX?) are the SLHA-block names associated to the mixing of the negatively (resp. positively) charged charginos that is to the mixing matrix \verb?UU? (resp. \verb?VV?)
\paragraph{Dirac fermions} Finally, concerning the declaration of Dirac fermions, the choice is left to the user to either provide both gauge eigenstates that form the Dirac fermion or only one gauge eigenstate letting thus {\fr} take care of the chirality projection. In the first case, the attribute for \verb?GaugeBasis? consists of a list of two bases with the convention that the first list refers to the left-handed component while the second one to the right-handed component. In both cases, the attributes for the options \verb?Value, BlockName? and \verb?MixingMatrix? take two lists for both chiralities following the same convention as above.\\
We profit of the Dirac fermions case to introduce the last option one can use in the declaration of mixings. As seen above, the relation for the mixing in the left-handed down-type quarks sector is given by
$$ d_L^0 = V_{\text{CKM}} . d_L$$
which means that if one wants to implement this mixing following the convention 
\begin{verbatim}
MassBasis = Value . GaugeBasis
\end{verbatim}
one would need the inverse of the CKM matrix $V_{\text{CKM}}^{-1}$. To circumvent this small problem, we introduced the option \verb?Inverse? which presence is only necessary when facing this kind of situation and whose attribute can be either the boolean \verb?True? or \verb?False? or an underscore if relevant.

\paragraph{Vacuum expectation values} In realistic new physics models, the ground state of the theory is non-trivial and neutral scalar fields must be shifted by their vacuum expectation value. To allow for a proper handling of the vacuum expectation values, it is asked to the user to gather them into the list \verb?M$vevs? following the structure
\begin{verbatim}
 M$vevs == { {field1, vev1}, {field2, vev2}, ...}
\end{verbatim}
where \verb?field1? and \verb?field2? refer to two gauge eigenstates and \verb?vev1? and \verb?vev2? to their respective vacuum expectation values which, on the other hand, have to be declared as any other parameter of {\fr} in the list \verb?M$Parameters?.

\subsection{Running the package} Once the model file adapted to allow for a proper handling of the mixings by our new routine, the user can load both {\fr} and his model as usual. If {\fr} detects the lists \verb?M$MixingsDescription? and \verb?M$vevs? it launches then some routines so that all the automatic declarations are accomplished. From the user side four main functions become available namely \verb?ComputeMassMatrix?, \verb?MixingSummary?, \verb?WriteASperGe? and \verb?RunASperGe?.\\

The function \verb?ComputeMassMatrix? computes the tree-level mass matrices of the model. It takes as argument the lagrangian and some options which are
\begin{itemize}
\item \verb?Mix? which takes as attribute either one label of mixing matrix or more. If present, {\fr} only computes the mass matrices associated to that/those label/labels, if not all mass matrices are computed.
\item \verb?Basis1 -> b1? and \verb?Basis2 -> b2? are two options that demand {\fr} to compute a mass matrix associated to the two bases \verb?b1? and \verb?b2? where we suppose the mass term in the lagrangian as follows 
$$ b_2^\dagger~M~b_1$$
\item \verb?ScreenOutput? is an option that takes as attribute a boolean. If set to \verb?True?, a whole set of information is printed to the screen while computing the mass matrices. By default, it is set to \verb?True?
\end{itemize}

The function \verb?MixingSummary? which takes as argument a mixing label, prints to the screen all the information associated to the label that is the gauge basis, the mass basis, the {\sc SLHA}-block, the symbol associated to the mixing matrix, the mass matrix itself and the mixing matrix\footnote{Some information might be not available depending on the options that were provided in the declaration of the mixing under consideration.}. \\

As indicated in the beginning of this section, we have developped a {\sc C++} package {\asperge} able to diagonalize numerically the mass matrices generated by {\fr} and to export the results in an {\sc SLHA}-compliant format. The source files for this code are generated automatically by {\fr}, some of them being model-dependent and the others model independent. 
To call this interface, it suffices to type in the {\mk} session where {\fr} is loaded the command 
\begin{verbatim} WriteASperGe[lag, options] \end{verbatim}
where \verb?options? stands for the options \verb?Mix? described above and the option \verb?Output? which takes as attribute the name of the output directory. Indeed, this command results in the creation of a directory whose name is either the model name plus the suffix \verb?_MD? or that provided as an attribute for the option \verb?Output?. This directory will contain all the source files, both model-dependent and model-independent, necessary for the {\asperge} package to diagonalize the mass matrices associated to the model under study. \\

Finally and before proceeding to the description of the content of the {\asperge} package, let us note that the routine \verb?RunASperGe[]? that takes no argument is able to run the {\asperge} package, read its ouptut and update the parameters in {\fr} accordingly.\\

The model-dependent files \verb?Parameters.cpp? included in the directory \verb?src? and its related header file \verb?Parameters.hpp? included in the sub-directory \verb?inc? contain all the information that allow to define the parameters of the model and the mass matrices that have been computed. The file \verb?main.cpp? which is at the root of the generated directory contains the source code to  initialize the mass matrices, to set the {\sc PDG} codes\footnote{As a convention, we have chosen to sort the {\sc PDG}-Ids in such a way that the lowest one corresponds to the lightest eigenstate of the mass matrix being digonalized.} associated with the mass eigenstates of the matrix and the command to launch the diagonalization. One last model-dependent file is created when launching the {\asperge} interface, namely \verb?externals.dat? in the sub-directory \verb?input?. This file is the {\sc SLHA}-file created by {\fr} from the model file. It contains all the numerical values for the external parameters, hence its name.\\

The model-independent files are all the other files present in the sub-directory \verb?src? together with their associated header files gathered in the sub-directory \verb?inc?. These files contain all the routines to read the input file, initialize all the parameters, diagonalize the mass matrices and generate the output. An important information to know when using the {\asperge} package is that the diagonalization routines use the {\sc GSL}-library functionalities \cite{GSL}. Though we allow for a non-standard installation, the library needs however to be installed.\\

To run the {\sc C++} code, it is then sufficient to issue the following commands in a {\sc Shell}:
\begin{verbatim}
cd <path-to-the-directory>
make
./ASperGe <input> <output> m1 m2 ...
\end{verbatim}
where \verb?<input>? and \verb?<output>? are the input and the output file and the optional sequence \verb? m1 m2 …? represents the names of the mixing matrices to be computed. The final command results in the creation of two files. The first one designated by \verb?<output>? contains the results of the diagonalization while the second one whose name is \verb?asperge.log? is a log file useful when debugging. Note that \verb?m1 m2 …? are not mandatory and if not provided, the package will just diagonalize all the mass matrices.\\

In our paper \cite{Alloul:2013fw}, we showed how one can achieve a successful implementation of the Standard Model, the Two-Higgs-Doublet Model, the Left-Right Symmetric Model (non-supersymmetric) and the Minimal Supersymmetric Model. Our results were compared with the litterature and turned out to be exact. We have also been able to reproduce all the mass matrices of the LRSUSY model studied in the chapter \ref{chap:lrsusy}. In the following subsection, we present an example of use of the package in the case of the Left-Right symmetric supersymmetric model.

\subsection{Example of use}\label{subsec:chain mixing}
To illustrate how the full machinery works, from the implementation of a mixing until obtaining a numerical spectrum let us get back to the example of the charginos in the left-right supersymmetric symmetric model described in the previous chapter:
\begin{verbatim}
Mix["char"] == { MassBasis   -> {{chmw[1],chmw[2],chmw[3],chmw[4],chmw[5],chmw[6]}, 
                                 {chpw[1],chpw[2],chpw[3],chpw[4],chpw[5],chpw[6]}}, 
                 GaugeBasis   -> {{wowlp,wowrp,h2w[1,2],h1w[1,2],h2Lw[3], h2Rw[3]}, 
                                  {wowlm,wowrm,h2w[2,1],h1w[2,1],h1Lw[3], h1Rw[3]}}
                 BlockName    -> {UMIX, VMIX}, 
                 MixingMatrix -> {UU, VV}
                },
\end{verbatim}
As stated above, the symbols \verb?wowlp,wowrp? and \verb?wowlm,wowrm? represent the fields $\tilde{W}_L^+, \tilde{W}_R^+$ and $\tilde{W}_L^-, \tilde{W}_R^-$ respectively. In order for {\fr} to handle properly this mixing it is mandatory to first declare the mixing between $\tilde{W}_{L,R}^{1}, \tilde{W}_{L,R}^2$ that gives rise to the charged winos and then declare the above mixing. This procedure we call ``chain-mixing" offers thus a flexible and efficient way to declare mixings going first from gauge eigenstates to $T^3$ eigenstates and then from the latter to mass eigenstates. For the mixings we are looking at, in addition to the mixing \verb?"char"?, we need thus the following lines in the model file:
\begin{verbatim}
Mix["2a"] == { MassBasis -> {wowl0,wowlp,wowlm}, GaugeBasis -> {wlow[3],wlow[1],wlow[2]}, 
               Value ->{{I,0,0}, {0,I/Sqrt[2],1/Sqrt[2]}, {0,I/Sqrt[2],-1/Sqrt[2]}} },
Mix["2b"] == { MassBasis -> {wowr0,wowrp,wowrm}, GaugeBasis -> {wrow[3],wrow[1],wrow[2]}, 
               Value ->{{I,0,0}, {0,I/Sqrt[2],1/Sqrt[2]}, {0,I/Sqrt[2],-1/Sqrt[2]}} }
\end{verbatim}
where the symbols \verb?wlow? and \verb?wrow? represent the winos $\tilde{W}_L$ and $\tilde{W}_R$ respectively. An important remark to keep in mind while proceeding to ``chain-mixing" declaration is to declare all fields involved in the different bases as regular fields.\\
This mixing implemented, we just need to add the information about the vacuum expectation values. This is done with the following line in the model file
\begin{verbatim}
M$vevs = { 
(*Bi-doublets vevs*)
{h1s[1,1], v1}, {h1s[2,2], vp1},  {h2s[1,1], vp2}, {h2s[2,2], v2}, 
(*vevs for matrix representations of triplets*)
{DeltaL10, v1L}, {DeltaL20, v2L}, {DeltaR10, v1R}, {DeltaR20, v2R}, 
(*``Left-handed" triplets*)
{h1Ls[1],v1L/Sqrt[2]}, {h1Ls[2],I*v1L/Sqrt[2]}, 
{h2Ls[1],v2L/Sqrt[2]}, {h2Ls[2],-I*v2L/Sqrt[2]},
(*``right-handed" triplets*)
{h1Rs[2],I*v1R/Sqrt[2]}, {h1Rs[1],v1R/Sqrt[2]},
{h2Rs[1],v2R/Sqrt[2]}, {h2Rs[2],-I*v2R/Sqrt[2]},
(*Singlet*)
{SPs, vs}
};
\end{verbatim}
where symbols \verb?h1s, h2s, h1Ls h2Ls, h1Rs, h2Rs, SPs? represent the scalar components of the superfields $ H_1, H_2, \delta_{1L}, \delta_{2L}, \delta_{1R}, \delta_{2R} \et S$ respectively. The symbols \verb?DeltaL10,DeltaL20, DeltaR10, DeltaR20? stand for the scalar components of the matrix representations $\Delta_{1L}, \Delta_{2L},\Delta_{1R}, \Delta_{2R}$ of the triplets\footnote{In this case also one has to use the ``chain-mixing" feature to first declare the relation between triplet and matrix form and then use the matrix form for further mixings.}. Note that in this declaration, we have the following relation among the vacuum expectation values
$$ \langle \Delta \rangle = \frac{\sigma^a}{\sqrt2} \langle \delta_a \rangle. $$
\begin{table}
\begin{center} 
\begin{tabular}{l | r }
  \hline\hline
  Parameter &  Scenario I.1 \\
  \hline
  \hline
  $M_1$ [GeV]    & 250  \\
  $M_{2L}$ [GeV] & 500  \\
  $M_{2R}$ [GeV] & 750  \\
  \hline
  $m_Z$ [GeV]    & 91.1876 \\
  $m_W$ [GeV]    & 80.399  \\
  $\alpha(m_Z)^{-1}$ & 127.9 \\
  \hline
  $v_R$ [GeV]       & 1000  \\
  $v_s$ [GeV]       & $10^5$\\
  $\tan\beta$       & 10 \\
  $\tan\tilde\beta$ & 1  \\
\hline
  $\lambda_L$ & 0.1 \\
  $\lambda_R$ & 0.1 \\
  $\lambda_s$ & 0.1 \\
  $\lambda_3$ & 0.1 \\
 \hline\hline
\end{tabular}
\caption{\footnotesize \label{tab: remember si1} Numerical values for the parameters in benchmark scenario of LRSUSY SI1}
\end{center}
\end{table}
Once the mixing declared properly, both {\fr} and the model file loaded in the current {\mk} session, it suffices to issue the command 
\begin{verbatim}
ComputeMassMatrix[lag]
\end{verbatim}
so that the mass matrix for charginos is computed. The result returned by {\fr} is that of eq.\eqref{mat: charginos} in chapter \ref{chap:lrsusy}, that is
\bea
\begin{pmatrix}
M_{2L} & 0 & \frac{g_L}{\sqrt2}\tilde{v}_2' & \frac{g_L}{\sqrt2}v_1 & -g_L v_{1L} & 0 \\
0 & M_{2R} & -\frac{g_R}{\sqrt2}v_2 & -\frac{g_R}{\sqrt2}\tilde{v}_1' & 0 & -g_R v_{1R} \\
\frac{g_L}{\sqrt2}v_2 & -\frac{g_R}{\sqrt2}\tilde{v}_2' & 0 & \tilde{\mu_3} & 0 & 0 \\
\frac{g_L}{\sqrt2}\tilde{v}_1' & -\frac{g_R}{\sqrt2}v_1 & \tilde{\mu}_3 & 0 & 0 & 0 \\
g_L v_{2L} & 0 & 0 & 0 & \tilde{\mu}_L & 0 \\
0 & g_R v_{2R} & 0 & 0 & 0 & \tilde{\mu}_R
\end{pmatrix}
\nonumber
\eea
To get the numerical spectrum of one of the scenarios of our phenomenological study presented in chapter \ref{chap:lrsusy}, one issues in the {\mk} session the command 
\begin{verbatim} WriteASperGe[lag, Output->LRSUSY] \end{verbatim}
edits the file \verb?LRSUSY/inputs/externals.dat? created automatically by {\fr} so that the input parameters are in agreement with those of the chosen scenario, compiles and runs
\begin{verbatim} ./ASperGe inputs/externals.dat output/out.dat \end{verbatim}
in a {\sc Shell}. The output of the computation, written in the file \verb?output/out.dat? can be further imported in the {\fr} session with the command 
\begin{verbatim} ReadLHAFile[Input->"LRSUSY/output/out.dat"] \end{verbatim}
in order, for example, to create an implementation of the model in the various {\sc MC}-tools interfaced with {\fr}.\\

In the case of the benchmark scenario SI.1 chosen in the previous chapter, for which we remind the setup in table \ref{tab: remember si1}, the output from {\sc ASperGe} is in agreement with the values presented figure \ref{fig:pure}.

\section{Future developments}\label{sec:spec gen future}
In this chapter, we have presented two packages that are running independently. The first one, focusing only on supersymmetric theories, provides all the necessary routines to extract analytically the two-loop renormalization group equations for any renormalizable supersymmetric theory implemented in {\fr}. The second one, extracts analytically the tree-level mass matrices for any lagrangian-based quantum field theory (be it supersymmetric or not) and coupled to the {\asperge} package provides a powerful tool to diagonalize numerically these mass matrices and use these results for further phenomenological investigations. \\

The final step before obtaining a fully working spectrum generator generator for supersymmetric theories is to extend the mass matrices extraction routine so that it generates them at the one-loop level and finally write a {\sc C++} routine able to run the RGEs, diagonalize the mass matrices, loop so that the spectrum is as accurate as possible and finally generate the output. We also plan to develop a routine in order to minimize the scalar potential and to be able to determine whether the spectrum is acceptable or not. The algorithm describing this can be found in the {\sc SuSpect}\cite{Djouadi:2002ze} or {\sc SoftSusy}\cite{Allanach:2001kg} manuals for example and is put in fig.\ref{fig:algo spec gen} for completeness.\\


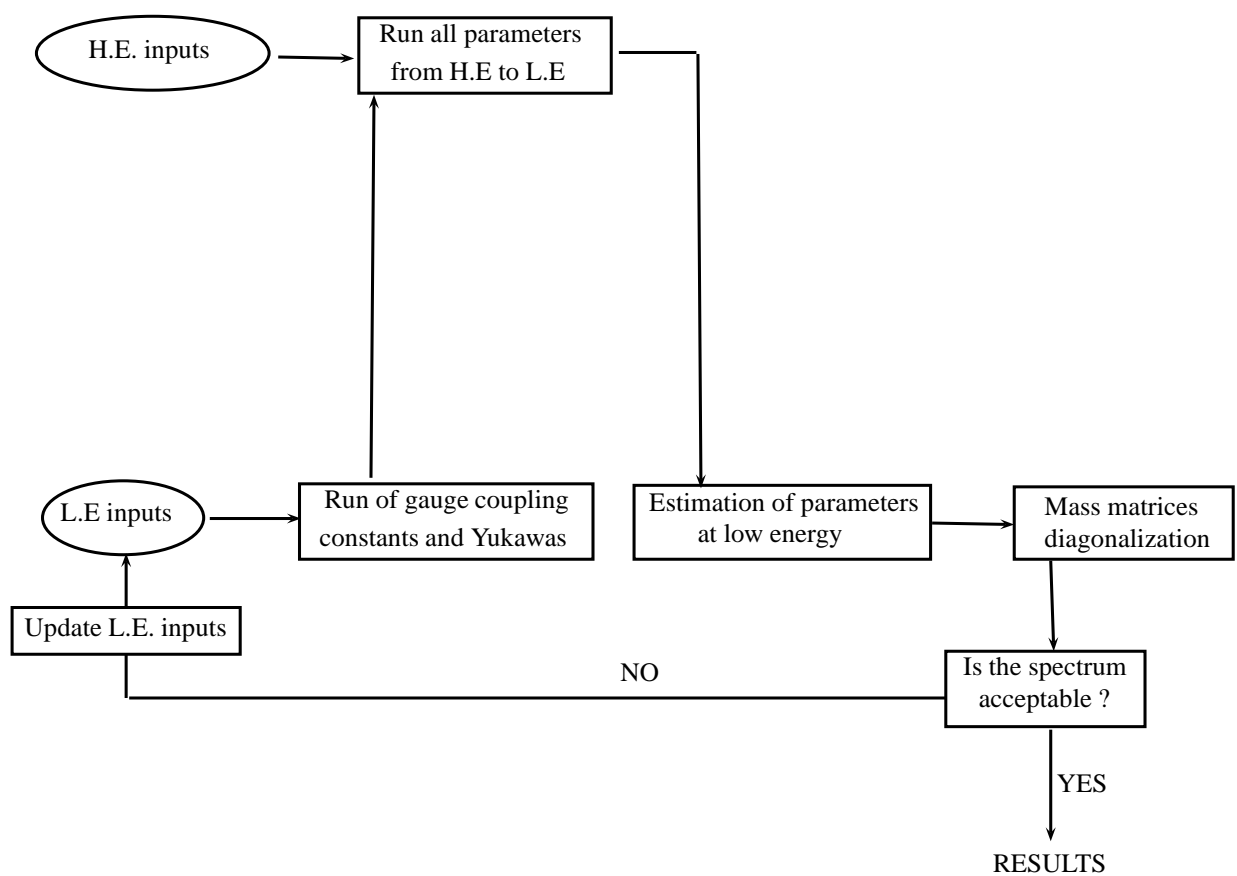
\begin{figure}
\scalebox{1} 
{
\begin{pspicture}(0,-6.3228126)(18.34,6.3028126)
\psframe[linewidth=0.04,dimen=outer](7.68,-1.1171875)(3.78,-2.1571875)
\psframe[linewidth=0.04,dimen=outer](7.92,5.0628123)(4.56,4.0228124)
\psframe[linewidth=0.04,dimen=outer](12.12,-1.1371875)(8.18,-2.1571875)
\psframe[linewidth=0.04,dimen=outer](16.1,-1.1571875)(13.18,-2.1571875)
\psellipse[linewidth=0.04,dimen=outer](1.48,-1.5871875)(1.08,0.51)
\psellipse[linewidth=0.04,dimen=outer](1.87,4.5728126)(1.55,0.51)
\psline[linewidth=0.04cm,arrowsize=0.05291667cm 2.0,arrowlength=1.4,arrowinset=0.4]{->}(2.62,-1.5971875)(3.82,-1.5971875)
\psline[linewidth=0.04cm,arrowsize=0.05291667cm 2.0,arrowlength=1.4,arrowinset=0.4]{->}(4.74,-1.0571876)(4.78,4.0228124)
\psline[linewidth=0.04cm,arrowsize=0.05291667cm 2.0,arrowlength=1.4,arrowinset=0.4]{->}(3.52,4.5228124)(4.52,4.5028124)
\psline[linewidth=0.04cm](8.0,4.5828123)(9.02,4.5828123)
\psline[linewidth=0.04cm,arrowsize=0.05291667cm 2.0,arrowlength=1.4,arrowinset=0.4]{->}(9.04,4.5628123)(9.08,-1.1971875)
\psline[linewidth=0.04cm,arrowsize=0.05291667cm 2.0,arrowlength=1.4,arrowinset=0.4]{->}(12.12,-1.6571875)(13.2,-1.6771874)
\psframe[linewidth=0.04,dimen=outer](14.94,-3.3371875)(12.28,-4.3771877)
\psline[linewidth=0.04cm,arrowsize=0.05291667cm 2.0,arrowlength=1.4,arrowinset=0.4]{->}(13.68,-2.1571875)(13.72,-3.3571875)
\psline[linewidth=0.04cm,arrowsize=0.05291667cm 2.0,arrowlength=1.4,arrowinset=0.4]{->}(13.68,-4.3971877)(13.68,-5.8771877)
\psline[linewidth=0.04cm](12.3,-3.9771874)(1.56,-3.9771874)
\psline[linewidth=0.04cm,arrowsize=0.05291667cm 2.0,arrowlength=1.4,arrowinset=0.4]{->}(1.52,-3.9771874)(1.52,-2.0571876)
\psframe[linewidth=0.04,dimen=outer,fillstyle=solid](3.04,-2.7571876)(0.0,-3.4171875)
\usefont{T1}{ptm}{m}{n}
\rput(1.512754,-3.0721874){Update L.E. inputs}
\usefont{T1}{ptm}{m}{n}
\rput(1.3901172,-1.5521874){L.E inputs}
\usefont{T1}{ptm}{m}{n}
\rput(5.7336717,-1.3721875){Run of gauge coupling}
\usefont{T1}{ptm}{m}{n}
\rput(5.687568,-1.8521875){constants and Yukawas}
\usefont{T1}{ptm}{m}{n}
\rput(1.8206445,4.5878124){H.E. inputs}
\usefont{T1}{ptm}{m}{n}
\rput(6.1806445,4.8078127){Run all parameters}
\usefont{T1}{ptm}{m}{n}
\rput(6.148877,4.3278127){from H.E to L.E}
\usefont{T1}{ptm}{m}{n}
\rput(10.17003,-1.4121875){Estimation of parameters}
\usefont{T1}{ptm}{m}{n}
\rput(9.94253,-1.8321875){ at low energy}
\usefont{T1}{ptm}{m}{n}
\rput(14.610732,-1.4321876){Mass matrices}
\usefont{T1}{ptm}{m}{n}
\rput(14.683232,-1.8921875){diagonalization}
\usefont{T1}{ptm}{m}{n}
\rput(13.644736,-3.5921874){Is the spectrum }
\usefont{T1}{ptm}{m}{n}
\rput(14.08877,-5.1121874){YES}
\usefont{T1}{ptm}{m}{n}
\rput(8.27542,-3.6321876){NO}
\usefont{T1}{ptm}{m}{n}
\rput(13.667363,-6.1721873){RESULTS}
\psline[linewidth=0.04cm,arrowsize=0.05291667cm 2.0,arrowlength=1.4,arrowinset=0.4]{->}(18.18,6.2828126)(18.32,-2.8571875)
\usefont{T1}{ptm}{m}{n}
\rput(13.59627,-4.0121875){acceptable ?}
\end{pspicture} 
}
\label{fig:algo spec gen}\caption{\footnotesize Algorithm explaining simply how to get a reasonable spectrum by combining RGEs runnings and mass matrices diagonalization. The main idea being to start from a set of input at both low energy (L.E.) and high energy (H.E.) and to first run the RGEs from L.E. to H.E., apply the H.E. input values and then run-back to L.E. to diagonalize the mass matrices. If the spectrum is acceptable ({\ie} no tachyons and check for electroweak symmetry breaking for example) the result is generated otherwise the input values at L.E. are re-calculated and the calculations start over}.
\end{figure}

\chapter{ Doubly charged  particles}\label{chap: doubly charged}

\section{Motivations}
At the Large Hadron Collider, searching for signals of new physics is for sure not a simple task, the background from Standard Model being very important especially in regions where the multiplicity of leptons is low. In addition to this, signals from ``new physics" processes can involve very low cross sections compared to those of the Standard Model. One is thus pushed to seek for configurations where the Standard Model predicts a low number of events to maximize the visibility of the deviation or for signatures that are not predicted by the Standard Model at all. In chapter \ref{chap:lrsusy} for example, we have concluded that the multilepton channel is certainly one of the best avenues to observe signals from a left-right symmetric supersymmetric model. We have also mentionned that a certain class of these models, where left-right symmetry is broken by Higgs fields belonging to the adjoint representation of $SU(2)_R$, predicts the existence of doubly-charged particles. The latter being absent from the field content of the Standard Model, their discovery would definitely point toward an extension of the Standard Model. But which one?\\

Indeed, these particles do not appear exclusively in left-right symmetric models but are also predicted in models like Little Higs models\cite{ArkaniHamed:2002qy,Lee:2005kd,Han:2005nk} or extra-dimensions\cite{Chen:2009gy,Csaki:2008qq,Kadosh:2010rm,delAguila:2010vg,delAguila:2010es}. The main difference between these different cases resides then in the representations to which belong the doubly-charged particles\footnote{Here we put apart all the other properties that might hint toward a particular model.}. For example, in the case of left-right symmetry the doubly-charged particles are scalar fields transforming as triplets under either $SU(2)_L$ or $SU(2)_R$; while in its supersymmetric version these doubly-charged scalar fields come together with their fermionic superpartners. Some other theories predict vector multiplets, {\ie} fields carrying a spin equal to one, where one of the components has a two-unit electric charge\footnote{The doubly-charged field can carry a higher spin and/or belong to a representation of $SU(2)$ other than $\singlet,\doublet \et \triplet$.}. It is thus easy to imagine that if a doubly-charged particle was to be discovered at the Large Hadron Collider, though this would dismiss many beyond the Standard Model theories, many others would survive. To compare with today's situation, the discovery of a Standard Model Higgs-like scalar particle by both the CMS and the ATLAS collaborations and the measurement of some of its couplings do not bring a clear answer on whether this is the Standard Model Higgs boson or not but it does constrain the parameter space of the various TeV scale models\footnote{In section \ref{sec:mssm}, we briefly discussed the consequences of such discovery on the MSSM's parameter space.}.\\

In the case of doubly-charged particles, a certain number of studies have already focused on the phenomenology these particles would induce if they were produced at the LHC (see for example \cite{Rentala:2011mr,Cuypers:1996ia,Meirose:2011cs}). These studies have however always been carried in the context of a precise model and thus do not bring any answer on how one could discriminate between various models. On the experimental side, both CMS and ATLAS experiments have led searches on these exotic particles. \\

The CMS collaboration focused only on the cases where the doubly-charged field is a fermion or a scalar. In both cases they find no excess over the Standard Model background and they use their data to deduce lower bounds. For the first case, {\ie} a doubly-charged fermion field, searches have been carried with both runs of 2011 at a center-of-mass energy of 7 TeV and a luminosity of 5 $fb^{-1}$ and 2012 one with a center-of-mass energy of 8 TeV and an integrated luminosity of 18.8 $fb^{-1}$. Assuming these particles to only couple to the photon and the Z-boson through $U(1)$ couplings and to have low decay rates (long-lived particles), they excluded such fermions up to 685 GeV \cite{Chatrchyan:2013oca}. As to the scalar fields, often called doubly-charged Higgs fields, searches were conducted assuming these particles to only decay into a pair of charged leptons and led them to exclude masses up to 453 GeV in the most pessimistic case\cite{Chatrchyan:2012ya}.\\ 

As to the ATLAS collaboration similar searches have been conducted. In the case of massive fermions\cite{Aad:2013pqd}, the exclusion limit, based on a 7 TeV center-of-mass energy run and with 4.4 $fb^{-1}$ of data, is set to 430 GeV while scalar fields are excluded up to 409 GeV. The latter exclusion limit is based on $4.7 fb^{-1}$ of data\cite{ATLAS:2012hi}. \\

The main message to keep from the above considerations is that these doubly-charged particles are far from being excluded yet, especially when they are promptly decaying particles and/or have different branching ratios than what was hypothesized. Moreover no study provided the key observables to analyze in order to discriminate between the various models predicting a doubly-charged particle. This chapter is exactly intended to this aim.\\

In this analysis, we adopt a model-independent approach extending the field content of the Standard Model minimally to include only a doubly-charged particle and its possible isospin partners. Considering scalar, fermion and vector fields lying in either the trivial, fundamental or adjoint representations of $SU(2)_L$, we construct effective Lagrangians with the only constraint to have gauge invariant interactions but keeping the number of new parameters minimum. We then compute analytically the various cross sections associated to the production of these new particles at the LHC running at a center-of-mass energy of 8 TeV together with the analytical expressions for their decay widths (whenever possible). Focusing only on final states with at least three charged leptons\footnote{In all this chapter ``lepton" will refer only to the charged light leptons, {\ie} the electron and the muon.} (this is in order to reduce the background from Standard Model), we deduce from the analytical formulas the masses above which such final states would be associated to cross sections smaller than $1 fb$, {\ie} hardly visible with a luminosity of 20 $fb^{-1}$. Finally, from these bounds we choose three different values for the mass of the new state\footnote{In this study, we consider all the components of the new multiplet to have the same mass so that they only decay into SM particles.} in each representation and perform a full Monte Carlo simulation in order to find the key observables to differentiate the various states.\\

The chapter is organized as follows. In the next three sections we construct our effective models, distinguishing between scalar, fermionic and vector fields and derive the analytic formulas for both the cross sections and decay widths. In these sections, we shall also present a numerical analysis based on the analytic formulas and derive the bounds on the masses for the signal to be observed at the LHC. In section \ref{sec: dc pheno}, we proceed to the Monte Carlo simulation and present some kinematic observables that can help us distinguishing between the various representations. Finally, concluding remarks are presented in section \ref{sec: dc conclusion}.\\

Before proceeding, we refer the reader to the chapter \ref{chap: SM} for the notations and conventions related to Standard Model's quantities and to the appendix \ref{annex: vertices} for the definition of the Feynman rules related to the vertices, propagators and more generally for all the useful definitions to carry out properly the computation of the cross-sections below. Finally, we have used a semi-automatic approach for the computation of the cross sections in the sense that Dirac matrices algebra that is needed to perform the latter calculations have been performed by the package {\sc FeynCalc}\cite{Mertig:1990an}.

\section{Scalar fields}\label{sec: db scalars}
Doubly-charged scalar fields appear in various models of particle physics and may be motivated by the seesaw mechanism for example. In this case they appear as components of a multiplet belonging to the representation $\triplet$ of $SU(2)_L$\cite{Arhrib:2012vp,Zee:1980ai,Cheng:1980qt,Gelmini:1980re}. The left-right symmetric models (either supersymmetric or not) are a good example as they predict scalar triplet fields for both $SU(2)_L \et SU(2)_R$. In the latter case, {\ie} when the doubly-charged scalar field belongs to an $SU(2)_R$ triplet, it transforms as a singlet under $SU(2)_L$. Doubly-charged scalar particles can also appear in Composite Higgs models like in the Littlest Higgs model introduced in \cite{ArkaniHamed:2002qy,Lee:2005kd,Han:2005nk}. In the latter, the Higgs scalar fields are pseudo-Goldstone bosons arising from the condensation of fermionic fields\cite{ArkaniHamed:2001ca}. Finally, a scalar doublet with hypercharge $3/2$ and thus a doubly-charged component can also be invoked to explain the smallness of neutrino masses \cite{Aoki:2011yk}.\\

In our simplified model, to take into account these different possibilities, we add to the Standard Model particle content three complex scalar fields $\phi$, $\Phi$ and $\pPhi$ belonging to the representation $\singlet, \doublet \et \triplet$ of $SU(2)_L$ respectively. We also choose their hypercharge so that the doubly-charged particle has the highest electric charge. Using the matrix representation for the triplet field (defined in eq.\ref{doubly charged}), we have:
$$ \phi \equiv \phi^{++},~~~~ \Phi^i = \begin{pmatrix} \Phi^{++} \\ \Phi^{+} \end{pmatrix}, ~~~~ \pPhi^i{}_j = \begin{pmatrix} \frac{\pPhi^+}{\sqrt2} & \pPhi^{++} \\ \pPhi^0 & - \frac{\pPhi^+}{\sqrt2} \end{pmatrix} $$
with the hypercharges set to
$$ Y_\phi = 2, ~~~~ Y_\Phi = \frac32, ~~~~ Y_{\pPhi} = 1$$
and the superscripts ``0", ``+" and ``++" stand for the electric charge of the various fields. The Lagrangian of the Standard model is then supplemented with kinetic terms for these new fields
\bea \label{eq: db kin scal} \lag_{kin} = D_\mu\phidagger D^\mu \phi + D_\mu \Phi^\dagger_i D^\mu \Phi^i + D_\mu \pPhi^\dagger_a D^\mu\pPhi^a + \dots, \eea
where covariant derivatives, dictated by gauge invariance, read simply
\bea \label{eq: db cov der}
	D_\mu \phi &=& \delm \phi - 2ig'B_\mu \phi,\\
	D_\mu \Phi^i &=& \delm \phi^i - i\frac32 g'B_\mu \Phi^i - i g \frac{(\sigma^a)^i{}_j}{2} \Phi^j W^a_\mu ,\\
	D_\mu \pPhi^a &=& \delm \pPhi^a - i g' B_\mu \pPhi^a + g \epsilon_{ab}{}^c \pPhi^c W_\mu^b .
\eea
The dots in the equation above stand for mass terms for these new fields. As a convention, we will use the letters from the beginning of the alphabet ($a,b,c$) to represent indices in the adjoint representation of $SU(2)_L$ while the letters from the middle of the alphabet ($i,j,k$) will stand for indices in the fundamental representation of the latter. \\

Due to their peculiar electric charge and keeping the extension of the Standard Model minimal, doubly-charged scalar fields are not allowed to decay into quarks. Setting $L$ ($l_R$, resp.) to be a left- (right-) handed Dirac field\footnote{Left- (right-) handed Dirac fermion stands for a four-component Dirac field whose left- (right-) handed part is zero. As indicated in chapter \ref{chap: SM}, this notation is particularly suited for cross section calculations. }, the Yukawa Lagrangian reads
\bea
\label{eq: db yuk scal}
	\lagr_{yuk} = \frac12 y^{(1)} \phi \bar{l}_R^c l_R + \frac{y^{(2)}}{\Lambda} \Phi^i \bar{L}^c_i \gamma_\mu D^\mu l^i_R + \frac12 y^{(3)} \pPhi^i{}_j \bar{L}^c_i L^j + {\rm h.c.} 
\eea
where the coupling $(y^{(i)})_{\big|i=1,2,3}$ is a $3 \times 3$ matrix in generation space. For clarity we have omitted flavor indices but let both $SU(2)_L$ and chirality indices. The superscript $c$ stands for charge conjugation. Finally, the four-component spinor product $\xi_R^c \lambda_L $ being equal to zero for any fermionic field, we have used a higher-dimensional operator suppressed by an effective scale $\Lambda$ for the yukawa coupling $y^{(2)} $of the doublet scalar field $\Phi$.

\subsection{Production cross-sections}
We proceed now to the computation of the cross-sections associated to the production of our new doubly-charged states. Starting from the Lagrangian in equations \eqref{eq: db kin scal} and \eqref{eq: db yuk scal}, we will restrain ourselves to the processes which will eventually lead to at least three leptons in the final state. Before proceeding, let us define some useful quantities:
\begin{itemize}
	\item $s_W \et c_W$ will denote the sine and the cosine of the Weinberg angle, respectively
	\item $\Gamma_p \et M_p$ are the total decay width and the mass associated to the particle $p$
	\item $\sh$ is the partonic center of mass energy, $x_p^2 = \frac{M_p^2}{\sh}$ and $\sh_p = \sh - M_p^2 + i\Gamma_p M_p $ the reduced kinematical variables associated to the particle $p$. 
	\item the electric charge $e_q$, the weak isospin number $T_{3q}$ and the coupling strengths to the $Z$-boson $L_q = 2 (T_{3q} - e_q s_W^2) \et R_q = -2e_q s_W^2$ associated to the quark $q$
	\item $CC \et NC$ will denote the charged and neutral currents, respectively.
	\item The K\"allen function $\lambda(x,y,z) = x^2 + y^2 + z^2 - 2xy -2xz - 2yz. $
\end{itemize}
Finally, the use of the numbers $1, 2 \et 3$  as subscripts will help to distinguish the quantities associated to the singlet, the doublet and the triplet cases.\\

From equation \eqref{eq: db kin scal}, we deduce that the multiplets can be produced either via neutral currents, that is a $Z$ boson or a photon decaying into a pair of doubly-charged particles, or charged currents in the case where the multiplet is a doublet or a triplet of $SU(2)_L$. In the latter case, the signal corresponds to the decay of a $W^\pm$ boson into a doubly-charged particle and its singly-charged isospin partner. The cross sections associated to the neutral current processes are
\be\bsp
 \frac{\d\sigma^{NC}_1}{\d t} = &\ 
    \frac{4 \pi \alpha^2 \sh}{9} \big[ 1 - 4x^2_{\phi^{++}}\big]^{\frac32} \bigg[
    \frac{e^2_q}{\sh^2} -
    \frac{ e_q (L_q+R_q) (\sh-M_Z^2)}{2 c_W^2 \sh |\sh_Z|^2} +
    \frac{L_q^2 + R_q^2}{8 c_W^4 |\sh_Z|^2}
  \bigg]  \ , \\
 \frac{\d\sigma^{NC}_2}{\d t} = &\ 
    \frac{4 \pi \alpha^2 \sh}{9} \big[ 1 - 4x^2_{\Phi^{++}}\big]^{\frac32} \bigg[  
    \frac{e^2_q}{\sh^2} +
    \frac{ e_q (1-4s_W^2) (L_q+R_q) (\sh-M_Z^2)}{8 c_W^2 s_W^2 \sh |\sh_Z|^2}+
    \frac{(1-4s_W^2)^2 (L_q^2 + R_q^2)}{128 c_W^4 s_W^4 |\sh_Z|^2}
  \bigg]  \ , \\
  \frac{\d\sigma^{NC}_3}{\d t} = &\
    \frac{4 \pi \alpha^2 \sh}{9} \big[ 1 - 4x^2_{\mathbf{\Phi}^{++}}\big]^{\frac32} \bigg[  
    \frac{e^2_q}{\sh^2} +
    \frac{ e_q (1-2s_W^2) (L_q+R_q) (\sh-M_Z^2)}{4 c_W^2 s_W^2 \sh |\sh_Z|^2} +
    \frac{(1-2s_W^2)^2 (L_q^2 + R_q^2)}{32 c_W^4 s_W^4 |\sh_Z|^2}
  \bigg]  \ .
\esp\label{eq:xsecscNC}\ee
Those associated with the charged current processes are
\be\bsp
 \frac{\d\sigma^{CC}_2}{\d t} = &\
   \frac{\pi \alpha^2 \sh}{72 s_W^4 |\sh_W|^2}  |V_{ij}^{\rm CKM}|^2\ \lambda^{\frac32}(1,x^2_{\Phi^{++}},x^2_{\Phi^+}) \ , \\
 \frac{\d\sigma^{CC}_3}{\d t} = &\
   \frac{\pi \alpha^2 \sh}{36 s_W^4 |\sh_W|^2}  |V_{ij}^{\rm CKM}|^2\ \lambda^{\frac32}(1,x^2_{\Phi^{++}},x^2_{\Phi^+}) \ ,
\esp\label{eq:xsecscCC}\ee
where $|V^{CKM}_{ij}|^2$ is the square of the module of element $i,j$ of the CKM matrix, as defined in chapter \ref{chap: SM}. These processes are of interest to our study as they lead to final states with at least three leptons, the doubly-charged scalar decaying into a pair of charged leptons and the singly charged into one charged lepton as can be deduced from the Yukawa Lagrangian \eqref{eq: db yuk scal}. In the case of the singlet and the doublet fields, these processes are the only ones that lead to such final states but in the case of the triplet field other production mechanisms and decay channels open when the neutral component acquires a vacuum expectation value ($v_\pPhi$). Indeed from the kinetic terms in equation \eqref{eq: db kin scal}, it is easy to see from the term
$$ D_\mu \pPhi^\dagger_a D^\mu\pPhi^a $$
that the vertices illustrated in figure \ref{fig: feyn diag trip} become possible when the neutral component $\pPhi^0$ acquires a vacuum expectation value 
$$ \pPhi^0 \rightarrow \frac{1}{\sqrt2} \big[ v_{\pPhi} + \mathbf{H}^0 + i \mathbf{A}^0 \big]$$
where $ \mathbf{H}^0 \et \mathbf{A}^0$ are its scalar and pseudo-scalar parts, respectively. These interactions may lead to final states with three charged leptons or more, after accounting for the leptonic decays of both the $Z$ and $W$ gauge bosons. The cross sections associated with the production of the charged components of the triplet scalar field together with a weak gauge boson are given by the following analytic formulas
\be\bsp
 \frac{\d\sigma^{W\mathbf\Phi^{++}}}{\d t} = &\ 
   \frac{\pi^2 \alpha^3 v^2_{\mathbf\Phi}}{18 s_W^6 \sh_W^2}  |V_{ij}^{\rm CKM}|^2\ \lambda^\frac12 (1,x^2_{\Phi^{++}},x^2_W)
      \bigg[ \frac{\lambda(1,x^2_{\Phi^{++}},x^2_W)}{x_W^2} + 12 \bigg]\ , \\ 
 \frac{\d\sigma^{Z\mathbf\Phi^+}}{\d t} = &\
   \frac{\pi^2 \alpha^3 v^2_{\mathbf\Phi}(1+s_W^2)^2}{72 s_W^6 c_W^2  \sh_W^2}  |V_{ij}^{\rm CKM}|^2\ \lambda^\frac12 (1,x^2_{\Phi^{++}},x^2_Z)
      \bigg[ \frac{\lambda(1,x^2_{\Phi^{++}},x^2_Z)}{x_Z^2} + 12 \bigg]\ .
\esp\label{eq:xsecscvev}\ee

\unitlength = 1mm
\begin{figure}
\begin{center}
\begin{fmffile}{dbchrgd1}
\begin{fmfgraph*}(30,30)
    \fmfleft{i1}
    \fmflabel{$\pPhi^{++}$}{i1}
	\fmfright{o1,o2}
	\fmflabel{$W^+_\mu$}{o1}
	\fmflabel{$W^+_\nu$}{o2}
	\fmf{dashes}{i1,v1}
    \fmflabel{$i\sqrt2\frac{e^2 v_{\pPhi}}{s^2_W} \eta^{\mu\nu}$}{v1}
    \fmf{photon}{o1,v1,o2}
\end{fmfgraph*}
\hskip3cm
\begin{fmfgraph*}(30,30)
    \fmfleft{i1}
    \fmflabel{$\pPhi^{+}$}{i1}
	\fmfright{o1,o2}
	\fmflabel{$W^+_\mu$}{o1}
	\fmflabel{$Z_\nu$}{o2}
	\fmf{dashes}{i1,v1}
    \fmflabel{$-i \frac{e^2 v_{\pPhi}(1+s_W^2)}{\sqrt2 s^2_Wc_W} \eta^{\mu\nu}$}{v1}
    \fmf{photon}{o1,v1,o2}
\end{fmfgraph*}

\end{fmffile}
\end{center}
\caption{\footnotesize \label{fig: feyn diag trip} Feynman diagram illustrating the interaction between the triplet scalar field and the weak gauge bosons $W \et Z$ after the neutral component $\pPhi^0$ acquired a vacuum expectation value.}
\end{figure}
Finally, we also consider the processes leading to the production of the CP even neutral scalar field $H^0$ as it can decay to final states with up to four charged light leptons through its couplings to the weak gauge bosons. The production processes are hence
$$ p~p \to H^0 ~ \mathbf\Phi^+ ,~~~ p~p \to H^0 ~ H^0 ~ Z, ~~~p~p \to H^0 ~ W^\pm $$
and the corresponding cross sections are given by the analytic formulas:
\be\bsp
    \frac{\d\sigma^{H^0\mathbf\Phi^{+}}}{\d t} = &\ |V_{ij}^{\rm CKM}|^2 \frac{\pi \alpha^2 \lambda^{3/2}(1,x_{\mathbf{\Phi^{+}}}^2,x_{\mathbf{H^{0}}}^2)}{72 s_W^4 |\hat{x}_Z|^2}\ ,\\
    \frac{\d\sigma^{H^0 Z}}{\d t} = &\ \frac{\pi^2 \alpha^3 v_{\pPhi}^2}{9c_W^4 s_W^6 \hat{s}|\hat{s}_Z|^2} \lambda^{\frac12}(1,x_{\mathbf{H}^0}^2,x_Z)^2 \Big[\frac{\lambda(1,x_{\mathbf{H}^0}^2,x_Z)}{x_Z^2} +12 \Big]\ ,\\
 \frac{\d\sigma^{W\mathbf H^{0}}}{\d t} = &\ \frac{\pi^2 \alpha^3 v^2_{\mathbf\Phi}}{36 s_W^6 \sh_W^2}  |V_{ij}^{\rm CKM}|^2\ \lambda^\frac12 (1,x^2_{\mathbf{H}^{0}},x^2_W)\bigg[ \frac{\lambda(1,x^2_{\mathbf{H}^{0}},x^2_W)}{x_W^2} + 12 \bigg]\ .
\esp\label{eq: db xsec trip neu}\ee

We draw the reader's attention to the fact that even if the vacuum expectation value $v_\pPhi$ is constrained to be small by the electroweak $\rho$-parameter as well as neutrinos masses; we, motivated by the case of left-right symmetric models where there is no real constraint on the value of the equivalent vev $v_L$ (see section \ref{sec:seesaw} of the previous chapter), decided to not constrain its value.

\subsection{Partial decay widths}
We now turn to the calculation of the partial decay widths of our new charged states. According to the Lagrangian of eq. \eqref{eq: db yuk scal}, they all may decay to a pair of same sign leptons with the associated decay rates
\be\bsp
  \Gamma_{1,\ell}^{++} = &\ \frac{M_{\phi^{++}}|y^{(1)}|^2}{32 \pi} \Big[1-2 x_\ell^2\Big] \sqrt{1-4 x_\ell^2}\ , \\
  \Gamma_{2,\ell}^{++} = &\ \frac{M_{\Phi^{++}} M_\ell^2 |y^{(2)}|^2}{8 \pi \Lambda^2} \Big[ 1-2 x_\ell^2\Big] \sqrt{1-4 x_\ell^2} \ , \\
  \Gamma_{3,\ell}^{++} = &\ \frac{M_{\mathbf{\Phi}^{++}} |y^{(3)}|^2}{32 \pi} \Big[1 - 2 x_\ell^2\Big] \sqrt{1 -4 x_\ell^2}\ .
\esp\label{eq:BRsc++1}\ee
The doubly-charged component of the triplet field, as seen above, can furthermore decay into a pair of $W$ gauge bosons with the width given by
\be\label{eq:BRsc++2}
  \Gamma_{3,WW}^{++} =  
    \frac{M^3_{\mathbf{\Phi}^{++}} \alpha^2 \pi v_{\mathbf{\Phi}}^2}{4 M_W^4 s_W^4} \sqrt{1 - 4 x_W^2} \Big[1 - 4 x_W^2 + 12 x_W^4\Big] \ .
\ee
The decays of the singly charged components of both the doublet and the triplet scalar fields are governed by the formulas below:
\be\bsp
  \Gamma_{2,\ell}^+ = &\ \frac{M_{\Phi^+} M_\ell^2 |y^{(2)}|^2}{16 \pi \Lambda ^2} \Big[1-x_\ell^2\Big]^2\ , \\
  \Gamma_{3,\ell}^+ = &\ \frac{M_{\mathbf{\Phi}^+} |y^{(3)}|^2}{32 \pi} \Big[1 - x_\ell^2\Big]^2\ , \\
  \Gamma_{3,WZ}^+ = &\ \frac{ M^3_{\mathbf{\Phi}^+} \alpha^2\pi v_{\mathbf{\Phi}}^2 (1 + s_W^2)^2}{8  c_W^2 s_W^4 M_Z^2 M_W^2 }
    \Big[\lambda(1,x^2_W,x^2_Z) + 12 x_Z^2 x_W^2\Big]  \sqrt{\lambda(1,x^2_W,x^2_Z)} \ .
\esp\label{eq:BRsc+1} \ee
Finally, the neutral component of the triplet multiplet is allowed to decay into a pair of weak gauge bosons of a pair of neutrinos and the corresponding decay widths are:
\be\bsp
  \Gamma_{3,Z Z}^0 = &\ \frac{\pi \alpha^2 M^3_{\mathbf{H}^0} v_{\mathbf{\Phi}}^2}{c_W^2 s_W^4 M^4_W}\sqrt{1-4x_W^2} \Big[1 - 4x_W^2 + 12 x_W^4\Big]\ ,\\
  \Gamma_{3,W W}^0 = &\  \frac{\pi \alpha^2 M^3_{\mathbf{H}^0} v_{\mathbf{\Phi}}^2}{4 s_W^4 M^4_Z}\sqrt{1-4x_Z^2} \Big[1 - 4x_Z^2 + 12 x_Z^4\Big]\ ,\\
  \Gamma_{3,\nu\nu}^{++} = &\ \frac{M_{\mathbf{H}^{0}} |y^{(3)}|^2}{16 \sqrt2 \pi} \ .
\esp\label{eq: db trip decay neutr}\ee

\subsection{Setup for the numerical analysis} Before moving into the analysis of the previous analytic results, let us first set the context.
\begin{itemize}
    \item First, in order to determine the multiplicity of charged leptons in the final state, we account for the decays of both the new states and those of the tau leptons and the weak gauge bosons taken from the Particle Data Group review \cite{Beringer:1900zz}. In the case of the tau lepton, its branching ratio into a charged light lepton $l$, and two neutrinos, denoted $\nu$ generically, is set to 
$$ {\rm BR}(\tau \to l ~2\nu) = 0.3521.$$
The $W$ gauge boson can decay into either an electron, a muon or a tau together with the associated neutrino. After accounting for the subsequent decay of the tau lepton into one of the charged leptons, we find that the branching ratio of the $W$ gauge boson into a charged light lepton is
$$ {\rm BR}(W \to l~\nu) = 0.261307 .$$
Finally, the branching ratio of the $Z$-boson into two light charged leptons is, after accounting for the subsequent decay of the tau lepton into lighter leptons,
$$ {\rm BR}(Z \to l^+ ~ l^-) = 0.073 .$$
In addition to this, one must also include the case where the $Z$-boson decays into a pair of tau leptons and one of them decays subsequently into a lighter lepton while the other one decays hadronically. This process is associated with the branching ratio
$$ {\rm BR}(Z \to \tau^+~\tau^- \to l^\pm ~ X) = 0.0156 $$
where $X$ stands for both hadrons and neutrinos. 
    \item The center-of-mass energy is set to $\sqrt{s}=8 {\rm TeV}$.
    \item Using the QCD factorization theorem, we convolve the partonic cross sections whose analytic expressions are given above with the universal parton densities $f_a \et f_b$ of partons $a \et b$ in order to obtain the hadronic cross sections. These parton densities depend on both the longitudinal momentum fractions of the two partons $x_{a,b} = \sqrt{\tau}e^{\pm y}$ and the unphysical factorization scale $\mu_F$. The hadronic cross sections read then
$$ \int_{4\frac{\tilde{M}^2}{s}}^1 d\tau \int_{-\frac{\ln\tau}{2}}^{\frac{\ln\tau}{2}} dy f_a(x_a,\mu_F^2) f_b(x_b,\mu_F^2) \sigma_p(x_ax_bs)$$
We employ the leading order set {\sc L1} of the {\sc CTEQ6} global parton density fit \cite{Pumplin:2002vw} which includes five light quarks and we identify the factorization scale to be the average mass of the produced final state particles $\tilde{M}$. 
    \item The CKM matrix elements are evaluated using the Wolfenstein parametrization. The corresponding four free parameters are set to 
$$ \lambda = 0.22535,~~~A = 0.811,~~~\bar{\rho} = 0.131,~~~\bar{\eta} = 0.345. $$
    \item Note that this setup will hold true when we discuss the analytic cross sections associated with the production of doubly-charged fermionic or vector fields.
\end{itemize}
Finally, we set all the Yukawa couplings $\big(y^{(i)}\big)_{\big | i=1,2,3}$ to be diagonal matrices with eigenvalues equal to~0.1; the energy scale $\Lambda$ suppressing the effective coupling of the doublet field is set to 1 TeV and the vacuum expectation value of the neutral component of the triplet Higgs field is taken to be equal to~100~GeV (see table \ref{tab: summary dc} for a summary of these values). As already mentionned, we do not allow the new states to decay into each other and consider hence the components of a same multiplet to be mass degenerate.
\begin{table}
\centering
\begin{tabular}{c| c}
\hline
\hline
Parameter & Numerical value \\
\hline
\hline
$y^{(1)},y^{(2)},y^{(3)}$ & $0.1\cdot \mathbbm{1}$\\
$\Lambda$ & 1 TeV \\
$v_\pPhi$ & 100 GeV\\
\hline
\end{tabular}
\caption{\label{tab: summary dc} Numerical values for the new parameters governing the production and the decays of the doubly-charged scalar fields. Here $ \mathbbm{1}$ stands for the $3\times 3$ indentity matrix.}
\end{table}

\subsection{Numerical analysis}
The numerical setup being defined, we can now turn to the analysis itself. Focusing only on final states with at least three light leptons, we want to know beyond which mass the number of events becomes too small or, equivalently, the cross section is lower than 1 fb. In figure \ref{fig: dc scalars} are presented the evolution of the cross sections for the scalar fields lying in the singlet (left panel), doublet (middle panel) and triplet (right panel) representations of $SU(2)_L$. 
\begin{figure}
\begin{tabular}{c c c}
\includegraphics[width=.32\columnwidth]{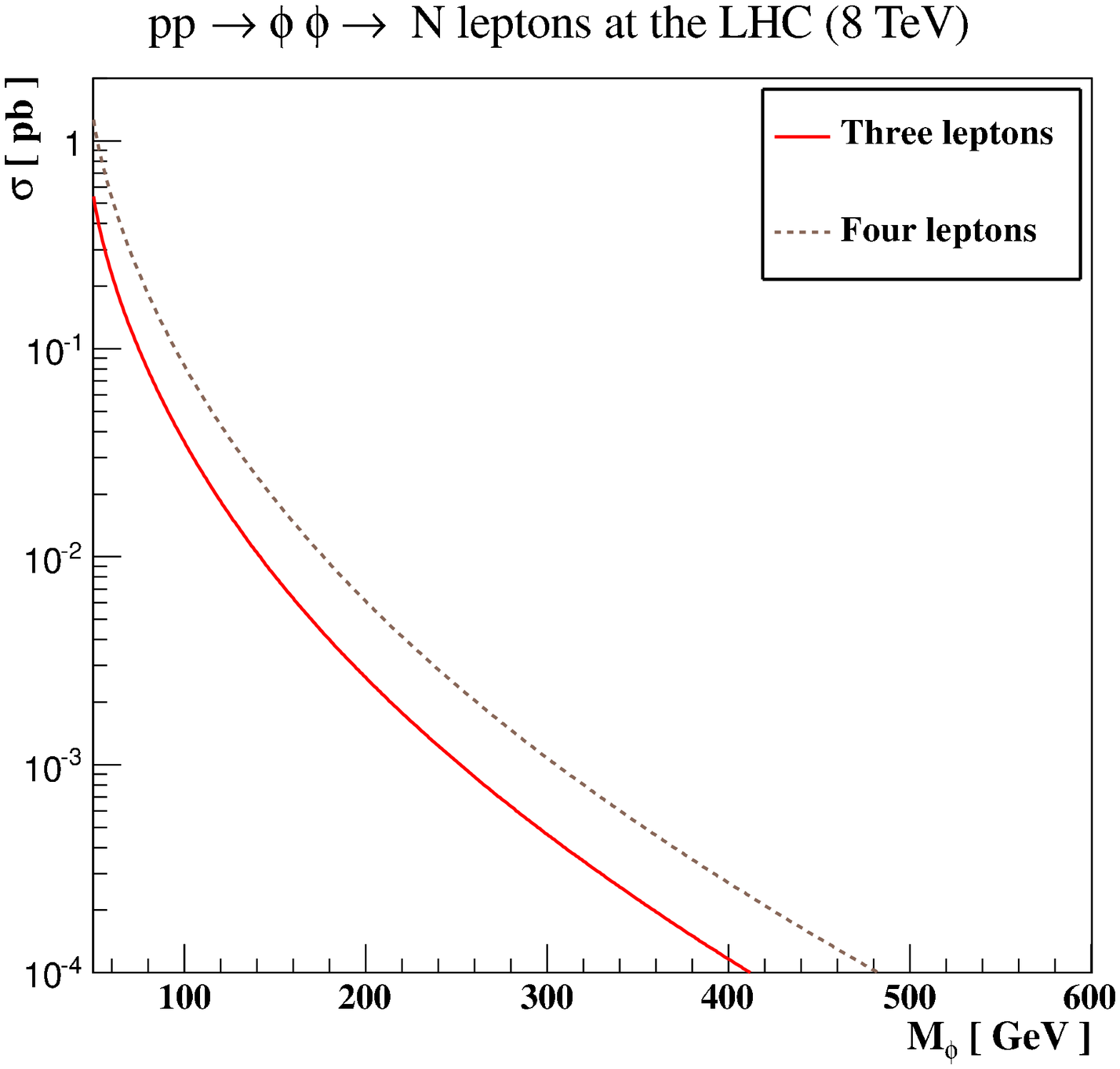} &\includegraphics[width=.32\columnwidth]{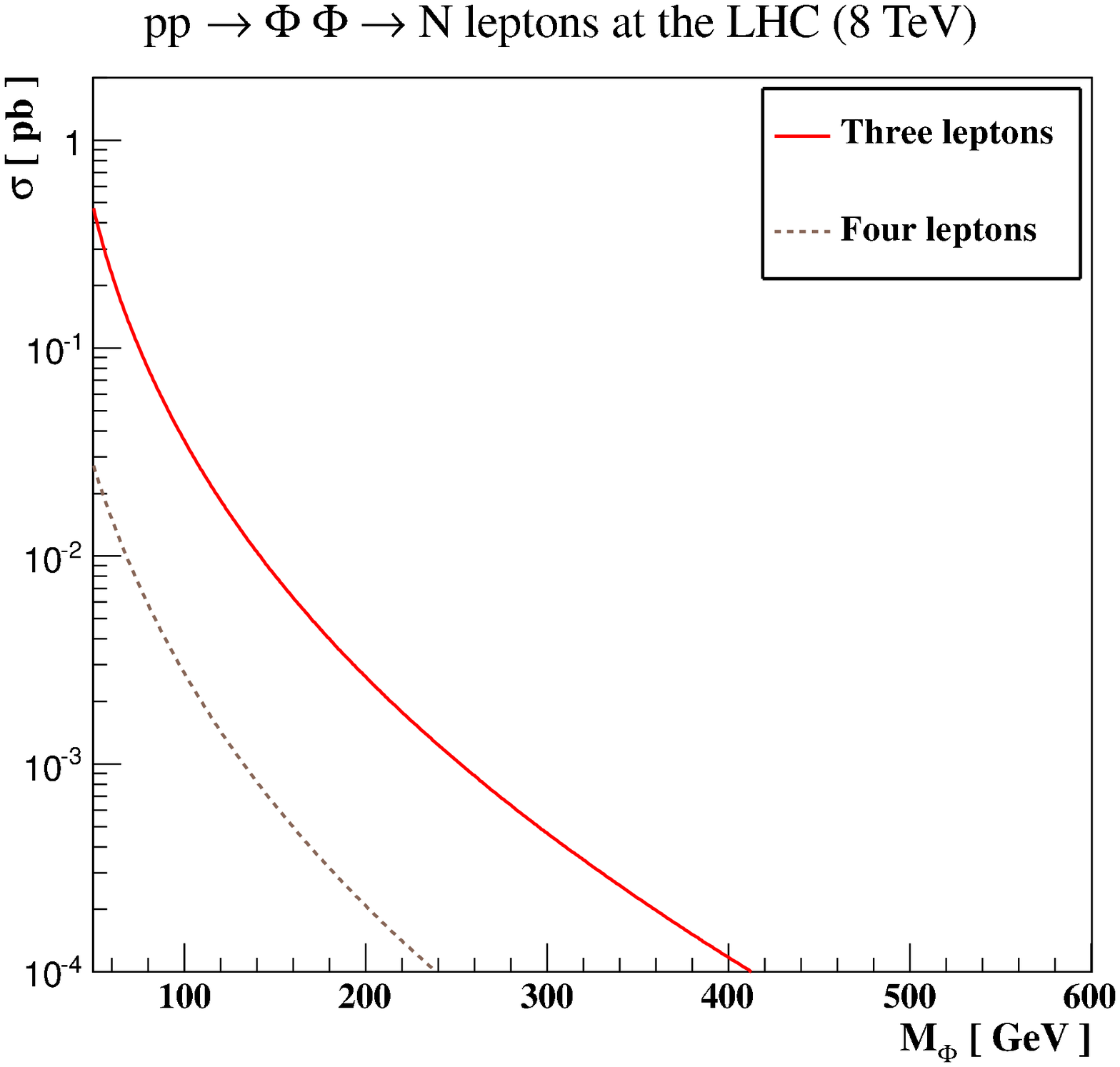} &\includegraphics[width=.32\columnwidth]{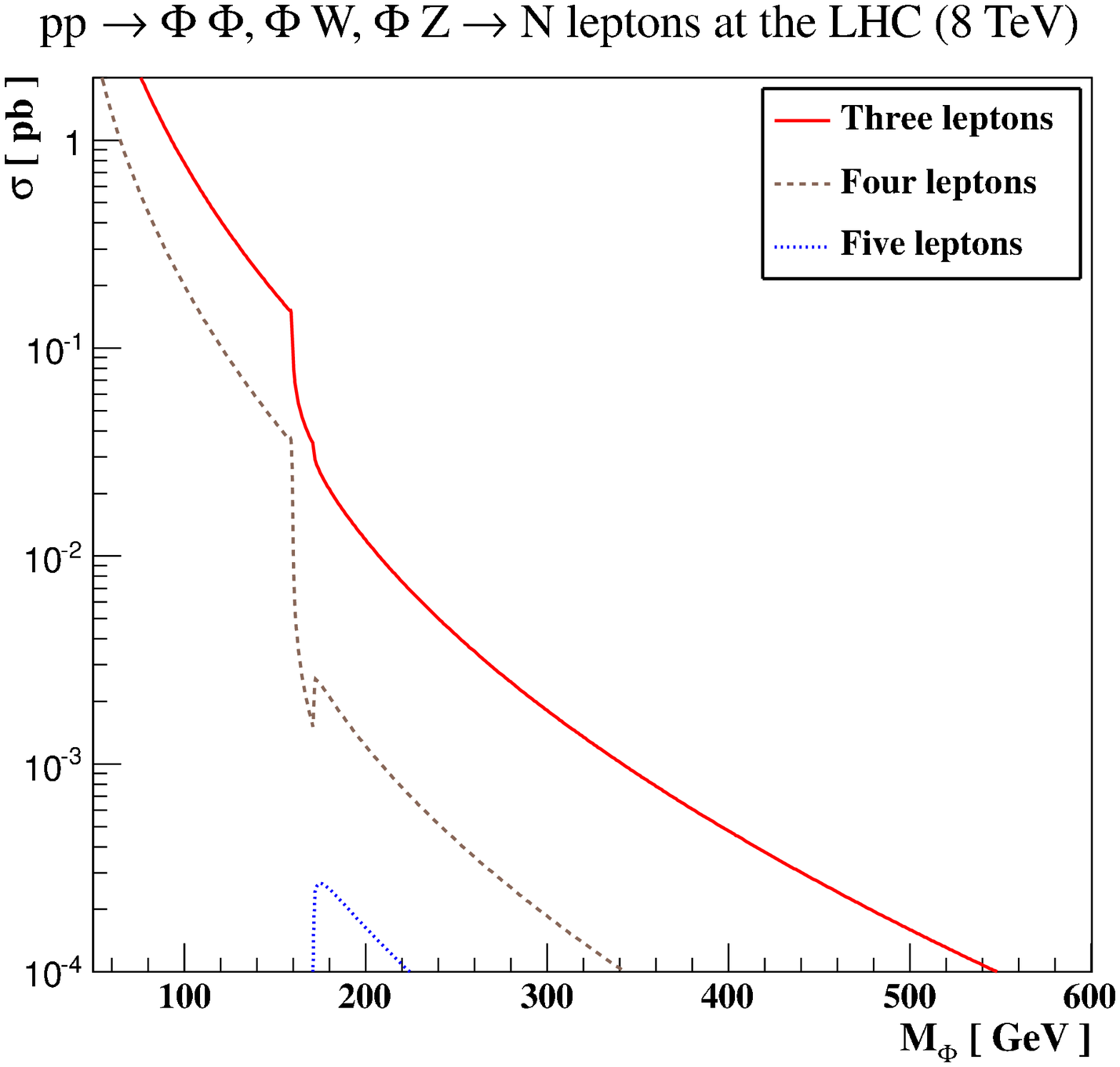}  
\end{tabular}
\caption{\label{fig: dc scalars} Evolution of the cross sections as a function of the mass of the new states for the production and decay into three leptons or more for the scalar fields lying in the trivial (left), fundamental (middle) and adjoint (right) representations of $SU(2)_L$.}
\end{figure}

\paragraph{Scalar singlet and doublet fields} In these cases thee scalar fields decay into either a pair of light leptons or into a pair of tau leptons. The latter decaying into lighter leptons with a frequency of $0.35$, the neutral currents leading to the pair production of these doubly charged scalars contribute to all the final states with a number of leptons ranging from 0 to 4. In the case of the singlet field, in order to have a cross section higher than the minimum one of $1{\rm  fb}$, the mass of the new field must be smaller than $\sim 330 {\rm GeV}$; while the doublet field $\Phi$ must be lighter than $\sim 250 {\rm GeV}$. 

\paragraph{Scalar triplet fields} In the case of the scalar triplet fields, the number of leptons in the final state can be as high as five. Indeed if one considers the charged current process leading to the production of the the doubly-charged state $\pPhi^{++}$ together with the singly charged one $\pPhi^-$. It is possible for both new particles to decay into the weak gauge bosons $Z$ and $W$ which subsequently decay into light charged leptons as illustrated in \eqref{eq: db sca 5l}
\bea \label{eq: db sca 5l} p~p \to \pPhi^{++} ~ \pPhi^{-} \to W^{+}~W^{+}~W^{+}~Z \to l^+~l^+~l^+~l^-~l^-~\nu .\eea
Though such signatures would be very clean as they are ``Standard Model free", they are associated with a maximum cross section of 0.02 fb which is impossible to observe in the context of the 2012 run of the LHC accomplished at a center-of-mass energy of 8 TeV and 20 ${\rm fb}^{-1}$ of accumulated data. Another interesting feature to observe in this case is when the decay into the weak gauge bosons becomes kinematically allowed. In the figures, this transition appears as a decrease in the cross section associated to the processes leading to three or four leptons in the final state while that associated to five-lepton final states becomes non zero. Actually this result depends on the size of the vacuum expectation value $v_\pPhi$; the larger value it has the more dominant the decays into weak gauge bosons become, as can be inferred from the decay widths presented in eq.\eqref{eq:BRsc++2} and \eqref{eq:BRsc++1}. This would lead to a decrease in the cross sections associated to three-lepton and four-lepton finale state because the $W$ and $Z$ bosons decay more often hadronically. In our case, we find that to have a measurable deviation at the LHC the doubly-charged scalar as a component of a triplet scalar multiplet must have a mass below $\sim$ 350 GeV. For completeness, curves drawn in figure \ref{fig: db sca trip} represent the contribution of each process (pair production, production of $\pPhi^{-}~\pPhi^{++}$ \dots) to three- and four- lepton signatures.
\begin{figure}
\centering
\begin{tabular}{c c}
\includegraphics[scale=0.3]{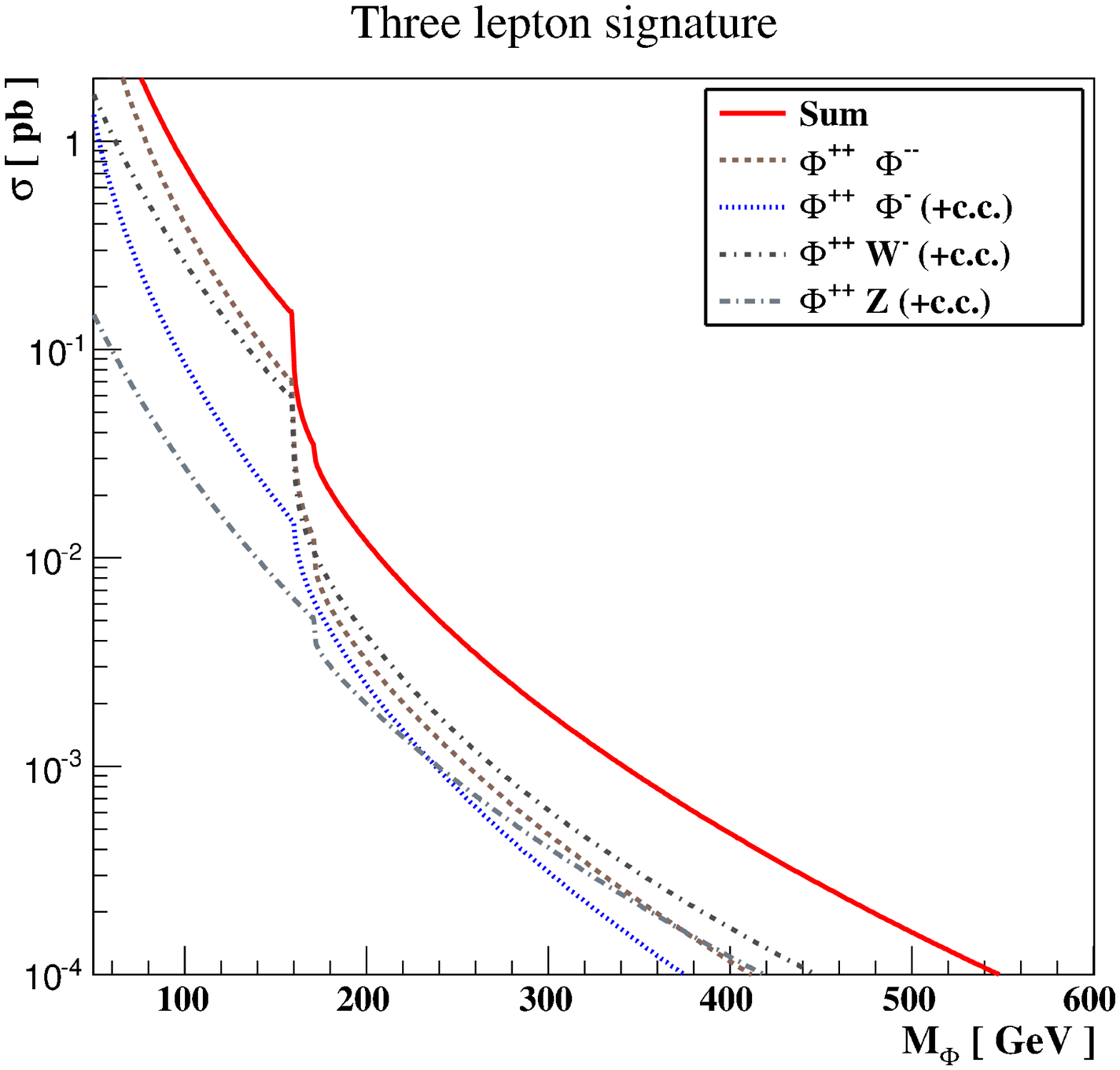} & \includegraphics[scale=0.3]{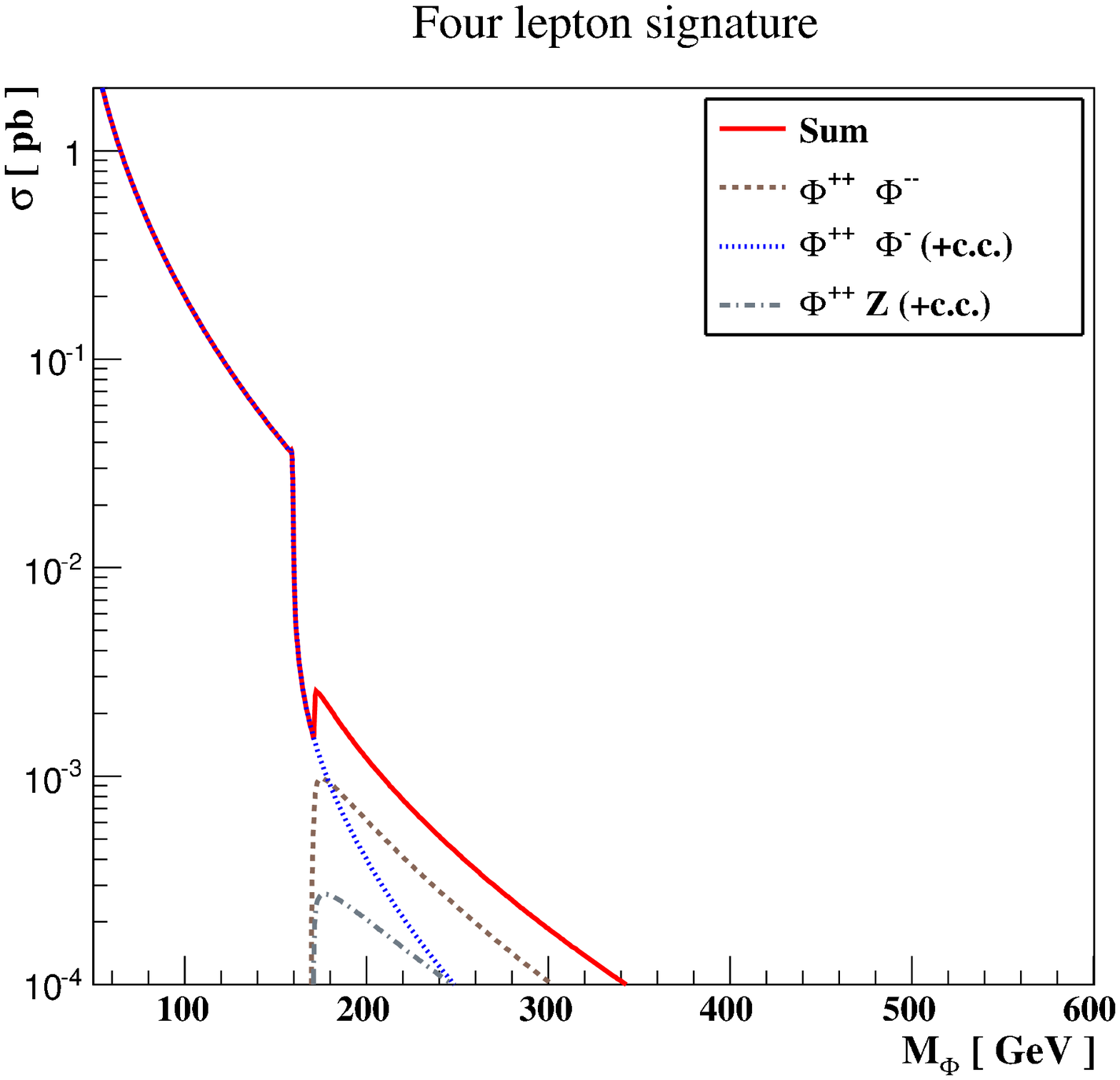}
\end{tabular}
\caption{\label{fig: db sca trip} Contribution of the various processes into three- and four-lepton final states in the case of a doubly-charged scalar belonging to a multiplet transforming as a triplet under $SU(2)_L$. We see that when the decays into the weak gauge bosons are open, the cross sections drop faster.}
\end{figure}

\section{Doubly-charged fermion fields}\label{sec: db fermions}
Doubly-charged fermions also appear in various TeV scale models of particle physics. The supersymmetric version of left-right symmetry predicts fermionic superpartners to the triplet scalar fields and thus doubly-charged fermionic fields transforming either as triplets or singlets under $SU(2)_L$. In extra-dimension models including the custodial symmetry, doubly-charged fermionic fields are also present but they appear as doublets under $SU(2)_L$ \cite{Chen:2009gy,Csaki:2008qq,Kadosh:2010rm,delAguila:2010vg,delAguila:2010es}.\\

Just like we did for the scalar case, we extend minimally the Standard Model Lagrangian to include a new fermionic multiplet lying in either the trivial, the fundamental or the adjoint representation of $SU(2)_L$. Their hypercharge are chosen so that the doubly-charged particle carries the highest electric charge. Our new fields are thus:

$$ \psi \equiv \psi^{++},~~~~ \Psi^i = \begin{pmatrix} \Psi^{++} \\ \Psi^{+} \end{pmatrix}, ~~~~ \pPsi^i{}_j = \begin{pmatrix} \frac{\pPsi^+}{\sqrt2} & \pPsi^{++} \\ \pPsi^0 & - \frac{\pPsi^+}{\sqrt2} \end{pmatrix} $$
with
$$ Y_\psi = 2, ~~~~ Y_\Psi = \frac32, ~~~~ Y_{\pPsi} = 1. $$

As usual, the kinetic terms for the above fields are expressed in terms of the covariant derivatives whose expression are identical to those from eqs.\eqref{eq: db cov der}
\be
  \lag_{\rm kin} = i \psibar \gamma^\mu D_\mu \psi + 
     i \Psibar_i \gamma^\mu D_\mu \Psi^i + 
     i \mathbf{\Psibar}_a \gamma^\mu D_\mu \mathbf\Psi^a\ + \dots.
\label{eq:Lfek}\ee
where, here also, we denote by dots the mass terms for the new fields. To allow for leptonic decays two choices are possible depending on the number of both new parameters and new fields we want to introduce. Keeping our extension of the Standard Model as minimal as possible, we restrain ourselves to introduce only one extra fermionic state that we denote $N$ at the cost of introducing non-renormalizable four-fermion interactions in the Yukawa Lagrangian whose strengths are set by the values of the parameters $ (G^{(i,j)})_{\big| i,j=1,2,3} $. These interactions are then supposed to be  suppressed by some cut-off energy scale $\Lambda$. Supposing the $N$ field to be a gauge singlet, as in sterile neutrino models, we write then the Yukawa interactions for the fields $\psi, \Psi \et \pPsi$ lying in the trivial, fundamental and adjoint representations of $SU(2)_L$, respectively
\bea
  \lag_{\rm F} &=&
  \frac{G^{(1,1)}}{2\Lambda^2} \big[\lbar^c_R l_R\big] \big[\Nbar P_L \psi\big] +
  \frac{G^{(1,2)}}{2\Lambda^2} \big[\lbar^c_R l_R\big] \big[\Nbar P_R \psi\big] 
+
  \frac{G^{(2,1)}}{\Lambda^2} \big[\lbar^c_R \Psi^i\big] \big[\Nbar L_i\big] +
  \frac{G^{(2,2)}}{\Lambda^2} \big[\lbar^c_R N\big] \big[\Lbar^{ic} \Psi_i\big]
\non \\ & +&
  \frac{G^{(3,1)}}{2\Lambda^2} \big[\Lbar^c_i L^j\big] \big[\Nbar P_L
    \mathbf\Psi^i{}_j\big] +
  \frac{G^{(3,2)}}{2\Lambda^2} \big[\Lbar^c_i L^j\big] \big[\Nbar P_R
    \mathbf\Psi^i{}_j\big]
 + {\rm h.c.} \ .
\label{eq:LfeF}\eea 
In these equations we have omitted all generation indices and let $P_L$ and $P_R$ to be the chirality projectors, that is
$$ P_L = \frac{1 - \gamma^5}{2} ~~~\et~~~  P_R = \frac{1 + \gamma^5}{2}. $$
The other drawback due to the non-renormalizable interactions we have allowed here, is that the new fields can only decay through prompt three-body decays into a pair of same sign leptons and a $N$ field. Such decay widths cannot be calculated analytically (no closed formulas exist yet) and to determine the branching ratios (only quantity really useful when considering decay chains) we will make use of the {\sc MadGraph 5} package\cite{Stelzer:1994ta,Maltoni:2002qb,Alwall:2007st,Alwall:2008pm,Alwall:2011uj,deAquino:2011ub}. \\

We now turn to the computation of the cross sections associated with the production of the new fermionic fields. The kinetic terms given in the Lagrangian of equation \eqref{eq:Lfek} only allow for neutral and charged current interactions. In the case of neutral currents, the production of a pair of doubly-charged fermion fields is given by the following formulas
\be\bsp
 \frac{\d\sigma^{NC}_1}{\d t} = &\ 
    \frac{16 \pi \alpha^2 \sh}{9} \big[1 + 2 x^2_{\psi^{++}}\big] 
    \sqrt{1 - 4x^2_{\psi^{++}}} \bigg[
    \frac{e^2_q}{\sh^2} -
    \frac{ e_q (L_q+R_q) (\sh-M_Z^2)}{2 c_W^2 \sh |\sh_Z|^2} +
    \frac{L_q^2 + R_q^2}{8 c_W^4 |\sh_Z|^2}
  \bigg]  \ , \\
 \frac{\d\sigma^{NC}_2}{\d t} = &\ 
    \frac{16 \pi \alpha^2 \sh}{9} \big[1 + 2 x^2_{\Psi^{++}}\big] 
    \sqrt{1 - 4x^2_{\Psi^{++}}} \bigg[  
    \frac{e^2_q}{\sh^2} +
    \frac{ e_q (1-4s_W^2) (L_q+R_q) (\sh-M_Z^2)}{8 c_W^2 s_W^2 \sh |\sh_Z|^2} +
    \frac{(1-4s_W^2)^2 (L_q^2 + R_q^2)}{128 c_W^4 s_W^4 |\sh_Z|^2}
  \bigg]  \ , \\
 \frac{\d\sigma^{NC}_3}{\d t} = &\ 
    \frac{16 \pi \alpha^2 \sh}{9} \big[1 + 2 x^2_{\mathbf \Psi^{++}}\big] \sqrt{1 - 
    4x^2_{\mathbf \Psi^{++}}} \bigg[
    \frac{e^2_q}{\sh^2} +
    \frac{ e_q (1-2s_W^2) (L_q+R_q) (\sh-M_Z^2)}{4 c_W^2 s_W^2 \sh |\sh_Z|^2} +
    \frac{(1-2s_W^2)^2 (L_q^2 + R_q^2)}{32 c_W^4 s_W^4 |\sh_Z|^2}
  \bigg]  \ , 
\esp\label{eq:xsecfe3NC}\ee

In the case where the $W^{\pm}$ gauge boson is the intermediate state, the only possible final states involving the new fermionic fields are those where the singly-charged component is produced together with the doubly-charged one. The cross sections for these processes read
\be\bsp
 \frac{\d\sigma^{CC}_2}{\d t} = &\ 
   \frac{\pi \alpha^2 \sh}{36 s_W^4 |\sh_W|^2}  |V_{ij}^{\rm CKM}|^2\ \sqrt{\lambda(1,x^2_{\Psi^{++}},x^2_{\Psi^+})} 
     \Big[ 1-(x_{\Psi^{++}}-x_{\Psi^+})^2\Big]   \Big[ 2+(x_{\Psi^{++}}+x_{\Psi^+})^2\Big] \ , \\
 \frac{\d\sigma^{CC}_3}{\d t} = &\ 
   \frac{\pi \alpha^2 \sh}{18 s_W^4 |\sh_W|^2}  |V_{ij}^{\rm CKM}|^2\ \sqrt{\lambda(1,x^2_{\Psi^{++}},x^2_{\Psi^+})} 
     \Big[ 1-(x_{\Psi^{++}}-x_{\Psi^+})^2\Big]   \Big[ 2+(x_{\Psi^{++}}+x_{\Psi^+})^2\Big] \ .
\esp\label{eq:xsecfe3CC}\ee

\subsection{Numerical analysis} 
As stated above, no closed formulas for three-body decays exist yet and we choose thus to use the {\sc MadGraph 5} package to compute the branching ratios, after having implemented the model using the {\sc UFO} interface of the {\sc FeynRules} package \cite{Degrande:2011ua}. We find that all three new doubly-charged states, that is the ones belonging to the singlet, the doublet and the triplet multiplets equally decay into electrons, muons and taus:
$$ {\rm BR}(F^{++} \to e^+~e^- ~ N)={\rm BR}(F^{++} \to \mu^+~\mu^- ~ N) = {\rm BR}(F^{++} \to \tau^+~\tau^- ~ N) \sim 0.33$$
where $F$ denotes any of the new fermionic fields. In the case of the doublet and the triplet multiplets, the singly charged state exhibits the same property; that is
$$ {\rm BR}(F^{+} \to e^+~\nu_e ~ N)={\rm BR}(F^{+} \to \mu^+~\nu_\mu ~ N) = {\rm BR}(F^{+} \to \tau^+~\nu_\tau^- ~ N) \sim 0.33.$$
In both previous equations, we have denoted by $F$ the new multiplets $\psi, \Psi \mbox{ or } \pPsi$. \\

The numerical setup being fixed as in section \ref{sec: db scalars} we furthermore set the couplings $G^{(i,j)}$ to a unique value of 
$$ G^{(i,j)} = 0.1\cdot \mathbbm{1}$$
where $\mathbbm{1}$ is the unit $3\times 3$ matrix while the energy scale for new physics is defined, as in the previous section, to a value of 1 TeV. As to the singlet field $N$ that we introduced in order for the decays of the new fermionic fields into SM leptons to be allowed, we set its mass to a value of 50 GeV, in agreement with the present experimental limits as given by the Particle Data Group review \cite{Beringer:1900zz}.\\

We find that in order to have sizeable effects on events with at least three charged light leptons, the new fermionic field must lie under approximately 555, 661 and 738 GeV when it transforms as a singlet, doublet or a triplet of $SU(2)_L$, respectively. Following the same approach than in the previous case, we  plot in figure \ref{fig: dc Ferm 3G} the cross sections for the various final states.

\begin{figure}
\begin{tabular}{c c c}
\includegraphics[width=.32\columnwidth]{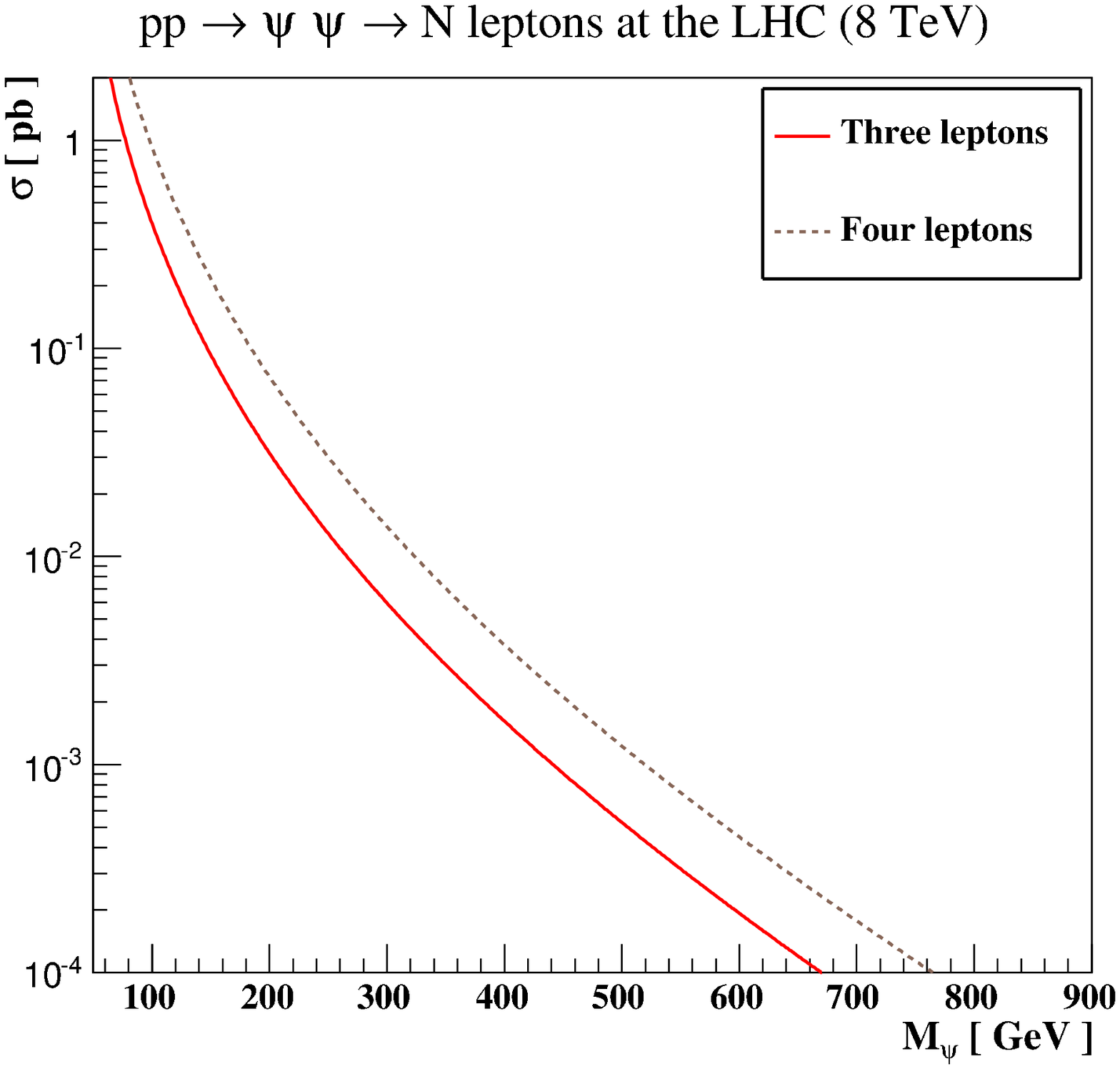} &\includegraphics[width=.32\columnwidth]{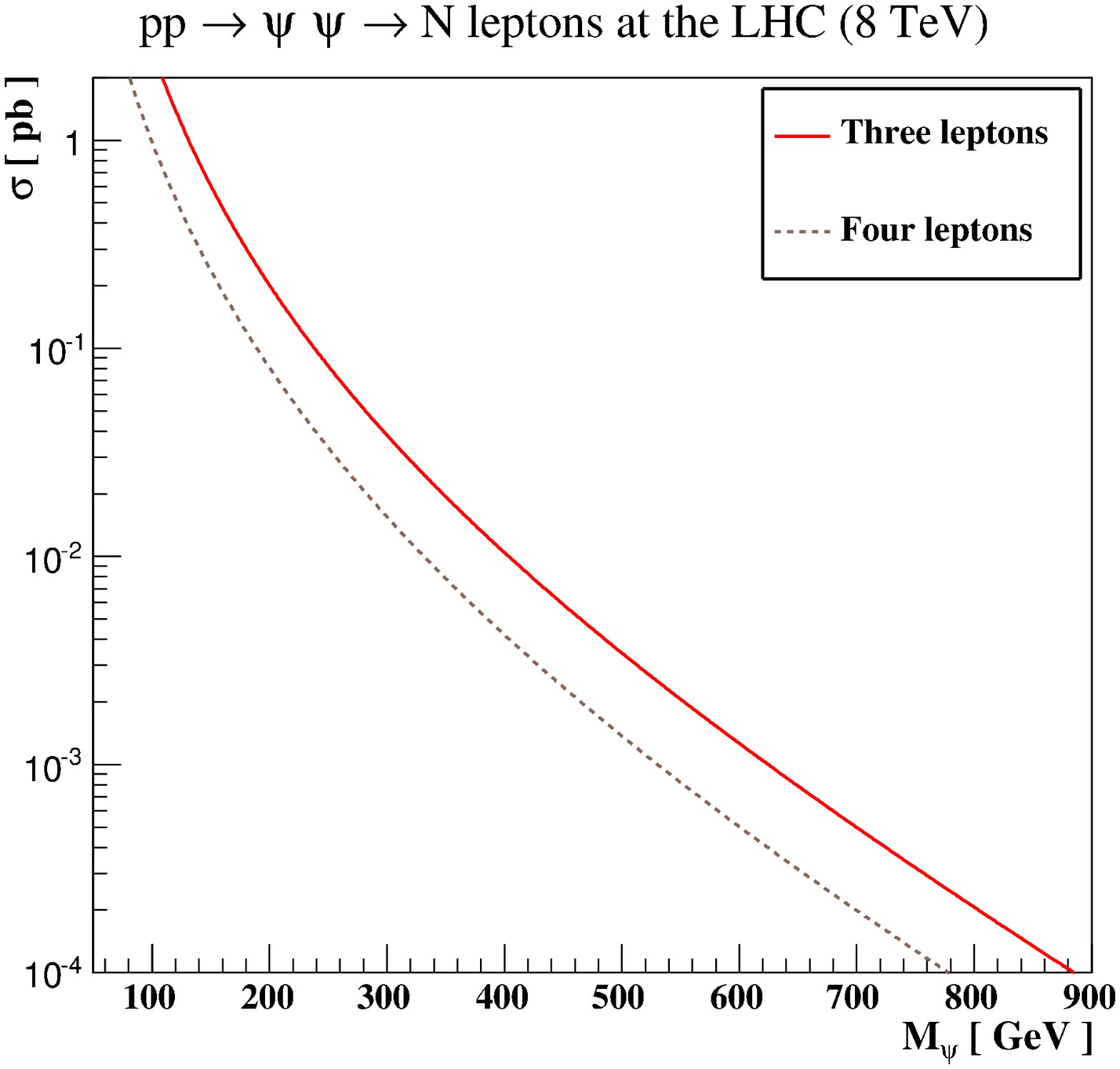} &\includegraphics[width=.32\columnwidth]{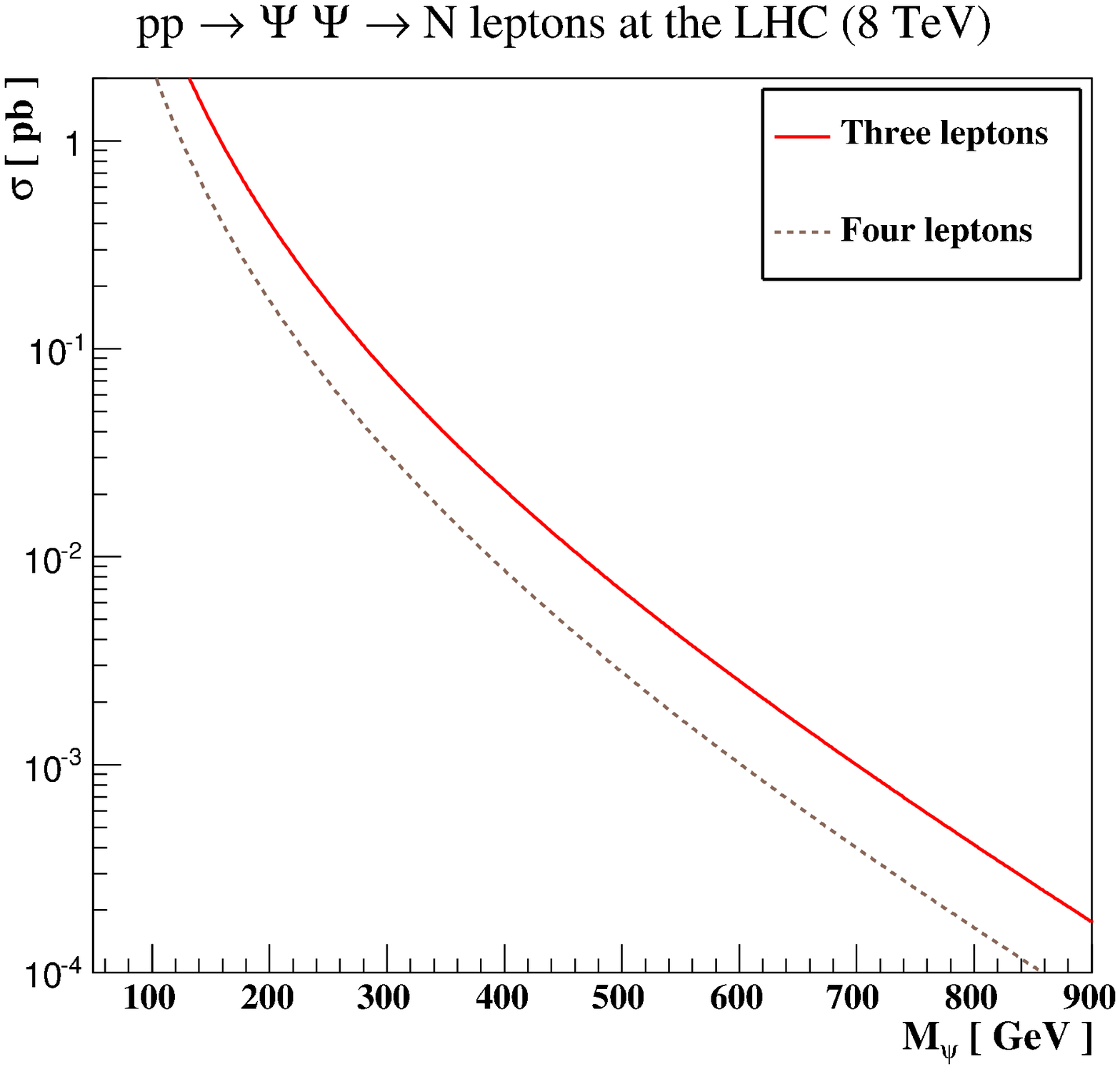}  
\end{tabular}
\caption{\label{fig: dc Ferm 3G} Evolution of the cross sections as a function of the mass of the new states for the production and decay into three leptons or more in the context of the extension of the Standard Model with a singlet (left panel), a doublet (middle panel) and a triplet (right panel) fermionic field. These cross sections hold when the fermionic fields do not mix with SM leptons.}
\end{figure}

\section{Doubly-charged fermion fields and a four generation Standard Model}
In the case where the field content of the Standard Model of particle physics is extended by a fermionic multiplet such as $\Psi \et \pPsi$ introduced in the previous section, it is possible to envisage a mixing in the lepton sector involving both SM's leptons and the singly-charged component of either $\Psi \mbox{ or } \pPsi$. Theoretically, this is made possible by the fact that after electroweak symmetry breaking, $\Psi^+ \et \pPsi^+$ have the same quantum numbers as the charged leptons. Denoting the fourth generation lepton $E'$ in the case where we have added the field $\Psi$ to the theory and $E''$ when the field content is extended with a triplet fermion field, the mixing matrices, whose elements are \textit{\`a priori} unknown read:
$$ \bpm  e' \\ \mu' \\ \tau' \\ E' \epm = M \bpm e \\ \mu \\ \tau \\ \Psi^+ \epm
  \qquad\text{and}\qquad
  \bpm e'' \\ \mu'' \\ \tau'' \\ E'' \epm =  M' \bpm e \\ \mu \\ \tau \\ \mathbf\Psi^+ \epm\ .$$

However, stringent experimental constraints exist from {\sc LEP}, the measurements of the muon anomalous magnetic moment, the leptonic flavor violating processes and from conversions nuclei. Taking into account these limits, we consider the mixing to happen only between the tau lepton and the new field with a maxiaml mixing angle:
\be
  \bpm e' \\ \mu' \\ \tau' \\ E' \epm = 
  \bpm 1 &0&0&0\\ 0&1&0&0\\ 0&0&c_\tau&s_\tau\\ 0&0&-s_\tau&c_\tau\epm
  \bpm e \\ \mu \\ \tau \\ \Psi^+ \epm
  \qquad\text{and}\qquad
  \bpm e'' \\ \mu'' \\ \tau'' \\ E'' \epm = 
  \bpm 1 &0&0&0\\ 0&1&0&0\\ 0&0&c_\tau&s_\tau\\ 0&0&-s_\tau&c_\tau\epm
  \bpm e \\ \mu \\ \tau \\ \mathbf\Psi^+ \epm\ .
  \label{eq: db mix lep}
\ee
with 
$$ s_\tau=0.99 $$

\subsection{Production cross-sections}
We now turn to the derivation of the analytic expressions for the cross sections associated to the production of the new states and leading to signatures with at least three leptons. As in the previous cases, such signatures can be obtained by pair producing doubly-charged fermions. In this case the cross sections are similar to those in eq.\eqref{eq:xsecfe3NC} for the doublet and the triplet fields:
\be\bsp
 \frac{\d\sigma^{NC}_2}{\d t} = &\ 
    \frac{16 \pi \alpha^2 \sh}{9} \big[1 + 2 x^2_{\Psi^{++}}\big] 
    \sqrt{1 - 4x^2_{\Psi^{++}}} \bigg[  
    \frac{e^2_q}{\sh^2} +
    \frac{ e_q (1-4s_W^2) (L_q+R_q) (\sh-M_Z^2)}{8 c_W^2 s_W^2 \sh |\sh_Z|^2} +
    \frac{(1-4s_W^2)^2 (L_q^2 + R_q^2)}{128 c_W^4 s_W^4 |\sh_Z|^2}
  \bigg]  \ , \\
 \frac{\d\sigma^{NC}_3}{\d t} = &\ 
    \frac{16 \pi \alpha^2 \sh}{9} \big[1 + 2 x^2_{\mathbf \Psi^{++}}\big] \sqrt{1 - 
    4x^2_{\mathbf \Psi^{++}}} \bigg[
    \frac{e^2_q}{\sh^2} +
    \frac{ e_q (1-2s_W^2) (L_q+R_q) (\sh-M_Z^2)}{4 c_W^2 s_W^2 \sh |\sh_Z|^2} +
    \frac{(1-2s_W^2)^2 (L_q^2 + R_q^2)}{32 c_W^4 s_W^4 |\sh_Z|^2}
  \bigg]  \ , 
\esp\label{eq:xsecfe4NC}\ee
In contrast to the case where the new fermions do not mix with the Standard Model lepton, neutral currents interactions producing a pair of singly-charged new fermion may lead to final states in which we are interested. Indeed, due to the mixing angle we have introduced in eq.\eqref{eq: db mix lep}, interactions between a $Z$-boson, a tau lepton and $E'$ or $E''$ are possible allowing the latter to cascade decay into a maximum of three charged leptons. We compute thus the cross sections associated to both the production of a pair of $E'$ ($E''$) and the production of  $E'$ ($E''$) together with a tau lepton:
\be\bsp
 \frac{\d\sigma_2^{E'E'}}{\d t} =&\
    \frac{4 \pi \alpha^2 \sh}{9} \sqrt{1 - 4x^2_{E'}} \bigg[ \big[1 + 2 x^2_{E'}\big] \frac{e^2_q}{\sh^2} -
    \frac{e_q (L_q+R_q)}{8c_W^2 s_W^2} \  \frac{\sh-M_Z^2}{ \sh |\sh_Z|^2}\ \big[1 + 2 x^2_{E'}\big] \big[2 + 4 s_W^2 - 3 s^2_\tau\big]\
      \\ &\ +
    \frac{(L_q^2+R_q^2)}{64 c_W^4 s_W^4} \  \frac{1}{ |\sh_Z|^2}\ \Big( \big[1 + 2 x^2_{E'}\big] \big[ 2(1+2 s_W^2)^2
      - 6 s_\tau^2 (1+2 s_W^2) \big] + s_\tau^4 (5+7x^2_{E'})\Big) \bigg]\ ,\\
  \frac{\d\sigma_2^{E'\tau}}{\d t} = &\
    \frac{\pi \alpha^2 (L_q^2+R_q^2) s_\tau^2 c_\tau^2}{288 c_W^4 s_W^4} \frac{\sh}{|\sh_Z|^2}\ \sqrt{\lambda(1,x_\tau^2,x_{E'}^2)} 
    \Big[ 5\big( 3 - 3 x_{E'}^2 -3 x_\tau^2- \lambda(1,x_\tau^2,x_{E'}^2) \big) + 24 x_{E'} x_\tau\Big] \ ,\\
 \frac{\d\sigma_3^{E'' E''}}{\d t} =&\ 
    \frac{4 \pi \alpha^2 \sh}{9} \sqrt{1 - 4x^2_{E'}} \bigg[ \big[1 + 2 x^2_{E'}\big] \frac{e^2_q}{\sh^2} -
    \frac{e_q (L_q+R_q)}{8c_W^2 s_W^2} \  \frac{\sh-M_Z^2}{ \sh |\sh_Z|^2}\ \big[1 + 2 x^2_{E'}\big] \big[4 s_W^2 - s_\tau^2\big]\
      \\& \ +
    \frac{(L_q^2+R_q^2)}{64 c_W^4 s_W^4} \  \frac{1}{ |\sh_Z|^2}\ \Big( \big[1 + 2 x^2_{E''}\big] \big[ 8 s_W^4
      - 4 s_\tau^2 s_W^2 \big] + s^4_\tau (1-x^2_{E''})\Big) \bigg]\ ,  \\
  \frac{\d\sigma_3^{E''\tau}}{\d t} = & \
    \frac{\pi \alpha^2 (L_q^2+R_q^2)s_\tau^2 c_\tau^2}{288 c_W^4 s_W^4} \frac{\sh}{|\sh_Z|^2}\ \sqrt{\lambda(1,x_\tau^2,x_{E'}^2)} 
    \Big[ 3\big( 1 - x_{E''}^2 - x_\tau^2\big) - \lambda(1,x_\tau^2,x_{E'}^2)  \Big] \ . \\
\esp \ee
Charged current interactions are also of interest to us, especially that now the new lepton field has a non vanishing branching ratio to final states with up to three leptons. Computing the cross sections associated with the latter processes for both the doublet and the triplet case, we find
\be\bsp
 \frac{\d\sigma_2^{\Psi^{++}E'}}{\d t} = &\ 
   \frac{\pi \alpha^2 \sh c_\tau^2}{36 s_W^4 |\sh_W|^2}  |V_{ij}^{\rm CKM}|^2\ \sqrt{\lambda(1,x^2_{\Psi^{++}},x^2_{E'})} 
     \Big[ 1-(x_{\Psi^{++}}-x_{E'})^2\Big]   \Big[ 2+(x_{\Psi^{++}}+x_{E'})^2\Big] \ , \\
 \frac{\d\sigma_2^{\Psi^{++} \tau}}{\d t} = &\ 
   \frac{\pi \alpha^2 \sh s_\tau^2}{36 s_W^4 |\sh_W|^2}  |V_{ij}^{\rm CKM}|^2\ \sqrt{\lambda(1,x^2_{\Psi^{++}},x^2_\tau)} 
     \Big[ 1-(x_{\Psi^{++}}-x_\tau)^2\Big]   \Big[ 2+(x_{\Psi^{++}}+x_\tau^2\Big] \ ,\\
 \frac{\d\sigma_3^{\mathbf{\Psi}^{++}E''} }{\d t} = &\ 
   \frac{\pi \alpha^2 \sh c_\tau^2}{18 s_W^4 |\sh_W|^2}  |V_{ij}^{\rm CKM}|^2\ \sqrt{\lambda(1,x^2_{\Psi^{++}},x^2_{E''})}
     \Big[ 1-(x_{\Psi^{++}}-x_{E''})^2\Big]   \Big[ 2+(x_{\Psi^{++}}+x_{E''})^2\Big] \ ,\\
  \frac{\d\sigma_3^{\mathbf{\Psi}^{++}\tau}}{\d t} = &\ 
   \frac{\pi \alpha^2 \sh s_\tau^2}{18 s_W^4 |\sh_W|^2}  |V_{ij}^{\rm CKM}|^2\ \sqrt{\lambda(1,x^2_{\Psi^{++}},x^2_\tau)} 
     \Big[ 1-(x_{\Psi^{++}}-x_\tau)^2\Big]   \Big[ 2+(x_{\Psi^{++}}+x_\tau)^2\Big] \ .
\esp\ee
In the case where the Standard Model is extended by a field belonging to the adjoint representation of $SU(2)_L$, there are two more candidate processes that might contribute to the signatures with free or low SM background. Indeed, the $E''$ field being allowed to decay to three leptons its production in association with the neutral component of $\pPsi$ becomes relevant to our study. The latter decaying to at most two charged light leptons, processes
$$ p~p \to \pPsi^0~\pPsi^0, ~~~~~ p~p \to E''~\pPsi^0 ~~~~~ \mbox{and} ~~~~ p~p \to \tau ~\pPsi^0 $$
may lead to five lepton signatures. The cross sections associated to these processes are 
\be \bsp
 \frac{\d\sigma_3^{\Psi^0\Psi^0} }{\d t} =&\ 
    \frac{\pi \alpha^2 (L_q^2+R_q^2)}{18 c_W^4 s_W^4} \frac{\sh}{|\sh_Z|^2}\ \sqrt{1-4x_{\mathbf{\Psi}^0}^2} 
    \Big[ 1 + 2 x_{\mathbf{\Psi}^0}^2 \Big] \ ,\\
 \frac{\d\sigma_3^{E'' \Psi^0}}{\d t} =&\ 
    \frac{\pi \alpha^2 \sh c_\tau^2}{18 s_W^4 |\sh_W|^2}  |V_{ij}^{\rm CKM}|^2\ \sqrt{\lambda(1,x^2_{\Psi^0},x^2_{E''})}
       \Big[ 1-(x_{\Psi^0}-x_{E''})^2\Big]   \Big[ 2+(x_{\Psi^0}+x_{E''})^2\Big] \ ,\\
 \frac{\d\sigma_3^{\tau\Psi^0} }{\d t} =&\ 
    \frac{\pi \alpha^2 \sh s_\tau^2}{18 s_W^4 |\sh_W|^2}  |V_{ij}^{\rm CKM}|^2\ \sqrt{\lambda(1,x^2_{\Psi^0},x^2_\tau)}
       \Big[ 1-(x_{\Psi^0}-x_\tau)^2\Big]   \Big[ 2+(x_{\Psi^0}+x_\tau)^2\Big] \ .
\esp \ee

\subsection{Partial decay widths}
Before analyzing the analytic formulas given above in order to determine the benchmark scenarios for our Monte Carlo study, we need to compute the decay widths associated to the various fields involved in the processes above. Starting with the doublet multiplet, we find that its doubly-charged component is allowed to decay exclusively to a $\tau$ lepton and a $W$ boson (we recall that the components of a same multiplet are mass degenerate). Hence, its total decay width reads
\be
  \Gamma^{++}_{2, \tau W} = \frac{M_{\Psi^{++}}^3 \alpha s_\tau^2}{8 M_W^2 s_W^2} 
    \sqrt{\lambda(1,x^2_W,x^2_\tau)} \bigg[ \lambda(1,x^2_W,x^2_\tau) - 3 x_W^2 \Big(x_W^2 - (1-x_\tau)^2\Big)\bigg] \ .
\label{eq:decfemix1}\ee
while that of the singly-charged component is the sum of the partial decay width into a tau lepton and a $Z$ boson and a tau-neutrino with a $W$ boson
$$ \Psi \to \tau ~ Z, ~~~~ \Psi \to \nu_\tau ~ W .$$
The analytic formulas for these two decay processes are 
\be\bsp
  \Gamma^+_{2, \tau Z} =&\ \frac{M_{E'}^3 \alpha s_\tau^2 c_\tau^2}{32 M_Z^2 c_W^2 s_W^2} 
    \sqrt{\lambda(1,x_Z^2,x_\tau^2)} 
     \bigg[ 5\lambda(1,x_Z^2,x_\tau^2) + 3 x_Z^2 \Big(5 (1 + x_\tau^2 - x_Z^2) - 8 x_\tau\Big) \bigg] \ ,  \\
  \Gamma^+_{2,\nu_\tau W} =&\ \frac{M_{E'}^3 \alpha s^2_\tau}{16 M_W^2 s_W^2} \Big[1-x_W^2\Big]^2 \Big[1+2x_W^2\Big]\ .
\esp\label{eq:decfemix2}\ee
\begin{figure}[!t]
\unitlength = 1mm
\begin{center}
\begin{fmffile}{dbchrgd6L}
\begin{fmfgraph*}(100,80)
	\fmfleft{i1,i2,i3,i4}
	\fmflabel{$u$}{i2}
	\fmflabel{$\bar{u}$}{i3}
	\fmfright{o1,o2,o3,o4,o5,o6,o7,o8,o9,o10}
	\fmflabel{$l^+$}{o1}
	\fmflabel{$l^-$}{o2}
	\fmflabel{$\nu_\tau$}{o3}
	\fmflabel{$\bar{\nu}_l$}{o4}
    \fmflabel{$l^-$}{o5}
	\fmflabel{$l^+$}{o6}
	\fmflabel{$l^-$}{o7}
	\fmflabel{$\bar{\nu}_\tau$}{o8}
	\fmflabel{$l^+$}{o9}
	\fmflabel{$\nu_l$}{o10}				
	\fmf{fermion,tension=2.5}{i2,v1,i3}
	\fmf{photon,tension=2.5,label=$\gamma^{\mu} / Z^\mu$}{v1,v2}
	\fmf{fermion,label=$E''^-$,label.side=right}{v2,v3}
	\fmf{fermion,label=$E''^+$}{v4,v2}
	\fmf{fermion,label=$Z$,label.side=left}{v3,v5}
	\fmf{fermion,label=$\tau^-$,label.side=left}{v3,v6}
	\fmf{fermion,label=$Z$,label.side=left}{v4,v7}
	\fmf{fermion,label=$\tau^+$,label.side=right}{v8,v4}
	\fmf{fermion}{o1,v5,o2}
	\fmf{fermion}{v6,o3}
	\fmf{fermion}{o4,v6,o5}
	
	\fmf{fermion}{o6,v7,o7}
	\fmf{fermion}{v8,o8}
	\fmf{fermion}{o9,v8,o10}
	
\end{fmfgraph*}
\end{fmffile}
\end{center}
\caption{\footnotesize\label{fig: feyn diag 6l} Feynman diagram associated to the process leading to a final state with six charged light leptons in the case where the fourth lepton $E$ originates from a fermion transforming either as a doublet or a triplet under $SU(2)_L$. In the latter case, though the cross section for pair producing the $E''$ fields can be as high as 0.46 pb for a mass of 100 GeV, the leptonic decays of the $Z$ bosons together with that of the $\tau$ lepton make this process have a tiny cross section which cannot be higher than $10^{-6} {\rm pb}$. In the former case where $E'$ originates from a doublet multiplet, the associated cross section can be as high as 0.16 fb when the mass is equal to 100 GeV, which though, not visible within the 2012 run of the LHC, can become important with both increasing luminosity and center-of-mass energy. }
\end{figure}
Finally, we also compute the decay widths associated to the components of the triplet field $\pPsi$. We find that the electrically non-neutral components have the same decay channels as $\Psi^{++}$ and $\Psi^+$
\be\bsp
  \Gamma^{++}_{3,\tau W} = &\ 
    \frac{M_{\mathbf{\psi}^{++}}^3 \alpha s_\tau^2}{4 M_W^2 s_W^2}\sqrt{\lambda(1,x^2_W, x_\tau^2)}
      \Big[ \lambda(1,x^2_W, x_\tau^2) + 3 x_W^2 \big( (1-x_\tau)^2-x_W^2\big)\Big] \ ,  \\
  \Gamma^+_{3,\tau Z} = &\ 
    \frac{ M_{E''}^3 \alpha s_\tau^2 c_\tau^2}{32 M_Z^2 c_W^2 s_W^2} \sqrt{\lambda(1,x^2_Z, x_\tau^2)}
      \Big[ \lambda(1,x^2_Z, x_\tau^2) + 3 x_Z^2 ( 1+x_\tau^2-x_Z^2) \Big] \ ,  \\
  \Gamma^+_{3,\nu_\tau W} =&\ \frac{M_{E''}^3 \alpha s_\tau^2}{16  M_W^2 s_W^2} \Big[1-x_W^2\Big]^2 \Big[1 + 2 x_W^2\Big] \ .
\esp\ee
As to the neutral field $\pPsi^0$, it only decays through the interactions induced by the covariant derivatives, that is 
$$ \pPsi^0 \to \tau^+ ~ W^- $$
where here also we have ignored the coupling into the $E'$ as not allowed kinematically in our setup. The total decay width for the neutral $\pPsi^0$ is hence
$$  \Gamma^0_{3, \tau W} =\frac{ M_{\mathbf{\psi}^0}^3 \alpha s_\tau^2}{4 M_W^2 s_W^2} \sqrt{\lambda(1,x^2_W, x_\tau^2)}  \Big[ \lambda(1,x^2_W, x_\tau^2) + 3 x_W^2 \big( (1-x_\tau)^2-x_W^2\big)\Big] .$$
\begin{figure}[!t]
\centering
\begin{tabular}{c c}
\includegraphics[width=.5\columnwidth]{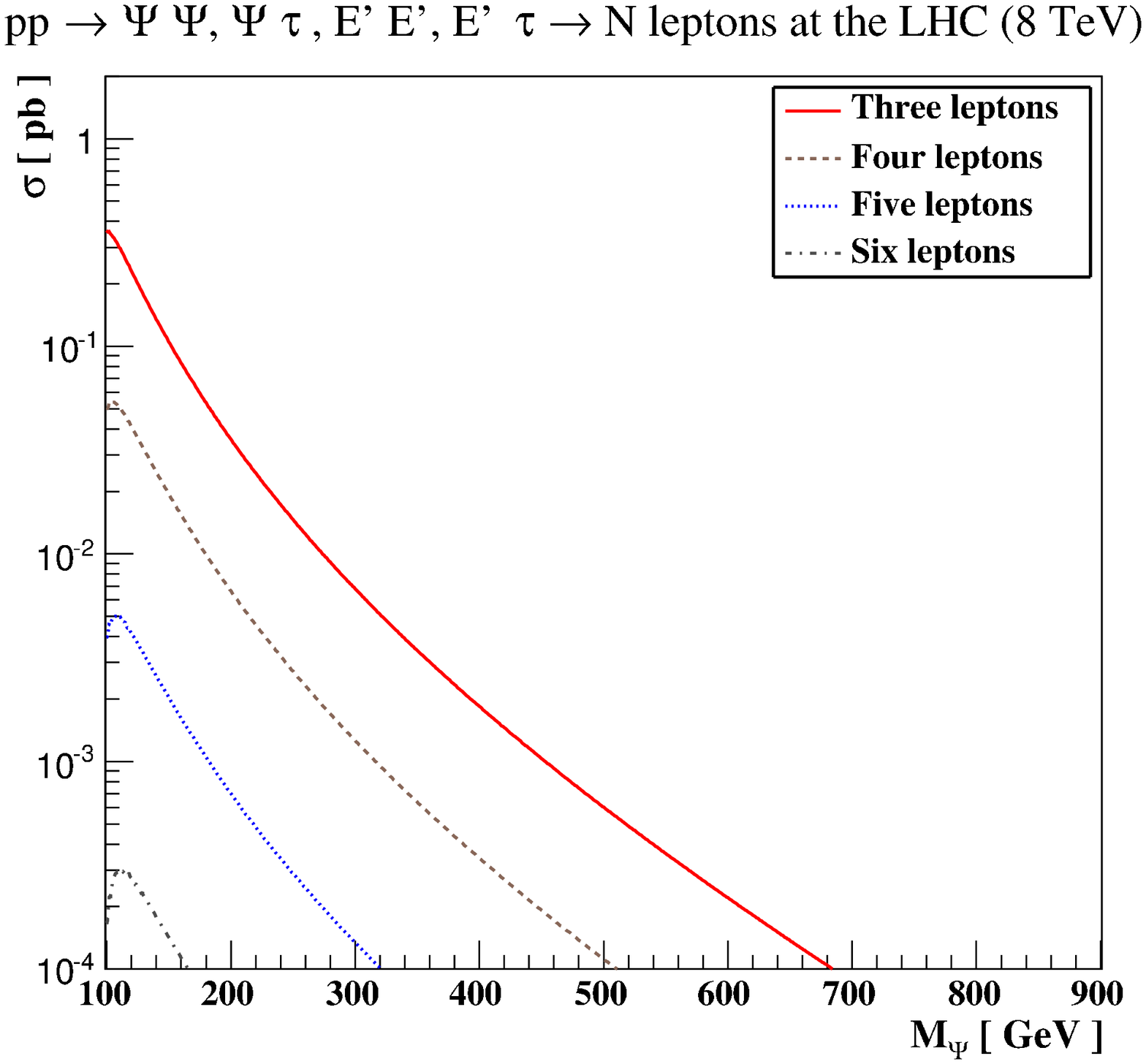} & \includegraphics[width=.5\columnwidth]{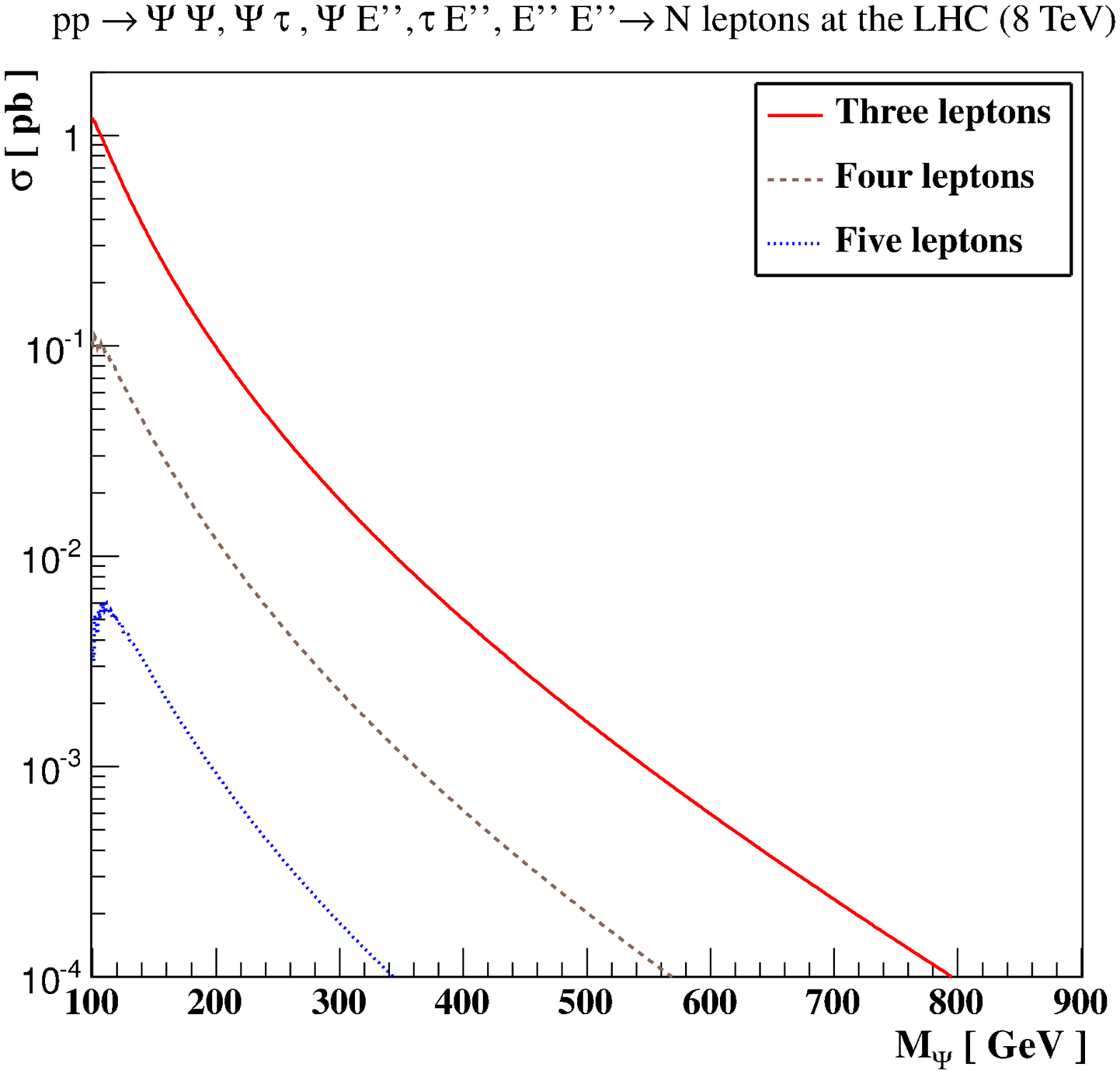}
\end{tabular}
\caption{\footnotesize\label{fig: db fermions 4g}Evolution of the cross sections for the model extending the SM with doubly-charged fermion fields lying in either the fundamental or adjoint representation of $SU(2)_L$ and whose singly-charged component mixes with SM leptons.}
\end{figure}

\subsection{Numerical analysis}
We are now ready to analyze the bahaviour of the cross sections when the mass of the new states is varied. In order to make the comparison easier between the two models, we gather in figure \ref{fig: db fermions 4g} the evolution of all cross sections given above for the new fermionic state when the mixing with the SM leptons is allowed. From the left corner to the right corner are presented succesively the cross sections for the multiplets lying in the fundamental and adjoint representation. A few remarks are in order here.
\begin{itemize}
    \item In the case where the new fermion that is added to the field content of the Standard Model transforms as a triplet under $SU(2)_L$, we find that though possible as outlined above, the cross section leading to a six lepton final state is of the order of $10^{-6}{\rm pb}$ in the best case, that is a mass of 100 GeV. This is because the only possible process leading to such a final state is that given by the feynman diagram of fig.\ref{fig: feyn diag 6l} which requires the leptonic decays of both two $Z$ bosons and two tau leptons. Hence, though the maximum cross section obtained for a mass of 100 GeV is around 0.46 pb, the tiny branching ratios into leptons of both the $Z$ and the tau leptons render it negligible. In figure \ref{fig: db fermions 4g}, the associated curve has been removed. 
    \item The six lepton signature in the case of the $SU(2)_L$ doublet field is, on the contrary, much more promising. Indeed, though under the limit of 1 fb which means that it is not observable within the 2012 run of the LHC and 20 ${\rm fb}^{-1}$ of accumulated data, with an increase in both the luminosity and the center-of-mass energy it can lead to sizeable effect in a region where the background from Standard Model is absent.
    \item Finally, the three lepton signature is the one where the number of events is the highest and we find that to have a cross section above 1 fb, the mass of the new states needs to be lower than  525 GeV in the case of the doublet field and 648 GeV in the case of the triplet field. 
\end{itemize}

\section{Vector fields}\label{sec: db vectors}
Finally, we turn to the last setup we want to consider, that is doubly-charged vectors lying in either the trivial, the fundamental or the adjoint representation of $SU(2)_L$ which can appear in theories with an extended gauge group \cite{Frampton:1989fu,Pal:1990xw,Pisano:1991ee,Frampton:1991wf} or in composite and technicolor models \cite{Farhi:1980xs,Harari:1982xy,Cabibbo:1983bk,Eichten:1984eu,Pancheri:1984sm,Biondini:2012ny}.\\

Following the same steps as before, we build an effective Lagrangian in order to allow for the new states to decay into the Standard Model fields. Guided only by gauge invariance and the constrain to keep the extension minimal, we define the new fields as 
\bea 
	V_\mu = V^{++}_\mu,~~~~~ \mV_\mu = \bpm \mV^{++}_\mu\\ \mV^{+}_\mu \epm, ~~~~ (\vV_\mu)^i{}_j = \bpm \frac{\vV^+_\mu}{\sqrt2} & \vV^{++}_\mu \\ \vV^0_\mu & - \frac{\vV^+_\mu}{\sqrt2} \epm .
\eea
and use a gauged version of the Proca Lagrangian to write their kinetic and gauge interactions terms  
\bea
\lagr_{kin} &=& -\frac12 \big[ D_\mu V_\nu^\dagger - D_\nu V_\mu^\dagger \big]\big[D^\mu V^\nu - D^\nu V^\mu\big] - \frac12 \big[D_\mu \mV_\nu^\dagger - D_\nu \mV_\mu^\dagger\big]\big[D^\mu\mV^\nu - D^\nu\mV^\mu\big]\n
&-& \frac12 \big[D_\mu \vV^\dagger_\nu - D_\nu \vV^\dagger_\mu \big]^a\big[D^\mu\vV^\nu - D^\nu\vV^\mu\big]_a,
\eea
where, following our convention, $a$ is an index in the adjoint representation of $SU(2)_L$. We also introduce the yukawa couplings $\Big(\tilde{g}^{(i)}\Big)_{\big|i=1,2,3}$ allowing for the new vector fields to decay into charged and/or neutral leptons:
\bea
\lagr_{yuk} = \frac{\tilde{g}^{(1)}}{\Lambda}V_\mu \bar{l}^c_R \sigma^{\mu\nu}D_\nu l_R + \tilde{g}^{(2)} \mV^i_\mu \bar{L}^c_i \gamma^\mu l_R + \frac{\tilde{g}^{(3)}}{\Lambda} (\vV_\mu)^i{}_j \bar{L}^c_i \sigma^{\mu\nu}D_\nu L^j.\label{db: vec lagr yuka}
\eea
In this equation, we have introduced $\sigma^{\mu\nu} = \frac{i}{4}[\gamma^\mu,\gamma^\nu]$ where $\gamma^\mu$ are the Dirac matrices, and the new physics energy scale $\Lambda$ to suppress the dimension-four operators $ \bar{l}^c_R \sigma^{\mu\nu}D_\nu l_R \et (\vV_\mu)^i{}_j \bar{L}^c_i \sigma^{\mu\nu}D_\nu L^j$ that identically vanish.\\

\subsection{Production cross sections}
Turning to the analytic calculations, we start by calculating the cross sections associated to processes leading potentially to signatures with more than two light charged leptons. We then compute the decay widths of the various fields and finally analyze numerically the behaviour of these cross sections.\\

From the Lagrangian we have introduced, the only production modes of thew states allowed in our model are those mediated by neutral and charged currents. In the first case, $q \bar{q}$ scattering (where $q$ is a quark) leads through an intermediate $Z$ boson, to the production of a pair of doubly-charged vector fields $$ q~\bar{q} \to V^{++} ~ V^{--},~~~~ q~\bar{q} \to \mV^{++} ~ \mV^{--},~~~~ q~\bar{q} \to \vV^{++} ~ \vV^{--} $$
to which we associate the cross sections:
\be\bsp
 \frac{\d\sigma^{NC}_1}{\d t} = &\ 
    \frac{4 \pi \alpha^2 \sh}{9} \sqrt{1 - 4x^2_{V^{++}}}\frac{1-x_{V^{++}}^2 - 12x_{V^{++}}^4}{x_{V^{++}}^2} \bigg[
    \frac{e^2_q}{\sh^2} -
    \frac{ e_q (L_q+R_q) (\sh-M_Z^2)}{2 c_W^2 \sh |\sh_Z|^2} +
    \frac{L_q^2 + R_q^2}{8 c_W^4 |\sh_Z|^2}
  \bigg]  \ , \\
 \frac{\d\sigma^{NC}_2}{\d t} = &\ 
  \frac{4 \pi \alpha^2 \sh}{9} \sqrt{1 - 4x^2_{\mV^{++}}}\frac{1-x_{\mV^{++}}^2 - 12x_{\mV^{++}}^4}{x_{\mV^{++}}^2} \bigg[
    \frac{e^2_q}{\sh^2} +
    \frac{ e_q (1-4s_W^2) (L_q+R_q) (\sh-M_Z^2)}{8 c_W^2 s_W^2 \sh |\sh_Z|^2} \
      \\& \ +
    \frac{(1-4s_W^2)^2 (L_q^2 + R_q^2)}{128 c_W^4 s_W^4 |\sh_Z|^2}
  \bigg]  \ , \\
 \frac{\d\sigma^{NC}_3}{\d t} = &\ 
  \frac{4 \pi \alpha^2 \sh}{9} \sqrt{1 - 4x^2_{\vV^{++}}}\frac{1-x_{\vV^{++}}^2 - 12x_{\vV^{++}}^4}{x_{\vV^{++}}^2} \bigg[
    \frac{e^2_q}{\sh^2} +
    \frac{ e_q (1-2s_W^2) (L_q+R_q) (\sh-M_Z^2)}{4 c_W^2 s_W^2 \sh |\sh_Z|^2} \
      \\& \ +
    \frac{(1-2s_W^2)^2 (L_q^2 + R_q^2)}{32 c_W^4 s_W^4 |\sh_Z|^2}
  \bigg]  \ , 
\esp\label{eq:xsecvecNC}\ee  
The other mode is triggered by the scattering of an up-type quark with an anti-down type quark leading, through an intermediate $W$ gauge boson, to the production of a singly-charged vector field together with its doubly-charged partner:
$$ u_i~\bar{d}_j \to V^{++} ~ V^{-},~~~~ u_i~\bar{d}_j \to \mV^{++} ~ \mV^{-},~~~~ u_i~\bar{d}_j \to \vV^{++} ~ \vV^{-} .$$
The analytic formulas associated to these processes read
\be\bsp
 \frac{\d\sigma^{CC}_2}{\d t} = &\ 
   \frac{\pi \alpha^2 \sh}{288 s_W^4 |\sh_W|^2}  |V_{ij}^{\rm CKM}|^2\ \sqrt{\lambda(1,x^2_{\mV^{++}},x^2_{\mV^+})} 
     \frac{\Big[ 1-(x_{\mV^{++}}-x_{\mV^+})^2\Big]   \Big[ 1-(x_{\mV^{++}}+x_{\mV^+})^2\Big]}{x_{\mV^{++}}^2x_{\mV^{+}}}\
      \\& \ \times \Big[\lambda(1,x_{\mV^{++}}^2,x_{\mV^+}^2) - 1 + 4 ( x_{\mV^+}^2 + x_{\mV^{++}}^2 + 3x_{\mV^+}^2x_{\mV^{++}}^2) \Big]
 \ , \\
 \frac{\d\sigma^{CC}_3}{\d t} = &\ 
   \frac{\pi \alpha^2 \sh}{144 s_W^4 |\sh_W|^2}  |V_{ij}^{\rm CKM}|^2\ \sqrt{\lambda(1,x^2_{\vV^{++}},x^2_{\vV^+})} 
     \frac{\Big[ 1-(x_{\vV^{++}}-x_{\vV^+})^2\Big]   \Big[ 1-(x_{\vV^{++}}+x_{\vV^+})^2\Big]}{x_{\vV^{++}}^2x_{\vV^{+}}}\
      \\& \ \times \Big[\lambda(1,x_{\vV^{++}}^2,x_{\vV^+}^2) - 1 + 4 ( x_{\vV^+}^2 + x_{\vV^{++}}^2 + 3x_{\vV^+}^2x_{\vV^{++}}^2) \Big]\ .
\esp\label{eq:xsecvecCC}\ee

\subsection{Partial decay widths}
In this simplified model, they are quiet simple as the only decay modes allowed for the new vector fields are dictated by the Yukawa Lagrangian given in equation \eqref{db: vec lagr yuka}. In other words, the doubly-charged components of the new multiplets only decay into a pair of same-sign leptons while the singly charged component decays into a charged lepton (tau or lighter) and the associated neutrino. The neutral component of the triplet vector field is not considered as it decays exclusively into a pair of neutrinos which translate to missing transverse energy at colliders. The associated decay widths for the various fields read
\be\bsp
 \Gamma_{1,\ell}^{++} = &\ \frac{M_{V^{++}}M_\ell^2(\tilde{g}^{(1)})^2}{96 \pi\Lambda^2} \Big[1-4 x_\ell^2\Big]^{3/2}\ , \\
 \Gamma_{2,\ell}^{++} = &\ \frac{M_{\mV^{++}} (\tilde{g}^{(2)})^2}{24 \pi} \Big[ 1-4 x_\ell^2\Big]^{3/2} \ , \\
 \Gamma_{3,\ell}^{++} = &\ \frac{M_{\vV^{++}}M_\ell^2 (\tilde{g}^{(3)})^2}{96 \pi \Lambda^2} \Big[1 - 4 x_\ell^2\Big]^{3/2}\ ,\\
 \Gamma_{2,\ell}^{+} = &\ \frac{M_{\mV^{++}}(\tilde{g}^{(2)})^2}{48\pi}\Big[2-3x_\ell^2+x_\ell^6\Big]\ ,\\
 \Gamma_{3,\ell}^{+} = &\ \frac{M_{\vV^{++}}M_\ell^2(\tilde{g}^{(3)})^2}{384\pi\Lambda^2}\Big[1-x_\ell^2\Big]^2\Big[2+x_\ell^2\Big]\ .
\esp\label{eq:decvec}\ee

\subsection{Numerical analysis}
For the numerical values, necessary to carry the calculations, we use the same setup as described in section \ref{sec: db scalars} and let the Yukawa couplings diagonal with the eigenvalues 0.1; the scale for the new physics to be equal to 1 TeV and we assume mass degeneracy inside the same multiplet. \\

From the above considerations we see that when the doubly-charged particle is a heavy vector field, one can expect at most four charged light leptons in the final state. Moreover, As can be deduced from figure \ref{fig: db vec}, where we have represented the evolution of the various cross sections leading to multilepton final states in the case of the singlet (left), doublet (middle) and triplet (right) vector fields, the doublet case is the one exhibiting the highest cross sections. Starting from well above 1 pb for a mass of 100 GeV, it reaches the 1 fb limit only for masses higher than 620 GeV while the singlet and triplet fields reach the same limit for the masses 392 GeV and 495 GeV respectively. 
\begin{figure}
\begin{tabular}{c c c}
\includegraphics[width=.32\columnwidth]{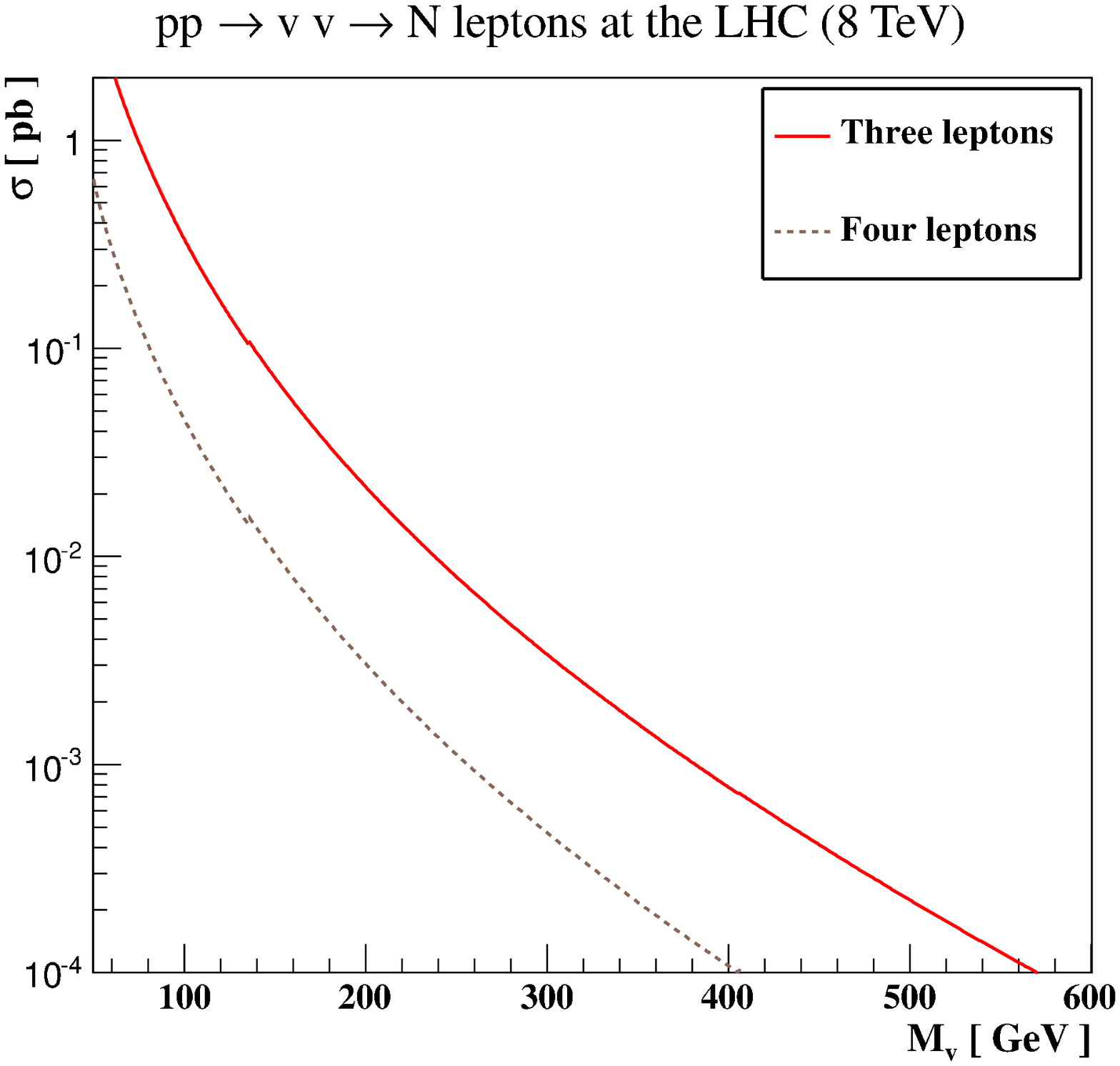} & \includegraphics[width=.32\columnwidth]{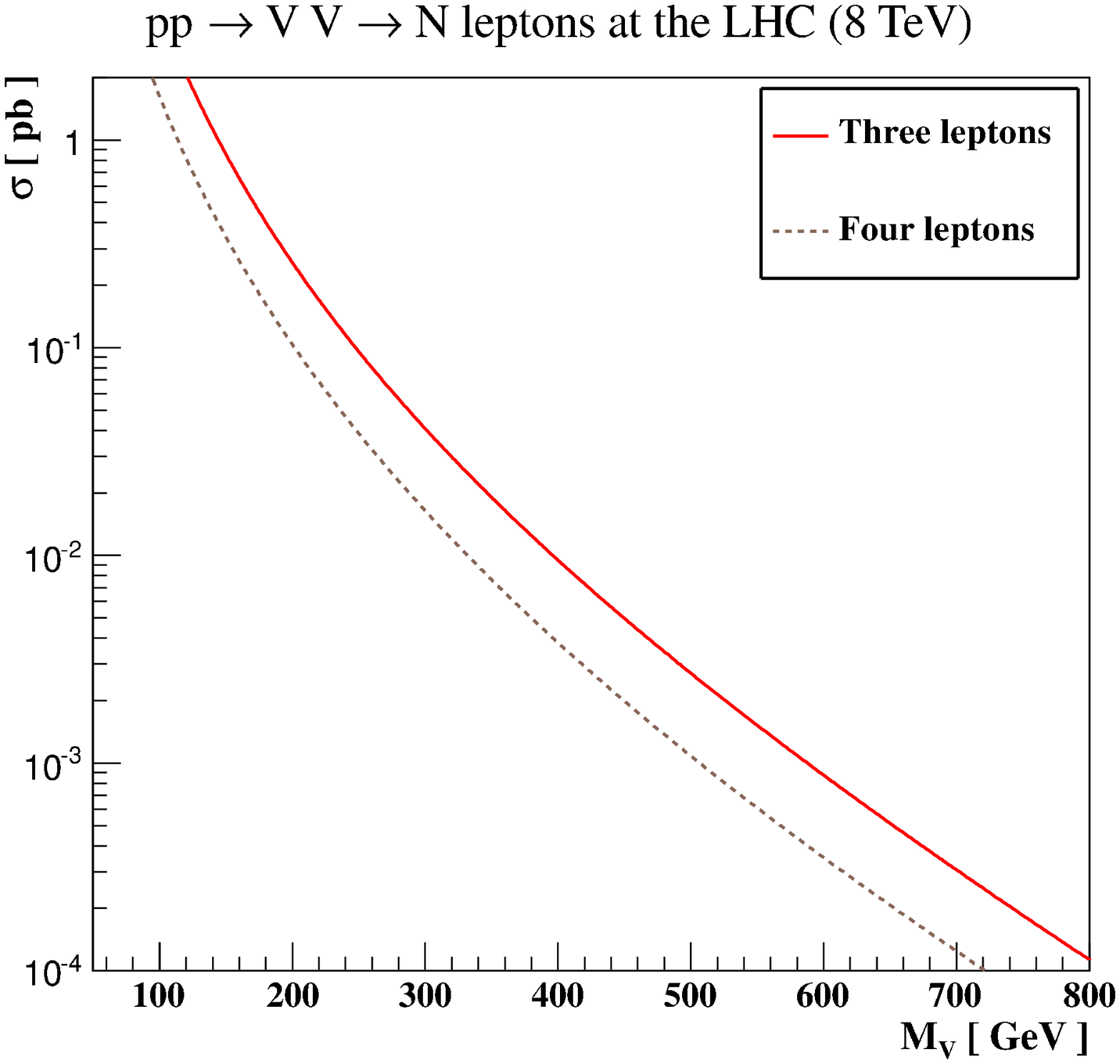} & \includegraphics[width=.32\columnwidth]{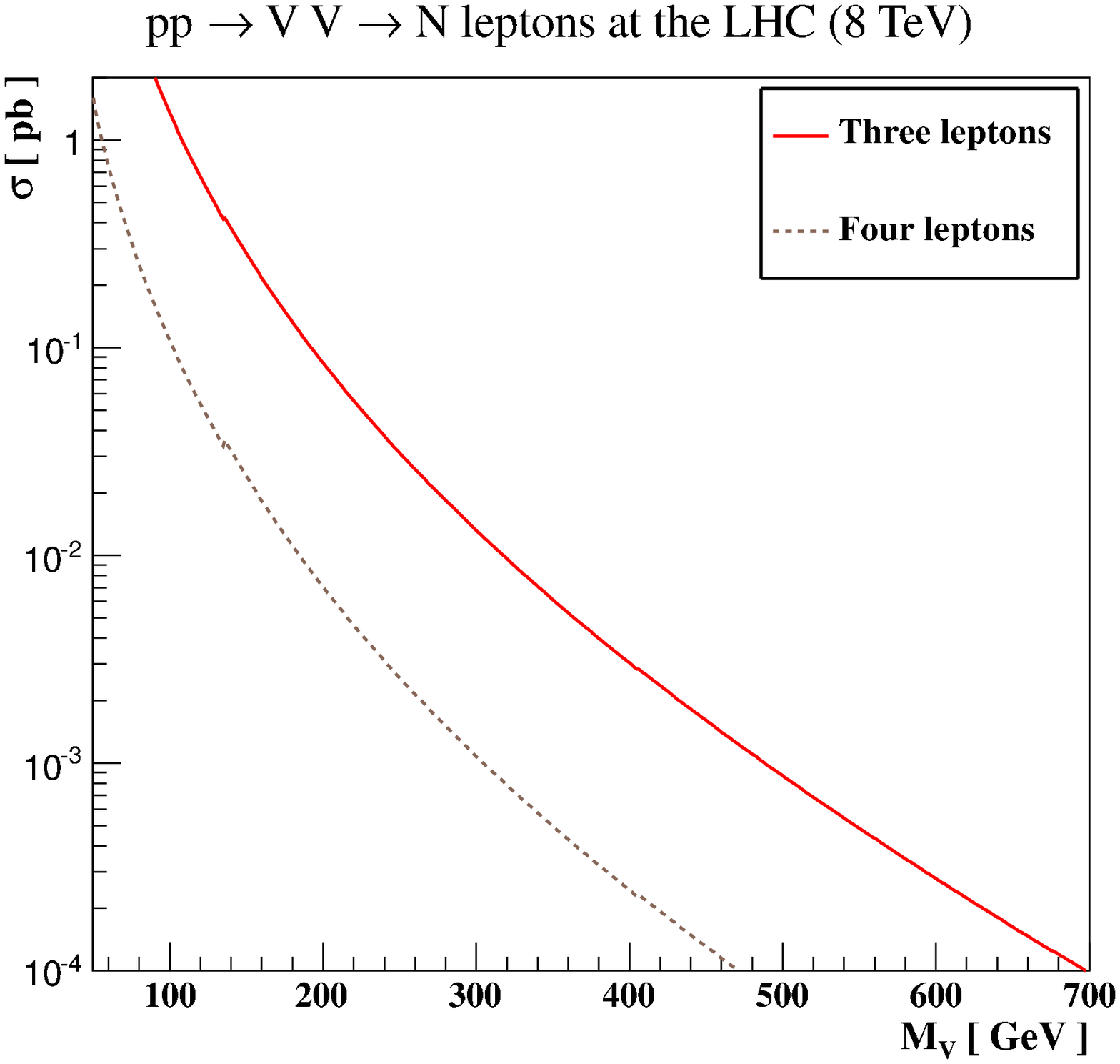}
\end{tabular}
\caption{\label{fig: db vec}Evolution of the cross sections for three and four lepton final states as a function of the mass of the new multiplet. From left to right are presented, succesively, the case where the vector field lies in the trivial, fundamental and adjoint representation of $SU(2)_L$. It is found that above a mass of 392 GeV for the singlet, 620 GeV for the doublet and 495 GeV for the triplet the cross section associated to three-lepton final states drops under the 1fb limit. }
\end{figure}

\section{Monte Carlo simulation}\label{sec: dc pheno}
\subsection{Summary of the analytical results}
The theorical study accomplished, we now want to investigate the kinematical distributions the various processes envisaged in the previous sections would induce. Focusing only on finale states with a minimum of three charged light leptons, we want to find the key observables that one might use to both know the kind of distributions such models would yield but also to emphasize on the differences between the various spin and isospin representations in order to distinguish them. Before proceeding let us first summarize the results from previous sections. \\

In table \ref{tab: db summary}, we summarize the various limits on the mass of the doubly-charged particle and its partners that we have derived together with the maximum number of leptons expected for each representation, ignoring however the six lepton final state induced by the fermion multiplet lying in the representation $\triplet$ of $SU(2)_L$. We recall also the numerical values we have used:
\begin{itemize}
    \item All Yukawa couplings are set to a unique value of $0.1\cdot \mathbbm{1}$.
    \item All new physics energy scales are set to 1 TeV.
    \item The new multiplets are considered mass degenerate forbidding thus decays inside a same multiplet
    \item In the case of leptons mixing with the singly-charged component of the doublet or triplet fermion fields, we have assumed the mixing large in order to comply with the experimental constraints
    $$ c_\tau = 0.01 .$$
    \item Finally, in the case of the scalar triplet field, the neutral component is allowed to have a vacuum expectation value of 100 GeV.
\end{itemize}
\begin{table}
\centering
\begin{tabular}{c|c|c|c||c|c|c|}
\cline{2-7}
& \multicolumn{3}{|c||}{Maximum mass [GeV]} & \multicolumn{3}{|c||}{Maximum number of leptons} \\
\cline{2-7}
& Singlet & Doublet & Triplet & Singlet & Doublet & Triplet \\
\hline
\hline
Scalars & 330 & 257 & 350 & 4 & 4 & 5\\
Fermions (3 Gen) & 555 & 661 & 738 & 4 & 4 & 4\\
Fermions (4 Gen) & - & 525 & 648 & - & 6 & 5 \\
Vectors & 392 & 619 & 495 & 4 & 4 & 4\\
\hline
\end{tabular}
\caption{\footnotesize\label{tab: db summary}In this table are presented the summary results for all simplified models considered above. First three columns under ``Maximum mass" correspond to the maximum mass the new multiplet can have in order to yield a cross section for three lepton final states at least equal to 1 fb. In the second part of the table are presented the maximum number of light charged leptons in the final state one can expect for each model. Six lepton final states in the case of the model in which the new multiplet belonging to the adjoint representation of $SU(2)_L$ mixes, through its singly charged component, with SM leptons has not been taken into account as the associated cross section is too low. Finally, the two sets of columns are disconnected, that is the limit on the mass only holds for three lepton final states.}
\end{table}

\subsection{Setup for the simulation}
In order to generate the Monte Carlo events necessary for the analysis, we follow almost the same pattern than in the case of the left-right symmetric supersymmetric model presented in chapter \ref{chap:lrsusy}. We start then by implmenting the various Lagrangians in {\sc FeynRules} in order to export the {\sc UFO} model file to {\sc MadGraph 5} and use, therefore, the latter to generate the parton level events. Contrary to the tool-chain used for the analysis of the left-right symmetric supersymmetric model, the parton level events generated by {\sc MadGraph 5} only contain SM particles in the final states, in other words the new multiplets introduced in our simplified models have been decayed. We then use the package {\sc Pythia 6} \cite{Sjostrand:2006za} to simulate the decays of the various unstable particles produced by the decays of the new multiplets (e.g. $W$ and $Z$ gauge bosons), parton showering and hadronization but let {\sc Tauola}\cite{Davidson:2010rw} handle the decays of the tau lepton. Finally, the package {\sc MadAnalysis 5} is used for both its interface to the {\sc FastJet} package in order to allow for a proper reconstruction of the various obects and also for the analysis facilities it provides. 
Finally, the selection criteria used in the LRSUSY study (see section \ref{sec:lrsusy pheno}) in order to ensure the reconstructed objects are well isolated and above the $p_T$ threshold are used here also. In addition, we only select events with at least three charged light leptons in the final state ensuring thus that the background from Standard Model, which is not simulated here, is under control.

\subsection{Differentiating between the various models}
From table \ref{tab: db summary}, we see that a five lepton state already helps to reduce the number of possible configurations as only three models predict such final states. In the case of final states with six charged light leptons, the possibilities are even more reduced as only remains the fermionic doublet field whose singly-charged component mixes with the tau lepton. Now the questions we want to address are:
\begin{itemize}
    \item How to characterize every configuration from some key kinematical distributions?
    \item How do these differences evolve with the mass of the various fields?
    \item Do those peculiar five and six lepton final states lead to sizeable effects with the rather conservative hypothesis of an ideal detector?
\end{itemize}

To answer the first two questions, we construct out of our models three benchmark scenarios taking the mass of the new multiplet to be equal to 100, 250 and 350 GeV successively. After analyzing various distributions we find that, in general, differences are enhanced with increasing mass but the number of events drops quickly. In the case of the scalar fields for example, the model leads to no visible event when the mass is set to 350 GeV, which is in agreement with table \ref{tab: db summary}.\\

The convention we adopt from now on is that, from the left-hand side to the right-hand side are represented successively the cases where the mass of the new multiplet is equal to 100, 250 and 350 GeV. From the top to the bottom of the figure, representations are varied taking successively the value singlet, doublet and triplet.
\paragraph{Angular distance $\Delta R$} Defined as 
$$ \Delta R = \sqrt{\Delta \phi^2 + \Delta \eta^2} $$
where $\phi$ stands for the azimuthal angle with respect to the beam direction and $\eta$ is the pseudo-rapidity; the angular distance between two objects is found to lead to signficant differences, especially in tails of the distributions. This is illustrated by the histograms in figure \ref{fig: db deltar} representing the evolution of this observable for the pair leading-lepton next-to-leading lepton (ordered by decreasing transverse momentum) when both the mass and the representation are varied.\\

For example, in the case of the scalar field, one remarks that events concentrate on rather small ranges of $\Delta R$ while the fermion cases have broader distributions. The vector fields, when their mass is equal to 100 GeV, lead to a peculiar distribution as after a sharp increase in the number of events there is a visible inflation of the curve before re-increasing again and reaching a peak around the value $\pi$. This behaviour however disappears when increasing the mass.\\

In some other cases, however, the differences are not that clear as can be seen in bottom-left corner of the figure. Therein, all models, apart from the vector field, exhibit the same behaviour. 

\begin{figure}
\vskip-1cm
\centering\begin{tabular}{c c c}
\includegraphics[width=.32\columnwidth]{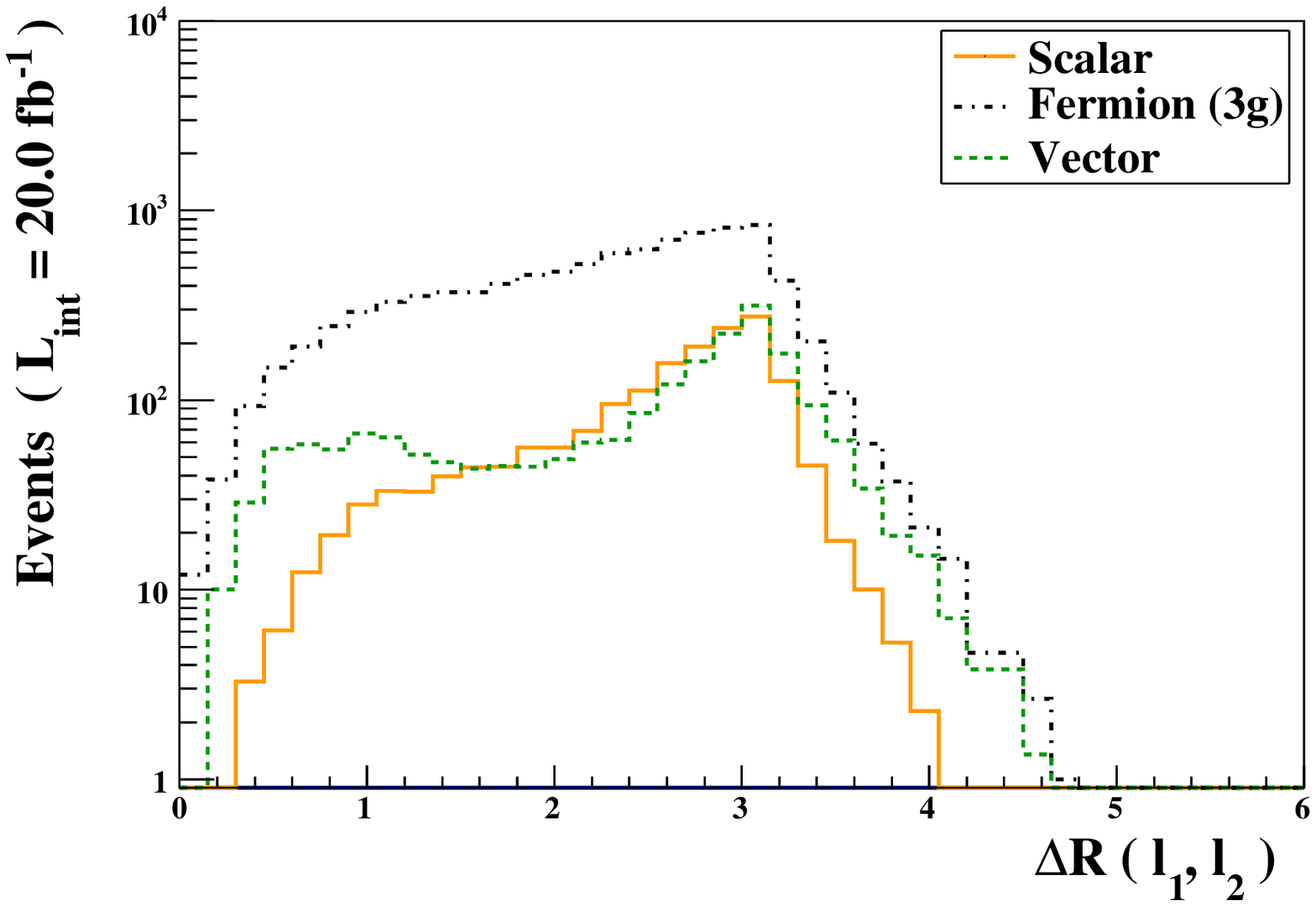} & \includegraphics[width=.32\columnwidth]{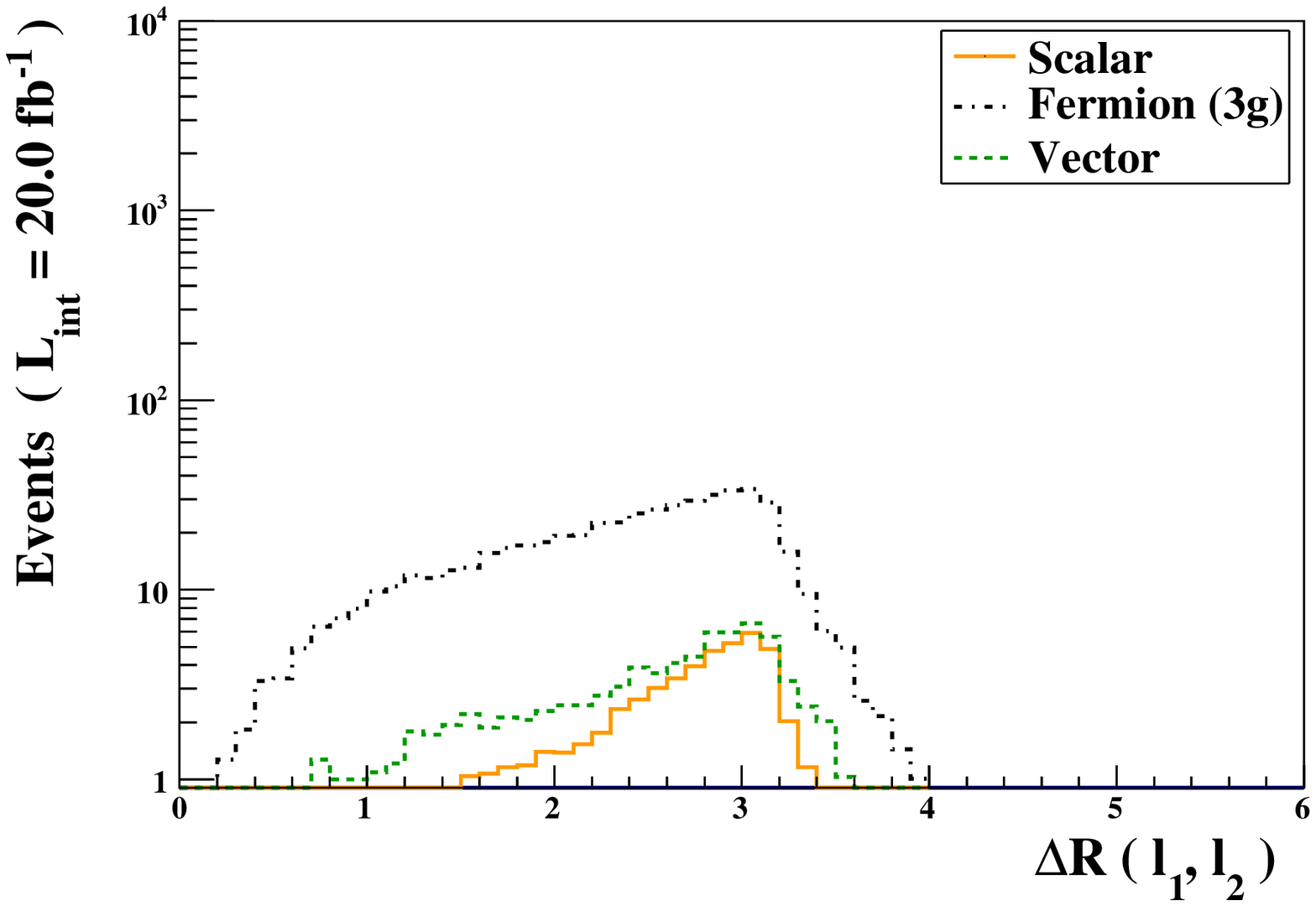} & \includegraphics[width=.32\columnwidth]{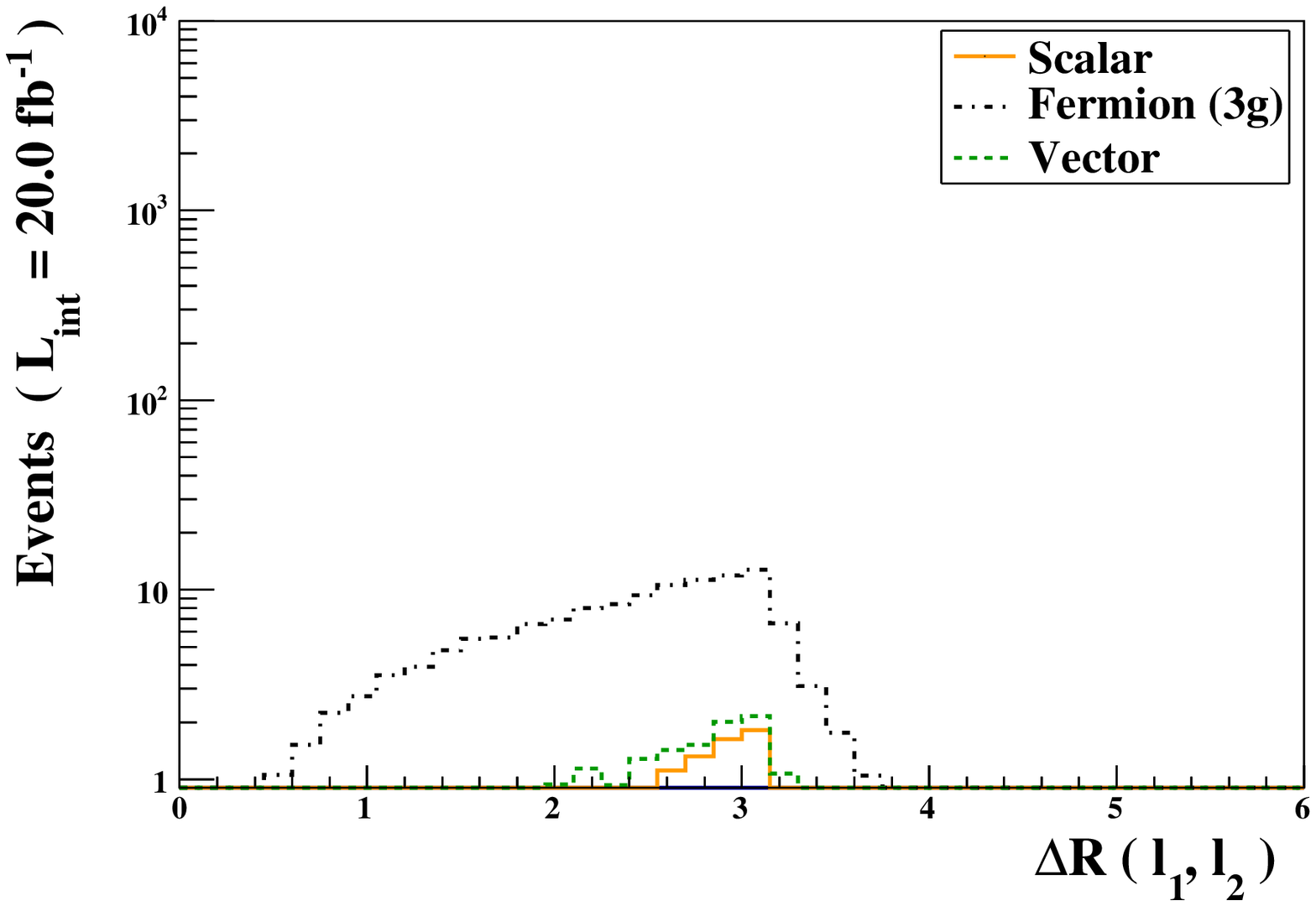}\\
\includegraphics[width=.32\columnwidth]{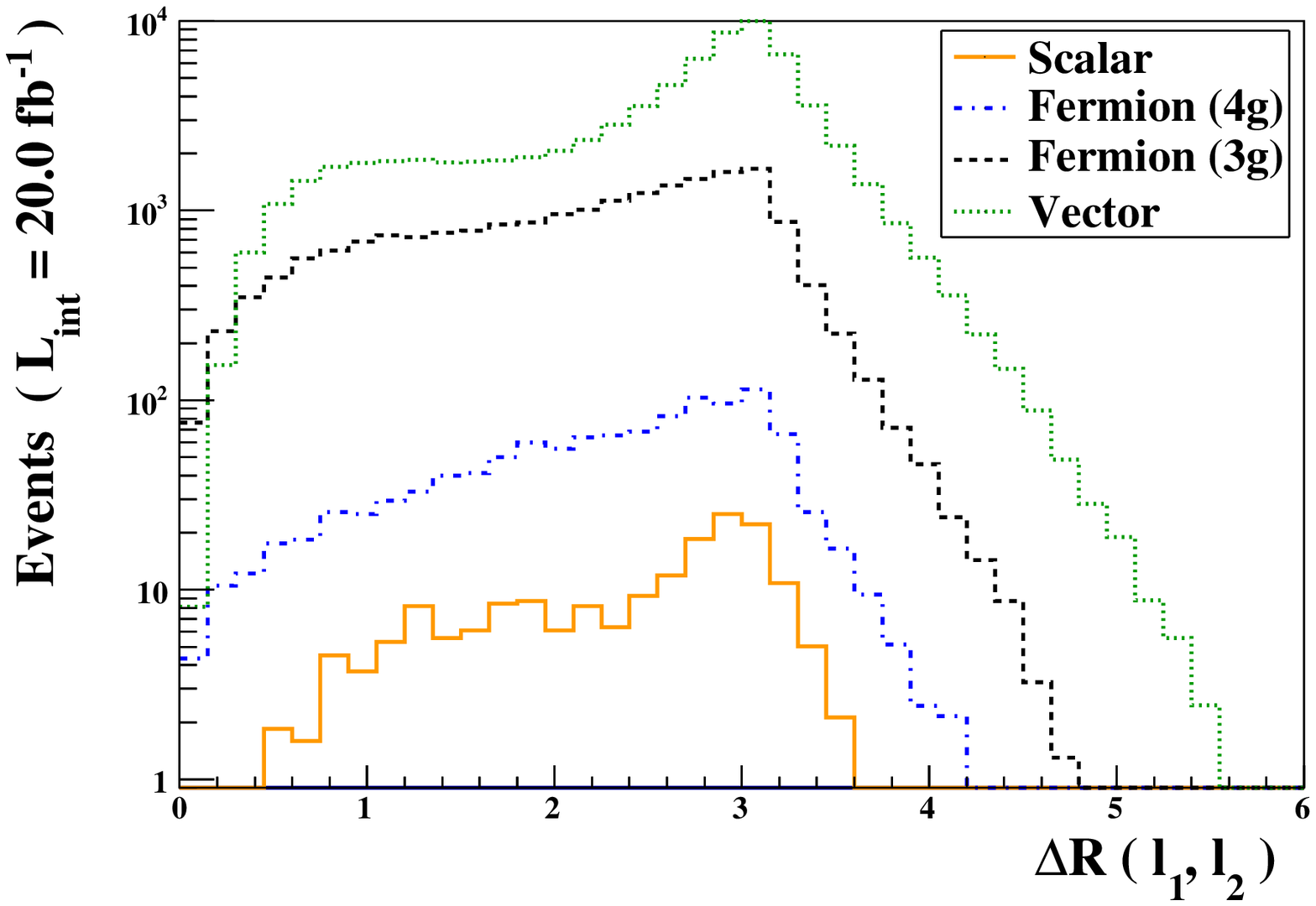} & \includegraphics[width=.32\columnwidth]{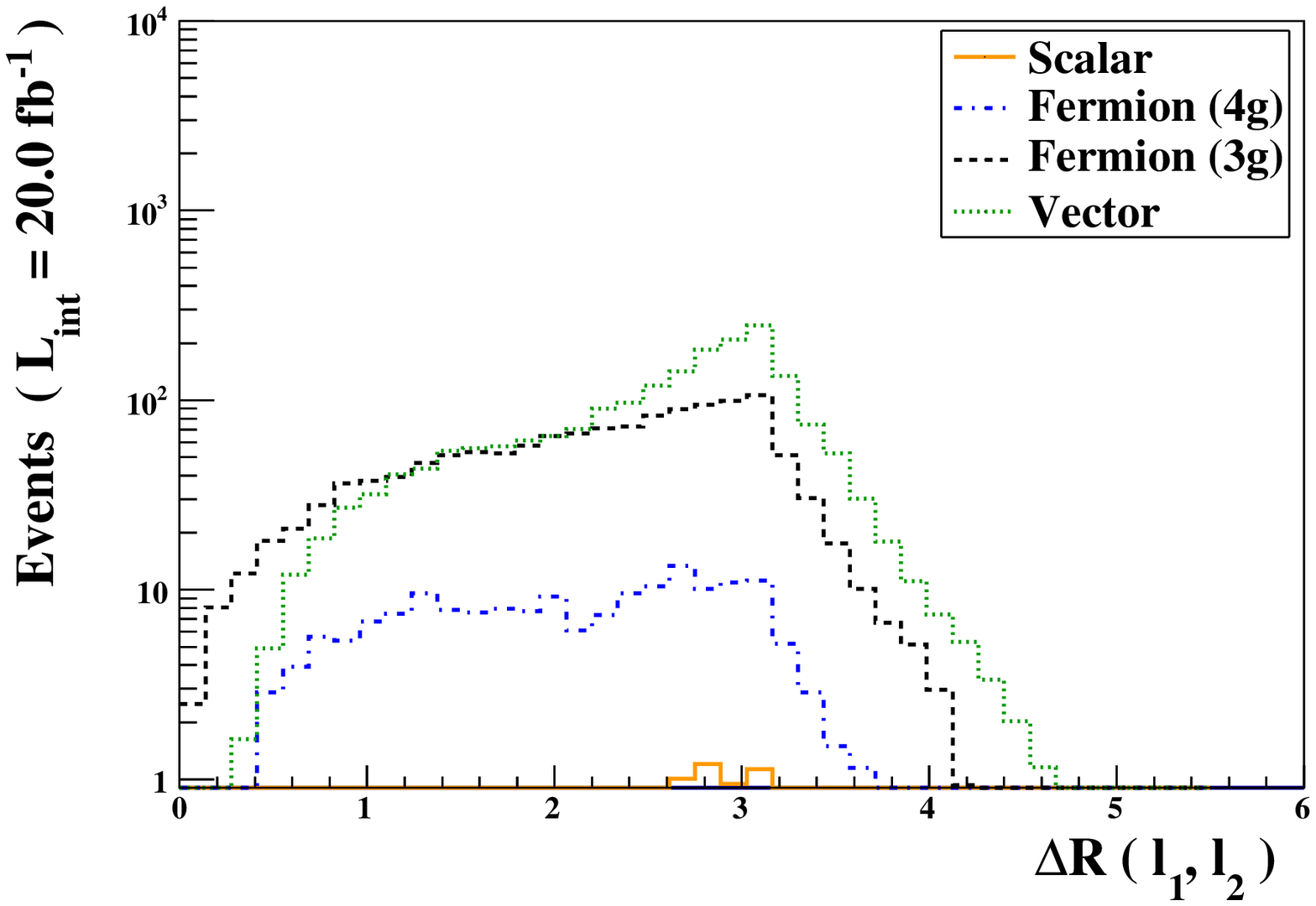} & \includegraphics[width=.32\columnwidth]{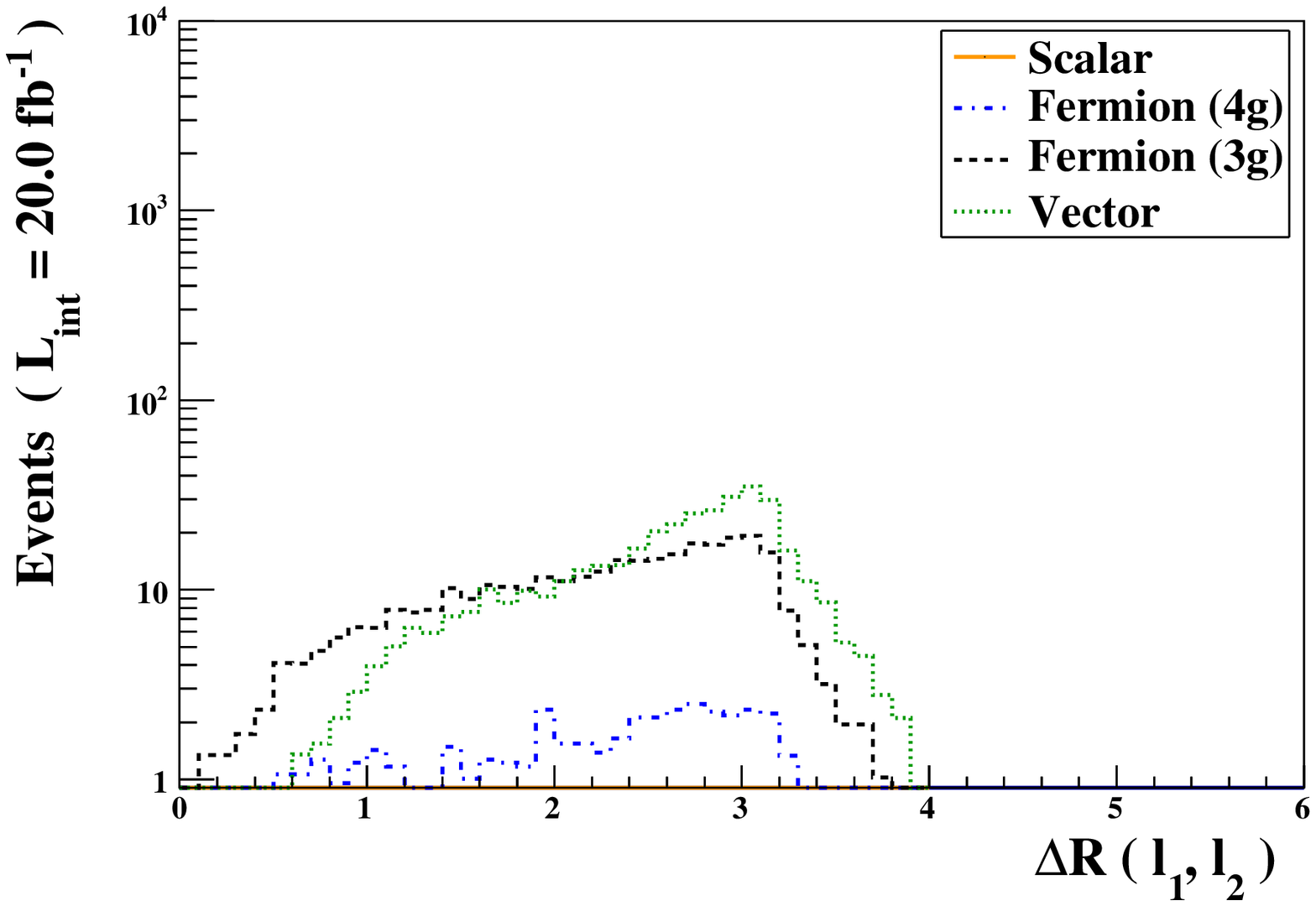}\\
\includegraphics[width=.32\columnwidth]{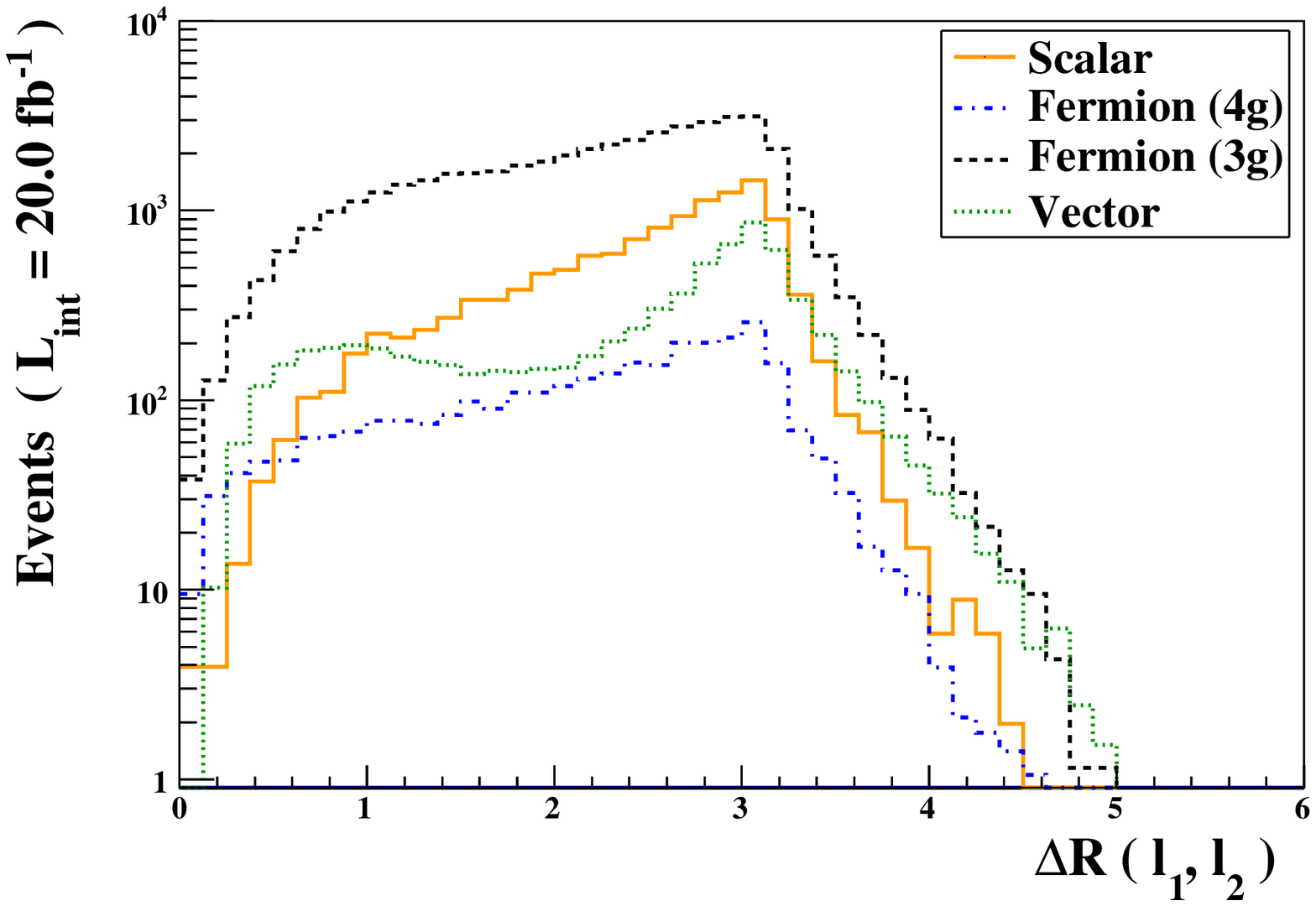} & \includegraphics[width=.32\columnwidth]{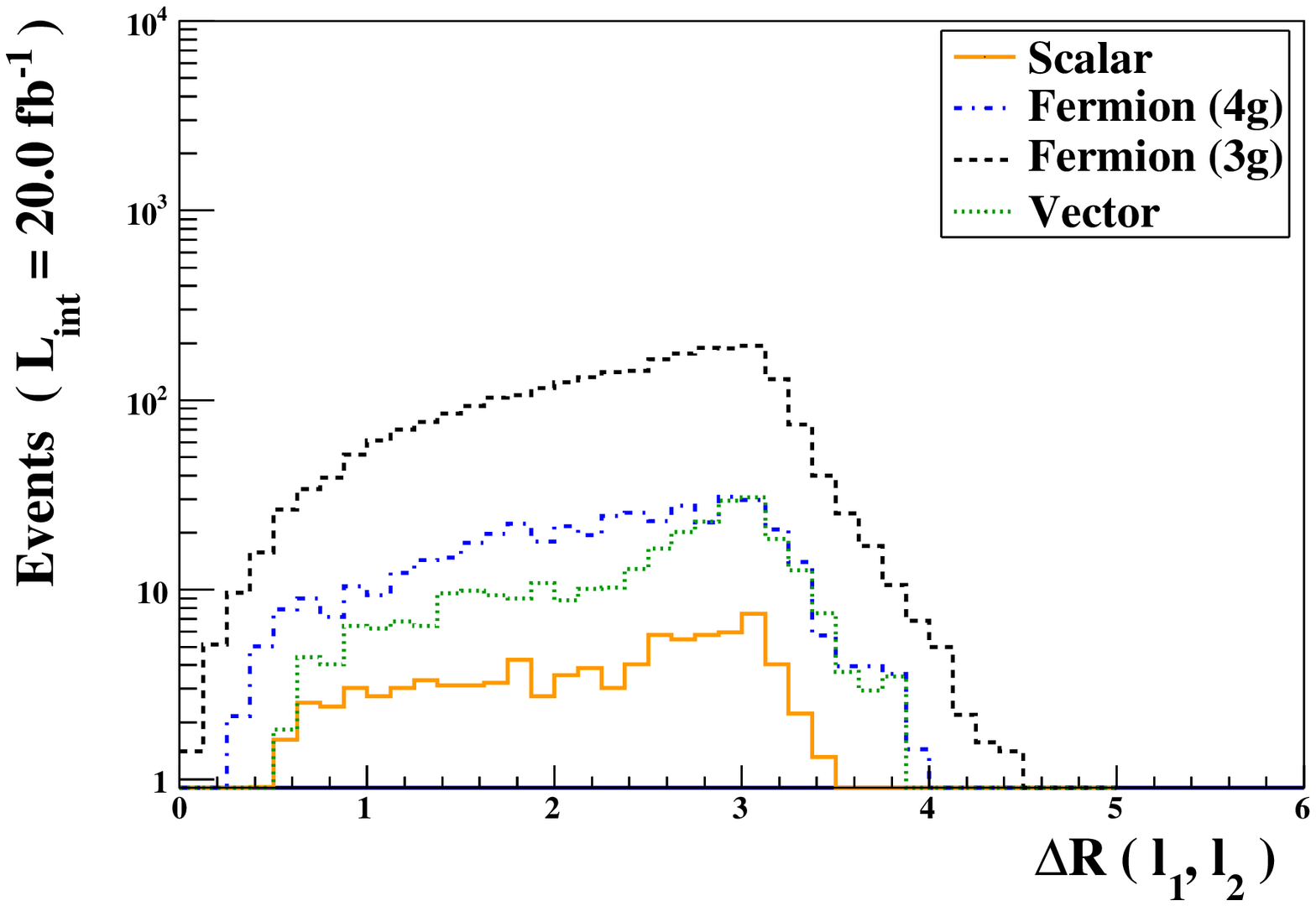} & \includegraphics[width=.32\columnwidth]{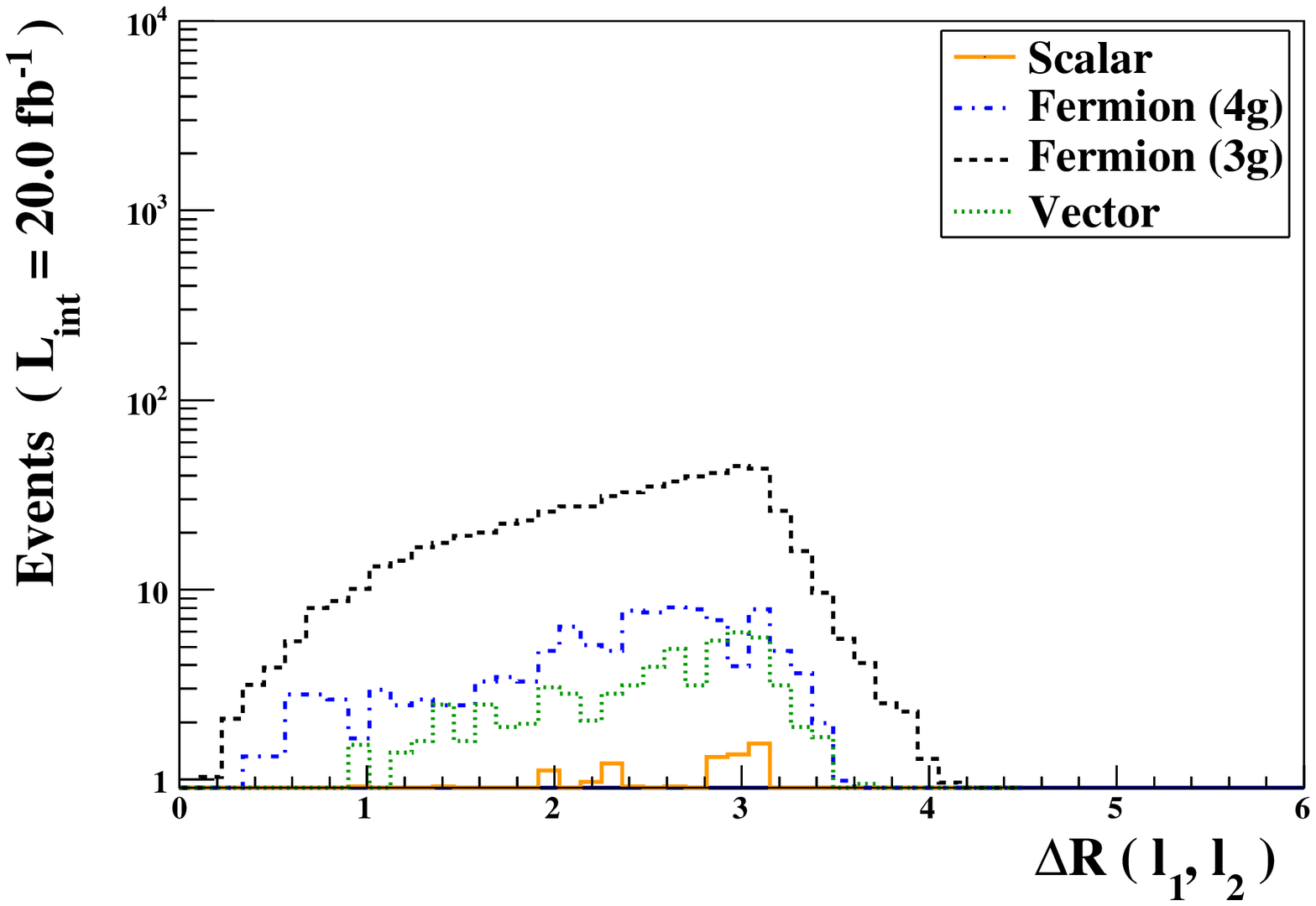}\\
\end{tabular}
\caption{\footnotesize\label{fig: db deltar} Evolution of the angular distance between the leading and next-to-leading leptons when both the mass (from left to right: 100, 250 and 350 GeV) and the isospin representation (from top to down: singlet, doublet and triplet under $SU(2)_L$) are varied. In these plots, ``Fermion (3g)" represents the simplified model where the Standard Model tau lepton does not mix with the new multiplet while ``Fermion (4g)" represents the other case where the mixing leads to four generations in the lepton sector. Apart from the peculiar behaviour of the vector field visible in the 100 GeV column (left) and which disappears for higher masses no real general behaviour can be extracted from this first serie of plots.}
\centering\begin{tabular}{c c c}
\includegraphics[width=.32\columnwidth]{Doublycharged/plots/PTL1_singlets_100.eps} & \includegraphics[width=.32\columnwidth]{Doublycharged/plots/PTL1_singlets_250.eps} & \includegraphics[width=.32\columnwidth]{Doublycharged/plots/PTL1_singlets_350.eps}\\
\includegraphics[width=.32\columnwidth]{Doublycharged/plots/PTL1_doublets_100.eps} & \includegraphics[width=.32\columnwidth]{Doublycharged/plots/PTL1_doublets_250.eps} & \includegraphics[width=.32\columnwidth]{Doublycharged/plots/PTL1_doublets_350.eps}\\
\includegraphics[width=.32\columnwidth]{Doublycharged/plots/PTL1_triplets_100.eps} & \includegraphics[width=.32\columnwidth]{Doublycharged/plots/PTL1_triplets_250.eps} & \includegraphics[width=.32\columnwidth]{Doublycharged/plots/PTL1_triplets_350.eps}\\
\end{tabular}
\caption{\footnotesize\label{fig: db ptl1} Evolution of the transverse momentum distribution when both the representation and the mass of the new states are varied.}
\end{figure}

\paragraph{Transverse momenta of the leptons} We now turn to the transverse momenta of the charged light leptons. We find that the transverse momenta of the three most energetic leptons lead to similar distributions, and choose to focus only on the leading one. Figure \ref{fig: db ptl1} illustrates the variation of the distribution associated to this observable when both the mass and the representation changes.\\

As a first general remark, one sees from the figure that all distributions increase sharply peaking approximately at the same value (for a given histogram), apart from the cases where the new field is a singlet and has a mass of 250 or 350 GeV. One remarks also that, when the new field transforms as a doublet under $SU(2)_L$ both the number of events and the tails are larger. The maximum being reached for a vector field whose mass is equal to 100 GeV where the tail extends up to values of the $p_T$ larger than 1 TeV. Finally, the doublet cases, {\ie} when the new fields belong to the fundamental representation of $SU(2)_L$, is the one where comparisons are made easier. Be it for a mass of 100 GeV, 250 GeV or 350, all the cases lead to quiet different distributions. The triplet cases, on the contrary, are more difficult to disentangle. 

\paragraph{Invariant mass} Another variable that might help us in answering the first two questions is the invariant mass of the pair leading - next-to-leading lepton. Following the same convention as above for the display of the histograms, we depict in figure \ref{fig: db ml1l2} the variation of the invariant mass as a function of the various parameters.
\begin{figure}
\centering\begin{tabular}{c c c}
\includegraphics[width=.32\columnwidth]{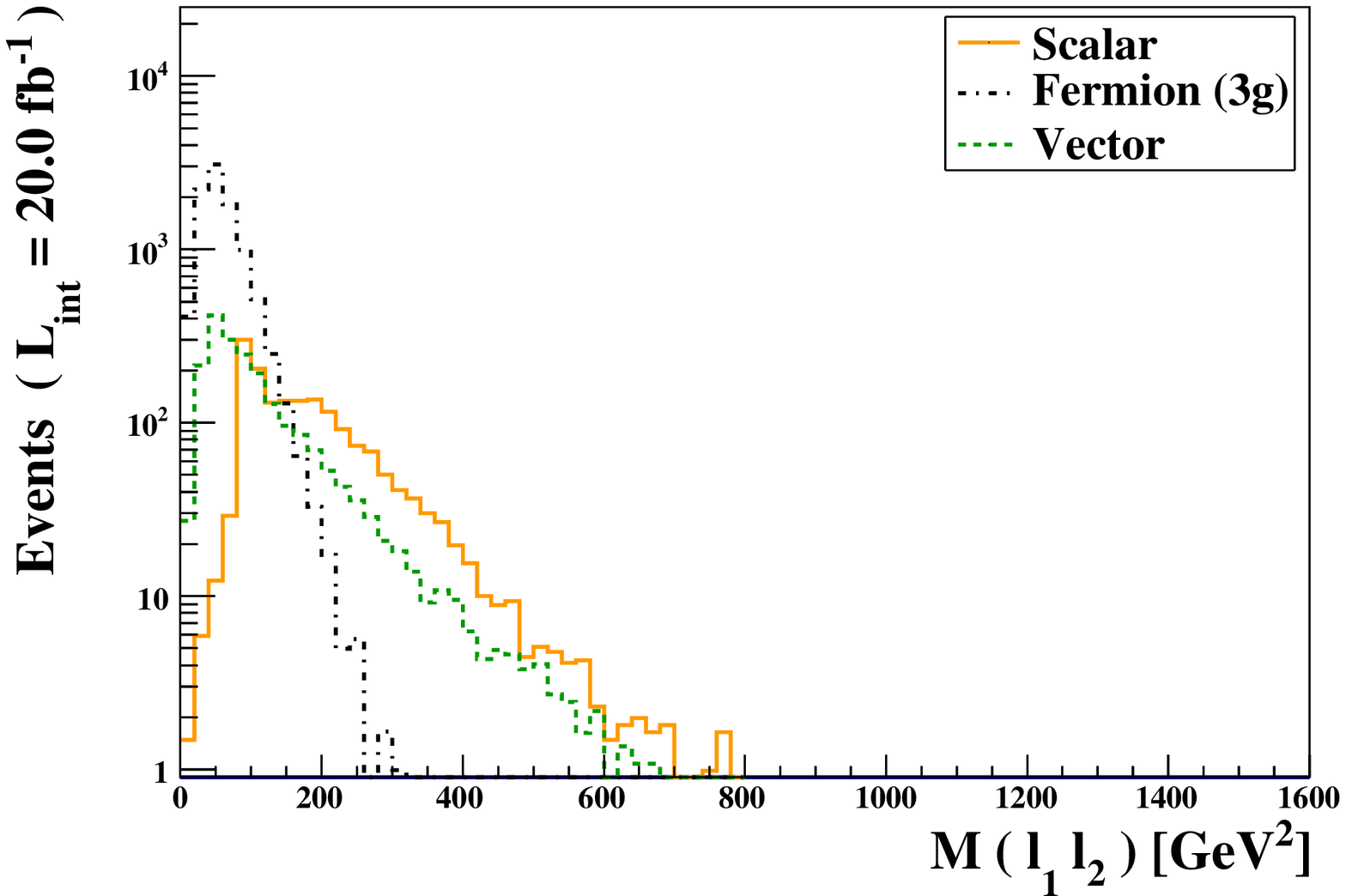} & \includegraphics[width=.32\columnwidth]{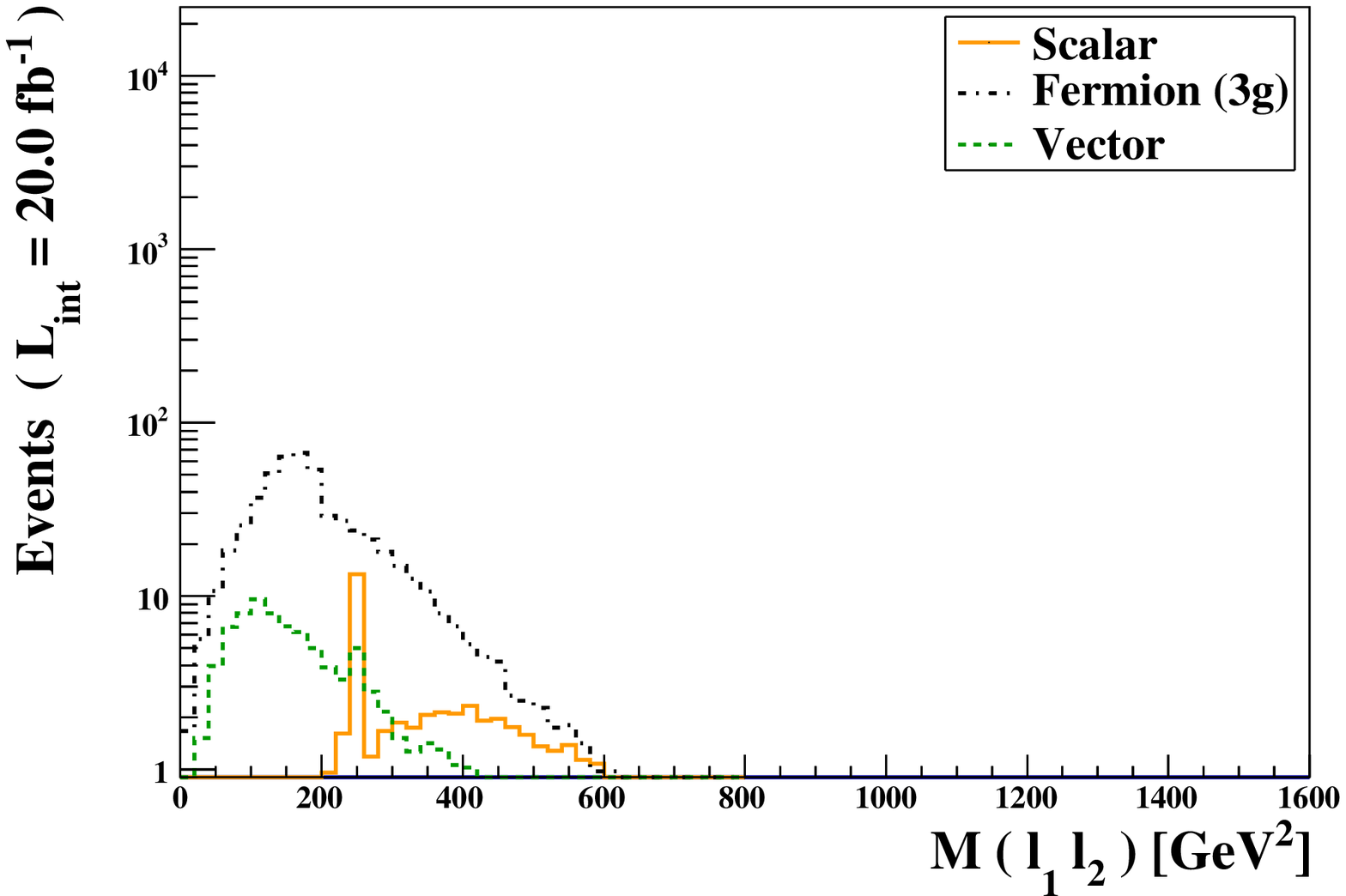} & \includegraphics[width=.32\columnwidth]{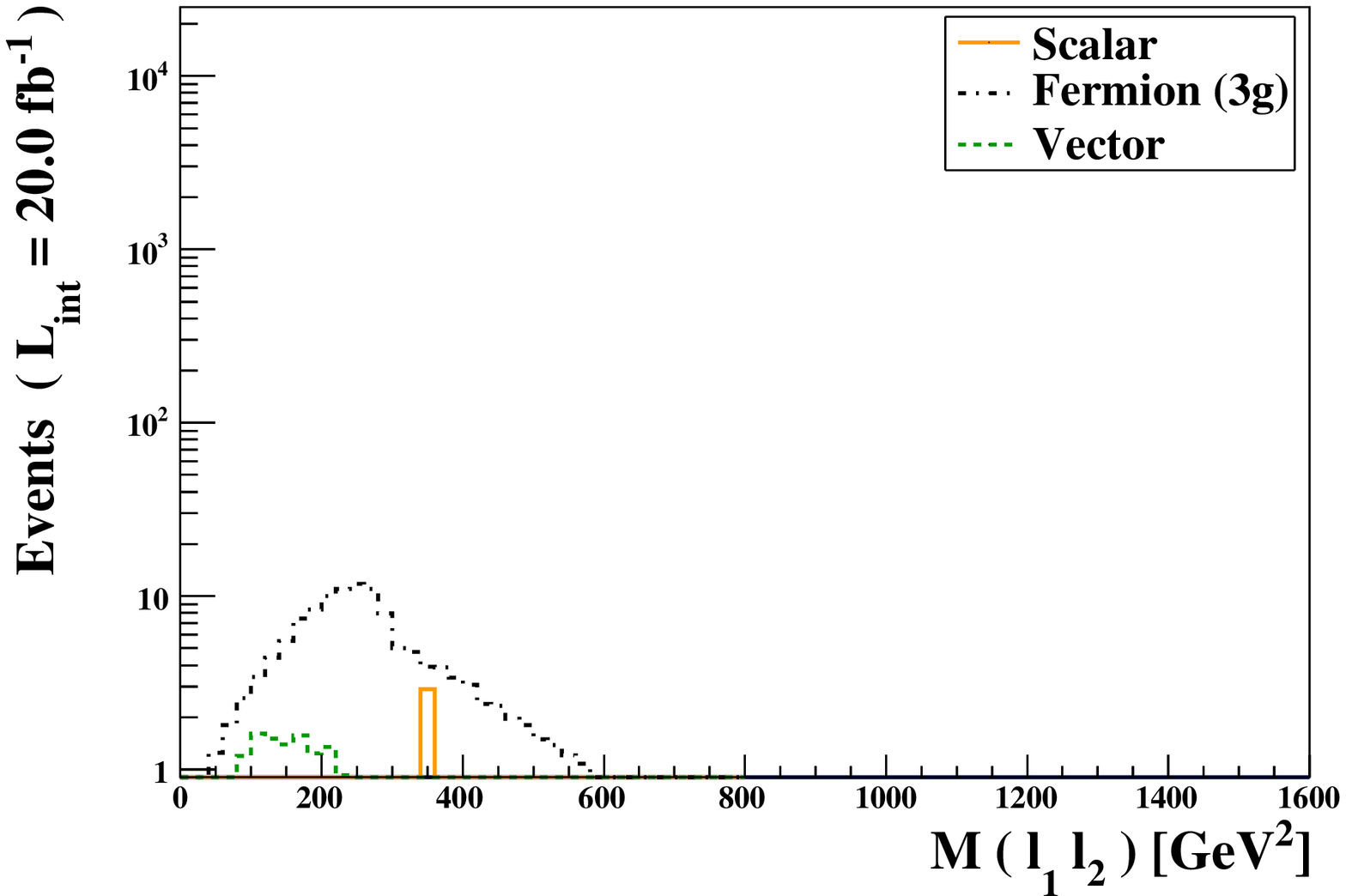}\\
\includegraphics[width=.32\columnwidth]{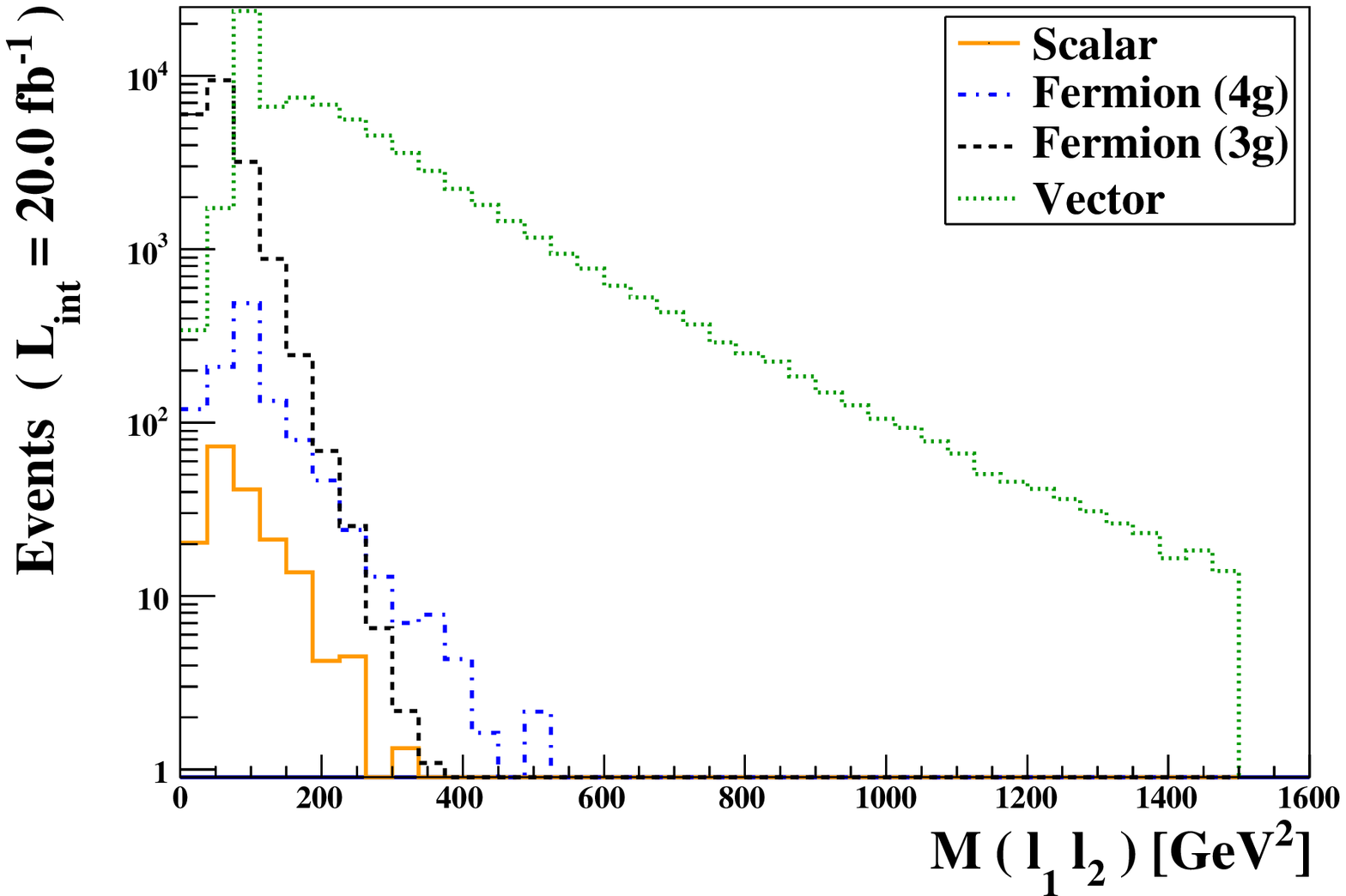} & \includegraphics[width=.32\columnwidth]{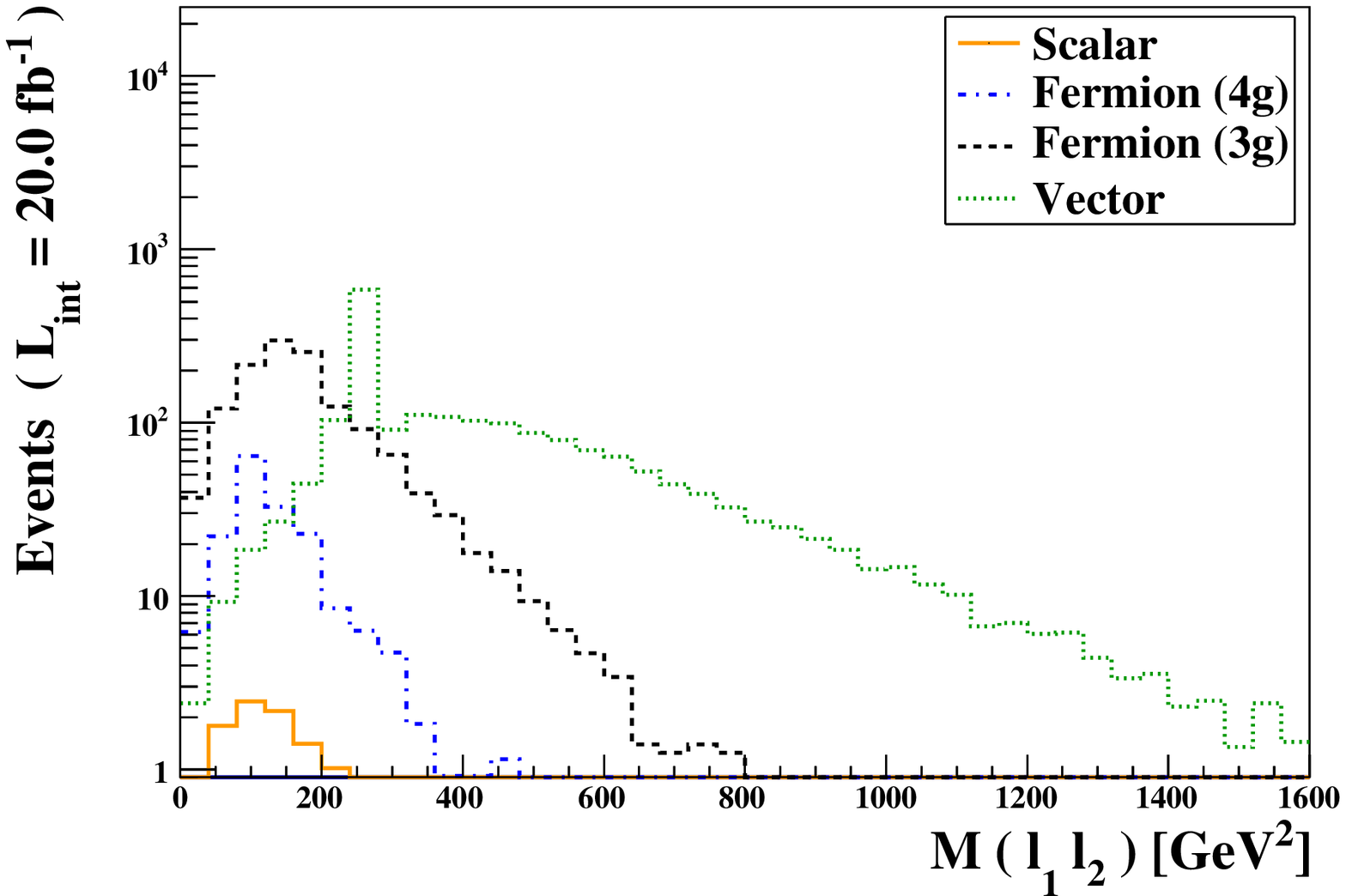} & \includegraphics[width=.32\columnwidth]{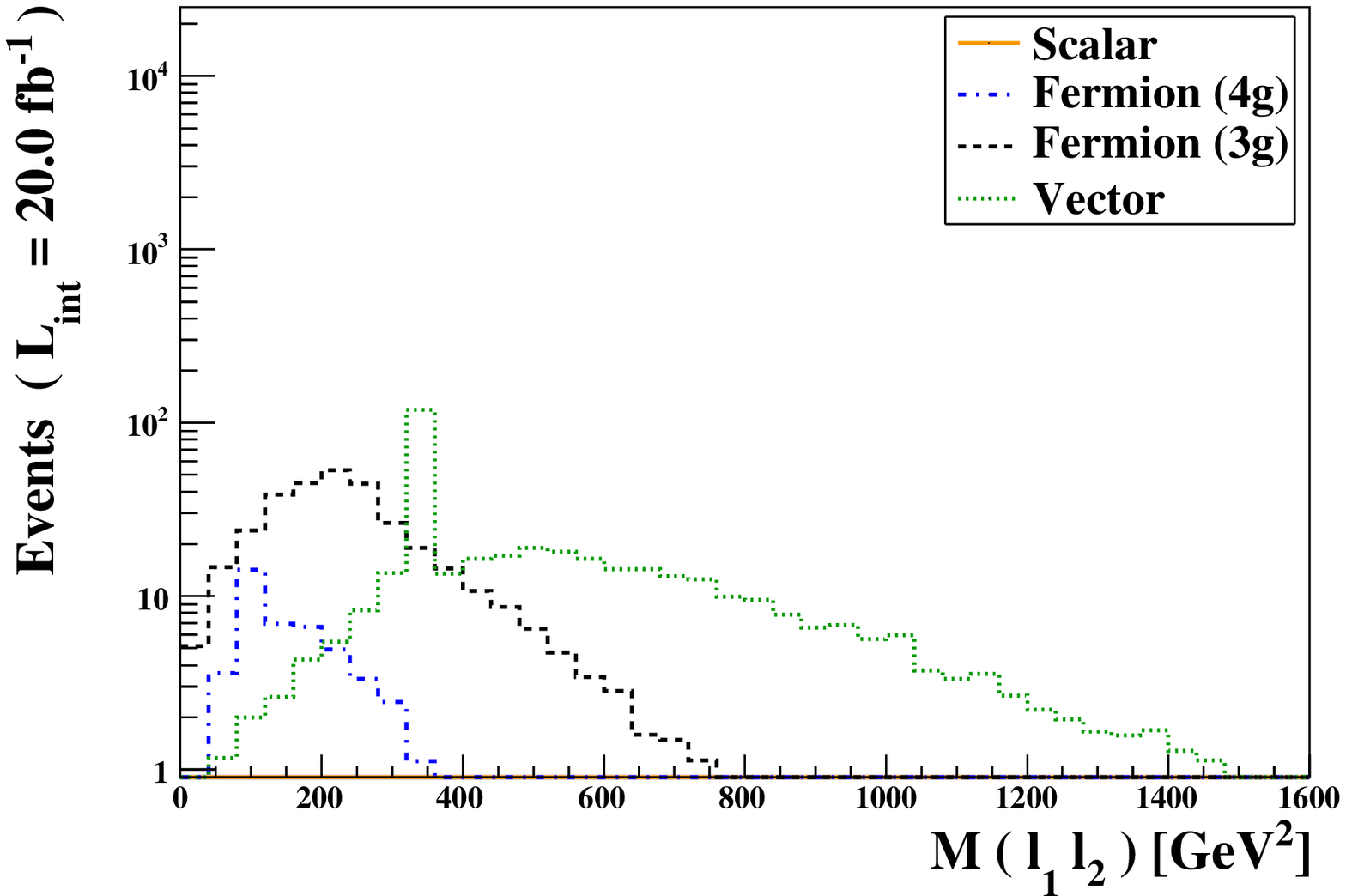}\\
\includegraphics[width=.32\columnwidth]{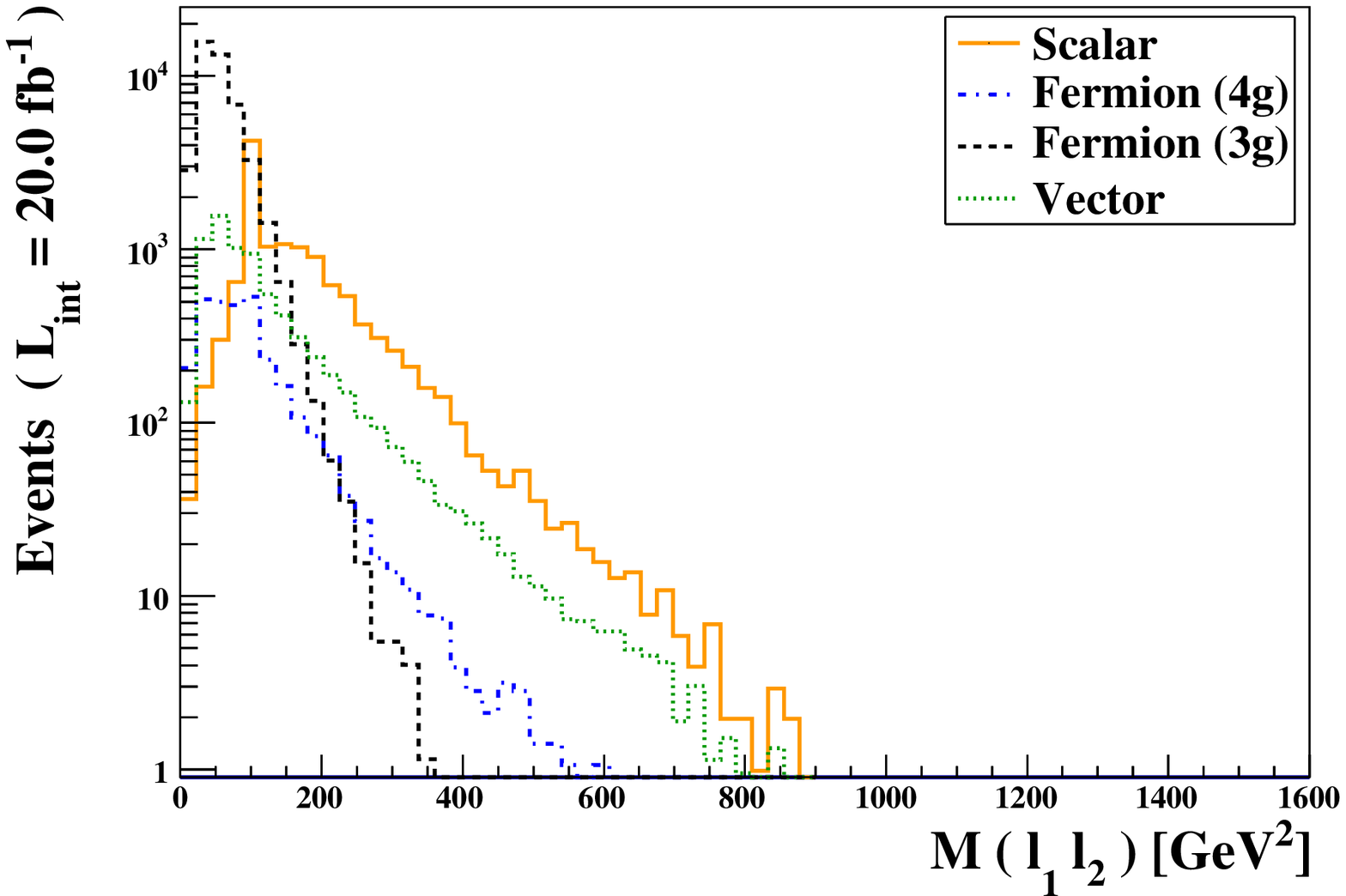} & \includegraphics[width=.32\columnwidth]{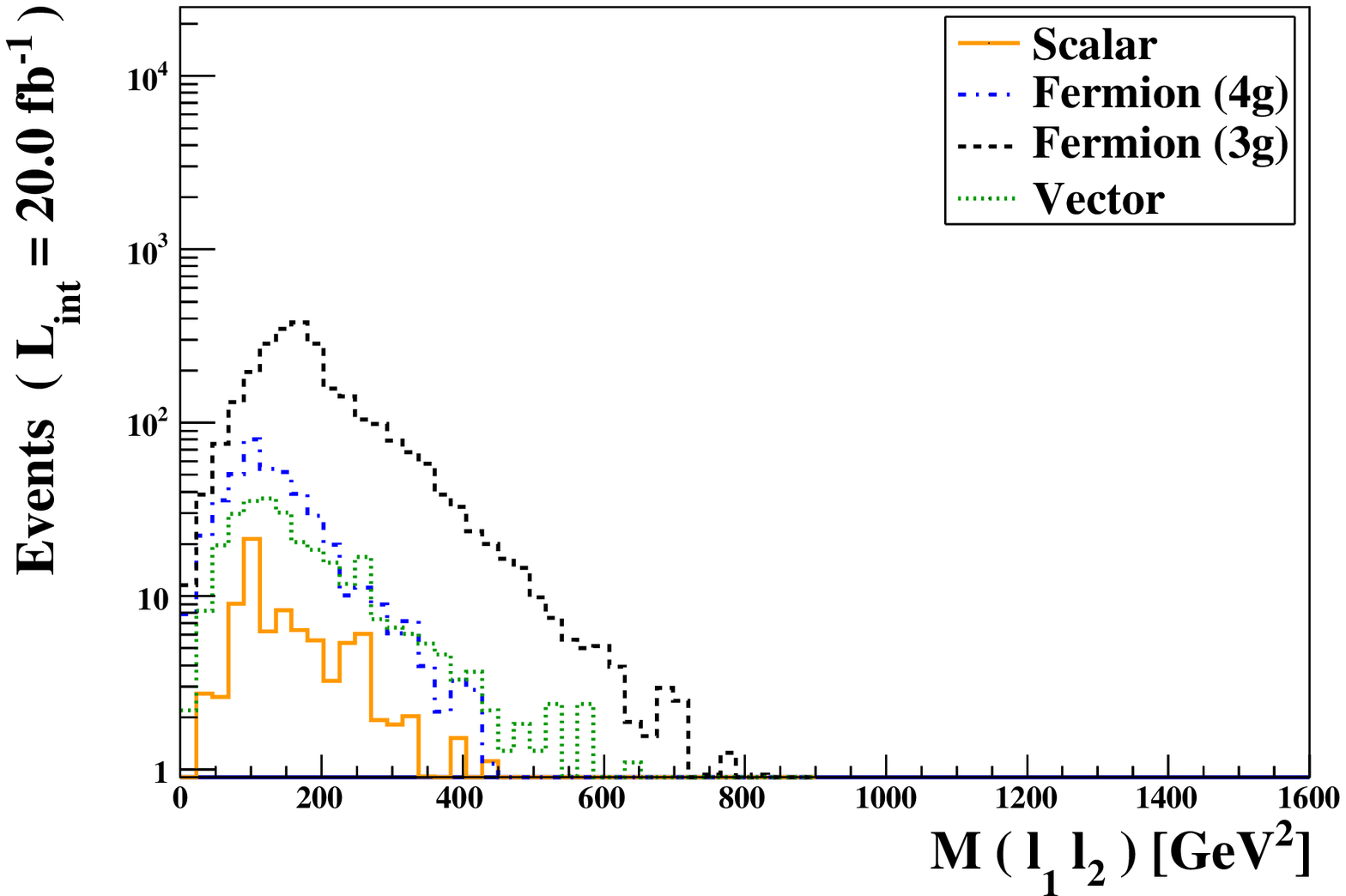} & \includegraphics[width=.32\columnwidth]{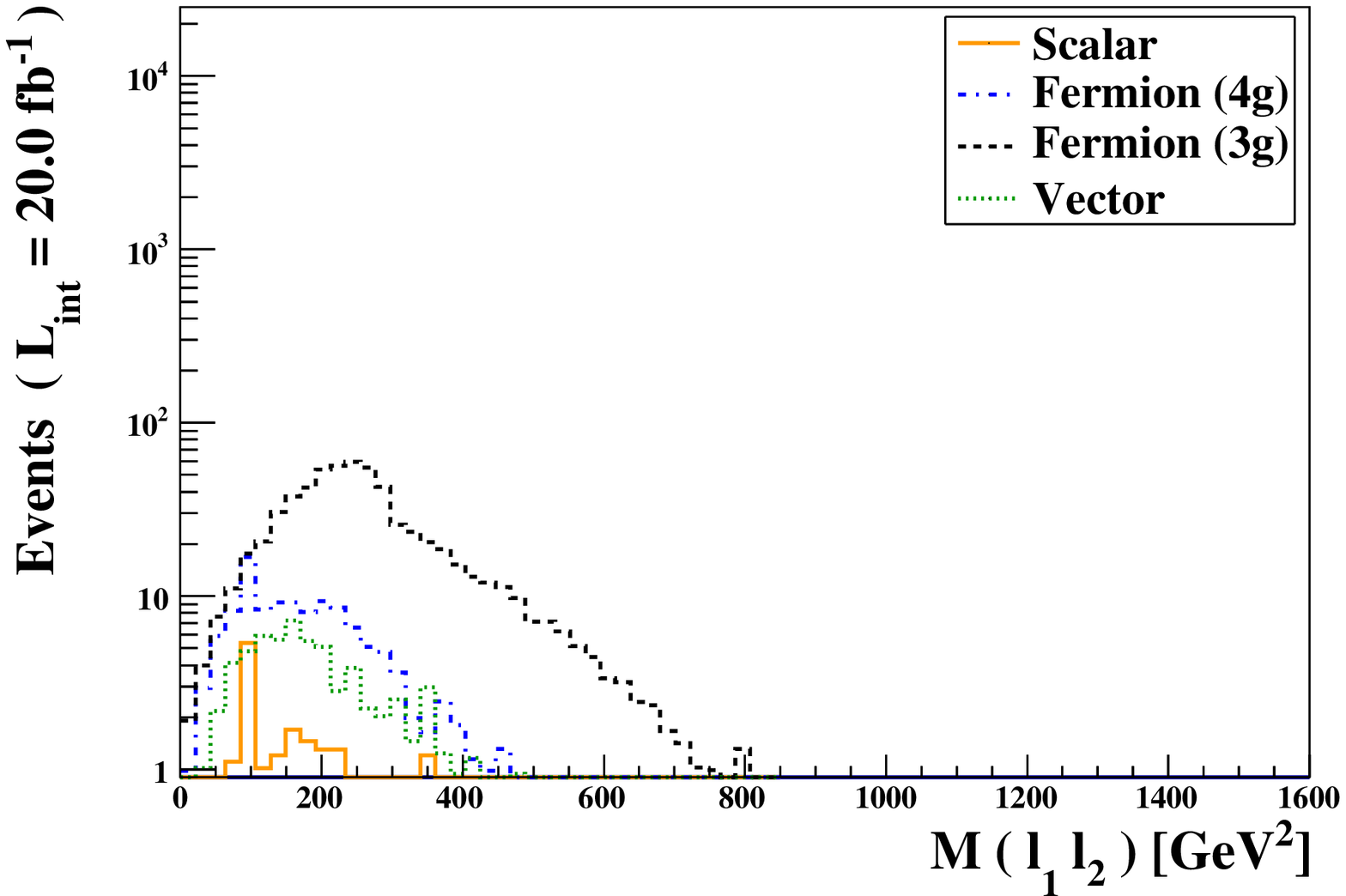}\\
\end{tabular}
\caption{\footnotesize\label{fig: db ml1l2} Evolution of the transverse momentum distribution when both the representation and the mass of the new states are varied.}
\end{figure}
Splitting the discussion line by line, we see that in the singlet case the differences between the various spin states become larger with the mass, the worst case scenario being when the new fields have a mass of 100 GeV. In the latter, however, the scalar and vector fields are quiet different from the fermion model with a three generation of leptons Standard Model. For higher masses, scalar fields are characterized by a clear peak around the mass of the new heavy doubly charged particle. \\

Turning to the doublet cases, we immediately remark the predominance of the vector fields amongst all others. Here again, we remark that differences become more visible with an increasing mass but when the new doublet field has a mass of 100 GeV it becomes impossible to distinguish between scalar and both fermionic fields. The triplet fields leading to the distributions depicted in the last line of figure \ref{fig: db ml1l2} seem to exhibit much less differences than in the other cases. More surprisingly, with an increasing mass the differences seem to reduce except maybe for the case where the new fermionic multiplet does not mix with the tau lepton whose tail is different from others'.

\paragraph{Five and six lepton final states} As indicated in table \ref{tab: db summary}, three models predict a fifth charged light lepton in the final state. The model with a new fermion transforming as a doublet and whose singly-charged component mixes with the tau lepton predicts even a sixth lepton. It was also found during the general discussions about the evolution of the various cross sections that these events could be associated with cross sections higher than 1 fb, for a given mass range, leading thus to potentially observable effects. The numerical simulation however shows that all of them but the one predicting a sixth lepton, have a fifth lepton whose transverse momentum is smaller than 10 GeV, not passing thus the selection criteria. The model with the fermion doublet and a four-generation SM has a small number of events surviving to the selection criteria but the $p_T$ remains very close to the threshold.

\section{Conclusion and outlook} \label{sec: dc conclusion}
In this chapter we have presented a systematic and model independent study on the production of doubly-charged particles at the LHC. Systematic because we have considered various types of doubly-charged particles varying their spin and the isospin representation to which they belong independently. The only constrain here having been to envisage only the most common representations allowing thus the doubly-charged particle to be a scalar, a fermion or a vector field transforming as a singlet, a doublet or a triplet under $SU(2)_L$. Model-independent because we did not consider a specific model for each case but rather constructed the Lagrangian only obeying to both constraints of gauge invariance and minimality, at the cost of having non-renormalizable terms in the Lagrangian, for some cases.\\

From the Lagrangians, we have derived the analytic expressions for both the cross sections associated with events leading eventually to multilepton final states (\textit{i.e.} at least three leptons) and the decay widths of the new fields. After a first numerical analysis, we have derived bounds on the masses of the new particles in order for these processes to be visible in the context of the 2012 run of the LHC. Finally we use these bounds and the fact that no real experimental limit exist on these doubly-charged particles to construct three scenarios from each simplified model in order to analyze, in the framework of a Monte Carlo simulation, the kind of observables one could consider to differentiate between various models of doubly-charged particles.\\

In this context we have explored various kinematic variables and found that one could focus on the angular distance between the leading and the subleading lepton, their transverse momentum and their invariant mass in order to distinguish between the various models. The combined analysis of these observables helping sometimes to disentangle some very close scenarios. Finally, though final states with five or six leptons are possible, we find that both cross sections and the momenta of the fifth and sixth lepton hardly pass the 1fb and 10 GeV thresholds respectively.\\

A few questions remain however unanswered, calling thus for extensions of this work. For example, one might wonder how these results are altered when moving from an ideal to a real detector? Another interesting question, calling for a broader collaboration, would be to use the {\sc CMS} and/or {\sc ATLAS} data in order to re-interpret them and eventually derive bounds holding for promptly decaying doubly-charged particles.

\chapter{Putting things into perspective}\label{chap: conclusion}
Considering the Standard Model of particle physics as an effective theory only valid at low energies is nowadays a common belief. Since early '70s, theorists have been trying to build new models in order to answer the open questions of the Standard Model. However, not a single experiment of those that have been conducted until now has been able to bring a clear answer to know which theory describes best the fundamental interactions. The Large Hadron Collider which focuses all the hopes has already started to reveal exciting results. The discovery of a Higgs boson of a mass of 126 GeV is a major breakthrough as it helps us in better understanding the electroweak symmetry breaking mechanism. Though further work is still needed to measure precisely the properties of this particle in order to know if it is the Standard Model Higgs boson or not, the mass of this particle puts stringent constraints on the models of new physics. In the context of supersymmetry, its minimal realization, the Minimal Supersymmetric Standard Model, has concentrated a lot of attention. An enormous phenomenological work has thus been performed and nowadays this model seems not to be in good shape, especially its constrained version. The discovery of the scalar field for example rises many questions as we have seen in chapter \ref{chap:susy}. On the other hand, non minimal supersymmetric models do not suffer from really stringent constraints and are thus not ruled out at all.\\

It is in this context that my first project has been conducted. The analysis presented in chapter \ref{chap:lrsusy} proposes to investigate in a \textit{top-down} approach the production of charginos and neutralinos at the Large Hadron Collider in the context of a left-right symmetric supersymmetric model. To this aim, we have we have built the model from some simple considerations, lifted some ambiguities in the definitions of some quantities relative to group theory and calculated the spectrum of the latter particles. In the second part of the work, we have built four scenarios representative of the Model and performed a full Monte Carlo simulation and a comparison with the Standard Model background. Only focusing on signatures where at least one charged light lepton is present, we have found that left-right symmetric supersymmetric models were very promising in final states with a multiplicity of leptons higher than two. For both the single lepton and dilepton signatures, the results were somehow mitigated as only one of our scenarios predicted enough events to be distinguishable from the background. We have also performed a comparison with the Minimal Supersymmetric Standard Model and found that these two models had very different phenomenologies which makes them easy to distringuish. \\

The technical details of the latter analysis show the importance of the automated tools in such analyses. The theoretical calculations we have made to extract the spectrum of the model for example, could not be carried analytically as the mass matrices for both the charginos and neutralinos were larger than $4\times4$. As to the Monte Carlo simulation in order to have a feeling of the signatures such models could lead at the Large Hadron Collider, it depends obviously on the use of computers. A \textit{top-down} approach can also rely, for a supersymmetric model like ours, on the use of the renormalization group equations. One then starts with some unification considerations at very high scale and then make use of the latter equations to know the low energy values of the parameters in the theory. Here also the use of a dedicated automated tool can help in gaining in efficiency and responsiveness. These considerations have motivated the second part of my work in which I have participated in the development of two automated tools allowing, for the first, to extract automatically the analytical expressions of the two-loop renormalization group equations associated with any supersymmetric renormalizable model. As to the second module, it allows to derive from a model that is implemented in {\sc FeynRules} the analytical expressions for the mass matrices and to generate automatically a {\sc C++} package able to diagonalize these mass matrices and return the spectrum together with the mixing matrices.\\

The last project I have contributed to is a phenomenological study of particles carrying a two-unit electric charge. This project has been initiated by the idea that such exotic particles that appear naturally in left-right symmetric models can also appear in the context of other different models. The idea was then to adopt a \textit{bottom-up} approach starting from the signature (a doubly-charged particle is produced) and to construct an effective theory able to describe the interactions of this particle. In order to be as general as possible, we have allowed the new exotic particle to be a scalar, a fermionic or a vector field lying in the trivial, fundamental or adjoint representation of $SU(2)_L$. This has defined our nine effective models. After analyzing the various production and decay modes of these particles, we have constructed three benchmark scenarios in order to find the key observables one could focus on in order to distinguish between the various doubly-charged particles. After analyzing the transverse momenta of the leptons, the invariant mass and the angular distance of the the pair leading - next-to-leading lepton, we have found that using only one of these variables could lead sometimes to ambiguous results while combining them enhances the discrimination power. \\

I hope that these phenomenological excursions have triggered in the reader's mind some interest in this kind of non-minimal models. Indeed, contrary to the famous models like the Minimal Supersymmetric Standard Model which parameter space is more and more constrained, non-minimal models are generally less constrained and may lead to very interesting phenomenology. Speaking of phenomenology, I think that several questions deserve some further investigations. For example, in the case of left-right symmetries, the investigation of the link between the high energy considerations (unification of gauge coupling constants, scalar masses universality$\dots$) and the low energy signatures deserves more attention. In the case of the doubly-charged particles,  an obvious extension of this work would be to re-analyze the data acquired by the Large Hadron Collider in order to see how constrained such particles could be. This would require obviously a broader collaboration with experimentalists.

\appendix
\chapter{Conventions}\label{annex:susy}
\section{Generalities}
In this appendix, we present the conventions that shall hold for the whole manuscript, except when stated otherwise. The Minkowski metric $\eta_{\mu \nu}$ and its inverse $\eta^{\mu \nu}$ are given by 
$$\eta_{\mu\nu} = \eta^{\mu\nu} = {\rm diag}(1,-1,-1,-1).$$
Lorentz indices will be denoted by Greek letters, gauge indices, that is, indices in the adjoint representation of a group $G$ by latin letters of the beginning of the alphabet $(a,b,\dots)$, while those of the middle of the alphabet $(i,j,\dots)$ will denote indices in the fundamental representation.\\

The Levi-Civita tensor is denoted $\epsilon$ and normalized to $\epsilon_{0123} = 1$. If we define
$$ \epsilon^{\alpha \beta \gamma \delta} = \epsilon_{\mu\nu\rho\sigma} \eta^{\alpha\mu} \eta^{\beta \nu} \eta{\gamma\rho}\eta{\delta\sigma} $$
then $\epsilon^{0123} = -1$.\\

Pauli $\sigma$ Matrices are
$$ \sigma^1 = \begin{pmatrix} 0 & 1 \\ 1 & 0 \end{pmatrix}, ~~ \sigma^2 = \begin{pmatrix} 0 & -i \\ i & 0 \end{pmatrix}, ~~ \sigma^3 = \begin{pmatrix} 1 & 0 \\ 0 & -1 \end{pmatrix} .$$
Care shall be taken to the position of the indices as $\sigma^i = - \sigma_i$. We also introduce 
$$\sigma^\mu = ( \sigma^0, \sigma^i),~~~ \bar{\sigma}^\mu = ( \sigma^0, -\sigma^i) $$
where $\sigma^0$ is a two by two identity matrix. We subsequently define the Dirac matrices $\gamma^\mu$ in the Weyl representation as
$$ \gamma^\mu = \begin{pmatrix} 0 & \sigma^\mu \\ \bar{\sigma}^\mu & 0 \end{pmatrix}.$$
The $\gamma_5$ matrix reads then
$$ \gamma_5 =i \gamma_0 \gamma_1 \gamma_2\gamma_3 = \begin{pmatrix} -1 & 0 \\ 0 & 1 \end{pmatrix}. $$
These matrices satisfy
$$ \{\gamma^\mu, \gamma^\nu\} = 2\eta^{\mu\nu}$$
and we define
$$ \gamma^{\mu\nu} = \frac13 [\gamma^\mu,\gamma^\nu]$$
where $[,]$ is the commutation operation.

\section{Two and four component spinors}
Weyl spinors are written following van der Waerden notations
$$ \lambda_L \to \lambda_\alpha, ~~~ \bar{\chi}_R \to \bar{\chi}^{\alphadot} $$
where $\lambda_L $ (respectively $\bar{\chi}_R$) is a left- (right-) handed spinor. Spinorial indices (noted with greek letters $\alpha, \beta, \dots$) are lowered and raised by the mean of the totally antisymmetric tensor of rank two $\epsilon$ that we define so that
$$ \epsilon_{1 2} = - \epsilon^{12} = 1,~~~ \epsilon_{\dot{1} \dot{2}} = - \epsilon^{\dot{1}\dot{2}} = 1,$$
and
$$ \epsilon_{\alpha \beta} \epsilon^{\beta \gamma} = \delta_\alpha{}^\gamma, ~~  \epsilon_{\dot{\alpha} \dot{\beta}} \epsilon^{\dot{\beta} \dot{\gamma}} = \delta_{\dot{\alpha}}{}^{\dot{\gamma}}$$
For the summation of the indices we take as a convention
$$ \lambda\cdot \lambda' = \lambda^\alpha \lambda'_\alpha, ~~~ \bar{\chi}\cdot\bar{\chi}' = \bar{\chi}_{\alphadot} \bar{\chi}'^{\alphadot}$$.
With these conventions, Dirac ($\psi_D$) and Majorana ($\psi_M$) spinors are defined as
$$ \psi_D = \begin{pmatrix} \lambda_\alpha \\ \bar{\chi}^{\alphadot} \end{pmatrix},~~~\psi_M = \begin{pmatrix} \lambda_\alpha \\ \bar{\lambda}^{\alphadot} \end{pmatrix}$$
with $\lambdabar_{\alphadot}$ is the complex conjugate of the Weyl spinor $\lambda^\alpha$. We also define the Dirac conjugate of a Dirac spinor as
$$ \bar{\psi}_D = \psi_D^\dagger\gamma_0 $$
and the charge conjugate spinor $\psi_D^c$ as
$$ \psi_D^c = C \bar{\psi}^t_D $$
where $C$ is the charge conjugation matrix defined as
$$ C = i \gamma^0 \gamma^2. $$
Finally, we define $\sigma^{\mu\nu} \et \bar{\sigma}^{\mu\nu}$ as
\bea 
\sigma^{\mu\nu}{}_\alpha{}^\beta = \frac{1}{4} ( \sigma^\mu{}_{\alpha\dot{\gamma}}\bar{\sigma}^{\nu}{}^{\dot{\gamma}\beta} - \sigma^\nu{}_{\alpha\dot{\gamma}} \bar{\sigma}^{\mu\dot{\gamma}\beta} )~~~\et~~~
\bar{\sigma}^{\mu\nu\dot{\alpha}}{}_{\dot{\beta}}= \frac{1}{4}(\bar{\sigma}^{\mu\dot{\alpha}\gamma}\sigma^{\nu}{}_{\gamma \dot{\beta}} - \bar{\sigma}^{\nu\dot{\alpha}\gamma}\sigma^{\mu}{}_{\gamma \dot{\beta}} )\nonumber
\eea
\subsection{Fundamental identities}
Let $\psi, \lambda \et \chi$ be three left-handed Weyl fermions. We have the following fundamental relations amongst these fields
\begin{align*}
\psi\cdot\lambda &= \lambda\cdot \psi ,~~~ (\psi\cdot\lambda)^\dagger = \psibar\cdot\lambdabar ;\\
\bar{\sigma}^{\mu\alphadot\alpha} &= \sigma^\mu{}_{\beta\betadot} \epsilon^{\alpha\beta}\epsilon^{\betadot\alphadot};\\
\psibar\sigmabar^\mu \lambda &= - \lambda \sigma^\mu \psibar ;\\
\psi\sigma^\mu\sigmabar^{\nu}\lambda &= \lambda\sigma^\nu\sigmabar^\mu\psi ;\\
\chi_\alpha (\psi\cdot\lambda) &= - \psi_\alpha (\chi\cdot\lambda) - \lambda_\alpha (\chi\cdot\psi)~~~(\rm Fierz~identity);\\
\frac12 \epsilon_{\mu\nu\rho\sigma}\sigma^{\rho\sigma} &= -i \sigma_{\mu\nu},~~\frac12 \epsilon_{\mu\nu\rho\sigma}\sigmabar^{\rho\sigma}=i\sigmabar_{\mu\nu};\\
\sigmabar^\mu\sigma^\nu\sigmabar^\rho + \sigmabar^\rho\sigma^\nu\sigmabar^\mu &= 2( \eta^{\mu\nu}\sigmabar^\rho + \eta^{\rho\nu}\sigmabar^\mu - \eta^{\mu\rho}\sigmabar^\nu);\\
\sigmabar^\mu\sigma^\nu\sigmabar^\rho - \sigmabar^\rho\sigma^\nu\sigmabar^\mu &= 2i\epsilon^{\mu\nu\rho\sigma}\sigmabar_\sigma ;\\
\sigma^\mu_{\alpha\alphadot}\sigmabar_{\mu}^{\betadot\beta} &= 2\delta_\alpha{}^\beta\delta_{\alphadot}{}^{\betadot},~~~\sigma^\mu{}_{\alpha\alphadot}\sigma_{\mu\beta\betadot} = 2\epsilon_{\alpha\beta}\epsilon{\alphadot\betadot}.
\end{align*}

To carry the calculations in superspace, we let 
$$\theta \et \thetabar $$
be the components of a Majorana spinor, define the derivation operations $\partial_\alpha \et \partial_{\alphadot}$ such that
$$\{ \partial_\alpha, \theta^\beta \} = \delta_\alpha{}^\beta, ~~~ \{ \bar{\partial}_{\alphadot}, \thetabar^{\dot{\beta}} \} = \delta_{\alphadot}{}^{\dot{\beta}}$$
the other anticommutators being equal to zero. The fundamental identities useful for the derivation of the various equations presented in chapter \ref{chap:susy} are
\begin{align*}
\theta^\alpha \theta^\beta &= -\frac12 \theta\cdot\theta \epsilon^{\alpha\beta},~~\bar{\theta}^{\alphadot}\bar{\theta}^{\betadot} = \frac12 \thetabar\cdot\thetabar \epsilon^{\alphadot\betadot},~~\theta^\alpha \bar{\theta}^{\alphadot} = \frac12 \theta \sigma^\mu\thetabar\sigmabar_{\mu}^{\alphadot \alpha} ;\\
\partial\cdot\partial (\theta\cdot\theta) &= -4,~~~ \bar{\partial}\cdot\bar{\partial} (\thetabar\cdot\thetabar) = -4;\\
(\psi\cdot\theta)~(\lambda\cdot\theta) &= -\frac12 (\theta\cdot\theta)~(\psi\lambda), ~~ (\theta\cdot\psi)~(\thetabar\cdot\lambdabar) = -\frac12(\theta\sigma^\mu\thetabar)~(\epsilonbar\sigmabar_\mu\lambda);\\
(\theta\sigma^\mu\lambdabar)~(\theta\cdot\psi) &= -\frac12 (\theta\cdot\theta)~( \psi\sigma^\mu\lambdabar),~~~ (\lambda\sigma^\mu\thetabar)~(\thetabar\cdot\psibar) = -\frac12(\thetabar\cdot\thetabar)~(\lambda\sigma^\mu\psibar );\\
(\lambda\sigma^\mu\thetabar)~(\theta\cdot\psi) &= -\frac12 (\theta\sigma^\mu\thetabar)~(\lambda\cdot\psi) - (\theta\sigma_\nu\thetabar)~( \lambda\sigma^{\mu\nu}\psi);\\
(\theta\sigma^\mu\lambdabar)~(\thetabar\cdot\psibar) &= -\frac12 (\theta\sigma^\mu\thetabar)~(\lambdabar\cdot\psibar) + (\theta\sigma_\nu\thetabar)~(\lambdabar\sigmabar^{\mu\nu} \psibar);\\
(\theta\sigma^\mu\thetabar)~(\theta\cdot\psi) &= -\frac12 (\theta\cdot\theta)~( \psi\sigma^\mu\thetabar);\\
(\theta\sigma^\mu\epsilonbar)~(\theta\sigma^\nu\thetabar) &= \frac12 (\theta\cdot\theta)~(\thetabar\cdot\epsilonbar)\eta^{\mu\nu} + (\theta\cdot\theta)~(\thetabar\sigmabar^{\mu\nu}\epsilonbar) ;\\
(\theta\sigma^\mu\thetabar)~(\theta\sigma^\nu\thetabar) &= \frac12(\theta\cdot\theta)~(\thetabar\cdot\thetabar)\eta^{\mu\nu};\\
\sigma^\mu_{\alpha\alphadot}\thetabar^{\alphadot} ~(\theta\sigma^\nu\thetabar) &= \frac12 (\thetabar\cdot\thetabar)~\sigma^{\mu}_{\alpha\alphadot}\sigmabar^{\alphadot\beta}\theta_{\beta};\\
\sigma^\mu_{\alpha\alphadot}\thetabar^{\alphadot} (\thetabar\cdot\lambdabar) &= -\frac12 (\thetabar\cdot\thetabar) \sigma^\mu_{\alpha\betadot}\lambdabar^{\betadot};\\
\theta^\alpha\sigma^\mu_{\alpha\alphadot}\sigmabar^{\mu\alphadot\beta} \sigma^{\rho}_{\beta\gammadot}\sigmabar^{\sigma\gammadot\gamma}\theta_{\gamma} &= (\theta\cdot\theta)~{\rm Tr}(\sigma^\mu\sigmabar^\nu\sigma^\rho\sigmabar^\sigma) = (\theta\cdot\theta)~(\eta^{\mu\nu}\eta^{\rho\sigma} + \eta^{\rho\nu}\eta^{\mu\sigma} - \eta^{\mu\rho}\eta^{\mu\sigma} - i\epsilon^{\mu\nu\rho\sigma});\\
\psi\sigma^{\mu\nu} &= - \lambda\sigma^{\mu\nu} \psi.
\end{align*}
In the above equations, the brackets denote that all spinorial indices are contracted, \textit{e.g.}
$$ (\theta\sigma^\mu\lambdabar) = \theta^\alpha \sigma^{\mu}_{\alpha\alphadot} \lambdabar^{\alphadot}.$$

\chapter{LRSUSY}\label{annex:lrsusy}
\section{Minimization of the scalar potential}
\bea
\frac{\partial V}{\partial v_{1L}} = 0 \!\!\!\!&=&\!\!\!\! v_{1L} \Bigg\{ -m_{\Delta_{1L}} - \mu_L^2 + g_{B-L}^2 \frac{ v_{2L}^2 + v_{2R}^2 - v_{1R}^2 - v_{1L}^2}{2} + g_{L}^2 \frac{ 2 v_{2L}^2 + v_1'^2 + v_2^2 - v_2'^2 -  v_1^2 - 2 v_{1L}^2 }{4} \n
  &&  - \lambda_L^2 \frac{  v_s^2 + v_{2L}^2}{2} - \sqrt{2}  \lambda_L  \mu_L v_s \Bigg\} \n
  && + v_{2L}\Bigg\{ \frac{\lambda_L}{2} \Big[ \cos\big(\alpha_1 +  \alpha_2\big) \lambda_3 v_1'  v_2' + \lambda_3 v_1  v_2 - \lambda_R  v_{1R}  v_{2R}   - 2\sqrt{2}  \mu_s  v_s  - 2B_L - \lambda_s v_s^2 - 2\xi_S \Big] - \frac{T_L v_s}{\sqrt{2}}\Bigg\};\n
\frac{\partial V}{\partial v_{2L}} = 0 \!\!\!\!&=&\!\!\!\! v_{2L}\Bigg\{- m_{\Delta_{2L}}  - \mu_L^2 + g_{B-L}^2 \frac{ v_{1R}^2 + v_{1L}^2  - v_{2L}^2 - v_{2R}^2}{2} + g_{L}^2 \frac{ v_1^2 + 2 v_{1L}^2 - v_2^2 - 2 v_{2L}^2 - v_1'^2 +  v_2'^2}{4}\n
&& -\lambda_L^2 \frac{ v_{1L}^2 +  v_s^2}{2} - \sqrt{2}  \lambda_L  \mu_L  v_s  \Bigg\}\n
&& + v_{1L} \Bigg\{ - B_L - \frac{T_L  v_s}{\sqrt2} 
+ \frac{\lambda_L}{2} \Big[ \lambda_3 v_1 v_2 + \cos\big( \alpha_1 +\alpha_2\big)\lambda_3 v_1' v_2' - \lambda_R v_{1R}  v_{2R}  - \frac{1}{\sqrt2} \mu_s v_s - \lambda_s  v_s^2 - 2\xi_S\Big]
\Bigg\};\n
\frac{\partial V}{\partial v_{1R}} = 0 \!\!\!\!&=&\!\!\!\! v_{1R}\Bigg\{  - m_{\Delta_{1R}} - \mu_R^2 g_{B-L}^2\frac{ - v_{1L}^2  - v_{1R}^2 + v_{2L}^2 + v_{2R}^2}{2}  + g_{R}^2 \frac{ v_1^2  - 2 v_{1R}^2  - v_2^2 +  2v_{2R}^2  -  v_1'^2 + v_2'^2}{4}\n
&& - \lambda_R^2 \frac{v_{2R}^2 + v_s^2}{2} - \sqrt{2} \lambda_R \mu_R \Bigg\} \n
&& + v_{2R}\Bigg\{ \frac{\lambda_R}{2} \Big[ \lambda_3 v_1  v_2 - \lambda_L  v_{1L}  v_{2L} + \cos\big(\alpha_1 +  \alpha_2\big)  \lambda_3  v_1'  v_2' - \mu_s \frac{v_s}{\sqrt2} - \lambda_s v_s^2- 2\xi_S \Big] - T_R \frac{v_s}{\sqrt{2}}- B_R \Bigg \};\n
\frac{\partial V}{\partial v_{2R}} = 0 \!\!\!\!&=&\!\!\!\! v_{2R}\Bigg\{- m_{\Delta_{2R}} - \mu_R^2 + g_{B-L}^2 \frac{ v_{1L}^2  + v_{1R}^2 -  v_{2L}^2 - v_{2R}^2}{2} + g_{R}^2 \frac{- v_1^2 + 2 v_{1R}^2  + v_2^2 - 2 v_{2R}^2 + v_1'^2 - v_2'^2}{4}\n
  && - \lambda_R^2 \frac{ v_{1R}^2 + v_s^2}{2}  - \sqrt{2} \lambda_R \mu_R v_{2R}  v_s \Bigg\} \n
  && + v_{1R}\Bigg\{ \frac{\lambda_R}{2} \Big[\lambda_3 v_1 v_2 - \lambda_L v_{1L} v_{2L} + \cos\big( \alpha_1 + \alpha_2\big) \lambda_3 v_1' v_2' - \frac{1}{\sqrt{2}} \mu_s v_s - \lambda_s v_s^2 - 2\xi_S \Big] -B_R  - T_R \frac{v_s}{\sqrt{2}}\nonumber;
\eea
\bea
\frac{\partial V}{\partial v_{1}} = 0 \!\!\!\!&=&\!\!\!\! v_1 \Bigg\{ -(m_{\Phi})^{11}  - \mu_3^2 + g_L^2 \frac{-v_1^2 -  2v_{1L}^2 + v_2^2 + 2v_{2L}^2 + v_1'^2 -  v_2'^2}{8} + g_{R}^2 \frac{-v_1^2 + 2v_{1R}^2 + v_2^2 - 2v_{2R}^2 + v_1'^2 -  v_2'^2}{8} \n
  && -\lambda_3^2 \frac{v_2^2 + v_s^2}{2} - \sqrt{2} \lambda_3 \mu_3 v_1 v_s \Bigg\} -  v_2'\frac{e^{i \alpha_2} (m_{\Phi})^{12} + e^{-i \alpha_2} (m_{\Phi})^{21} }{2}\n
  && + v_2 \Bigg\{ \frac{\lambda_3}{2} \Big[\lambda_L v_{1L}  v_{2L} + \lambda_R  v_{1R} v_{2R} - \cos\big(\alpha_1 + \alpha_2\big) \lambda_3 v_2'v_1' + \frac{1}{\sqrt{2}} \mu_s v_s + \lambda_s v_s^2 + 2\xi_S\Big] + B_3 + \frac{T_3 v_s}{\sqrt{2}}\Bigg\}   ;\n
\frac{\partial V}{\partial v_{2}} = 0 \!\!\!\!&=&\!\!\!\! v_2 \Bigg\{-(m_{\Phi})^{22} - \mu_3^2 + g_{L}^2 \frac{v_1^2 + 2 v_{1L}^2 - v_2^2 - 2 v_{2L}^2 - v_1'^2 -  v_1'^2 + v_2'^2 }{8} + g_{R}^2 \frac{v_1^2 - 2 v_{1R}^2 - v_2^2 + 2 v_{2R}^2 + v_2'^2}{8}\n
  && -\lambda_3^2 \frac{v_1^2 - v_s^2}{2} - \sqrt{2} \lambda_3  \mu_3 v_s \Bigg\} -  v_1' \frac{e^{-i\alpha_1} (m_{\Phi})^{12} + e^{i \alpha_1} (m_{\Phi})^{21}}{2}  \n
  && + v_1\Bigg\{ \frac{\lambda_3}{2} \Big[ 2 \xi_S + \lambda_L v_{1L} v_{2L} + \lambda_R v_{1R} v_{2R} + \cos\big( \alpha_1 + \alpha_2 \big) \lambda_3 v_1' v_2' + \frac{1}{\sqrt{2}} \mu_s v_s + \lambda_s v_s^2 \Big]+ B_3 + T_3\frac{v_s}{\sqrt{2}} \Bigg\};\n
\frac{\partial V}{\partial v_{1}'} = 0 \!\!\!\!&=&\!\!\!\! v_1' \Bigg\{- (m_{\Phi}^2)^{11} - \mu_3^2 + g_L^2 \frac{v_1^2 + 2 v_{1L}^2 - v_2^2 - 2 v_2L^2 - v_1'^2 + v_2'^2}{8} + g_R^2 \frac{v_1^2 - 2 v_{1R}^2 - v_2^2 + 2 v_{2R}^2 - v_1'^2 + v_2'^2}{8} \n
  && -\lambda_3^2 \frac{ v_2'^2 + v_s^2}{2} - \sqrt{2} \lambda_3 \mu_3 v_s \Bigg\} - v_2 \frac{e^{- i \alpha_1} (m_{\Phi}^2)^{12} + e^{i \alpha_1} (m_{\Phi}^2)^{21}}{2} \n
  && + v_2' \cos\big(\alpha_1 + \alpha_2 \big)  \Bigg\{
     - \frac{1}{2}\lambda_3\Big[ \lambda_3 v_1 v_2 + \lambda_L v_{1L} v_{2L} + \lambda_R v_{1R} v_{2R} + \frac{1}{\sqrt2}\mu_s v_s + \lambda_s v_s^2 +2 \xi_s \Big] +  B_3 + \frac{1}{\sqrt2} T_3 v_s  \Bigg\};\n
\frac{\partial V}{\partial v_{2}'} = 0 \!\!\!\!&=&\!\!\!\! v_2'\Bigg\{- (m_{\Phi}^2)^{22} - \mu_3^2 - g_L^2 \frac{v_1^2 - 2 v_{1L}^2 + v_2^2 + 2 v_{2L}^2 + v_1'^2 - v_2'^2}{8} - g_R^2 \frac{v_1^2 + 2 v_{1R}^2 + v_2^2 - 2 v_{2R}^2 + v_1'^2 - v_2'^2}{8}\n
  && - \lambda_3^2 \frac{v_1'^2 + v_s^2}{2} - \sqrt{2} \lambda_3 \mu_3  v_s \Bigg\} - v_1\frac{e^{i \alpha_2} (m_{\Phi}^2)^{12} + e^{-i \alpha_2} (m_{\Phi}^2)^{21} }{2} \n
  && +  v_1' \cos\big(\alpha_1 + \alpha_2\big) \Bigg\{\frac{1}{2}\lambda_3 \Big[ - \lambda_3 v_1 v_2  + \lambda_L v_{1L} v_{2L} + \lambda_R v_{1R} v_{2R} + \frac{1}{\sqrt{2}} \mu_s v_s +\lambda_s v_s^2 + 2\xi_s \Big] + B_3 +  T_3\frac{v_s}{\sqrt{2}}\Bigg\};\n
\frac{\partial V}{\partial \alpha_1} = 0 \!\!\!\!&=&\!\!\!\! v_1'v_2'\sin\big(\alpha_1 + \alpha_2\big) \Bigg\{ \frac{1}{2}\lambda_3 \Big[\lambda_3 v_1 v_2 - \lambda_L v_{1L} v_{2L} - \lambda_R v_{1R} v_{2R} - 2\sqrt{2} \mu_s v_s - \lambda_s v_s^2 - 2\xi_s \Big]
- B_3 - \frac{1}{\sqrt{2}} T_3 v_s\Bigg\}\n
&& - i \frac{v_2 v_1'}{2} \big( e^{i \alpha_1} (m_{\Phi}^2)^{21} - e^{-i \alpha_1} (m_{\Phi}^2)^{12} \big);\n
\frac{\partial V}{\partial \alpha_2} = 0 \!\!\!\!&=&\!\!\!\! v_1' v_2'\sin\big(\alpha_1 + \alpha_2\big) \Bigg\{ \frac{1}{2}\lambda_3 \Big[\lambda_3 v_1 v_2 - \lambda_L v_{1L} v_{2L} - \lambda_R v_{1R} v_{2R} - 2\sqrt{2}\mu_s  v_s - \lambda_s v_s^2 - 2 \xi_s \Big] - B_3 - T_3  v_s \frac{1}{\sqrt{2}}\Bigg\}\n
&& -i \frac{v_1 v_2'}{2} \big( e^{i \alpha_2} (m_{\Phi}^2)^{12}  - (m_{\Phi}^2)^{21} e^{-i \alpha_2} \big).
\eea

\chapter{Doubly-charged particles}{\label{annex: vertices}}
Keeping the same notations as in chapter \ref{chap: doubly charged}, we list here all the Feynman rules we have used to carry properly the calculations. For commodity, the output has been generated by the {\sc TeXInterface} included in {\sc FeynRules} but we need to set the notations. Let $a,b \et c$ be three fields involved in a vertex with the coupling $i\lambda$, This will be written as\\
\begin{minipage}{1.5in}
\begin{displaymath}
\left(
\begin{array}{cc}
 a & 1 \\
 b & 2 \\
 c & 3 \\
\end{array}
\right)
\end{displaymath}\vskip0.2cm
\end{minipage}
\begin{minipage}{6in}
\begin{sloppypar}\begin{flushleft}$
i\lambda
$\end{flushleft}\end{sloppypar}
\end{minipage}\\
which corresponds to the following vertex where the arrows correspond to the direction of the momenta.
\begin{center}
\begin{fmffile}{convfr}
\begin{fmfgraph*}(30,30)
    \fmfleft{i1}
    \fmflabel{$a$}{i1}
	\fmfright{o1,o2}
	\fmflabel{$b$}{o1}
	\fmflabel{$c$}{o2}
	\fmf{fermion,label=1}{i1,v1}
    \fmflabel{$i\lambda $}{v1}
    \fmf{fermion,label=2}{o1,v1}
    \fmf{fermion,label=3}{o2,v1}
\end{fmfgraph*}
\end{fmffile}
\end{center}
The numbers on the lines label the momenta and correspond to the numbers appearing in the notation above. Therefore if the particle $a$ carried the momentum $p_1$ and the vertex depended on it we would have instead of $i\lambda$ something like
$$ i \lambda f(p_1)$$
where $f(p_1)$ is a function of the momentum $p_1$ of the particle $a$. Therefore, in the following all the indices will carry a number in subscript. We will denote by $s_i$ spin indices and $\mu_i$ Lorentz indices.

\section{Feynman rules for doubly-charged scalars}
\subsection{Singlet scalar field case}
\begin{minipage}{1.5in}
\begin{displaymath}
\left(
\begin{array}{cc}
 \gamma& 1 \\
 \phi^{++} & 2 \\
 \phi^{--} & 3 \\
\end{array}
\right)
\end{displaymath}\vskip0.2cm
\end{minipage}
\begin{minipage}{6in}
\begin{sloppypar}\begin{flushleft}$
2 i e (p_2{}^{\mu_1}- p_3{}^{\mu_1})
$\end{flushleft}\end{sloppypar}
\end{minipage}
\begin{minipage}{1.5in}
\begin{displaymath}
\left(
\begin{array}{cc}
 l^+& 1 \\
 l^+& 2 \\
 \phi^{++} & 3 \\
\end{array}
\right)
\end{displaymath}\vskip0.2cm
\end{minipage}
\begin{minipage}{6in}
\begin{sloppypar}\begin{flushleft}$
i y^{(1)}  \frac{1+\gamma^5}{2}
$\end{flushleft}\end{sloppypar}
\end{minipage}
\begin{minipage}{1.5in}
\begin{displaymath}
\left(
\begin{array}{cc}
 \phi^{++} & 1 \\
 \phi^{--} & 2 \\
 Z & 3 \\
\end{array}
\right)
\end{displaymath}\vskip0.2cm
\end{minipage}
\begin{minipage}{6in}
\begin{sloppypar}\begin{flushleft}$
-\frac{2 i e s_W }{c_W}(p_1{}^{\mu_3}-p_2{}^{\mu_3})
$\end{flushleft}\end{sloppypar}
\end{minipage}

\subsection{Doublet scalar field case}
\begin{minipage}{1.5in}
\begin{displaymath}
\left(
\begin{array}{cc}
 \gamma& 1 \\
 \Phi^{++}& 2 \\
 \Phi^{--} & 3 \\
\end{array}
\right)
\end{displaymath}\vskip0.2cm
\end{minipage}
\begin{minipage}{6in}
\begin{sloppypar}\begin{flushleft}$
2 i e (p_2{}^{\mu_1}- p_3{}^{\mu_1})
$\end{flushleft}\end{sloppypar}
\end{minipage}
\begin{minipage}{1.5in}
\begin{displaymath}
\left(
\begin{array}{cc}
 \Phi^{-} & 1 \\
 \Phi^{++}& 2 \\
 W^{+} & 3 \\
\end{array}
\right)
\end{displaymath}\vskip0.2cm
\end{minipage}
\begin{minipage}{6in}
\begin{sloppypar}\begin{flushleft}$
-\frac{i e}{\sqrt{2} s_W}(p_1{}^{\mu_3} + p_2{}^{\mu_3})
$\end{flushleft}\end{sloppypar}
\end{minipage}
\begin{minipage}{1.5in}
\begin{displaymath}
\left(
\begin{array}{cc}
 \Phi^{++}& 1 \\
 \Phi^{--} & 2 \\
 Z & 3 \\
\end{array}
\right)
\end{displaymath}\vskip0.2cm
\end{minipage}
\begin{minipage}{6in}
\begin{sloppypar}\begin{flushleft}$
\frac{i c_W e }{2 s_W}( p_1{}^{\mu_3} - p_2{}^{\mu_3}) -\frac{3 i e s_W }{2 c_W}(p_1{}^{\mu_3}-p_2{}^{\mu_3})
$\end{flushleft}\end{sloppypar}
\end{minipage}
\begin{minipage}{1.5in}
\begin{displaymath}
\left(
\begin{array}{cc}
 l^+& 1 \\
 l^+& 2 \\
 \Phi^{++}& 3 \\
\end{array}
\right)
\end{displaymath}\vskip0.2cm
\end{minipage}
\begin{minipage}{6in}
\begin{sloppypar}\begin{flushleft}$
-\frac{y}{\Lambda}\Big[ \slashed{p}_1.\frac{1+\gamma^5}{2}-\slashed{p}_2.\frac{1-\gamma^5}{2}\Big]
$\end{flushleft}\end{sloppypar}
\end{minipage}
\begin{minipage}{1.5in}
\begin{displaymath}
\left(
\begin{array}{cc}
 \gamma& 1 \\
 \Phi^{+} & 2 \\
 \Phi^{-} & 3 \\
\end{array}
\right)
\end{displaymath}\vskip0.2cm
\end{minipage}
\begin{minipage}{6in}
\begin{sloppypar}\begin{flushleft}$
i e (p_2{}^{\mu_1}-p_3{}^{\mu_1})
$\end{flushleft}\end{sloppypar}
\end{minipage}
\begin{minipage}{1.5in}
\begin{displaymath}
\left(
\begin{array}{cc}
 \Phi^{+} & 1 \\
 \Phi^{-} & 2 \\
 Z & 3 \\
\end{array}
\right)
\end{displaymath}\vskip0.2cm
\end{minipage}
\begin{minipage}{6in}
\begin{sloppypar}\begin{flushleft}$
\frac{i c_W e }{2 s_W}\Big[p_2{}^{\mu_3} - p_1{}^{\mu_3}\Big] + \frac{3 i e s_W }{2 c_W}\Big[p_2{}^{\mu_3} - p_1{}^{\mu_3}\Big]
$\end{flushleft}\end{sloppypar}
\end{minipage}
\begin{minipage}{1.5in}
\begin{displaymath}
\left(
\begin{array}{cc}
 l^+& 1 \\
\nu_L & 2 \\
 \Phi^{+} & 3 \\
\end{array}
\right)
\end{displaymath}\vskip0.2cm
\end{minipage}
\begin{minipage}{6in}
\begin{sloppypar}\begin{flushleft}$
\frac{y^{(2)} }{\Lambda}\slashed{p}_1.\frac{1+\gamma^5}{2}
$\end{flushleft}\end{sloppypar}
\end{minipage}


\subsection{Triplet scalar field case}
\begin{minipage}{1.5in}
\begin{displaymath}
\left(
\begin{array}{cc}
 \gamma& 1 \\
 \pPhi^{++} & 2 \\
 \pPhi^{--}  & 3 \\
\end{array}
\right)
\end{displaymath}\vskip0.2cm
\end{minipage}
\begin{minipage}{6in}
\begin{sloppypar}\begin{flushleft}$
2 i e (p_2{}^{\mu_1}- p_3{}^{\mu_1})
$\end{flushleft}\end{sloppypar}
\end{minipage}
\begin{minipage}{1.5in}
\begin{displaymath}
\left(
\begin{array}{cc}
 l^+& 1 \\
 l^+& 2 \\
 \pPhi^{++} & 3 \\
\end{array}
\right)
\end{displaymath}\vskip0.2cm
\end{minipage}
\begin{minipage}{6in}
\begin{sloppypar}\begin{flushleft}$
i y^{(3)}  \frac{1-\gamma^5}{2}
$\end{flushleft}\end{sloppypar}
\end{minipage}
\begin{minipage}{1.5in}
\begin{displaymath}
\left(
\begin{array}{cc}
 \pPhi^{-}  & 1 \\
 \pPhi^{++} & 2 \\
 W^{+} & 3 \\
\end{array}
\right)
\end{displaymath}\vskip0.2cm
\end{minipage}
\begin{minipage}{6in}
\begin{sloppypar}\begin{flushleft}$
\frac{i e }{s_W}\Big[p_1{}^{\mu_3}- p_2{}^{\mu_3}\Big]
$\end{flushleft}\end{sloppypar}
\end{minipage}
\begin{minipage}{1.5in}
\begin{displaymath}
\left(
\begin{array}{cc}
 \pPhi^{++} & 1 \\
 W^{+} & 2 \\
 W^{+} & 3 \\
\end{array}
\right)
\end{displaymath}\vskip0.2cm
\end{minipage}
\begin{minipage}{6in}
\begin{sloppypar}\begin{flushleft}$
-\frac{i \sqrt{2} e{}^2 v{}_{\pPhi }}{s_W^2} \eta_{\mu_2\mu_3}
$\end{flushleft}\end{sloppypar}
\end{minipage}
\begin{minipage}{1.5in}
\begin{displaymath}
\left(
\begin{array}{cc}
 \pPhi^{++} & 1 \\
 \pPhi^{--}  & 2 \\
 Z & 3 \\
\end{array}
\right)
\end{displaymath}\vskip0.2cm
\end{minipage}
\begin{minipage}{6in}
\begin{sloppypar}\begin{flushleft}$
\frac{i c_W e }{s_W}\Big[ p_1{}^{\mu_3}- p_2{}^{\mu_3}\Big]+\frac{i e s_W }{c_W} \Big[p_2{}^{\mu_3}- p_1{}^{\mu_3}\Big]
$\end{flushleft}\end{sloppypar}
\end{minipage}
\begin{minipage}{1.5in}
\begin{displaymath}
\left(
\begin{array}{cc}
 \gamma& 1 \\
 \pPhi^{+}  & 2 \\
 \pPhi^{-}  & 3 \\
\end{array}
\right)
\end{displaymath}\vskip0.2cm
\end{minipage}
\begin{minipage}{6in}
\begin{sloppypar}\begin{flushleft}$
i e \Big[ p_2{}^{\mu_1}- p_3{}^{\mu_1}\Big]
$\end{flushleft}\end{sloppypar}
\end{minipage}
\begin{minipage}{1.5in}
\begin{displaymath}
\left(
\begin{array}{cc}
 l^+& 1 \\
\nu_L & 2 \\
 \pPhi^{+}  & 3 \\
\end{array}
\right)
\end{displaymath}\vskip0.2cm
\end{minipage}
\begin{minipage}{6in}
\begin{sloppypar}\begin{flushleft}$
\frac{i y^{(3)} }{\sqrt{2}} \frac{1-\gamma^5}{2}
$\end{flushleft}\end{sloppypar}
\end{minipage}
\begin{minipage}{1.5in}
\begin{displaymath}
\left(
\begin{array}{cc}
 \gamma& 1 \\
 \pPhi^{+}  & 2 \\
 W^{+} & 3 \\
\end{array}
\right)
\end{displaymath}\vskip0.2cm
\end{minipage}
\begin{minipage}{6in}
\begin{sloppypar}\begin{flushleft}$
\frac{i e{}^2 v{}_{\pPhi } }{\sqrt{2} s_W}\eta_{\mu_1\mu_3}
$\end{flushleft}\end{sloppypar}
\end{minipage}
\begin{minipage}{1.5in}
\begin{displaymath}
\left(
\begin{array}{cc}
 \pPhi^{0} & 1 \\
 \pPhi^{+}  & 2 \\
 W^{+} & 3 \\
\end{array}
\right)
\end{displaymath}\vskip0.2cm
\end{minipage}
\begin{minipage}{6in}
\begin{sloppypar}\begin{flushleft}$
-\frac{i e  }{\sqrt{2} s_W}\Big[p_1{}^{\mu_3}- p_2{}^{\mu_3}\Big]
$\end{flushleft}\end{sloppypar}
\end{minipage}
\begin{minipage}{1.5in}
\begin{displaymath}
\left(
\begin{array}{cc}
 \pPhi^{+}  & 1 \\
 \pPhi^{-}  & 2 \\
 Z & 3 \\
\end{array}
\right)
\end{displaymath}\vskip0.2cm
\end{minipage}
\begin{minipage}{6in}
\begin{sloppypar}\begin{flushleft}$
-\frac{i e s_W }{c_W}\Big[p_1{}^{\mu_3}-p_2{}^{\mu_3}\Big]
$\end{flushleft}\end{sloppypar}
\end{minipage}
\begin{minipage}{1.5in}
\begin{displaymath}
\left(
\begin{array}{cc}
 \pPhi^{+}  & 1 \\
 W^{+} & 2 \\
 Z & 3 \\
\end{array}
\right)
\end{displaymath}\vskip0.2cm
\end{minipage}
\begin{minipage}{6in}
\begin{sloppypar}\begin{flushleft}$
i e{}^2v{}_{\pPhi }\Big[\frac{1+s_W^2 }{\sqrt{2}c_W s_W^2}\Big]\eta_{\mu_2\mu_3}
$\end{flushleft}\end{sloppypar}
\end{minipage}
\begin{minipage}{1.5in}
\begin{displaymath}
\left(
\begin{array}{cc}
\nu_L & 1 \\
\nu_L & 2 \\
 \pPhi^{0} & 3 \\
\end{array}
\right)
\end{displaymath}\vskip0.2cm
\end{minipage}
\begin{minipage}{6in}
\begin{sloppypar}\begin{flushleft}$
-\frac{i y^{(3)} }{\sqrt{2}} \frac{1-\gamma^5}{2}
$\end{flushleft}\end{sloppypar}
\end{minipage}
\begin{minipage}{1.5in}
\begin{displaymath}
\left(
\begin{array}{cc}
 \bar{\nu}_L & 1 \\
 \bar{\nu}_L & 2 \\
 \pPhi^{0} & 3 \\
\end{array}
\right)
\end{displaymath}\vskip0.2cm
\end{minipage}
\begin{minipage}{6in}
\begin{sloppypar}\begin{flushleft}$
-\frac{i y^{(3)} }{\sqrt{2}} \frac{1+\gamma^5}{2}
$\end{flushleft}\end{sloppypar}
\end{minipage}
\begin{minipage}{1.5in}
\begin{displaymath}
\left(
\begin{array}{cc}
 \pPhi^{0} & 1 \\
 W^{-}& 2 \\
 W^{+} & 3 \\
\end{array}
\right)
\end{displaymath}\vskip0.2cm
\end{minipage}
\begin{minipage}{6in}
\begin{sloppypar}\begin{flushleft}$
\frac{i e^2 v_{\pPhi } }{s_W^2}\eta_{\mu_2\mu_3}
$\end{flushleft}\end{sloppypar}
\end{minipage}
\begin{minipage}{1.5in}
\begin{displaymath}
\left(
\begin{array}{cc}
 \pPhi^{0} & 1 \\
 Z & 2 \\
 Z & 3 \\
\end{array}
\right)
\end{displaymath}\vskip0.2cm
\end{minipage}
\begin{minipage}{6in}
\begin{sloppypar}\begin{flushleft}$
\frac{2 i e{}^2v_{\pPhi } }{s_W^2 c_W{}^2} \eta_{\mu_2\mu_3}
$\end{flushleft}\end{sloppypar}
\end{minipage}

\section{Feynman rules for doubly-charged Fermions with a four generation SM}

\subsection{Doublet fermion field case}
\begin{minipage}{1.5in}
\begin{displaymath}
\left(
\begin{array}{cc}
 \Psi^{--} & 1 \\
\Psi^{++}& 2 \\
 \gamma& 3 \\
\end{array}
\right)
\end{displaymath}\vskip0.2cm
\end{minipage}
\begin{minipage}{6in}
\begin{sloppypar}\begin{flushleft}$
2 i e \gamma_{s_1s_2}^{\mu_3}
$\end{flushleft}\end{sloppypar}
\end{minipage}
\begin{minipage}{1.5in}
\begin{displaymath}
\left(
\begin{array}{cc}
E'^- & 1 \\
\Psi^{++}& 2 \\
 W^{+} & 3 \\
\end{array}
\right)
\end{displaymath}\vskip0.2cm
\end{minipage}
\begin{minipage}{6in}
\begin{sloppypar}\begin{flushleft}$
\frac{i c_\tau e }{\sqrt{2} s_W}\gamma_{s_1s_2}^{\mu_3}
$\end{flushleft}\end{sloppypar}
\end{minipage}
\begin{minipage}{1.5in}
\begin{displaymath}
\left(
\begin{array}{cc}
\tau^{-} & 1 \\
\Psi^{++}& 2 \\
 W^{+} & 3 \\
\end{array}
\right)
\end{displaymath}\vskip0.2cm
\end{minipage}
\begin{minipage}{6in}
\begin{sloppypar}\begin{flushleft}$
\frac{i e s_{\tau }}{\sqrt{2} s_W} \gamma_{s_1s_2}^{\mu_3}
$\end{flushleft}\end{sloppypar}
\end{minipage}
\begin{minipage}{1.5in}
\begin{displaymath}
\left(
\begin{array}{cc}
 \Psi^{--} & 1 \\
\Psi^{++}& 2 \\
 Z & 3 \\
\end{array}
\right)
\end{displaymath}\vskip0.2cm
\end{minipage}
\begin{minipage}{6in}
\begin{sloppypar}\begin{flushleft}$
ie\frac{ 1- 4s_W^2 }{2 s_wc_W}\gamma_{s_1s_2}^{\mu_3}
$\end{flushleft}\end{sloppypar}
\end{minipage}
\begin{minipage}{1.5in}
\begin{displaymath}
\left(
\begin{array}{cc}
E'^+  & 1 \\
E'^- & 2 \\
 \gamma& 3 \\
\end{array}
\right)
\end{displaymath}\vskip0.2cm
\end{minipage}
\begin{minipage}{6in}
\begin{sloppypar}\begin{flushleft}$
-i e \gamma_{s_1s_2}^{\mu_3}
$\end{flushleft}\end{sloppypar}
\end{minipage}
\begin{minipage}{1.5in}
\begin{displaymath}
\left(
\begin{array}{cc}
 \tau^{+} & 1 \\
E'^- & 2 \\
 \gamma& 3 \\
\end{array}
\right)
\end{displaymath}\vskip0.2cm
\end{minipage}
\begin{minipage}{6in}
\begin{sloppypar}\begin{flushleft}$
2i c_\tau e s_{\tau }  \gamma_{s_1s_2}^{\mu_3}
$\end{flushleft}\end{sloppypar}
\end{minipage}
\begin{minipage}{1.5in}
\begin{displaymath}
\left(
\begin{array}{cc}
E'^+ & 1 \\
E'^- & 2 \\
 Z & 3 \\
\end{array}
\right)
\end{displaymath}\vskip0.2cm
\end{minipage}
\begin{minipage}{6in}
\begin{sloppypar}\begin{flushleft}$
i ec_\tau^2\frac{ 1+ 2s_W^2 }{2 c_W} \gamma_{s_1s_2}^{\mu_3} -i e s_\tau^2\gamma^{\mu_3}. \Big[ \frac{ 1-2s_W^2 }{2 s_W} \frac{1-\gamma^5}{2} -\frac{ s_W }{c_W}\frac{1+\gamma^5}{2}\Big]
$\end{flushleft}\end{sloppypar}
\end{minipage}
\begin{minipage}{1.5in}
\begin{displaymath}
\left(
\begin{array}{cc}
 \tau^{+} & 1 \\
E'^- & 2 \\
 Z & 3 \\
\end{array}
\right)
\end{displaymath}\vskip0.2cm
\end{minipage}
\begin{minipage}{6in}
\begin{sloppypar}\begin{flushleft}$
i ec_\tau s_\tau\frac{ 1+ 2s_W^2 }{2 c_W} \gamma_{s_1s_2}^{\mu_3} -i e s_\tau c_\tau\gamma^{\mu_3}. \Big[ \frac{ 1-2s_W^2 }{2 s_W} \frac{1-\gamma^5}{2} -\frac{ s_W }{c_W}\frac{1+\gamma^5}{2}\Big]
$\end{flushleft}\end{sloppypar}
\end{minipage}
\begin{minipage}{1.5in}
\begin{displaymath}
\left(
\begin{array}{cc}
\bar{\nu}_\tau & 1 \\
E'^- & 2 \\
 W^{-}& 3 \\
\end{array}
\right)
\end{displaymath}\vskip0.2cm
\end{minipage}
\begin{minipage}{6in}
\begin{sloppypar}\begin{flushleft}$
-\frac{i e s_{\tau }}{\sqrt{2} s_W}\gamma^{\mu_3}.\frac{1-\gamma^5}{2}
$\end{flushleft}\end{sloppypar}
\end{minipage}

\subsection{Triplet fermion field case}

\begin{minipage}{1.5in}
\begin{displaymath}
\left(
\begin{array}{cc}
 \pPsi^{--} & 1 \\
 \pPsi^{++}& 2 \\
 \gamma& 3 \\
\end{array}
\right)
\end{displaymath}\vskip0.2cm
\end{minipage}
\begin{minipage}{6in}
\begin{sloppypar}\begin{flushleft}$
2 i e \gamma_{s_1s_2}^{\mu_3}
$\end{flushleft}\end{sloppypar}
\end{minipage}
\begin{minipage}{1.5in}
\begin{displaymath}
\left(
\begin{array}{cc}
E''^- & 1 \\
 \pPsi^{++}& 2 \\
 W^{+} & 3 \\
\end{array}
\right)
\end{displaymath}\vskip0.2cm
\end{minipage}
\begin{minipage}{6in}
\begin{sloppypar}\begin{flushleft}$
-\frac{i c_\tau e }{s_W}\gamma_{s_1s_2}^{\mu_3}
$\end{flushleft}\end{sloppypar}
\end{minipage}
\begin{minipage}{1.5in}
\begin{displaymath}
\left(
\begin{array}{cc}
 \pPsi^{++}& 1 \\
\tau^{-} & 2 \\
 W^{+} & 3 \\
\end{array}
\right)
\end{displaymath}\vskip0.2cm
\end{minipage}
\begin{minipage}{6in}
\begin{sloppypar}\begin{flushleft}$
\frac{i e s_{\tau }}{s_W} \gamma_{s_1s_2}^{\mu_3}
$\end{flushleft}\end{sloppypar}
\end{minipage}
\begin{minipage}{1.5in}
\begin{displaymath}
\left(
\begin{array}{cc}
 \pPsi^{--} & 1 \\
 \pPsi^{++}& 2 \\
 Z & 3 \\
\end{array}
\right)
\end{displaymath}\vskip0.2cm
\end{minipage}
\begin{minipage}{6in}
\begin{sloppypar}\begin{flushleft}$
i e \frac{1-s_W^2}{s_Wc_W} \gamma_{s_1s_2}^{\mu_3}
$\end{flushleft}\end{sloppypar}
\end{minipage}
\begin{minipage}{1.5in}
\begin{displaymath}
\left(
\begin{array}{cc}
 E''^+ & 1 \\
E''^- & 2 \\
 \gamma& 3 \\
\end{array}
\right)
\end{displaymath}\vskip0.2cm
\end{minipage}
\begin{minipage}{6in}
\begin{sloppypar}\begin{flushleft}$
-i e \gamma_{s_1s_2}^{\mu_3}
$\end{flushleft}\end{sloppypar}
\end{minipage}
%
\begin{minipage}{1.5in}
\begin{displaymath}
\left(
\begin{array}{cc}
E''^- & 1 \\
 \pPsi^{0} & 2 \\
 W^{-}& 3 \\
\end{array}
\right)
\end{displaymath}\vskip0.2cm
\end{minipage}
\begin{minipage}{6in}
\begin{sloppypar}\begin{flushleft}$
\frac{i c_\tau e }{s_W}\gamma_{s_1s_2}^{\mu_3}
$\end{flushleft}\end{sloppypar}
\end{minipage}
\begin{minipage}{1.5in}
\begin{displaymath}
\left(
\begin{array}{cc}
E''^- & 1 \\
 \pPsi^{++}& 2 \\
 W^{+} & 3 \\
\end{array}
\right)
\end{displaymath}\vskip0.2cm
\end{minipage}
\begin{minipage}{6in}
\begin{sloppypar}\begin{flushleft}$
-\frac{i c_\tau e }{s_W}\gamma_{s_1s_2}^{\mu_3}
$\end{flushleft}\end{sloppypar}
\end{minipage}
\begin{minipage}{1.5in}
\begin{displaymath}
\left(
\begin{array}{cc}
 E''^+ & 1 \\
E''^- & 2 \\
 Z & 3 \\
\end{array}
\right)
\end{displaymath}\vskip0.2cm
\end{minipage}
\begin{minipage}{6in}
\begin{sloppypar}\begin{flushleft}$
\frac{i c_\tau^2 e s_W }{c_W}\gamma_{s_1s_2}^{\mu_3}- ies_\tau^2 \gamma^{\mu_3}.\Big[\frac{ 1- 2s_W^2 }{2 s_W c_W} \frac{1-\gamma^5}{2} -\frac{s_W }{c_W}\frac{1+\gamma^5}{2}\Big]
$\end{flushleft}\end{sloppypar}
\end{minipage}
\begin{minipage}{1.5in}
\begin{displaymath}
\left(
\begin{array}{cc}
 \tau^{+} & 1 \\
E''^- & 2 \\
 Z & 3 \\
\end{array}
\right)
\end{displaymath}\vskip0.2cm
\end{minipage}
\begin{minipage}{6in}
\begin{sloppypar}\begin{flushleft}$
\frac{i c_\tau s_\tau e s_W }{c_W}\gamma_{s_1s_2}^{\mu_3}- ies_\tau c_\tau \gamma^{\mu_3}.\Big[\frac{ 1- 2s_W^2 }{2 s_W c_W} \frac{1-\gamma^5}{2} -\frac{s_W }{c_W}\frac{1+\gamma^5}{2}\Big]
$\end{flushleft}\end{sloppypar}
\end{minipage}
\begin{minipage}{1.5in}
\begin{displaymath}
\left(
\begin{array}{cc}
\bar{\nu}_\tau & 1 \\
E''^- & 2 \\
 W^{-}& 3 \\
\end{array}
\right)
\end{displaymath}\vskip0.2cm
\end{minipage}
\begin{minipage}{6in}
\begin{sloppypar}\begin{flushleft}$
-\frac{i e s_{\tau }}{\sqrt{2} s_W} \gamma^{\mu_3}.\frac{1-\gamma^5}{2}
$\end{flushleft}\end{sloppypar}
\end{minipage}
\begin{minipage}{1.5in}
\begin{displaymath}
\left(
\begin{array}{cc}
 \pPsi^{0} & 1 \\
\tau^{-} & 2 \\
 W^{-}& 3 \\
\end{array}
\right)
\end{displaymath}\vskip0.2cm
\end{minipage}
\begin{minipage}{6in}
\begin{sloppypar}\begin{flushleft}$
-\frac{i e s_{\tau } }{s_W}\gamma_{s_1s_2}^{\mu_3}
$\end{flushleft}\end{sloppypar}
\end{minipage}
\begin{minipage}{1.5in}
\begin{displaymath}
\left(
\begin{array}{cc}
 \pPsi^{0} & 1 \\
 \pPsi^{0} & 2 \\
 Z & 3 \\
\end{array}
\right)
\end{displaymath}\vskip0.2cm
\end{minipage}
\begin{minipage}{6in}
\begin{sloppypar}\begin{flushleft}$
-i e \frac{ 1- 2s_W^2 }{c_W} \gamma_{s_1s_2}^{\mu_3}
$\end{flushleft}\end{sloppypar}
\end{minipage}


\section{Feynman rules for doubly-charged vectors}

\subsection{Singlet vector field case}
\begin{minipage}{1.5in}
\begin{displaymath}
\left(
\begin{array}{cc}
 \gamma& 1 \\
 V^{++}& 2 \\
 V^{--}& 3 \\
\end{array}
\right)
\end{displaymath}\vskip0.2cm
\end{minipage}
\begin{minipage}{6in}
\begin{sloppypar}\begin{flushleft}$
2 i e \Big[p_2{}^{\mu_3} \eta_{\mu_1\mu_2}-p_3{}^{\mu_2} \eta_{\mu_1\mu_3}- ( p_2 -p_3)^{\mu_1} \eta_{\mu_2\mu_3}\Big]
$\end{flushleft}\end{sloppypar}
\end{minipage}
\begin{minipage}{1.5in}
\begin{displaymath}
\left(
\begin{array}{cc}
 V^{++}& 1 \\
 V^{--}& 2 \\
 Z & 3 \\
\end{array}
\right)
\end{displaymath}\vskip0.2cm
\end{minipage}
\begin{minipage}{6in}
\begin{sloppypar}\begin{flushleft}$
\frac{2 i e s_W}{c_W}\Big[ p_2{}^{\mu_1} \eta_{\mu_2\mu_3}- p_1{}^{\mu_2} \eta_{\mu_1\mu_3}+( p_1 - p_2)^{\mu_3} \eta_{\mu_1\mu_2}\Big]
$\end{flushleft}\end{sloppypar}
\end{minipage}
\begin{minipage}{1.5in}
\begin{displaymath}
\left(
\begin{array}{cc}
 l^+& 1 \\
 l^+& 2 \\
 V^{++}& 3 \\
\end{array}
\right)
\end{displaymath}\vskip0.2cm
\end{minipage}
\begin{minipage}{6in}
\begin{sloppypar}\begin{flushleft}$
\frac{i \tilde{g}^{(1)} }{4 \Lambda}\Big[ \gamma^{\mu_3}( \slashed{p}_2 -  \slashed{p}_1) + (\slashed{p}_1 - \slashed{p}_2)\gamma^{\mu_3}\Big]\frac{1+\gamma^5}{2}
$\end{flushleft}\end{sloppypar}
\end{minipage}

\subsection{Doublet vector field case}
\begin{minipage}{1.5in}
\begin{displaymath}
\left(
\begin{array}{cc}
 W^{+} & 1 \\
\mV^{-}  & 2 \\
\mV^{++}  & 3 \\
\end{array}
\right)
\end{displaymath}\vskip0.2cm
\end{minipage}
\begin{minipage}{6in}
\begin{sloppypar}\begin{flushleft}$
+\frac{i e }{\sqrt{2} s_W}\Big[p_3{}^{\mu_2} \eta_{\mu_1\mu_3} - p_2{}^{\mu_3} \eta_{\mu_1\mu_2} + (p_2 - p_3)^{\mu_1} \eta_{\mu_2\mu_3}\Big]
$\end{flushleft}\end{sloppypar}
\end{minipage}
\begin{minipage}{1.5in}
\begin{displaymath}
\left(
\begin{array}{cc}
 \gamma& 1 \\
\mV^{++}  & 2 \\
\mV^{--}  & 3 \\
\end{array}
\right)
\end{displaymath}\vskip0.2cm
\end{minipage}
\begin{minipage}{6in}
\begin{sloppypar}\begin{flushleft}$
2 i e \Big[ p_2{}^{\mu_3} \eta_{\mu_1\mu_2}- p_3{}^{\mu_2} \eta_{\mu_1\mu_3}- ( p_2 - p_3){}^{\mu_1} \eta_{\mu_2\mu_3}\Big]
$\end{flushleft}\end{sloppypar}
\end{minipage}
\begin{minipage}{1.5in}
\begin{displaymath}
\left(
\begin{array}{cc}
\mV^{++}  & 1 \\
\mV^{--}  & 2 \\
 Z & 3 \\
\end{array}
\right)
\end{displaymath}\vskip0.2cm
\end{minipage}
\begin{minipage}{6in}
\begin{sloppypar}\begin{flushleft}$
\Big[ \frac{i c_W e }{2 s_W} - \frac{3 i e s_W}{2 c_W}\Big] (p_2 - p_1)^{\mu_3} \eta_{\mu_1\mu_2}  + 
\Big[ \frac{i c_W e }{2 s_W}-\frac{3 i e s_W}{2 c_W}\Big] ( p_1{}^{\mu_2} \eta_{\mu_1\mu_3} - p_2{}^{\mu_1} \eta_{\mu_2\mu_3})
$\end{flushleft}\end{sloppypar}
\end{minipage}
\begin{minipage}{1.5in}
\begin{displaymath}
\left(
\begin{array}{cc}
 l^+& 1 \\
 l^+& 2 \\
\mV^{++}  & 3 \\
\end{array}
\right)
\end{displaymath}\vskip0.2cm
\end{minipage}
\begin{minipage}{6in}
\begin{sloppypar}\begin{flushleft}$
i \tilde{g}^{(2)}  \gamma^{\mu_3}.\gamma^5_{s_1s_2}
$\end{flushleft}\end{sloppypar}
\end{minipage}
\begin{minipage}{1.5in}
\begin{displaymath}
\left(
\begin{array}{cc}
 l^+& 1 \\
\nu_L & 2 \\
 \mV^{+} & 3 \\
\end{array}
\right)
\end{displaymath}\vskip0.2cm
\end{minipage}
\begin{minipage}{6in}
\begin{sloppypar}\begin{flushleft}$
-i \tilde{g}^{(2)}  \gamma^{\mu_3}.\frac{1+\gamma^5}{2}
$\end{flushleft}\end{sloppypar}
\end{minipage}

\subsection{Triplet vector field case}
\begin{minipage}{1.5in}
\begin{displaymath}
\left(
\begin{array}{cc}
 \gamma& 1 \\
 \vV^{++}& 2 \\
 \vV^{--} & 3 \\
\end{array}
\right)
\end{displaymath}\vskip0.2cm
\end{minipage}
\begin{minipage}{6in}
\begin{sloppypar}\begin{flushleft}$
2 i e \Big[ p_2{}^{\mu_3} \eta_{\mu_1\mu_2}- p_3{}^{\mu_2} \eta_{\mu_1\mu_3} - ( p_2 - p_3)^{\mu_1} \eta_{\mu_2\mu_3}\Big]
$\end{flushleft}\end{sloppypar}
\end{minipage}
\begin{minipage}{1.5in}
\begin{displaymath}
\left(
\begin{array}{cc}
 \vV^{-} & 1 \\
 \vV^{++}& 2 \\
 W^{+} & 3 \\
\end{array}
\right)
\end{displaymath}\vskip0.2cm
\end{minipage}
\begin{minipage}{6in}
\begin{sloppypar}\begin{flushleft}$
+\frac{i e }{s_W}\Big[ p_1{}^{\mu_2} \eta_{\mu_1\mu_3} - p_2{}^{\mu_1} \eta_{\mu_2\mu_3} + ( p_2 - p_1)^{\mu_3} \eta_{\mu_1\mu_2}\Big]
$\end{flushleft}\end{sloppypar}
\end{minipage}
\begin{minipage}{1.5in}
\begin{displaymath}
\left(
\begin{array}{cc}
 \vV^{++}& 1 \\
 \vV^{--} & 2 \\
 Z & 3 \\
\end{array}
\right)
\end{displaymath}\vskip0.2cm
\end{minipage}
\begin{minipage}{6in}
\begin{sloppypar}\begin{flushleft}$
\Big[\frac{i e s_W}{c_W} - \frac{i c_W e }{s_W}\ \Big] ( p_1- p_2)^{\mu_3} \eta_{\mu_1\mu_2}+
\Big[\frac{i e s_W}{c_W} - \frac{i c_W e }{s_W}\Big]\Big[  - p_1{}^{\mu_2} \eta_{\mu_1\mu_3} + p_2{}^{\mu_1} \eta_{\mu_2\mu_3}\Big]
$\end{flushleft}\end{sloppypar}
\end{minipage}
\begin{minipage}{1.5in}
\begin{displaymath}
\left(
\begin{array}{cc}
 l^+& 1 \\
 l^+& 2 \\
 \vV^{++}& 3 \\
\end{array}
\right)
\end{displaymath}\vskip0.2cm
\end{minipage}
\begin{minipage}{6in}
\begin{sloppypar}\begin{flushleft}$
+\frac{i \tilde{g}^{(3)} }{4 \Lambda}\Big[ \gamma^{\mu_3}(\slashed{p}_2 - \slashed{p}_1) + (\slashed{p}_1 - \slashed{p}_2)\gamma^{\mu_3}\Big].\frac{1-\gamma^5}{2}
$\end{flushleft}\end{sloppypar}
\end{minipage}
\begin{minipage}{1.5in}
\begin{displaymath}
\left(
\begin{array}{cc}
 \gamma& 1 \\
 \vV^{+} & 2 \\
 \vV^{-} & 3 \\
\end{array}
\right)
\end{displaymath}\vskip0.2cm
\end{minipage}
\begin{minipage}{6in}
\begin{sloppypar}\begin{flushleft}$
i e \Big[ p_2{}^{\mu_3} \eta_{\mu_1\mu_2} - p_3{}^{\mu_2} \eta_{\mu_1\mu_3} - ( p_2 + p_3)^{\mu_1} \eta_{\mu_2\mu_3}\Big]
$\end{flushleft}\end{sloppypar}
\end{minipage}
\begin{minipage}{1.5in}
\begin{displaymath}
\left(
\begin{array}{cc}
 \vV^{+} & 1 \\
 \vV^{-} & 2 \\
 Z & 3 \\
\end{array}
\right)
\end{displaymath}\vskip0.2cm
\end{minipage}
\begin{minipage}{6in}
\begin{sloppypar}\begin{flushleft}$
\frac{i e s_W}{c_W}\Big[ (p_1 - p_2)^{\mu_3} \eta_{\mu_1\mu_2} - p_1{}^{\mu_2} \eta_{\mu_1\mu_3} - p_2{}^{\mu_1} \eta_{\mu_2\mu_3}\Big]
$\end{flushleft}\end{sloppypar}
\end{minipage}
\begin{minipage}{1.5in}
\begin{displaymath}
\left(
\begin{array}{cc}
 l^+& 1 \\
\nu_L & 2 \\
 \vV^{+} & 3 \\
\end{array}
\right)
\end{displaymath}\vskip0.2cm
\end{minipage}
\begin{minipage}{6in}
\begin{sloppypar}\begin{flushleft}$
\frac{i\tilde{g}^{(3)}}{2 \sqrt{2} \Lambda}\Big[ ( p_1 - p_2)^{\mu_3}  - \gamma^{\mu_3} ( \slashed{p}_1  -\slashed{p}_2)\Big]\frac{1-\gamma^5}{2}
$\end{flushleft}\end{sloppypar}
\end{minipage}


\bibliographystyle{utphys}
\bibliography{biblio}{}
\end{document}